\documentclass[journal,final,twocolumn,10pt,twoside]{IEEEtran}

\usepackage{cite}
\usepackage{graphicx}
\usepackage{picinpar}
\usepackage{amsmath}
\usepackage{subfig}
\usepackage[export]{adjustbox} 
\usepackage{enumerate}

\usepackage{colortbl}
\usepackage{tikz}
\usepackage{xcolor}
\colorlet{linkequation}{green}

\usepackage{pdflscape}
\usepackage{url}

\usepackage{amssymb}
\usepackage{mathdots}
\usepackage{upgreek}
\usepackage{breqn}
\usepackage{hyperref}

\newcommand{\bglo}[1]{(#1)}

\newcommand{\eref}[1]{Eq.~(\ref{#1})}

\newcommand{\fref}[1]{Fig.~\ref{#1}}

\newcommand{\tref}[1]{Table~\ref{#1}}
\newcommand{\sref}[1]{Section~\ref{#1}}

\newcommand{\ket}[1]{| #1 \rangle}

\newcommand*{\cref}[1]{%
  \begingroup
    \hypersetup{
      linkcolor=linkequation,
      linkbordercolor=linkequation,
    }%
    \ref{#1}%
  \endgroup
}

\begin{document}
\title{A Survey on Quantum Channel Capacities}
\author{Laszlo Gyongyosi,$^{1,2,3,*}$~\IEEEmembership{Member,~IEEE}, Sandor Imre,$^{2}$~\IEEEmembership{Senior Member,~IEEE}, and Hung Viet Nguyen,$^{1}$~\IEEEmembership{Member,~IEEE}\\
$^{1}$School of Electronics and Computer Science, University of Southampton, Southampton SO17 1BJ, UK\\
$^{2}$Department of Networked Systems and Services, Budapest University of Technology and Economics, Budapest, H-1117 Hungary\\
$^{3}$MTA-BME Information Systems Research Group, Hungarian Academy of Sciences, Budapest, H-1051 Hungary
\thanks{This work was partially supported by the European Research Council through the Advanced Fellow Grant, in part by the Royal Society’s Wolfson Research Merit Award, in part by the Engineering and Physical Sciences Research Council under Grant EP/L018659/1, by the Hungarian Scientific Research Fund - OTKA K-112125, and by the National Research Development and Innovation Office of Hungary (Project No. 2017-1.2.1-NKP-2017-00001).}
\thanks{*Email: \href{mailto:l.gyongyosi@soton.ac.uk}{l.gyongyosi@soton.ac.uk}}
}

\maketitle
\vspace{-0.5cm}
\begin{abstract}
Quantum information processing exploits the quantum nature of information. It offers fundamentally new solutions in the field of computer science and extends the possibilities to a level that cannot be imagined in classical communication systems. For quantum communication channels, many new capacity definitions were developed in comparison to classical counterparts. A quantum channel can be used to realize classical information transmission or to deliver quantum information, such as quantum entanglement. Here we review the properties of the quantum communication channel, the various capacity measures and the fundamental differences between the classical and quantum channels.
\end{abstract}


\section{Introduction}
\label{sec1}
\label{Introduction}
According to Moore's Law [\cref{Moore65}], the physical limitations of classical semiconductor-based technologies could be reached within the next few years. We will then step into the age of quantum information. When first quantum computers become available on the shelf, today's encrypted information will not remain secure. Classical computational complexity will no longer guard this information. Quantum communication systems exploit the quantum nature of information offering new possibilities and limitations for engineers when designing protocols. Quantum communication systems face two major challenges. 

First, available point-to-point communication link should be connected on one hand to cover large distances an on the other hand to reach huge number of users in the form of a network. Thus, the quantum Internet [\cref{Kimble08}], [\cref{Lloyd04}] requires quantum repeaters and quantum switches/routers. Because of the so called \textit{no-cloning theorem} [\cref{Wootters82}], which is the simple consequence of the postulates of the quantum mechanics, the construction of these network entities proves to be very hard [\cref{VanMeter14}].

The other challenge -- this paper focuses on -- is the amount of information which can be transmitted over quantum channels, i.e. the capacity. The capacity of a communication channel describes the capability of the channel for delivering information from the sender to the receiver, in a faithful and recoverable way. Thanks to Shannon we can calculate the capacity of classical channels within the frames of classical information theory\footnote{$  $Quantum Shannon theory has deep relevance concerning the information transmission and storage in quantum systems. It can be regarded as a natural generalization of classical Shannon theory$  $.$    $ Classical information theory represents an orthogonality-restricted case of quantum information theory.} [\cref{Shannon48}]. However, the different capacities of quantum channels have been discovered just in the `90s, and there are still many open questions about the different capacity measures. 

Many new capacity definitions exist for quantum channels in comparison to a classical communication channel. In the case of a classical channel, we can send only classical information while quantum channels extend the possibilities, and besides the classical information we can deliver entanglement-assisted classical information, private classical information, and of course, quantum information [\cref{Bennett02}], [\cref{Devetak05}]. On the other hand, the elements of classical information theory cannot be applied in general for quantum information --in other words, they can be used only in some special cases. There is no general formula to describe the capacity of every quantum channel model, but one of the main results of the recent researches was a simplified picture in which various capacities of a quantum channel (i.e., the classical, private, quantum) are all non-additive [\cref{Imre13}]. 

In possession of admitted capacity definitions they have to be calculated for various channel models. Channels behave in very different ways in free-space or in optical fibers and these two main categories divides into many subclasses and special cases [\cref{Giovannetti04}], [\cref{Giovannetti14}], [\cref{Zhang12}]. 

Since capacity shows only the theoretically achievable transmission rate and gives no construction rules how to reach or near them, therefore quantum channel/error correction coding has similar importance from practical implementation point of view as in case of classical information theory [\cref{Gaitan08}].

This paper is organized as follows. In \sref{sec2}, preliminaries are summarized. In \sref{sec3}, we study the classical information transmission capability of quantum channels. In \sref{sec4}, we discuss the quantum capacity. Numerical examples are included in \sref{sec5}. \sref{sec6} focuses on the practical implementations of quantum channels. Finally, \sref{sec7} concludes the paper. Supplementary material is included in the Appendix. 

\section{Preliminaries}
\label{sec2}

\subsection{Applications and Gains of Quantum Communications}
Before discussing the modeling, characteristics and capacities of quantum channels we present their potential to improve state-of the-art communication and computing systems.

We highlight the fact that from application point of view the concept of ‘channel’ can represent any medium possessing an input to receive information and an output to give back a modified version of this information. This simplified definition highlights the fact that not only an optical fiber, a copper cable or a free-space link can be regarded as channel but a computer memory, too.

Quantum communication systems are capable of providing absolute randomness, absolute security, of improving transmission quality as well as of bearing much more information in comparison to the current classical binary based systems. Moreover, when the benefits of quantum computing power are properly employed, the quantum based solutions are capable of supporting the execution of tasks much faster  or beyond the capability of the current binary based systems [\cref{Denchev16}]. The appealing gains and the associated application scenarios that we may expect from quantum communications are as follows. 

The general existence of a qubit $\psi$ in a superposition state (see the next sub-sections of \sref{sec2}) of two pure quantum states $\ket{0}$ and $\ket{1}$ can be represented by
\begin{eqnarray}
\label{qbit}
\ket{\psi} &=& \alpha \ket{0} + \beta\ket{1},\   
\end{eqnarray}
where $\alpha$ and $\beta$ are complex number. If a qubit $\psi$ is measured by $\ket{0}$ and $\ket{1}$ bases, the measurement result is randomly obtained in the state of $\ket{0}$ or $\ket{1}$ with the corresponding probability of $|\alpha|^2$ or $|\beta|^2$.  This random nature of quantum measure have been favourably used for providing high quality random number generator [\cref{Ivanova17}, \cref{Khanmohammadi15}], [\cref{Ma16}]. It is important to note that along with the measurement randomness, no-cloning theorem [\cref{Wootters82}] of qubit says that it is not possible to clone a qubit. This characteristics allow quantum based solutions to  support absolute security, to which there have been abundant examples of quantum based solutions  [\cref{Ghilen17}], [\cref{Li17}], [\cref{Liao17}], [\cref{Zhang17}], [\cref{Wu16}] where a popular example of mature applications is quantum key distribution (QKD) [\cref{Bennett14}], [\cref{Bennett83}].

Quantum entanglement is a unique characteristic of quantum mechanics, which is another valuable foundation for provisioning the absolute secure communication. Let us consider a two qubit system $\sigma$ represented by
\begin{eqnarray}
\label{2qbit}
\ket{\sigma} &=& \alpha_{00} \ket{0}\ket{0} + \alpha_{01} \ket{0}\ket{1} + \alpha_{10} \ket{1}\ket{0} + \alpha_{11} \ket{1}\ket{1},\   
\end{eqnarray}
where  $\alpha_{00}, \alpha_{01}, \alpha_{10}, \alpha_{11}$ are complex numbers having $|\alpha_{00}|^2 + |\alpha_{01}|^2 +  |\alpha_{10}| + |\alpha_{11}|^2 =1$. If the system $\sigma$ is prepared in one of the four states (see Appendix), for example  
\begin{eqnarray}
\label{2qbit}
\ket{\sigma} &=& \alpha_{00} \ket{0}\ket{0} +  \alpha_{11} \ket{1}\ket{1},\   
\end{eqnarray}
where $|\alpha_{00}|^2 + |\alpha_{11}|^2 =1$, the measurement result of the two qubits is in either $\ket{00}$ or $\ket{11}$ state. In this state, the two qubits are entangled, meaning that having the measurement result of either of the two is sufficient to  know the measurement result of the other. As a result, if the two entangled qubits are separated  in the distance, for example 144 km terrestial distance [\cref{Fedrizzi09}] or earth-station to satellite 1200 km distance [\cref{Yin17}], information can be secretly transmitted over two locations, where there exists entanglement between the two locations. The entanglement based transmission can be employed for transmitting classical bits by using the superdense coding protocol [\cref{Abeyesinghe06}], [\cref{Babar13}], [\cref{Hsieh10}] or for transmitting qubits using the quantum teleportation protocol [\cref{Bennett05}], [\cref{Hellemans04}].   

Classical channels handle classical information i.e. orthogonal (distinguishable) basis states while quantum channels may deliver superposition states (linear combination of basis states). Of course, since quantum mechanic is more complete than classical information theory classical information and classical channels can be regarded as special cases of quantum information and channels. Keeping in mind the application scenarios, there is a major difference between classical and quantum information. Human beings due to their limited senses can perceive only classical information; therefore measurement is needed to perform conversion between the quantum and classical world.

From the above considerations, quantum channels can be applied in several different ways for information transmission. If classical information is encoded to quantum states, the quantum channel delivers this information between its input and output and finally a measurement device converts the information back to the classical world. In many practical settings, quantum channels are used to transfer classical information only. 

The most discussed practical application of this approach is QKD. Optical fiber based [\cref{Idquantique}], [\cref{Jouguet13}], [\cref{Korzh15}], [\cref{Takesue15}] ground-ground [\cref{Zeilinger12}] and ground-space [\cref{Liao17sk}] systems have already been demonstrated. These protocols – independently whether they are first-generation single photon systems or second-generation multi photon solutions – exchange classical sequences between Alice and Bob over the quantum channel being encoded in non-orthogonal quantum states. Since the no-cloning theorem [\cref{Imre05}], [\cref{Wootters82}] makes no possible to copy (to eavesdrop) the quantum states without error, symmetric ciphering keys can be established for both parties. In this case quantum channel is used to create a new quality instead of improving the performance of classical communication.

Furthermore, quantum encoding can improve the transmission rates of certain channels. For example the well-known bit-flip channel inverts the incoming bit value by probability $p$ and leaves it unchanged by $(1-p)$. Classically this type of channel can not transmit any information at all if $p=0.5$ even if we apply redundancy for error correction. However, if classical bits are encoded into appropriate quantum bits one-by-one, i.e., no redundancy is used, the information will be delivered without error. This means that quantum communication improves the classical information transmission capability of the bit-flip channel form 0 to the maximum 1. The different models of classical information transmission over a quantum channel will be detailed in \sref{sec3} (particulary in \sref{sec3c}-\sref{sec3g}).

The second approach applies quantum channels to deliver quantum information and this information is used to improve the performance of classical communication systems. The detailed discussion of the transmission of quantum information is the subject of \sref{sec4}. These protocols exploit over-quantum-channel-shared entangled states, i.e. entanglement assisted communications is considered. In case of quantum superdense coding [\cref{Bennett92}], [\cref{Berces15}], [\cref{Imre05}] we assume that Alice and Bob have already shared an entangled Bell-pair, such as $\left| {{\beta }_{00}} \right\rangle $  (see Appendix), expressed as
\begin{equation} \label{6)} 
 \begin{array}{l}
\begin{split}
\left|\left.{\beta }_{00}\right\rangle \right.\mathrm{=}\frac{\mathrm{1}}{\sqrt{\mathrm{2}}}\left(\left|\left.00\right\rangle \right.\mathrm{+}\left|\left.\mathrm{11}\right\rangle \right.\right). 
\end{split}
\end{array}
\end{equation} 

When Alice wants to communicate with Bob, she encodes two classical bits into the half pair she possesses and sends this quantum bit to Bob over the quantum channel. Finally, Bob leads his own qubit together the received one to a measuring device which decodes the original two classical bits. Practically 2 classical bits have been transferred at the expense of 1 quantum bit, i.e., the entanglement assisted quantum channels can outperform classical ones.

Another practical example of this approach is distributed medium access control. In this case a classical communication channel is supported by pre-shared entanglement. It is well-known that WiFi and other systems can be derived from the Slotted Aloha protocol [\cref{Abramson70}] widely used as a reference. Slotted Aloha can deliver $[0.5/e,1/e]$ packets in average in each timeslot if the number of nodes is known for everyone, and optimal access strategy is used by everyone. This is because of collisions and unused timeslots. Practically the size of the population can be only estimated which decreases the efficiency. However, if special entangled $\ket{w_n}$ states are generated as
\begin{equation}\label{aaaa}
\ket{w_n}=\frac{1}{\sqrt{n}}\sum^{n}_{i=1}\ket{2^{(n-i)}}.
\end{equation}
and used to coordinate the channel access in a distributed way the timeslot usage will improve to 100\% and there is no need to know the number of users.

Further important application scenarios are related to quantum computers where quantum information has to be delivered between modules over quantum connections. Similarly quantum memories are practically quantum channels – of course with different characteristics compared to communication channels – which store and read back quantum information.

\subsection{Privacy and Performance Gains of Quantum Channels}
Due to the inherent no-cloning theory of quantum mechanics, the random nature of quantum measurement as well as to the unique entanglement phenomenon of quantum mechanics, secure communications can be guaranteed by quantum communications. The private classical capacity of a quantum channel is detailed in \sref{sec3c}. 

Moreover, quantum communications using quantum channels is capable of carrying much more information in comparison to the current classical binary based systems.  Let us have a closer look at \eref{qbit}, where obviously one qubit contains superpositioned $2^1$ distinct states or values, which is equivalent to at least 2 bits. In the case of using two qubits in \eref{2qbit},  $2^2$ distinct states or values are simultaneously conveyed by two qubits, meaning at least $2^2\times 2$ bits are carried by 2 qubits. Generally, $n$ qubits can carry up to $2^n$ states, which corresponds to $2^n\times n$ bits. The superposition nature of qubits leads to the advent of powerful quantum computing, which is in some cases proved be 100 millions times faster than the classical computer [\cref{Denchev16}]. Moreover, in theory quantum computer is capable of providing the computing power that is beyond the capability of its classical counterpart. Importantly, in order to realise such supreme computing power, the crucial part is quantum communications, which has to be used for transmitting qubits within the quantum processor as well as between distributed quantum processors.

Additionally, quantum receivers [\cref{Becerra14}] relying on quantum communications principle has proved to outperform classical homodyne or heterodyne receiver in the context of optical communications. For the sake of brevity, please allow us to refer interested readers to the references [\cref{Becerra14}], [\cref{Tsujino11}].

\subsection{Communication over a Quantum Channel}
Communication through a quantum channel cannot be described by the results of classical information theory; it requires the generalization of classical information theory by quantum perception of the world. In the general model of communication over a quantum channel $\mathcal{N}$, the encoder encodes the message in some coded form, and the receiver decodes it, however in this case, the whole communication is realized through a quantum system. 

The information sent through quantum channels is carried by quantum states, hence the encoding is fundamentally different from any classical encoder scheme. The encoding here means the preparation of a quantum system, according to the probability distribution of the classical message being encoded. Similarly, the decoding process is also different: here it means the measurement of the received quantum state. The properties of quantum communication channel, and the fundamental differences between the classical and quantum communication channel cannot be described without the elements of quantum information theory. 

The model of the quantum channel represents the physically allowed transformations which can occur on the sent quantum system. The result of the channel transformation is another quantum system, while the quantum states are represented by matrices. The physically allowed channel transformations could be very different; nevertheless they are always \textit{Completely Positive Trace Preserving} (CPTP) transformations (trace: the sum of the elements on the main diagonal of a matrix). The trace preserving property therefore means that the corresponding density matrices (density matrix: mathematical description of a quantum system) at the input and output of the channel have the same trace. 

 The input of a quantum channel is a quantum state, which encodes information into a physical property. The quantum state is sent through a quantum communication channel, which in practice can be implemented e.g. by an optical-fiber channel, or by a wireless quantum communication channel. To extract any information from the quantum state, it has to be measured at the receiver's side. The outcome of the measurement of the quantum state (which might be perturbed) depends on the transformation of the quantum channel, since it can be either totally probabilistic or deterministic. In contrast to classical channels, a quantum channel transforms the information coded into quantum states, which can be e.g. the spin state of the particle, the ground and excited state of an atom, or several other physical approaches. The classical capacity of a quantum channel has relevance if the goal is transmit classical information in a quantum state, or would like to send classical information privately via quantum systems (private classical capacity). The quantum capacity has relevance if one would like to transmit quantum information such as superposed quantum states or quantum entanglement. 

First, we discuss the process of transmission of information over a quantum channel. Then, the interaction between quantum channel output and environment will be described. 

\subsubsection{The Quantum Channel Map}
From algebraic point of view, quantum channels are linear CPTP maps, while from a geometrical viewpoint, the quantum channel $\mathcal{N}$ is an affine transformation. While, from the algebraic view the transformations are defined on density matrices, in the geometrical approach, the qubit transformations are also interpretable via the Bloch sphere (a geometrical representation of the pure state space of a qubit system) as Bloch vectors (vectors in the Bloch sphere representation). Since, density matrices can be expressed in terms of Bloch vectors, hence the map of a quantum channel $\mathcal{N}$ also can be analyzed in the geometrical picture. 

To preserve the condition for a density matrix $\rho $, the noise on the quantum channel $\mathcal{N}$ must be trace-preserving (TP), i.e., 
\begin{equation} \label{1.1)} 
Tr\left(\rho \right)\mathrm{=}Tr\left(\mathcal{N}\left(\rho \right)\right),                                            
\end{equation} 
and it must be Completely Positive (CP), i.e., for any identity map \textit{I}, the map $I\mathrm{\otimes}\mathcal{N}$ maps a semi-positive Hermitian matrix to a semi-positive Hermitian matrix. 
\begin{center}
\begin{figure}[htbp]
\vspace{-0.6cm}
\begin{center}
\includegraphics[angle = 0,width=0.9\linewidth]{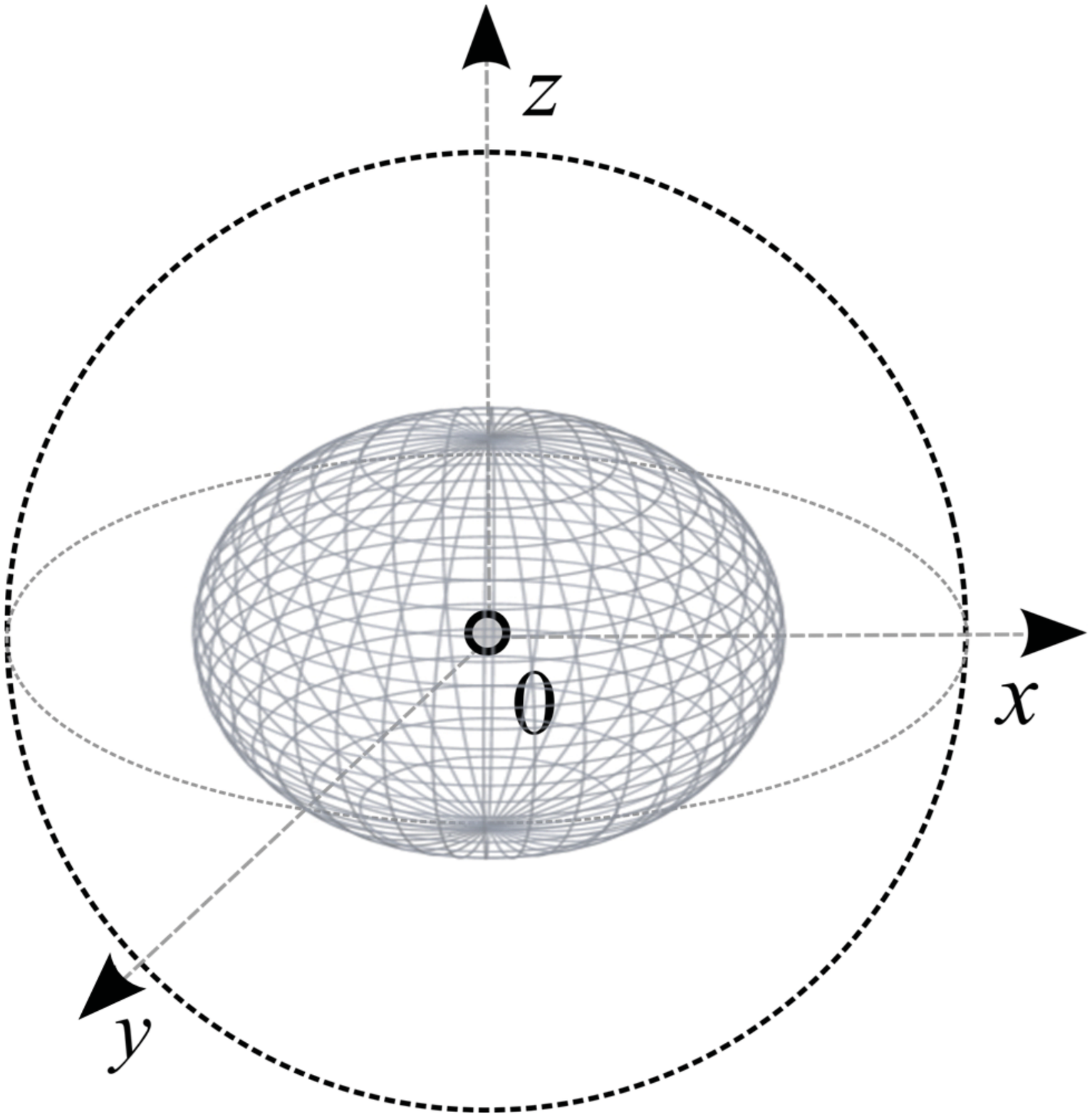}
\caption{Geometrical picture of a noisy qubit quantum channel on the Bloch sphere [Imre13].} 
 \label{fig1_1}
 \end{center}
\end{figure}
\end{center}

For a unital quantum channel $\mathcal{N}$, the channel map transforms the \textit{I} identity transformation to the \textit{I} identity transformation, while this condition does not hold for a non-unital channel. 
To express it, for a unital quantum channel, we have 
\begin{equation} \label{1.2)} 
\mathcal{N}\left(I\right)\mathrm{=}I,              
\end{equation} 
while for a non-unital quantum channel, 
\begin{equation} \label{1.3)} 
\mathcal{N}\left(I\right)\mathrm{\ne }I.              
\end{equation} 

Focusing on a qubit channel, the image of the quantum channel's linear transform is an \textit{ellipsoid} on the Bloch sphere, as it is depicted in \fref{fig1_1}. For a unital quantum channel, the center of the geometrical interpretation of the channel ellipsoid is equal to the center of the Bloch sphere. This means that a unital quantum channel preserves the average of the system states. On the other hand, for a non-unital quantum channel, the center of the channel ellipsoid will differ from the center of the Bloch sphere. The main difference between unital and non-unital channels is that the non-unital channels do not preserve the average state in the center of the Bloch sphere. It follows from this that the numerical and algebraic analysis of non-unital quantum channels is more complicated than in the case of unital ones. While unital channels shrink the Bloch sphere in different directions with the center preserved, non-unital quantum channels shrink both the original Bloch sphere and move the center from the origin of the Bloch sphere. This fact makes our analysis more complex, however, in many cases, the physical systems cannot be described with unital quantum channel maps. Since the unital channel maps can be expressed as the convex combination of the basic unitary transformations, the unital channel maps can be represented in the Bloch sphere as different rotations with shrinking parameters. On the other hand, for a non-unital quantum map, the map cannot be decomposed into a convex combination of unitary rotations [\cref{Imre13}]. 

\subsubsection{Steps of the Communication}
The transmission of information through classical channels and quantum channels differs in many ways. If we would like to describe the process of information transmission through a quantum communication channel, we have to introduce the three main phases of quantum communication. In the first phase, the sender, Alice, has to encode her information to compensate the noise of the channel $\mathcal{N}$ (i.e., for error correction), according to properties of the physical channel - this step is called \textit{channel coding. }After the sender has encoded the information into the appropriate form, it has to be put on the quantum channel, which transforms it according to its channel map - this second phase is called the \textit{channel evolution}. The quantum channel $\mathcal{N}$ conveys the quantum state to the receiver, Bob; however this state is still a superposed and probably \textit{mixed} (according to the noise of the channel) quantum state. To extract the information which is encoded in the state, the receiver has to make a measurement - this \textit{decoding process} (with the error correction procedure) is the third phase of the communication over a quantum channel. 

\begin{center}
\begin{figure}[htbp]
\begin{center}
\includegraphics[angle = 0,width=1\linewidth]{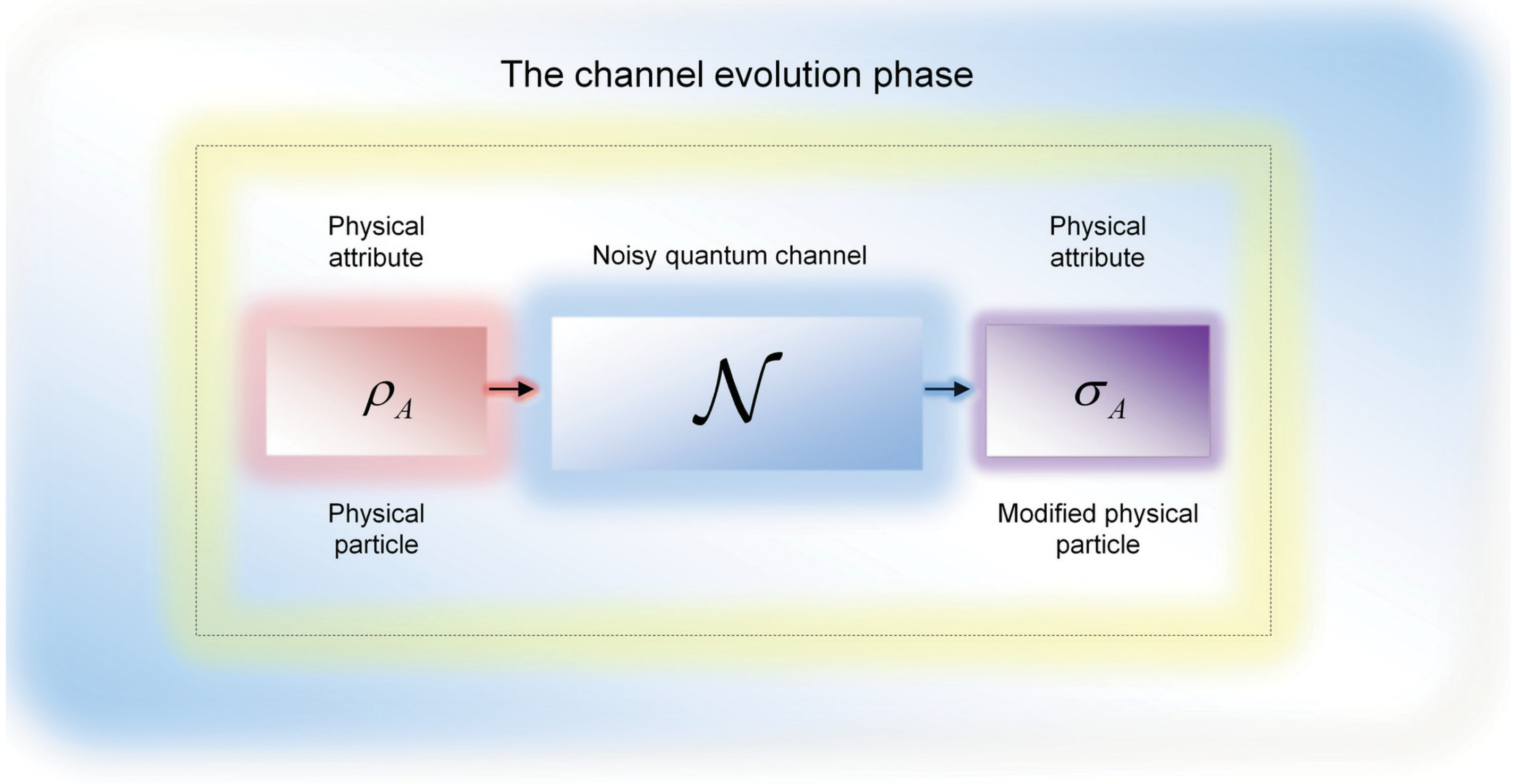}
\caption{The channel evolution phase.} 
 \label{fig1_3}
 \end{center}
\end{figure}
\end{center}

The channel transformation represents the noise of the quantum channel. Physically, the quantum channel is the medium, which moves the particle from the sender to the receiver. The noise disturbs the state of the particle, in the case of a half-spin particle, it causes spin precession. The channel evolution phase is illustrated in \fref{fig1_3}.

Finally, the measurement process responsible for the decoding and the extraction of the encoded information. The previous phase determines the success probability of the recovery of the original information. If the channel $\mathcal{N}$ is completely noisy, then the receiver will get a maximally mixed quantum state. The output of the measurement of a maximally mixed state is completely undeterministic: it tells us nothing about the original information encoded by the sender. On the other hand, if the quantum channel $\mathcal{N}$ is completely noiseless, then the information which was encoded by the sender can be recovered with probability 1: the result of the measurement will be completely deterministic and completely correlated with the original message. In practice, a quantum channel realizes a map which is in between these two extreme cases. A general quantum channel transforms the original pure quantum state into a mixed quantum state, - but not into a maximally mixed state - which makes it possible to recover the original message with a high - or low - probability, depending on the level of the noise of the quantum channel $\mathcal{N}$.

\subsection{Formal Model}

\begin{center}
\begin{figure}[htbp]
\vspace{-0.45cm}
\begin{center}
\includegraphics[angle = 0,width=1\linewidth]{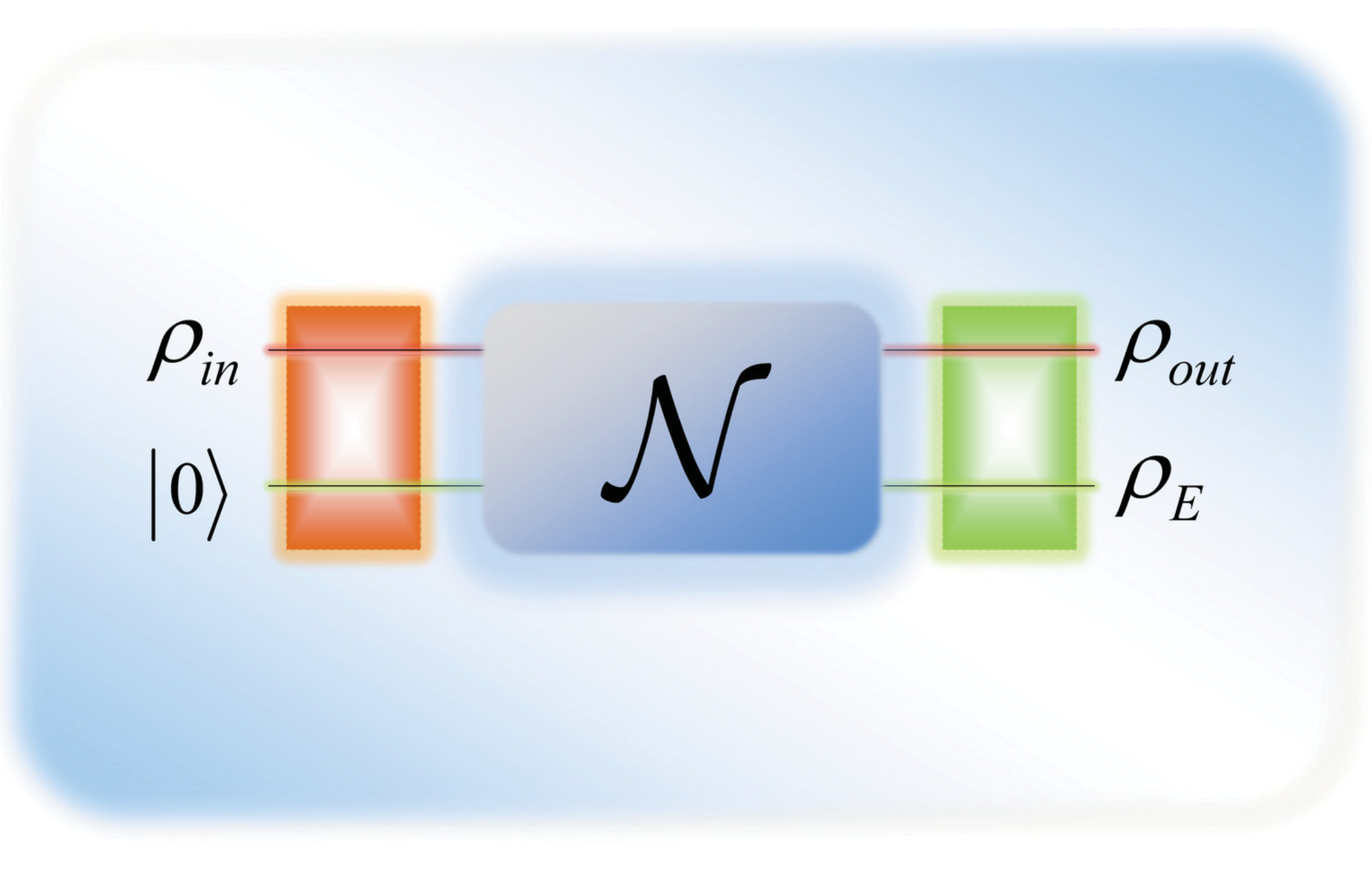}
\caption{The general model of transmission of information over a noisy quantum channel.} 
 \label{fig1_5}
 \end{center}
\end{figure}
\end{center}

As shown in \fref{fig1_5}, the information transmission through the quantum channel $\mathcal{N}$ is defined by the ${\rho }_{in}$ input quantum state and the initial state of the environment ${\rho }_E\mathrm{=}\left|\left.0\right\rangle \right.\left\langle \left.0\right|\right.$. In the initial phase, the environment is assumed to be in the pure state $\left|\left.0\right\rangle \right.$. The system state which consist of the input quantum state ${\rho }_{in}$ and the environment ${\rho }_E\mathrm{=}\left|\left.0\right\rangle \right.\left\langle \left.0\right|\right.$, is called the\textit{ composite state} ${\rho }_{in}\mathrm{\otimes }{\rho }_E$. 

If the quantum channel $\mathcal{N}$ is used for information transmission, then the state of the composite system changes unitarily, as follows: 
\begin{equation} \label{1.4)} 
U\left({\rho }_{in}\mathrm{\otimes }{\rho }_E\right)U^{\mathrm{\dagger }},                                                                     
\end{equation} 
where $U$ is a unitary transformation, and $U^{\mathrm{\dagger }}U\mathrm{=}I$. After the quantum state has been sent over the quantum channel $\mathcal{N}$, the ${\rho }_{out}$ output state can be expressed as: 
\begin{equation} \label{1.5)} 
\mathcal{N}\left({\rho }_{in}\right)\mathrm{=}{\rho }_{out}\mathrm{=}Tr_E\left[U\left({\rho }_{in}\mathrm{\otimes }{\rho }_E\right)U^{\mathrm{\dagger }}\right],                                                           
\end{equation} 
where $Tr_E$ traces out the environment \textit{E} from the joint state. Assuming the environment \textit{E} in the pure state $\left|\left.0\right\rangle \right.$, ${\rho }_E\mathrm{=}\left|\left.0\right\rangle \right.\left\langle \left.0\right|\right.$, the $\mathcal{N}\left({\rho }_{in}\right)$ noisy evolution of the channel $\mathcal{N}$ can be expressed as:
\begin{equation} \label{1.6)} 
\mathcal{N}\left({\rho }_{in}\right)\mathrm{=}{\rho }_{out}\mathrm{=}Tr_EU{\rho }_{in}\mathrm{\otimes }\left|\left.0\right\rangle \right.\left\langle \left.0\right|\right.U^{\mathrm{\dagger }},                                                            
\end{equation} 
while the post-state ${\rho }_E$ of the environment after the transmission is 
\begin{equation} \label{1.7)} 
{\rho }_E\mathrm{=}Tr_BU{\rho }_{in}\mathrm{\otimes }\left|\left.0\right\rangle \right.\left\langle \left.0\right|\right.U^{\mathrm{\dagger }},                                                                         
\end{equation} 
where $Tr_B$ traces out the output system \textit{B}. In general, the \textit{i}-th input quantum state ${\rho }_i$ is prepared with probability $p_i$, which describes the ensemble $\left\{p_i,{\rho }_i\right\}$. The average of the \textit{input} quantum system is 
\begin{equation} \label{1.8)} 
{\sigma }_{in}\mathrm{=}\sum_i{p_i{\rho }_i},                                                             
\end{equation} 
The average (or the mixture) of the \textit{output} of the quantum channel is denoted by 
\begin{equation} \label{ZEqnNum172273} 
{\sigma }_{out}\mathrm{=}\mathcal{N}\left({\sigma }_{in}\right)\mathrm{=}\sum_i{p_i\mathcal{N}\left({\rho }_i\right)}.                                                                 
\end{equation}

\subsection{Quantum Channel Capacity}

 The capacity of a communication channel describes the capability of the channel for sending information from the sender to the receiver, in a faithful and recoverable way. The perfect ideal communication channel realizes an identity map. For a quantum communication channel, it means that the channel can transmit the quantum states perfectly. Clearly speaking, the capacity of the quantum channel measures the closeness to the ideal identity transformation \textit{I}. 

To describe the information transmission capability of the quantum channel $\mathcal{N}$, we have to make a distinction between the various capacities of a quantum channel. The encoded quantum states can carry classical messages or quantum messages. In the case of classical messages, the quantum states encode the output from a \textit{classical information source}, while in the latter the source is a \textit{quantum information source}. 

\begin{center}
\begin{figure}[htbp]
\vspace{-0.5cm}
\begin{center}
\includegraphics[angle = 0,width=1\linewidth]{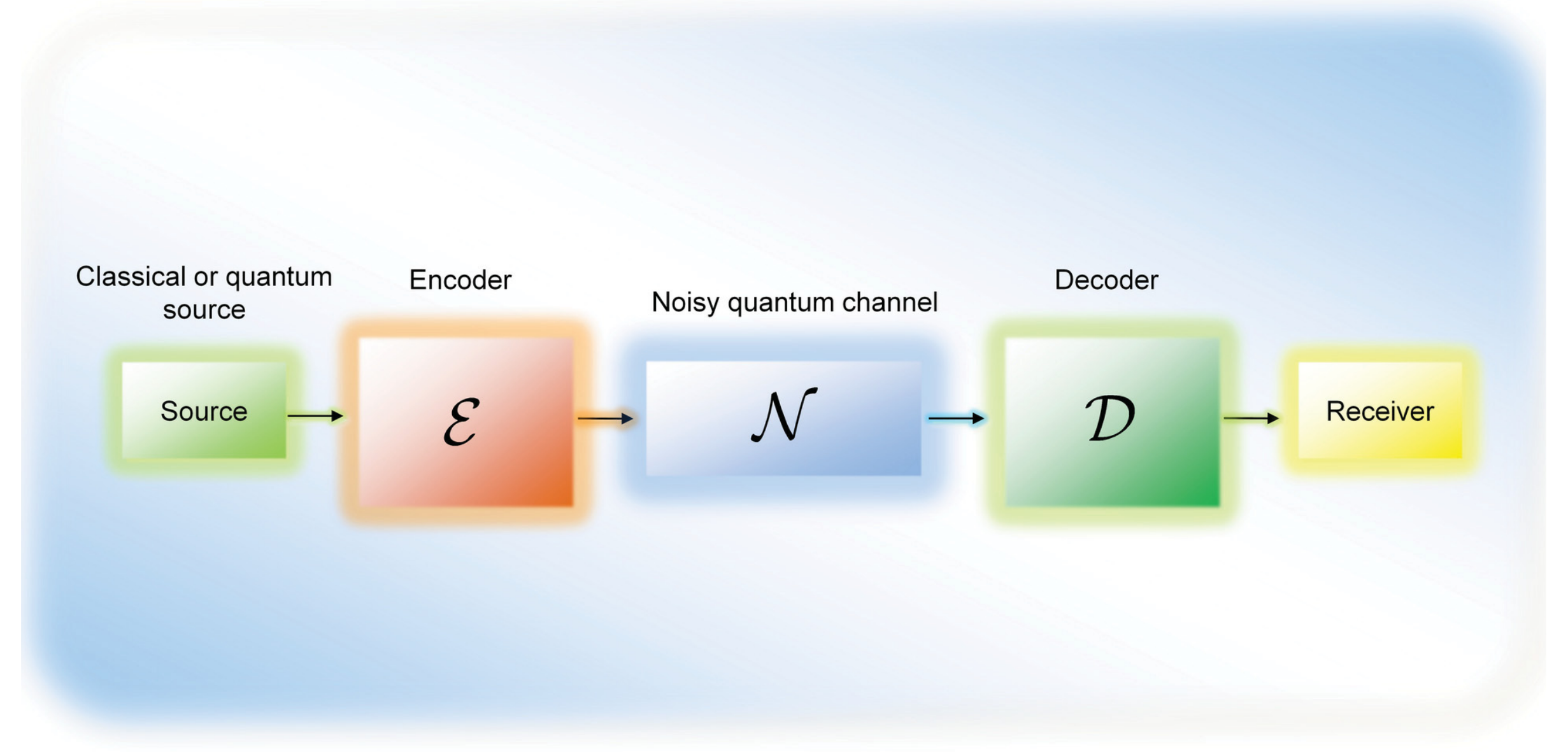}
\caption{Communication over a noisy quantum channel.} 
 \label{fig1_6}
 \end{center}
\end{figure}
\end{center}

On one hand for classical communication channel \textit{N}, only one type of capacity measure can be defined, on the other hand for a quantum communication channel $\mathcal{N}$ a number of different types of quantum channel capacities can be applied, with different characteristics. There are plenty of open questions regarding these various capacities. In general, the \textit{single-use} capacity of a quantum channel is not equal to the \textit{asymptotic} capacity of the quantum channel (As we will see later, it also depends on the type of the quantum channel). The asymptotic capacity gives us the amount of information which can be transmitted in a reliable form using the quantum channel infinitely many times. The encoding and the decoding functions mathematically can be described by the operators $\mathcal{E}$ and $\mathcal{D}$, realized on the blocks of quantum states. These superoperators describe unitary transformations on the input states together with the environment of the quantum system. The model of communication through noisy quantum channel $\mathcal{N}$ with encoding, delivery and decoding phases is illustrated in \fref{fig1_6}. 

We note, in our paper we will use the terms \textit{classical quantity} and \textit{quantum quantity} with relation to the quantum channel $\mathcal{N}$ as follows: 
\begin{enumerate}
\item \textit{classical quantity}: it is a measure of the classical transmission capabilities of a quantum channel. (See later: Holevo information, quantum mutual information, etc., in \sref{sec3})
\item \textit{quantum quantity}: it is a measure of the quantum transmission capabilities of a quantum channel (See later: quantum coherent information,etc., in \sref{sec4})
\end{enumerate}
 If we mention classical quantity we will do this with relation to the quantum channel $\mathcal{N}$, i.e., for example the Holevo information is also not a typical' classical quantity since it is describes a quantum system not a classical one, but with relation to the quantum channel we can use the \textit{classical }mark.
 The historical background with the description of the most relevant works can be found in the Related Work part of each section. For detailed information see [\cref{Imre13}].

\subsection{Definitions}
Quantum information theory also has relevance to the discussion of the capacity of quantum channels and to information transmission and storage in quantum systems. As we will see in this section, while the transmission of product states can be described similar to classical information, on the other hand, the properties of quantum entanglement cannot be handled by the elements of classical information theory. Of course, the elements of classical information theory can be viewed as a subset of the larger and more complex quantum information theory [\cref{Zhang16}]. 

First, we summarize the basic definitions and formulas of quantum information theory. We introduce the reader to the description of a noisy quantum channel, purification, isometric extension, Kraus representation and the von Neumann entropy. Next, we describe the encoding of quantum states and the meaning of Holevo information, the quantum mutual information and quantum conditional entropy. 

\subsubsection{Discussion}
 Before starting the discussion on various capacities of quantum channels and the related consequences we summarize the basic definitions and formulas of quantum information theory intended to represent the information stored in quantum states. Those readers who are familiar with density matrices, entropies etc. may run through this section.

The world of quantum information processing (QIP) is describable with the help of quantum information theory (QIT), which is the main subject of this section. We will provide an overview of the most important differences between the compressibility of classical bits and quantum bits, and between the capacities of classical and quantum communication channels. To represent classical information with quantum states, we might use pure orthogonal states. In this case there is no difference between the compressibility of classical and quantum bits.

Similarly, a quantum channel can be used with pure orthogonal states to realize classical information transmission, or it can be used to transmit non-orthogonal states or even quantum entanglement. Information transmission also can be approached using the question, whether the input consists of unentangled or entangled quantum states. This leads us to say that for quantum channels many new capacity definitions exist in comparison to a classical communication channel.

Quantum information theory also has relevance to the discussion of the capacity of quantum channels and to information transmission and storage in quantum systems. While the transmission of product states can be described similar to classical information, on the other hand, the properties of quantum entanglement cannot be handled by the elements of classical information theory. Of course, the elements of classical information theory can be viewed as a subset of the larger and more complex quantum information theory.

Before we would start to our introduction to quantum information theory, we have to make a clear distinction between quantum information theory and quantum information processing. Quantum information theory is rather a generalization of the elements and functions of classical information theory to describe the properties of quantum systems, storage of information in quantum systems and the various quantum phenomena of quantum mechanics. While quantum information theory aims to provide a stable theoretical background, quantum information processing is a more general and rather experimental field: it answers what can be achieved in engineering with the help of quantum information. Quantum information processing includes the computing, error-correcting schemes, quantum communication protocols, field of communication complexity, etc.

The character of classical information and quantum information is significantly different. There are many phenomena in quantum systems which cannot be described classically, such as entanglement, which makes it possible to store quantum information in the correlation of quantum states. Entangled quantum states are named to EPR states after Einstein, Podolsky and Rosen, or Bell states, after J. Bell. Quantum entanglement was discovered in the 1930s, and it may still yield many surprises in the future. Currently it is clear that entanglement has many classically indescribable properties and many new communication approaches based on it. Quantum entanglement plays a fundamental role in advanced quantum communications, such as teleportation, quantum cryptography etc.

The elements of quantum information theory are based on the laws of quantum mechanics. The main results of quantum information processing were laid down during the end of the twentieth century, the most important results being stated by Feynman, Bennett, DiVincenzo, Devetak, Deutsch, Holevo, Lloyd, Schumacher, Shor and many more. After the basic concepts of quantum information processing had been stated, researchers started to look for efficient quantum error correction schemes and codes, and started to develop the theoretical background of fault-tolerant quantum computation. The main results from this field were presented by Bennett, Schumacher, Gottesman, Calderbank, Preskill, Knill, and Kerckhoff. On the other hand, there are still many open questions about quantum computation. The theoretical limits of quantum computers were discovered by Bennett, Bernstein, Brassard and Vazirani: quantum computers can provide at best a quadratic reduction in the complexity of search-based problems, hence if we give an NP-complete problem to quantum computer, it still cannot solve it. Recently, the complexity classes of quantum information processing have been investigated, and many new classes and lower bounds have been found.

By the end of the twentieth century, many advanced and interesting properties of quantum information theory had been discovered, and many possible applications of these results in future communication had been developed. One of the most interesting revealed connections was that between quantum information theory and the elements of geometry. The space of quantum states can be modeled as a convex set which contains points with different probability distributions, and the geometrical distance between these probability distributions can be measured by the elementary functions of quantum information theory, such as the von Neumann entropy or the quantum relative entropy function. The connection between the elements of quantum information theory and geometry leads us to the application of advanced computational geometrical algorithms to quantum space, to reveal the still undiscovered properties of quantum information processing, such as the open questions on the capacities of the quantum channels or their additivity properties. The connection between the Hilbert space of quantum states and the geometrical distance can help us to reveal the fantastic properties of quantum bits and quantum state space.

Several functions have been defined in quantum information theory to describe the statistical distances between the states in the quantum space: one of the most important is the quantum relative entropy function which plays a key role in the description of entanglement, too. This function has many different applications, and maybe this function plays the most important role in the questions regarding the capacity of quantum channels. The possible applications of the quantum relative entropy function have been studied by Schumacher and Westmoreland and by Vedral.

Quantum information theory plays fundamental role in the description of the data transmission through quantum communication channels. At the dawn of this millennium new problems have arisen, whose solutions are still not known, and which have opened the door to many new promising results such as the superactivation of zero-capacity quantum channels in 2008, and then the superactivation of the zero-error capacities of the quantum channels in 2009 and 2010. One of the earliest works on the capacities of quantum communication channels was published in the early 1970s. Along with other researchers, Holevo was showed that there are many differences between the properties of classical and quantum communication channels, and illustrated this with the benefits of using entangled input states. Later, he also stated that quantum communication channels can be used to transmit both classical and quantum information. Next, many new quantum protocols were developed, such as teleportation or superdense coding. After Alexander Holevo published his work, about thirty years later, he, with Benjamin Schumacher and Michael Westmoreland presented one of the most important result in quantum information theory, called the \textit{Holevo-Schumacher-Westmoreland} (HSW) theorem [\cref{Holevo98}], [\cref{Schumacher97}]. The HSW-theorem is a generalization of the classical noisy channel coding theorem from classical information theory to a noisy quantum channel. The HSW theorem is also called the \textit{product-state} classical channel capacity theorem of a noisy quantum channel. The understanding of the classical capacity of a quantum channel was completed by 1997 by Schumacher and Westmoreland, and by 1998 by Holevo, and it has tremendous relevance in quantum information theory, since it was the first to give a mathematical proof that a noisy quantum channel can be used to transmit classical information in a reliable form. The HSW theorem was a very important result in the history of quantum information theory, on the other hand it raised a lot of questions regarding the transmission of classical information over general quantum channels. 

The quantum capacity of a quantum channel was firstly formulated by Seth Lloyd in 1996, then by Peter Shor in 2002, finally it was completed by Igor Devetak in 2003, - the result is known as the \textit{LSD channel capacity} [\cref{Devetak03a}], [\cref{Lloyd97}], [\cref{Shor02}]. While the classical capacity of a quantum channel is described by the maximum of quantum mutual information and the Holevo information, the quantum capacity of the quantum channels is described by a completely different correlation measure: called the \textit{quantum coherent information}. The concept of quantum coherent information plays a fundamental role in the computation of the quantum capacity which measures the asymptotic quantum capacity of the quantum capacity in general.  For the complete historical background with the references see the Related Works.

\subsubsection{Density Matrix and Trace Operator}

 In this section we introduce a basic concept of quantum information theory, called the \textit{density matrix}.

 Before we start to discuss the density matrix, we introduce some terms. An $n\mathrm{\times }n$  square matrix \textit{A} is called \textit{positive-semidefinite} if $\left\langle  \psi  | \left. A \right|\psi  \right\rangle $ is a non-negative real number for every vector $\left|\left.\psi \right\rangle \right.$. If $A\mathrm{=}A^{\mathrm{\dagger }}$, i.e., \textit{A} has Hermitian matrix and the $\left\{{\lambda }_{\mathrm{1}},{\lambda }_{\mathrm{2}}\mathrm{,\dots }{\lambda }_n\right\}$ eigenvalues of \textit{A} are all non-negative real numbers then it is positive-semidefinite. This definition has important role in quantum information theory, since \textit{every density matrix is positive-semidefinite. }It means, for any vector $\left|\left.\varphi \right\rangle \right.$ the positive-semidefinite property says that
\begin{equation} \label{ZEqnNum374970} 
\left\langle \left.\varphi \right|\right.\rho \left|\left.\varphi \right\rangle \right.\mathrm{=}\sum^n_{i\mathrm{=1}}{p_i\left\langle \varphi \mathrel{\left|\vphantom{\varphi  {\psi }_i}\right.\kern-\nulldelimiterspace}{\psi }_i\right\rangle \left\langle {\psi }_i\mathrel{\left|\vphantom{{\psi }_i \varphi }\right.\kern-\nulldelimiterspace}\varphi \right\rangle }\mathrm{=}\sum^n_{i\mathrm{=1}}{p_i{\left|\left\langle \varphi \mathrel{\left|\vphantom{\varphi  {\psi }_i}\right.\kern-\nulldelimiterspace}{\psi }_i\right\rangle \right|}^{\mathrm{2}}}\mathrm{\ge }\mathrm{0}.                                
\end{equation} 
In \eqref{ZEqnNum374970} we used, the density matrix is denoted by $\rho $, and it describes the system by the classical probability weighted sum of possible states
\begin{equation} \label{ZEqnNum711440} 
\rho \mathrm{=}\sum_i{p_i}\left|\left.{\psi }_i\right\rangle \right.\left\langle \left.{\psi }_i\right|\right.,                                                                            
\end{equation} 
where $\left|\left.{\psi }_i\right\rangle \right.$ is the \textit{i}-th system state occurring with classical probability $p_i$. As can be seen, this density matrix describes the system as a probabilistic mixture of the possible known states the so called \textit{pure states}. For pure state $\left|\left.\psi \right\rangle \right.$ the density matrix is $\rho \mathrm{=}\left|\left.\psi \right\rangle \right.\left\langle \left.\psi \right|\right.$ and the rank of the matrix is equal to one. Trivially, classical states e.g. $\left|\left.0\right\rangle \right.$ and $\left|\left.\mathrm{1}\right\rangle \right.$ are pure, however, if we know that our system is prepared to the \textit{superposition} $\frac{\mathrm{1}}{\sqrt{\mathrm{2}}}\left(\left|\left.0\right\rangle \right.\mathrm{+}\left|\left.\mathrm{1}\right\rangle \right.\right)$ then this state is pure, too. Clearly speaking, while superposition is a quantum linear combination of orthonormal basis states weighted by probability amplitudes, mixed states are classical linear combination of pure superpositions (quantum states) weighted by classical probabilities.

The density matrix contains all the possible information that can be extracted from the quantum system. It is possible that two quantum systems possess the same density matrices: in this case, these quantum systems are called indistinguishable, since it is not possible to construct a measurement setting, which can distinguish between the two systems.

 The density matrix $\rho $ of a simple pure quantum system which can be given in the state vector representation $\left|\left.\psi \right\rangle \right.\mathrm{=}\alpha \left|\left.0\right\rangle \right.\mathrm{+}\beta \left|\left.\mathrm{1}\right\rangle \right.$ can be expressed as the outer product of the \textit{ket} and \textit{bra} vectors, where bra is the transposed complex conjugate of ket, hence for $\left|\left.\psi \right\rangle \right.\mathrm{=}\left[ \begin{array}{l}
\alpha  \\ 
\beta  \end{array}
\right],\left\langle \left.\psi \right|\right.\mathrm{=}\left[ \begin{array}{ll}
{\alpha }^{\mathrm{*}} & {\beta }^{\mathrm{*}} \end{array}
\right]$ the density matrix is
\begin{equation} \label{2.3)} 
\begin{split}
\rho \mathrm{=}\left|\left.\psi \right\rangle \right.\left\langle \left.\psi \right|\right.&\mathrm{=}\left[ \begin{array}{l}
\alpha  \\ 
\beta  \end{array}
\right]\left[ \begin{array}{ll}
{\alpha }^{\mathrm{*}} & {\beta }^{\mathrm{*}} \end{array}
\right]\\&\mathrm{=}\left[ \begin{array}{ll}
\alpha {\alpha }^{\mathrm{*}} & \alpha {\beta }^{\mathrm{*}} \\ 
{\alpha }^{\mathrm{*}}\beta  & \beta {\beta }^{\mathrm{*}} \end{array}
\right]\mathrm{=}\left[ \begin{array}{ll}
{\left|\alpha \right|}^{\mathrm{2}} & \alpha {\beta }^{\mathrm{*}} \\ 
{\alpha }^{\mathrm{*}}\beta  & {\left|\beta \right|}^{\mathrm{2}} \end{array}
\right].                                                    
\end{split}
\end{equation} 
The density matrix $\rho \mathrm{=}\sum^n_{i\mathrm{=1}}{p_i\left|\left.{\psi }_i\right\rangle \right.\left\langle \left.{\psi }_i\right|\right.}$ contains the probabilistic mixture of different pure states, which representation is based on the fact that the mixed states can be decomposed into weighted sum of pure states [\cref{Watrous06}].

To reveal important properties of the density matrix, we introduce the concept of the \textit{trace operation}. The trace of a density matrix is equal to the sum of its diagonal entries. For an $n\mathrm{\times }n$ square matrix $A$, the $Tr$ trace operator is defined as
\begin{equation} \label{2.4)} 
Tr\left(A\right)\mathrm{=}a_{\mathrm{11}}\mathrm{+}a_{\mathrm{22}}\mathrm{+\dots +}a_{nn}\mathrm{=}\sum^n_{i\mathrm{=1}}{a_{ii}},                                                          
\end{equation} 
where $a_{ii}$ are the elements of the main diagonal. The trace of the matrix \textit{A} is also equal to the sum of the \textit{eigenvalues} of its matrix. The eigenvalue is the factor by which the \textit{eigenvector} changes if it is multiplied by the matrix \textit{A}, for each eigenvectors. The \textit{eigenvectors} of the square matrix \textit{A} are those non-zero vectors, whose direction remain the same to the original vector after multiplied by the matrix \textit{A}. It means, the eigenvectors remain proportional to the original vector. For square matrix\textit{ A}, the non-zero vector $\boldsymbol{\mathrm{v}}$ is called \textit{eigenvector} of \textit{A}, if there is a scalar $\lambda $ for which
\begin{equation} \label{2.5)} 
A\boldsymbol{\mathrm{v}}\mathrm{=}\lambda \boldsymbol{\mathrm{v}},                                                 
\end{equation} 
where $\lambda $ is the \textit{eigenvalue} of \textit{A} corresponding to the eigenvector $\boldsymbol{\mathrm{v}}$.

The trace operation gives us the sum of the eigenvalues of positive-semidefinite \textit{A}, for each eigenvectors, hence $Tr\left(A\right)\mathrm{=}\sum^n_{i\mathrm{=1}}{{\lambda }_i}$, and $Tr\left(A^k\right)\mathrm{=}\sum^n_{i\mathrm{=1}}{{\lambda }^k_i}$.                                                                \label{2.6)}
 Using the eigenvalues, the \textit{spectral decomposition} of density matrix $\rho $ can be expressed as
\begin{equation} \label{2.7)} 
\rho \mathrm{=}\sum_i{{\lambda }_i}\left|\left.{\varphi }_i\right\rangle \right.\left\langle \left.{\varphi }_i\right|\right. ,                                                          
\end{equation} 
where $\left|\left.{\varphi }_i\right\rangle \right.$ are orthonormal vectors.

 The trace is a linear map, hence for square matrices \textit{A} and \textit{B}
\begin{equation} \label{2.8)} 
Tr\left(A\mathrm{+}B\right)\mathrm{=}Tr\left(A\right)\mathrm{+}Tr\left(B\right),                                                                  
\end{equation} 
and
\begin{equation} \label{2.9)} 
Tr\left(sA\right)\mathrm{=}sTr\left(A\right),                                                               
\end{equation} 
where \textit{s} is a scalar. Another useful formula, that for $m\mathrm{\times }n$ matrix \textit{A }and $n\mathrm{\times }m$ matrix \textit{B},
\begin{equation} \label{2.10)} 
Tr\left(AB\right)\mathrm{=}Tr\left(BA\right),                                                            
\end{equation} 
which holds for any matrices \textit{A} and \textit{B} for which the product matrix \textit{AB} is a square matrix, since
\begin{equation} \label{2.11)} 
Tr\left(AB\right)\mathrm{=}\sum^m_{i\mathrm{=1}}{\sum^n_{j\mathrm{=1}}{A_{ij}B_{ji}}}\mathrm{=}Tr\left(BA\right).                                                           
\end{equation} 
Finally, we mention that the trace of a matrix \textit{A} and the trace of its transpose $A^T$ are equal, hence
\begin{equation} \label{2.12)} 
Tr\left(A\right)\mathrm{=}Tr\left(A^T\right).                                                   
\end{equation} 
If we take the conjugate transpose $A^{\mathrm{*}}$ of the $m\mathrm{\times }n$ matrix \textit{A}, then we will find that
\begin{equation} \label{2.13)} 
Tr\left(A^{\mathrm{*}}A\right)\mathrm{\ge }\mathrm{0},                                      
\end{equation} 
which will be denoted by $\left\langle A,A\right\rangle $ and it is called the \textit{inner product}. For matrices \textit{A} and \textit{B}, the inner product $\left\langle A,B\right\rangle \mathrm{=}Tr\left(B^{\mathrm{*}}A\right)$, which can be used to define the angle between the two vectors. The inner product of two vectors will be zero if and only if the vectors are orthogonal.

 As we have seen, the trace operation gives the sum of the eigenvalues of matrix \textit{A}, this property can be extended to the density matrix, hence for each eigenvectors ${\lambda }_i$ of density matrix $\rho $
\begin{equation} \label{2.14)} 
Tr\left(\rho \right)\mathrm{=}\sum^n_{i\mathrm{=1}}{{\lambda }_i}.                                                              
\end{equation} 
Now, having introduced the \textit{trace} operation, we apply it to a density matrix. If we have an \textit{n-}qubit system being in the state $\rho \mathrm{=}\sum^n_{i\mathrm{=1}}{p_i\left|\left.{\psi }_i\right\rangle \right.\left\langle \left.{\psi }_i\right|\right.}$, then
\begin{equation} \label{ZEqnNum355630} 
\begin{split}
Tr\left( \sum\limits_{i=1}^{n}{{{p}_{i}}\left| {{\psi }_{i}} \right\rangle \left\langle  {{\psi }_{i}} \right|} \right)&=\sum\limits_{i=1}^{n}{{{p}_{i}}Tr\left( \left| {{\psi }_{i}} \right\rangle \left\langle  {{\psi }_{i}} \right| \right)}\\
&=\sum\limits_{i=1}^{n}{{{p}_{i}}\left( \left\langle  {{\psi }_{i}} | {{\psi }_{i}} \right\rangle  \right)=1},                              
\end{split}
\end{equation} 
where we exploited the relation for unit-length vectors $\left|\left.{\psi }_i\right\rangle \right.$
\begin{equation} \label{2.16)} 
\left\langle  {{\psi }_{i}} | {{\psi }_{i}} \right\rangle \equiv 1.
\end{equation} 
Thus the trace of any density matrix is equal to one
\begin{equation} \label{2.17)} 
Tr\left(\rho \right)\mathrm{=1}.                                           
\end{equation} 
The trace operation can help to distinguish \textit{pure} and \textit{mixed} states since for a given \textit{pure} state $\rho $
\begin{equation} \label{2.18)} 
Tr\left({\rho }^{\mathrm{2}}\right)\mathrm{=1},                                      
\end{equation} 
while for a \textit{mixed} state $\sigma $,
\begin{equation} \label{2.19)} 
Tr\left({\sigma }^{\mathrm{2}}\right)\mathrm{<1}.                                                              
\end{equation} 
where $Tr\left({\rho }^{\mathrm{2}}\right)\mathrm{=}\sum^n_{i\mathrm{=1}}{{\lambda }^{\mathrm{2}}_i}$ and $Tr\left({\sigma }^{\mathrm{2}}\right)\mathrm{=}\sum^n_{i\mathrm{=1}}{{\omega }^{\mathrm{2}}_i}$, where ${\omega }_i$ are the eigenvalues of density matrix $\sigma $.

 Similarly, for a pure \textit{entangled} system ${\rho }_{EPR}$
\begin{equation} \label{ZEqnNum618270} 
Tr\left({\rho }^{\mathrm{2}}_{EPR}\right)\mathrm{=1},                                      
\end{equation} 
while for any mixed subsystem ${\sigma }_{EPR}$ of the entangled state (i.e., for a half-pair of the entangled state), we will have
\begin{equation} \label{ZEqnNum469343} 
Tr\left({\sigma }^{\mathrm{2}}_{EPR}\right)\mathrm{<1}.                                      
\end{equation} 
The density matrix also can be used to describe the effect of a unitary transform on the probability distribution of the system. The probability that the whole quantum system is in $\left|\left.{\psi }_i\right\rangle \right.$ can be calculated by the trace operation. If we apply unitary transform $U$ to the state $\rho \mathrm{=}\sum^n_{i\mathrm{=1}}{p_i\left|\left.{\psi }_i\right\rangle \right.\left\langle \left.{\psi }_i\right|\right.}$, the effect can be expressed as follows:
\begin{equation} \label{2.22)} 
\sum^n_{i\mathrm{=1}}{p_i\left(U\left|\left.{\psi }_i\right\rangle \right.\right)}\left(\left\langle \left.{\psi }_i\right|\right.U^{\mathrm{\dagger }}\right)\mathrm{=}U\left(\sum^n_{i\mathrm{=1}}{p_i\left|\left.{\psi }_i\right\rangle \right.\left\langle \left.{\psi }_i\right|\right.}\right)U^{\mathrm{\dagger }}\mathrm{=}U\rho U^{\mathrm{\dagger }}. 
\end{equation} 
If the applied transformation is not unitary, a more general operator denoted by $G$ is introduced, and with the help of this operator the transform can be written as
\begin{equation} \label{2.23)} 
G\left(\rho \right)\mathrm{=}\sum^n_{i\mathrm{=1}}{A_i}\rho A^{\mathrm{\dagger }}_i\mathrm{=}\sum^n_{i\mathrm{=1}}{A_i}\left(p_i\left|\left.{\psi }_i\right\rangle \right.\left\langle \left.{\psi }_i\right|\right.\right)A^{\mathrm{\dagger }}_i,                                                        
\end{equation} 
where $\sum^n_{i\mathrm{=1}}{A_i}A^{\mathrm{\dagger }}_i\mathrm{=}I,$ for every matrices $A_i$. In this sense, operator $G$ describes the physically admissible or \textit{Completely Positive Trace Preserving} (CPTP)\textit{ }operations. The application of a CPTP operator $G$ on density matrix $\rho $ will result in a matrix $G\left(\rho \right)$, which in this case is still a density matrix.

 Now we can summarize the two most important properties of density matrices:
\begin{enumerate}
\item \textit{The density matrix }$\rho $\textit{ is a positive-semidefinite matrix, see }\eqref{ZEqnNum374970}\textit{.}
\item \textit{The trace of any density matrix }$\rho $\textit{ is equal to 1, see }\eqref{ZEqnNum355630}\textit{.}
The properties of a quantum measurement are as follows.
\end{enumerate}

\subsubsection{Quantum Measurement}

 Now, let us turn to measurements and their relation to density matrices. Assuming a projective measurement device, defined by measurement operators - i.e., projectors $\{$$P_j$$\}$. The projector $P_j$ is a Hermitian matrix, for which $P_j\mathrm{=}P^{\mathrm{\dagger }}_j$ and $P^{\mathrm{2}}_j\mathrm{=}P_j$. According to the \textit{3${}^{rd}$ Postulate of Quantum Mechanic}s the trace operator can be used to give the probability of outcome \textit{j} belonging to the operator $P_j$ in the following way
\begin{equation} \label{2.24)} 
\mathrm{Pr}\left[\left.j\right|P_j\rho \right]\mathrm{=}Tr\left(P_j\rho P^{\mathrm{\dagger }}_j\right)\mathrm{=}Tr\left(P^{\mathrm{\dagger }}_jP_j\rho \right)\mathrm{=}Tr\left(P_j\rho \right).                                                 
\end{equation} 
After the measurement, the measurement operator $P_j$ leaves the system in a post measurement state
\begin{equation} \label{2.25)} 
{\rho }_j\mathrm{=}\frac{P_j\left[\sum^n_{i\mathrm{=1}}{p_i\left|\left.{\psi }_i\right\rangle \right.\left\langle \left.{\psi }_i\right|\right.}\right]P_j}{Tr\left(P_j\left[\sum^n_{i\mathrm{=1}}{p_i\left|\left.{\psi }_i\right\rangle \right.\left\langle \left.{\psi }_i\right|\right.}\right]P_j\right)}\mathrm{=}\frac{P_j\rho P_j}{Tr\left(P_j\rho P_j\right)}\mathrm{=}\frac{P_j\rho P_j}{Tr\left(P_j\rho \right)}.                                               
\end{equation} 
If we have a pure quantum state $\left|\left.\psi \right\rangle \right.\mathrm{=}\alpha \left|\left.0\right\rangle \right.\mathrm{+}\beta \left|\left.\mathrm{1}\right\rangle \right.$, where $\alpha \mathrm{=}\left\langle 0\mathrel{\left|\vphantom{0 \psi }\right.\kern-\nulldelimiterspace}\psi \right\rangle $ and $\beta \mathrm{=}\left\langle \mathrm{1}\mathrel{\left|\vphantom{\mathrm{1} \psi }\right.\kern-\nulldelimiterspace}\psi \right\rangle $. Using the trace operator, the measurement probabilities of $\left|\left.0\right\rangle \right.$ and $\left|\left.\mathrm{1}\right\rangle \right.$ can be expressed as
\begin{equation} \label{2.26)} 
 \begin{array}{l}
\mathrm{Pr}\left[\left.j\mathrm{=0}\right|\psi \right]\mathrm{=}Tr\left(P_j\rho \right)\mathrm{=}Tr\mathrm{(}\left|\left.0\right\rangle \right.\underbrace{\left\langle \left.0\right|\right.\left|\left.\psi \right\rangle \right.}_{\left\langle 0\mathrel{\left|\vphantom{0 \psi }\right.\kern-\nulldelimiterspace}\psi \right\rangle }\left\langle \left.\psi \right|\right.\mathrm{)} \\ 
\mathrm{=}\left\langle 0\mathrel{\left|\vphantom{0 \psi }\right.\kern-\nulldelimiterspace}\psi \right\rangle Tr\left(\left|\left.0\right\rangle \right.\left\langle \left.\psi \right|\right.\right)\mathrm{=}\left\langle 0\mathrel{\left|\vphantom{0 \psi }\right.\kern-\nulldelimiterspace}\psi \right\rangle \left\langle \psi \mathrel{\left|\vphantom{\psi  0}\right.\kern-\nulldelimiterspace}0\right\rangle  \\ 
\mathrm{=}\left\langle 0\mathrel{\left|\vphantom{0 \psi }\right.\kern-\nulldelimiterspace}\psi \right\rangle {\left(\left\langle 0\mathrel{\left|\vphantom{0 \psi }\right.\kern-\nulldelimiterspace}\psi \right\rangle \right)}^{\mathrm{*}}\mathrm{=}\alpha \mathrm{\cdot }{\alpha }^{\mathrm{*}}\mathrm{=}{\left|\alpha \right|}^{\mathrm{2}}, \end{array}
\end{equation} 
and
\begin{equation} \label{2.27)} 
 \begin{array}{l}
\mathrm{Pr}\left[\left.j\mathrm{=1}\right|\psi \right]\mathrm{=}Tr\left(P_j\rho \right)\mathrm{=}Tr\mathrm{(}\left|\left.\mathrm{1}\right\rangle \right.\underbrace{\left\langle \left.\mathrm{1}\right|\right.\left|\left.\psi \right\rangle \right.}_{\left\langle \mathrm{1}\mathrel{\left|\vphantom{\mathrm{1} \psi }\right.\kern-\nulldelimiterspace}\psi \right\rangle }\left\langle \left.\psi \right|\right.\mathrm{)} \\ 
\mathrm{=}\left\langle \mathrm{1}\mathrel{\left|\vphantom{\mathrm{1} \psi }\right.\kern-\nulldelimiterspace}\psi \right\rangle Tr\left(\left|\left.\mathrm{1}\right\rangle \right.\left\langle \left.\psi \right|\right.\right)\mathrm{=}\left\langle \mathrm{1}\mathrel{\left|\vphantom{\mathrm{1} \psi }\right.\kern-\nulldelimiterspace}\psi \right\rangle \left\langle \psi \mathrel{\left|\vphantom{\psi  \mathrm{1}}\right.\kern-\nulldelimiterspace}\mathrm{1}\right\rangle  \\ 
\mathrm{=}\left\langle \mathrm{1}\mathrel{\left|\vphantom{\mathrm{1} \psi }\right.\kern-\nulldelimiterspace}\psi \right\rangle {\left(\left\langle \mathrm{1}\mathrel{\left|\vphantom{\mathrm{1} \psi }\right.\kern-\nulldelimiterspace}\psi \right\rangle \right)}^{\mathrm{*}}\mathrm{=}\beta \mathrm{\cdot }{\beta }^{\mathrm{*}}\mathrm{=}{\left|\beta \right|}^{\mathrm{2}}, \end{array}
\end{equation} 
in accordance with our expectations. Let us assume we have an \textit{orthonormal} basis $M\mathrm{=}\left\{\left|\left.x_{\mathrm{1}}\right\rangle \right.\left\langle \left.x_{\mathrm{1}}\right|\right.\mathrm{,\dots ,}\left|\left.x_n\right\rangle \right.\left\langle \left.x_n\right|\right.\right\}$ and an arbitrary (i.e., non-diagonal) density matrix $\rho $. The set of Hermitian operators $P_i\mathrm{=}\left\{\left|\left.x_i\right\rangle \right.\left\langle \left.x_i\right|\right.\right\}$ satisfies the \textit{completeness relation}, where $P_i\mathrm{=}\left|\left.x_i\right\rangle \right.\left\langle \left.x_i\right|\right.$ is the projector over $\left|\left.x_i\right\rangle \right.$, i.e., quantum measurement operator $M_i\mathrm{=}\left|\left.x_i\right\rangle \right.\left\langle \left.x_i\right|\right.$ is a valid measurement operator. The measurement operator $M_i$ projects the input quantum system $\left|\left.\psi \right\rangle \right.$ to the pure state $\left|\left.x_i\right\rangle \right.$ from the orthonormal basis $M\mathrm{=}\left\{\left|\left.x_{\mathrm{1}}\right\rangle \right.\left\langle \left.x_{\mathrm{1}}\right|\right.\mathrm{,\dots ,}\left|\left.x_n\right\rangle \right.\left\langle \left.x_n\right|\right.\right\}$. Now, the probability that the quantum state $\left|\left.\psi \right\rangle \right.$ is after the measurement in basis state $\left|\left.x_i\right\rangle \right.$ can be expressed as
\begin{equation} \label{2.28)} 
\begin{split}
&\left\langle \left.\psi \right|M^{\mathrm{\dagger }}_iM_i\mathrel{\left|\vphantom{\left.\psi \right|M^{\mathrm{\dagger }}_iM_i \psi }\right.\kern-\nulldelimiterspace}\psi \right\rangle \mathrm{=}\left\langle \left.\psi \right|P_i\mathrel{\left|\vphantom{\left.\psi \right|P_i \psi }\right.\kern-\nulldelimiterspace}\psi \right\rangle \\&\mathrm{=}\left(\sum^n_{j\mathrm{=}\mathrm{1}}{x^{\mathrm{*}}_j\left\langle \left.x_j\right|\right.}\right)\left|\left.x_i\right\rangle \right.\left\langle \left.x_i\right|\right.\left(\sum^n_{l\mathrm{=1}}{\left|\left.x_l\right\rangle \right.x_l}\right)\mathrm{=}{\left|x_i\right|}^{\mathrm{2}}.                           
\end{split}
\end{equation} 
In the computational basis  $\left\{\left|\left.x_{\mathrm{1}}\right\rangle \right.\mathrm{,\dots ,}\left|\left.x_n\right\rangle \right.\right\}$, the state of the quantum system after the measurement can be expressed as
\begin{equation} \label{2.29)} 
\rho \mathrm{'}\mathrm{=}\sum^n_{i\mathrm{=1}}{p_i\left|\left.x_i\right\rangle \right.\left\langle \left.x_i\right|\right.},                                                         
\end{equation} 
and the matrix of the quantum state $\rho \mathrm{'}$ will be \textit{diagonal} in the computational basis $\left\{\left|\left.x_i\right\rangle \right.\right\}$, and can be given by
\begin{equation} \label{2.30)} 
\rho \mathrm{'}\mathrm{=}\left[ \begin{array}{llll}
p_{\mathrm{1}} & 0 & \mathrm{\dots } & 0 \\ 
0 & p_{\mathrm{2}} & 0 & \mathrm{\vdots } \\ 
\mathrm{\vdots } & \mathrm{\vdots } & \mathrm{\ddots } & 0 \\ 
0 & 0 & 0 & p_n \end{array}
\right].                                                                     
\end{equation} 
To illustrate it, let assume we have an initial (not diagonal) density matrix in the computational basis $\left\{\left|\left.0\right\rangle \right.,\left|\left.\mathrm{1}\right\rangle \right.\right\}$ e.g. $\left|\left.\psi \right\rangle \right.\mathrm{=}\alpha \left|\left.0\right\rangle \right.\mathrm{+}\beta \left|\left.\mathrm{1}\right\rangle \right.$ with $p\mathrm{=}{\left|\alpha \right|}^{\mathrm{2}}$ and  $\mathrm{1-}p\mathrm{=}{\left|\beta \right|}^{\mathrm{2}}$ as
\begin{equation} \label{ZEqnNum308462} 
\rho \mathrm{=}\left|\left.\psi \right\rangle \right.\left\langle \left.\psi \right|\right.\mathrm{=}\left[ \begin{array}{ll}
{\left|\alpha \right|}^{\mathrm{2}} & \alpha {\beta }^{\mathrm{*}} \\ 
{\alpha }^{\mathrm{*}}\beta  & {\left|\beta \right|}^{\mathrm{2}} \end{array}
\right], 
\end{equation} 
and we have orthonormal basis $M\mathrm{=}\left\{\left|\left.0\right\rangle \right.\left\langle \left.0\right|\right.,\left|\left.\mathrm{1}\right\rangle \right.\left\langle \left.\mathrm{1}\right|\right.\right\}$. In this case, the after-measurement state can be expressed as
\begin{equation} \label{ZEqnNum990791} 
\rho \mathrm{'}\mathrm{=}p\left|\left.0\right\rangle \right.\left\langle \left.0\right|\right.\mathrm{+}\left(\mathrm{1-}p\right)\left|\left.\mathrm{1}\right\rangle \right.\left\langle \left.\mathrm{1}\right|\right.\mathrm{=}\left[ \begin{array}{ll}
{\left|\alpha \right|}^{\mathrm{2}} & 0 \\ 
0 & {\left|\beta \right|}^{\mathrm{2}} \end{array}
\right]\mathrm{=}\left[ \begin{array}{ll}
p & 0 \\ 
0 & \mathrm{1-}p \end{array}
\right]. 
\end{equation} 
As it can be seen, the matrix of $\rho \mathrm{'}$ is a diagonal matrix in the computational basis $\left\{\left|\left.0\right\rangle \right.,\left|\left.\mathrm{1}\right\rangle \right.\right\}$. Eq. \eqref{ZEqnNum308462} and \eqref{ZEqnNum990791} highlights the difference between quantum superpositions (probability amplitude weighted sum) and classical probabilistic mixtures of quantum states. 

 Now, let us see the result of the measurement on the input quantum system $\rho $
\begin{equation} \label{ZEqnNum388676} 
M\left(\rho \right)\mathrm{=}\sum^{\mathrm{1}}_{j\mathrm{=0}}{M_j}\rho M^{\mathrm{\dagger }}_j\mathrm{=}M_0\rho M^{\mathrm{\dagger }}_0\mathrm{+}M_{\mathrm{1}}\rho M^{\mathrm{\dagger }}_{\mathrm{1}}. 
\end{equation} 
For the measurement operators $M_0\mathrm{=}\left|\left.0\right\rangle \right.\left\langle \left.0\right|\right.$ and $M_{\mathrm{1}}\mathrm{=}\left|\left.\mathrm{1}\right\rangle \right.\left\langle \left.\mathrm{1}\right|\right.$ the completeness relation holds
\begin{equation} \label{2.34)} 
\begin{split}
\begin{array}{l}
\sum^{\mathrm{1}}_{j\mathrm{=0}}{M_j}M^{\mathrm{\dagger }}_j \mathrm{=}\left|\left.0\right\rangle \right.\left\langle \left.0\right|\right.\left|\left.0\right\rangle \right.\left\langle \left.0\right|\right.\mathrm{+}\left|\left.\mathrm{1}\right\rangle \right.\left\langle \left.\mathrm{1}\right|\right.\left|\left.\mathrm{1}\right\rangle \right.\left\langle \left.\mathrm{1}\right|\right.\\\mathrm{=}\left|\left.0\right\rangle \right.\left\langle \left.0\right|\right.\mathrm{+}\left|\left.\mathrm{1}\right\rangle \right.\left\langle \left.\mathrm{1}\right|\right.\mathrm{=}\left[ \begin{array}{ll}
\mathrm{1} & 0 \\ 
0 & \mathrm{1} \end{array}
\right]\mathrm{=}I. \end{array}
\end{split}
\end{equation} 
Using input system $\rho \mathrm{=}\left|\left.\psi \right\rangle \right.\left\langle \left.\psi \right|\right.$, where $\left|\left.\psi \right\rangle \right.\mathrm{=}\alpha \left|\left.0\right\rangle \right.\mathrm{+}\beta \left|\left.\mathrm{1}\right\rangle \right.$, the state after the measurement operation is
\begin{equation} \label{2.35)} 
 \begin{array}{l}
M\left(\rho \right)\mathrm{=}\sum^{\mathrm{1}}_{j\mathrm{=0}}{M_j}\rho M^{\mathrm{\dagger }}_j \\ 
\mathrm{=}\left|\left.0\right\rangle \right.\left\langle \left.0\right|\right.\rho \left|\left.0\right\rangle \right.\left\langle \left.0\right|\right.\mathrm{+}\left|\left.\mathrm{1}\right\rangle \right.\left\langle \left.\mathrm{1}\right|\right.\rho \left|\left.\mathrm{1}\right\rangle \right.\left\langle \left.\mathrm{1}\right|\right. \\ 
\mathrm{=}\left|\left.0\right\rangle \right.\left\langle \left.0\right|\right.\left|\left.\psi \right\rangle \right.\left\langle \left.\psi \right|\right.\left|\left.0\right\rangle \right.\left\langle \left.0\right|\right.\mathrm{+}\left|\left.\mathrm{1}\right\rangle \right.\left\langle \left.\mathrm{1}\right|\right.\left|\left.\psi \right\rangle \right.\left\langle \left.\psi \right|\right.\left|\left.\mathrm{1}\right\rangle \right.\left\langle \left.\mathrm{1}\right|\right. \\ 
\mathrm{=}\left|\left.0\right\rangle \right.\left\langle 0\mathrel{\left|\vphantom{0 \psi }\right.\kern-\nulldelimiterspace}\psi \right\rangle \left\langle 0\mathrel{\left|\vphantom{0 \psi }\right.\kern-\nulldelimiterspace}\psi \right\rangle \left\langle \left.0\right|\right.\mathrm{+}\left|\left.\mathrm{1}\right\rangle \right.\left\langle \mathrm{1}\mathrel{\left|\vphantom{\mathrm{1} \psi }\right.\kern-\nulldelimiterspace}\psi \right\rangle \left\langle \mathrm{1}\mathrel{\left|\vphantom{\mathrm{1} \psi }\right.\kern-\nulldelimiterspace}\psi \right\rangle \left\langle \left.\mathrm{1}\right|\right. \\ 
\mathrm{=}{\left|\left\langle 0\mathrel{\left|\vphantom{0 \psi }\right.\kern-\nulldelimiterspace}\psi \right\rangle \right|}^{\mathrm{2}}\left|\left.0\right\rangle \right.\left\langle \left.0\right|\right.\mathrm{+}{\left|\left\langle \mathrm{1}\mathrel{\left|\vphantom{\mathrm{1} \psi }\right.\kern-\nulldelimiterspace}\psi \right\rangle \right|}^{\mathrm{2}}\left|\left.\mathrm{1}\right\rangle \right.\left\langle \left.\mathrm{1}\right|\right. \\ 
\mathrm{=}{\left|\alpha \right|}^{\mathrm{2}}\left|\left.0\right\rangle \right.\left\langle \left.0\right|\right.\mathrm{+}{\left|\beta \right|}^{\mathrm{2}}\left|\left.\mathrm{1}\right\rangle \right.\left\langle \left.\mathrm{1}\right|\right.\mathrm{=}p\left|\left.0\right\rangle \right.\left\langle \left.0\right|\right.\mathrm{+1-}p\left|\left.\mathrm{1}\right\rangle \right.\left\langle \left.\mathrm{1}\right|\right.. \end{array}
\end{equation} 
As we have found, after the measurement operation $M\left(\rho \right)$, the \textit{off-diagonal} entries will have zero values, and they \textit{have no relevance}. As follows, the initial input system $\rho \mathrm{=}\left|\left.\psi \right\rangle \right.\left\langle \left.\psi \right|\right.$ after operation \textit{M} becomes
\begin{equation} \label{ZEqnNum801705} 
\rho \mathrm{=}\left[ \begin{array}{ll}
{\left|\alpha \right|}^{\mathrm{2}} & \alpha {\beta }^{\mathrm{*}} \\ 
{\alpha }^{\mathrm{*}}\beta  & {\left|\beta \right|}^{\mathrm{2}} \end{array}
\right]{{\stackrel{M}{\rightarrow}}}\rho \mathrm{'}\mathrm{=}\left[ \begin{array}{ll}
{\left|\alpha \right|}^{\mathrm{2}} & 0 \\ 
0 & {\left|\beta \right|}^{\mathrm{2}} \end{array}
\right]. 
\end{equation}

\paragraph{Orthonormal Basis Decomposition}

 Let assume we have orthonormal basis $\left\{\left|\left.b_{\mathrm{1}}\right\rangle \right.,\left|\left.b_{\mathrm{2}}\right\rangle \right.\mathrm{,\dots ,}\left|\left.b_n\right\rangle \right.\right\}$, which basis can be used to rewrite the quantum system $\left|\left.\psi \right\rangle \right.$ in a unique decomposition
\begin{equation} \label{ZEqnNum973996} 
\left|\left.\psi \right\rangle \right.\mathrm{=}b_{\mathrm{1}}\left|\left.b_{\mathrm{1}}\right\rangle \right.\mathrm{+}b_{\mathrm{2}}\left|\left.b_{\mathrm{2}}\right\rangle \right.\mathrm{+\dots +}b_n\left|\left.b_n\right\rangle \right.\mathrm{=}\sum^n_{i\mathrm{=1}}{b_i\left|\left.b_i\right\rangle \right.},                                                   
\end{equation} 
with complex $b_i$. Since $\left\langle \psi \mathrel{\left|\vphantom{\psi  \psi }\right.\kern-\nulldelimiterspace}\psi \right\rangle \mathrm{=1}$, we can express it in the form
\begin{equation} \label{ZEqnNum154263} 
\left\langle \psi \mathrel{\left|\vphantom{\psi  \psi }\right.\kern-\nulldelimiterspace}\psi \right\rangle \mathrm{=}\sum^n_{i\mathrm{=1}}{\sum^n_{j\mathrm{=1}}{b^{\mathrm{*}}_ib_j\left\langle b_i\mathrel{\left|\vphantom{b_i b_j}\right.\kern-\nulldelimiterspace}b_j\right\rangle }}\mathrm{=}\sum^n_{i\mathrm{=1}}{{\left|b_i\right|}^{\mathrm{2}}\mathrm{=}}\mathrm{1},                                                           
\end{equation} 
where $b^{\mathrm{*}}_i$ is the complex conjugate of \textit{probability amplitude} $b_i$, thus ${\left|b_i\right|}^{\mathrm{2}}$ is the \textit{probability} $p_i$ of measuring the quantum system $\left|\left.\psi \right\rangle \right.$ in the given basis state $\left|\left.b_i\right\rangle \right.$, i.e.,
\begin{equation} \label{2.39)} 
p_i\mathrm{=}{\left|b_i\right|}^{\mathrm{2}}.                                                  
\end{equation} 
Using \eqref{ZEqnNum711440}, \eqref{ZEqnNum973996} and \eqref{ZEqnNum154263} the density matrix of quantum system $\left|\left.\psi \right\rangle \right.$ can be expressed as
\begin{equation} \label{ZEqnNum456813} 
 \begin{array}{l}
\rho \mathrm{=}{\left|b_{\mathrm{1}}\right|}^{\mathrm{2}}\left|\left.b_{\mathrm{1}}\right\rangle \right.\left\langle \left.b_{\mathrm{1}}\right|\right.\mathrm{+}{\left|b_{\mathrm{2}}\right|}^{\mathrm{2}}\left|\left.b_{\mathrm{2}}\right\rangle \right.\left\langle \left.b_{\mathrm{2}}\right|\right.\mathrm{+\dots }\mathrm{+}{\left|b_n\right|}^{\mathrm{2}}\left|\left.b_n\right\rangle \right.\left\langle \left.b_n\right|\right. \\ 
\mathrm{=}\sum^n_{i\mathrm{=1}}{{\left|b_i\right|}^{\mathrm{2}}\left|\left.b_i\right\rangle \right.\left\langle \left.b_i\right|\right.}\mathrm{=}\sum^n_{i\mathrm{=1}}{p_i\left|\left.b_i\right\rangle \right.\left\langle \left.b_i\right|\right..} \end{array}
\end{equation} 
This density matrix is a diagonal matrix with the probabilities in the diagonal entries
\begin{equation} \label{ZEqnNum335306} 
\rho \mathrm{=}\left[ \begin{array}{llll}
p_{\mathrm{1}} & \mathrm{\cdots } & 0 & 0 \\ 
0 & p_{\mathrm{2}} & 0 & \mathrm{\vdots } \\ 
\mathrm{\vdots } & 0 & \mathrm{\ddots } & 0 \\ 
0 & \mathrm{\cdots } & 0 & p_n \end{array}
\right].                                                                          
\end{equation} 
The diagonal property of density matrix \eqref{ZEqnNum456813} in \eqref{ZEqnNum335306} can be checked, since the elements of the matrix can be expressed as
\begin{equation} \label{2.42)} 
\begin{split}
{{\rho }_{ij}}&=\left\langle  {{b}_{i}} | \left. \rho  \right|{{b}_{j}} \right\rangle \\
&=\left\langle  {{b}_{i}} \right|\left( \sum\limits_{l=1}^{n}{{{p}_{i}}\left| {{b}_{i}} \right\rangle \left\langle  {{b}_{i}} \right|} \right)\left| {{b}_{j}} \right\rangle =\sum\limits_{l=1}^{n}{{{p}_{l}}\left\langle  {{b}_{i}} | {{b}_{l}} \right\rangle }\left\langle  {{b}_{l}} | {{b}_{j}} \right\rangle ,                                        
\end{split}
\end{equation} 
where $\sum^n_{l\mathrm{=1}}{p_i\mathrm{=1}}$.

\paragraph{The Projective and POVM Measurement}

 The \textit{projective measurement} is also known as the \textit{von Neumann measurement} is formally can be described by the Hermitian operator $\mathcal{Z}$, which has the spectral decomposition
\begin{equation} \label{ZEqnNum183918} 
\mathcal{Z}\mathrm{=}\sum_m{{\lambda }_mP_m}.                                                                                      
\end{equation} 
where $P_m$ is a projector to the eigenspace of $\mathcal{Z}$ with eigenvalue ${\lambda }_m$. For the projectors
\begin{equation} \label{2.44)} 
\sum_m{P_m\mathrm{=}I},                                                  
\end{equation} 
and they are pairwise orthogonal. The measurement outcome \textit{m} corresponds to the eigenvalue ${\lambda }_m$, with measurement probability
\begin{equation} \label{2.45)} 
\Pr \left[ m\left| \psi  \right\rangle  \right]=\left\langle  \psi  | \left. {{P}_{m}} \right|\psi  \right\rangle .
\end{equation} 
When a quantum system is measured in an orthonormal basis $\left|\left.m\right\rangle \right.$, then we make a projective measurement with projector $P_m\mathrm{=}\left|\left.m\right\rangle \right.\left\langle \left.m\right|\right.$, thus \eqref{ZEqnNum183918} can be rewritten as
\begin{equation} \label{2.46)} 
\mathcal{Z}\mathrm{=}\sum_m{mP_m}.                                                                          
\end{equation} 
The $\mathcal{P}$ \textit{POVM }(Positive Operator Valued Measurement)\textit{ }is intended to select among the non-orthogonal states ${\left\{\left|\left.{\psi }_i\right\rangle \right.\right\}}^m_{i\mathrm{=1}}$ and defined by a \textit{set} of POVM operators ${\left\{{\mathcal{M}}_i\right\}}^{m\mathrm{+1}}_{i\mathrm{=1}}$, where
\begin{equation} \label{2.47)} 
{\mathcal{M}}_i\mathrm{=}{\mathcal{Q}}^{\mathrm{\dagger }}_i{\mathcal{Q}}_i,                                                  
\end{equation} 
and since we are not interested in the post-measurement state the exact knowledge about measurement operator ${\mathcal{Q}}_i$ is not required. For POVM operators ${\mathcal{M}}_i$ the completeness relation holds,
\begin{equation} \label{2.48)} 
\sum_i{{\mathcal{M}}_i}\mathrm{=}I.                                                  
\end{equation} 
For the POVM the probability of a given outcome \textit{n} for the state $\left|\left.\psi \right\rangle \right.$ can be expressed as
\begin{equation} \label{2.49)} 
\Pr \left[ i\left| \psi  \right\rangle  \right]=\left\langle  \psi  | \left. {{\mathcal{M}}_{i}} \right|\psi  \right\rangle .
\end{equation} 
The POVM also can be imagined as a `black-box', which outputs a number from 1 to \textit{m} for the given input quantum state $\psi $, using the set of operators
\begin{equation} \label{2.50)} 
\left\{{\mathcal{M}}_{\mathrm{1}}\mathrm{,\dots ,}{\mathcal{M}}_m,{\mathcal{M}}_{m\mathrm{+1}}\right\},                                                                     
\end{equation} 
where $\left\{{\mathcal{M}}_{\mathrm{1}}\mathrm{,\dots ,}{\mathcal{M}}_m\right\}$ are responsible to distinguish \textit{m} different typically non-orthogonal states i.e., if we observe $i\mathrm{\in }\left[\mathrm{1,}m\right]$ on the display of the measurement device we can be sure, that the result is correct. However, because unknown non-orthogonal states can not be distinguished with probability 1, we have to introduce an extra measurement operator, ${\mathcal{M}}_{m\mathrm{+1}}$, as the price of the distinguishability of the \textit{m} different states and if we obtain \textit{m+}1 as measurement results we can say nothing about $\left|\left.\psi \right\rangle \right.$. This operator can be expressed as
\begin{equation} \label{2.51)} 
{\mathcal{M}}_{m\mathrm{+1}}\mathrm{=}I\mathrm{-}\sum^m_{i\mathrm{=1}}{{\mathcal{M}}_i}.                                                                  
\end{equation} 
Such ${\mathcal{M}}_{m\mathrm{+1}}$ can be always constructed if the states in ${\left\{\left|\left.{\psi }_n\right\rangle \right.\right\}}^m_{n\mathrm{=1}}$ are linearly independent. We note, we will omit listing operator ${\mathcal{M}}_{m\mathrm{+1}}$ in further parts of the paper. The POVM measurement apparatus will be a key ingredient to distinguish quantum codewords with zero-error, and to reach the zero-error capacity of quantum channels.

The POVM can be viewed as the most general formula from among of any possible measurements in quantum mechanics. Therefore the effect of a projective measurement can be described by POVM operators, too. Or with other words, the projective measurements are the special case POVM measurement [\cref{Imre05}]. The elements of the POVM are not necessarily orthogonal, and the number of the elements can be larger than the dimension of the Hilbert space which they are originally used in.

\subsection{Geometrical Interpretation of the Density Matrices}

 While the \textit{wavefunction} representation is the full physical description of a quantum system in the space-time, the tensor product of multiple copies of two dimensional Hilbert spaces is its discrete version, with discrete finite-dimensional Hilbert spaces. The geometrical representation also can be extended to analyze the geometrical structure of the transmission of information though a quantum channel, and it also provides a very useful tool to analyze the capacities of different quantum channel models.

As it has been mentioned, the Bloch sphere is a geometrical conception, constructed to represent two-level quantum systems in a more expressive way than is possible with algebraic tools. The Bloch sphere has unit radius and is defined in a three-dimensional real vector space. The pure states are on the surface of the Bloch sphere, while the mixed states are in the interior of the original sphere. In the Bloch sphere representation, the state of a single qubit $\left|\left.\psi \right\rangle \right.\mathrm{=}\alpha \left|\left.0\right\rangle \right.\mathrm{+}\beta \left|\left.\mathrm{1}\right\rangle \right.$ can be expressed as
\begin{equation} \label{2.52)} 
\left|\left.\psi \right\rangle \right.\mathrm{=}e^{i\delta }\left(\mathrm{cos}\frac{\theta }{\mathrm{2}}\left|\left.0\right\rangle \right.\mathrm{+}e^{i\varphi }\mathrm{sin}\frac{\theta }{\mathrm{2}}\left|\left.\mathrm{1}\right\rangle \right.\right),                                                           
\end{equation} 
where $\delta $ is the global phase factor, which can be ignored from the computations, hence the state $\left|\left.\psi \right\rangle \right.$ in the terms of the angle $\theta $ and $\varphi $ can be expressed as
\begin{equation} \label{2.53)} 
\left|\left.\psi \right\rangle \right.\mathrm{=cos}\frac{\theta }{\mathrm{2}}\left|\left.0\right\rangle \right.\mathrm{+}e^{i\varphi }\mathrm{sin}\frac{\theta }{\mathrm{2}}\left|\left.\mathrm{1}\right\rangle \right.. 
\end{equation} 
The Bloch sphere is a very useful tool, since it makes possible to describe various, physically realized one-qubit quantum systems, such as the photon polarization, spins or the energy levels of an atom. Moreover, if we would like to compute the various channel capacities of the quantum channel, the geometrical expression of the channel capacity also can be represented by the Bloch sphere. Before we would introduce the geometrical calculation of the channel capacities, we have to start from the geometrical interpretation of density matrices. The density matrix $\rho $ can then be expressed using the Pauli matrices (a set of three complex matrices which are Hermitian and unitary) ${\sigma }_X\mathrm{=}\left[ \begin{array}{ll}
0 & \mathrm{1} \\ 
\mathrm{1} & 0 \end{array}
\right]$, ${\sigma }_Y\mathrm{=}\left[ \begin{array}{ll}
0 & \mathrm{-}i \\ 
i & 0 \end{array}
\right]$ and ${\sigma }_Z\mathrm{=}\left[ \begin{array}{ll}
\mathrm{1} & 0 \\ 
0 & \mathrm{-}\mathrm{1} \end{array}
\right]$ as
\begin{equation} \label{2.54)} 
\rho \mathrm{=}\frac{\mathrm{1+}r_X{\sigma }_X\mathrm{+}r_Y{\sigma }_Y\mathrm{+}r_Z{\sigma }_Z}{\mathrm{2}},                                                                         
\end{equation} 
where  $\boldsymbol{\mathrm{r}}\mathrm{=}\left(r_X,r_Y,r_Z\right)\mathrm{=}\left(\mathrm{sin}\theta \mathrm{cos}\phi \mathrm{,sin}\theta \mathrm{sin}\phi \mathrm{,cos}\theta \right)$ is the Bloch vector, $\left\|\left(r_X,r_Z,r_Y\right)\right\|\mathrm{\le }\mathrm{1}$, and $\sigma \mathrm{=}{\left({\sigma }_X,{\sigma }_Y,{\sigma }_Z\right)}^T$. In the vector representation, the previously shown formula can be expressed as
\begin{equation} \label{2.55)} 
\rho \mathrm{=}\frac{\mathrm{1+}\boldsymbol{\mathrm{r}}\sigma }{\mathrm{2}}.                                                              
\end{equation} 
In conclusion, every state can be expressed as linear combinations of the Pauli matrices and according to these Pauli matrices every state can be interpreted as a point in the three-dimensional real vector space. If we apply a unitary transformation $U$ to the density matrix $\rho $, then it can be expressed as
\begin{equation} \label{2.56)} 
\rho \mathrm{\to }\rho \mathrm{'}\mathrm{=}U\rho U^{\mathrm{\dagger }}\mathrm{=}\frac{\mathrm{1+}U\boldsymbol{\mathrm{r}}\sigma U^{\mathrm{\dagger }}}{\mathrm{2}}\mathrm{=}\frac{\mathrm{1+}U\boldsymbol{\mathrm{r}}U^{\mathrm{\dagger }}\sigma }{\mathrm{2}}, 
\end{equation} 
and $\boldsymbol{\mathrm{r}}\mathrm{'}\mathrm{=}U\boldsymbol{\mathrm{r}}U^{\mathrm{\dagger }}$ realizes a unitary transformation on $\boldsymbol{\mathrm{r}}$ as a rotation.

A density matrix $\rho $ can be expressed in a `weighted form' of density matrices ${\rho }_{\mathrm{1}}$ and ${\rho }_{\mathrm{2}}$ as follows:
\begin{equation} \label{2.57)} 
\rho \mathrm{=}\gamma {\rho }_{\mathrm{1}}\mathrm{+}\left(\mathrm{1-}\gamma \right){\rho }_{\mathrm{2}},                                                                
\end{equation} 
where $\mathrm{0}\mathrm{\le }\gamma \mathrm{\le }\mathrm{1}$, and ${\rho }_{\mathrm{1}}$ and ${\rho }_{\mathrm{2}}$ are pure states, and lie on a line segment connecting the density matrices in the Bloch sphere representation. Using probabilistic mixtures of the pure density matrices, any quantum state which lies between the two states can be expressed as a convex combination
\begin{equation} \label{2.58)} 
\rho \mathrm{=}p{\rho }_{\mathrm{1}}\mathrm{+}\left(\mathrm{1-}p\right){\rho }_{\mathrm{2}}, \mathrm{0}\mathrm{\le }p\mathrm{\le }\mathrm{1}.                                                                 
\end{equation} 
This remains true for an arbitrary number of quantum states, hence this result can be expressed for arbitrary number of density matrices. Mixed quantum states can be represented as \textit{statistical mixtures} of pure quantum states. The statistical representation of a pure state is unique. On the other hand we note that the decomposition of a mixed quantum state is not unique.
In the geometrical interpretation a pure state $\rho $ is on the surface of the Bloch sphere, while the mixed state $\sigma $ is inside. A maximally mixed quantum state, $\sigma \mathrm{=}\frac{\mathrm{1}}{\mathrm{2}}I$, can be found in the center of the Bloch sphere. The mixed state can be expressed as probabilistic mixture of pure states $\left\{{\rho }_{\mathrm{1}},{\rho }_{\mathrm{2}}\right\}$ and $\left\{{\rho }_{\mathrm{3}},{\rho }_{\mathrm{4}}\right\}$. As it has been stated by von Neumann, the \textit{decomposition of a mixed state is not unique}, since it can be expressed as a mixture of $\left\{{\rho }_{\mathrm{1}},{\rho }_{\mathrm{2}}\right\}$ or equivalently of $\left\{{\rho }_{\mathrm{3}},{\rho }_{\mathrm{4}}\right\}$.

 One can use a pure state $\rho $ to recover mixed state $\sigma $ from it, after the effects of environment ($E$) are traced out. With the help of the partial trace operator, Bob, the receiver, can decouple the environment from his mixed state, and the original state can be recovered by discarding the effects of the environment. If Bob's state is a \textit{probabilistic mixture} $\sigma \mathrm{=}\sum_i{p_i}\left|\left.{\varphi }_i\right\rangle \right.\left\langle \left.{\varphi }_i\right|\right.$, then a global pure \textit{purification} state $\left|\left.\mathrm{\Psi }\right\rangle \right.$ exists, which from Bob's state can be expressed as
\begin{equation} \label{2.59)} 
\sigma \mathrm{=}Tr_{E}\left|\left.\mathrm{\Psi }\right\rangle \right.\left\langle \left.\mathrm{\Psi }\right|\right..                                                                                  
\end{equation} 
Note, density matrix $\sigma $ can be recovered from $\left|\left.\mathrm{\Psi }\right\rangle \right.$ after discarding the environment. The decoupling of the environment can be achieved with the $Tr_{E}$ operator. For any unitary transformation of the environment, the pure state $\left|\left.\mathrm{\Psi }\right\rangle \right.$ is a unique state.

 We have seen, that the decomposition of mixed quantum states into pure quantum states is not unique, hence for example, it can be easily verified by the reader, that the decomposition of a mixed state $\sigma \mathrm{=}\frac{\mathrm{1}}{\mathrm{2}}\left(\left|\left.0\right\rangle \right.\left\langle \left.0\right|\right.\mathrm{+}\left|\left.\mathrm{1}\right\rangle \right.\left\langle \left.\mathrm{1}\right|\right.\right)$ can be made with pure states $\left\{\left|\left.0\right\rangle \right.,\left|\left.\mathrm{1}\right\rangle \right.\right\}$, and also can be given with pure states $\left\{\frac{\mathrm{1}}{\sqrt{\mathrm{2}}}\left(\left|\left.0\right\rangle \right.\mathrm{+}\left|\left.\mathrm{1}\right\rangle \right.\right),\frac{\mathrm{1}}{\sqrt{\mathrm{2}}}\left(\left|\left.0\right\rangle \right.\mathrm{-}\left|\left.\mathrm{1}\right\rangle \right.\right)\right\}$. Here, we have just changed the basis from rectilinear to diagonal, and we have used just pure states - and it resulted in the same mixed quantum state.

\subsection{Channel System Description}

\begin{center}
\begin{figure*}[htbp]
\begin{center}
\includegraphics[angle = 0,width=0.7\linewidth]{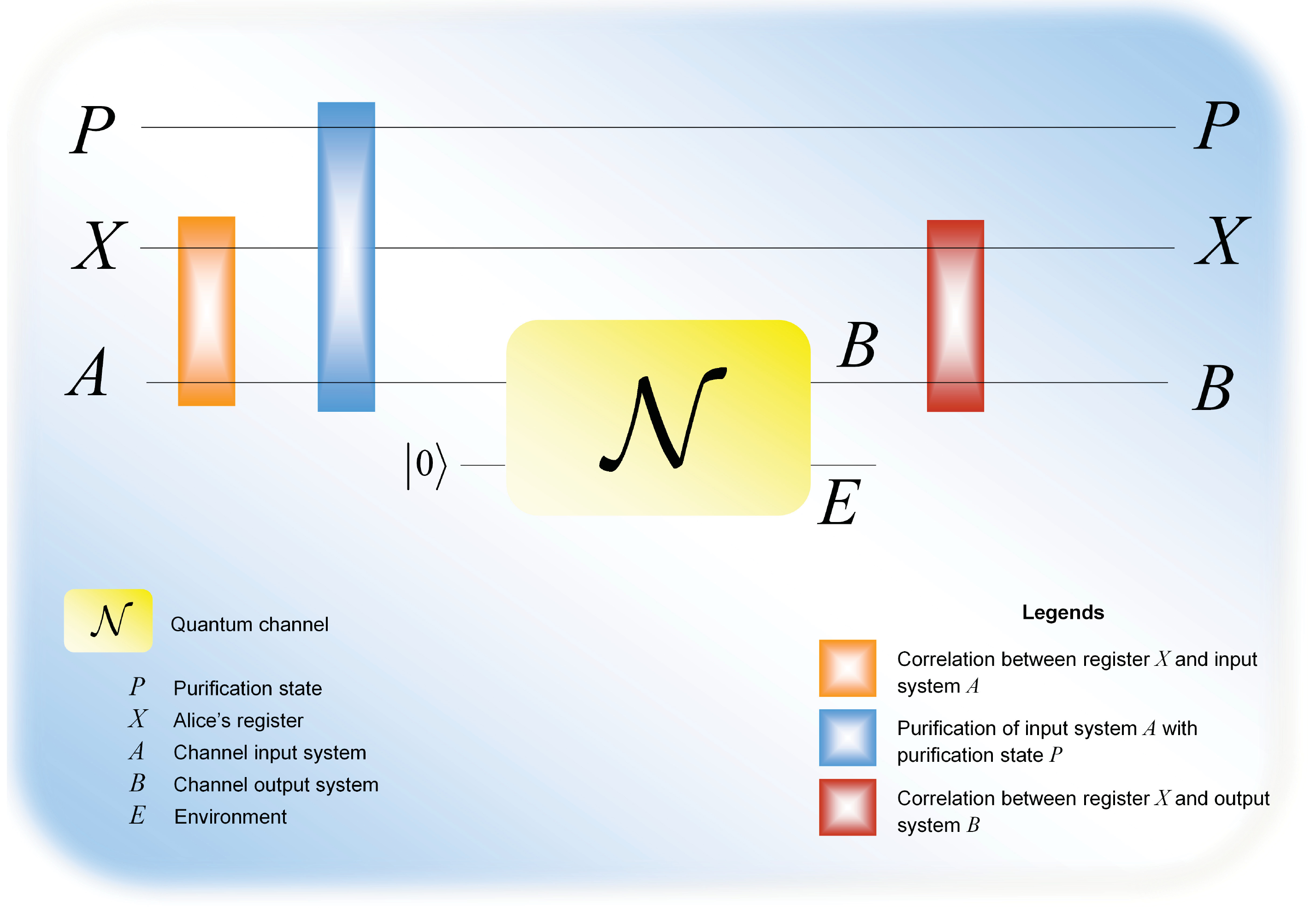}
\caption{Detailed model of a quantum communication channel exposing the interaction with the environment. Alice's register is denoted by \textit{X}, the input system is \textit{A} while \textit{P} is the purification state. The environment of the channel is denoted by \textit{E}, the output of the channel is \textit{B}. The quantum channel has positive classical capacity if and only if the channel output system \textit{B} will be correlated with Alice's register \textit{X}.} 
 \label{fig2_5}
 \end{center}
\end{figure*}
\end{center}

 If we are interested in the origin of noise (randomness) in the quantum channel the model should be refined in the following way: Alice's register \textit{X}, the purification state \textit{P}, channel input \textit{A}, channel output \textit{B}, and the environment state \textit{E}. The input system \textit{A} is described by a quantum system ${\rho }_x$, which occurs on the input with probability $p_X\left(x\right)$. They together form an ensemble denoted by ${\left\{p_X\left(x\right),{\rho }_x\right\}}_{x\mathrm{\in }X}$, where \textit{x} is a classical variable from the register \textit{X}. In the preparation process, Alice generates pure states ${\rho }_x$ according to random variable \textit{x}, i.e., the input density operator can be expressed as ${\rho }_x\mathrm{=}\left|\left.x\right\rangle \right.\left\langle \left.x\right|\right.$, where the classical states ${\left\{\left|\left.x\right\rangle \right.\right\}}_{x\mathrm{\in }X}$ form an orthonormal basis. According to the elements of Alice's register \textit{X}, the input system can be characterized by the quantum system 
\begin{equation} \label{2.60)} 
{\rho }_A\mathrm{=}\sum_{x\mathrm{\in }X}{p_X\left(x\right){\rho }_x\mathrm{=}}\sum_{x\mathrm{\in }X}{p_X\left(x\right)\left|\left.x\right\rangle \right.\left\langle \left.x\right|\right..} 
\end{equation} 
The system description is illustrated in \fref{fig2_5}. 

 The system state ${\rho }_x$ with the corresponding probability distribution $p_X\left(x\right)$ can be indentified by a set of measurement operators $M\mathrm{=}{\left\{\left|\left.x\right\rangle \right.\left\langle \left.x\right|\right.\right\}}_{x\mathrm{\in }X}$. If the density operators ${\rho }_x$ in ${\rho }_A$ are mixed, the probability distribution $p_X\left(x\right)$ and the classical variable \textit{x} from the register \textit{X} cannot be indentified by the measurement operators $M\mathrm{=}{\left\{\left|\left.x\right\rangle \right.\left\langle \left.x\right|\right.\right\}}_{x\mathrm{\in }X}$, since the system state ${\rho }_x$ is assumed to be a mixed or in a non-orthonormal state. Alice's register \textit{X} and the quantum system \textit{A} can be viewed as a tensor product system as 
\begin{equation} \label{2.61)} 
{\left\{p_X\left(x\right),\left|\left.x\right\rangle \right.{\left\langle \left.x\right|\right.}_X\mathrm{\otimes }{\rho }^x_A\right\}}_{x\mathrm{\in }X},                                
\end{equation} 
where the classical variable \textit{x} is correlated with the quantum system ${\rho }_x$, using orthonormal basis ${\left\{\left|\left.x\right\rangle \right.\right\}}_{x\mathrm{\in }X}$. Alice's register \textit{X} represents a classical variable, the channel input system is generated corresponding to the register \textit{X} in the form of a quantum state, and it is described by the density operator ${\rho }^x_A$. The input system \textit{A} with respect to the register \textit{X}, is described by the density operator 
\begin{equation} \label{2.62)} 
{\rho }_{XA}\mathrm{=}\sum_{x\mathrm{\in }X}{p_X\left(x\right)}\left|\left.x\right\rangle \right.{\left\langle \left.x\right|\right.}_X\mathrm{\otimes }{\rho }^x_A,                      
\end{equation} 
where ${\rho }^x_A\mathrm{=}\left|\left.{\psi }_x\right\rangle \right.{\left\langle \left.{\psi }_x\right|\right.}_A$ is the density matrix representation of Alice's input state ${\left|\left.{\psi }_x\right\rangle \right.}_A$.

\subsubsection{Purification}

 The \textit{purification} gives us a new viewpoint on the noise of the quantum channel. Assuming Alice's side \textit{A }and Alice's register \textit{X}, the spectral decomposition of the density operator ${\rho }_A$ can be expressed as 
\begin{equation} \label{ZEqnNum854084} 
{{\rho }_{A}}=\sum\limits_{x}{{{p}_{X}}\left( x \right)\left| x \right\rangle {{\left\langle  x \right|}_{A}}},
\end{equation} 
where $p_X\left(x\right)$ is the probability of variable \textit{x} in Alice's register \textit{X}. The $\left\{p_X\left(x\right),\left|\left.x\right\rangle \right.\right\}$ together is called an ensemble, where $\left|\left.x\right\rangle \right.$ is a quantum state according to classical variable \textit{x}. Using the set of orthonormal basis vectors ${{\left\{ {{\left| x \right\rangle }_{P}} \right\}}_{x\in X}}$ of the purification system \textit{P}, the purification of \eqref{ZEqnNum854084} can be given in the following way:
\begin{equation} \label{ZEqnNum381325} 
{\left|\left.\varphi \right\rangle \right.}_{PA}\mathrm{=}\sum_x{\sqrt{p_X\left(x\right)}{\left|\left.x\right\rangle \right.}_P{\left|\left.x\right\rangle \right.}_A.} 
\end{equation} 
From the purified system state ${\left|\left.\varphi \right\rangle \right.}_{PA}$, the original system state ${\rho }_A$ can be expressed with the partial trace operator (see Appendix) $Tr_P\left(\mathrm{\cdot }\right)$, which operator traces out the purification state from the system 
\begin{equation} \label{ZEqnNum494152} 
{\rho }_A\mathrm{=}Tr_P\left(\left|\left.\varphi \right\rangle \right.{\left\langle \left.\varphi \right|\right.}_{PA}\right).                                                
\end{equation} 
From joint system \eqref{ZEqnNum381325} and the purified state \eqref{ZEqnNum494152}, one can introduce a new definition. The \textit{extension} of ${\rho }_A$ can be given as
\begin{equation} \label{2.66)} 
{\rho }_A\mathrm{=}Tr_P\left({\omega }_{PA}\right),                                                       
\end{equation} 
where ${\omega }_{PA}$ is the joint system of purification state \textit{P} and channel input \textit{A } [\cref{Wilde11}], which represents a noisy state.

\subsubsection{Isometric Extension}

 \textit{Isometric extension} has utmost importance, because it helps us to understand what happens between the quantum channel and its environment whenever a quantum state is transmitted from Alice to Bob. Since the channel and the environment together form a closed physical system the isometric extension of the quantum channel $\mathcal{N}$ is the \textit{unitary representation} of the channel
\begin{equation} \label{2.67)} 
\mathcal{N}\mathrm{:}U_{A\mathrm{\to }BE},                                                                       
\end{equation} 
enabling the `one-sender and two-receiver' view: beside Alice the sender, both Bob and the environment of the channel are playing the receivers. In other words, the output of the noisy quantum channel $\mathcal{N}$ can be described only after the environment of the channel is traced out 
\begin{equation} \label{ZEqnNum611748} 
{\rho }_B\mathrm{=}Tr_E\left(U_{A\mathrm{\to }BE}\left({\rho }_A\right)\right)\mathrm{=}\mathcal{N}\left({\rho }_A\right).                                                              
\end{equation}

\subsubsection{Kraus Representation}

 The map of the quantum channel can also be expressed by means of a special tool called the \textit{Kraus Representation}. For a given input system ${\rho }_A$ and quantum channel $\mathcal{N}$, this representation can be expressed as 
\begin{equation} \label{2.69)} 
\mathcal{N}\left({\rho }_A\right)\mathrm{=}\sum_i{N_i}{\rho }_AN^{\mathrm{\dagger }}_i,                                                                                       
\end{equation} 
where $N_i$ are the Kraus operators, and $\sum_i{N^{\mathrm{\dagger }}_iN_i}\mathrm{=}I$. The isometric extension of $\mathcal{N}$ by means of the \textit{Kraus Representation} can be expressed as 
\begin{equation} \label{ZEqnNum440296} 
{\rho }_B\mathrm{=}\mathcal{N}\left({\rho }_A\right)\mathrm{=}\sum_i{N_i}{\rho }_AN^{\mathrm{\dagger }}_i\mathrm{\to }U_{A\mathrm{\to }BE}\left({\rho }_A\right)\mathrm{=}\sum_i{N_i}\mathrm{\otimes }{\left|\left.i\right\rangle \right.}_E.                                          
\end{equation} 
The action of the quantum channel $\mathcal{N}$ on an operator $\left|\left.k\right\rangle \right.\left\langle \left.l\right|\right.$, where $\left\{\left|\left.k\right\rangle \right.\right\}$ form an orthonormal basis also can be given in operator form using the Kraus operator $N_{kl}\mathrm{=}\mathcal{N}\left(\left|\left.k\right\rangle \right.\left\langle \left.l\right|\right.\right)$. By exploiting the property $UU^{\mathrm{\dagger }}\mathrm{=}P_{BE}$, for the input quantum system ${\rho }_A$
\begin{equation} \label{2.71)} 
\begin{split}
\begin{array}{l}
{\rho }_B\mathrm{=} U_{A\mathrm{\to }BE}\left({\rho }_A\right)\mathrm{=}U{\rho }_AU^{\mathrm{\dagger }} \\ 
\mathrm{=}\left(\sum_i{N_i}\mathrm{\otimes }{\left|\left.i\right\rangle \right.}_E\right){\rho }_A\left(\sum_j{N^{\mathrm{\dagger }}_j}\mathrm{\otimes }{\left\langle \left.j\right|\right.}_E\right)\\\mathrm{=}\sum_{i,j}{N_i}{\rho }_AN^{\mathrm{\dagger }}_j\mathrm{\otimes }\left|\left.i\right\rangle \right.{\left\langle \left.j\right|\right.}_E. \end{array}
\end{split}
\end{equation} 
If we trace out the environment, we get the equivalence of the two representations
\begin{equation} \label{ZEqnNum458981} 
{\rho }_B\mathrm{=}Tr_E\left(U_{A\mathrm{\to }BE}\left({\rho }_A\right)\right)\mathrm{=}\sum_i{N_i}{\rho }_AN^{\mathrm{\dagger }}_i.                                   
\end{equation}

\subsubsection{The von Neumann Entropy}

 Quantum information processing exploits the quantum nature of information. It offers fundamentally new solutions in the field of computer science and extends the possibilities to a level that cannot be imagined in classical communication systems. On the other hand, it requires the generalization of classical information theory through a quantum perception of the world. As Shannon entropy plays fundamental role in classical information theory, the von Neumann entropy does the same for quantum information. The von Neumann entropy $\mathrm{S}\left(\rho \right)$ of quantum state $\rho $ can be viewed as an extension of classical entropy for quantum systems. It measures the information of the quantum states in the form of the uncertainty of a quantum state. The classical Shannon entropy $H\left(X\right)$ of a variable \textit{X }with probability distribution $p\left(X\right)$ can be defined as 
\begin{equation} \label{2.73)} 
H\left(X\right)\mathrm{=-}\sum_{x\mathrm{\in }X}{p\left(x\right)}\mathrm{log}\left(p\left(x\right)\right), 
\end{equation} 
with $\mathrm{1}\mathrm{\le }H\left(X\right)\mathrm{\le }\mathrm{log}\left(\left|X\right|\right)$, where $\left|X\right|$ is the cardinality of the set \textit{X}. 

 The von Neumann entropy 
\begin{equation} \label{2.74)} 
\mathrm{S}\left(\rho \right)\mathrm{=-}Tr\left(\rho \mathrm{log}\left(\rho \right)\right) 
\end{equation} 
measures the information contained in the quantum system $\rho $. Furthermore $\mathrm{S}\left(\rho \right)$ can be expressed by means of the Shannon entropy for the eigenvalue distribution 
\begin{equation} \label{2.75)} 
\mathrm{S}\left(\rho \right)\mathrm{=}H\left(\lambda \right)\mathrm{=-}\sum^d_{i\mathrm{=1}}{{\lambda }_i\mathrm{log}\left({\lambda }_i\right)}, 
\end{equation} 
where \textit{d} is the level of the quantum system and ${\lambda }_i$ are the eigenvalues of density matrix $\rho $.

\subsubsection{The Holevo Quantity}

 The \textit{Holevo bound} determines the amount of information that can be extracted from a single qubit state. If Alice sends a quantum state ${\rho }_i$ with probability $p_i$ over an ideal quantum channel, then at Bob's receiver a mixed state
\begin{equation} \label{2.76)} 
{\rho }_B\mathrm{=}{\rho }_A\mathrm{=}\sum_i{p_i{\rho }_i} 
\end{equation} 
appears. Bob constructs a measurement $\left\{M_i\right\}$ to extract the information encoded in the quantum states. If he applies the measurement to ${\rho }_A$, the probability distribution of Bob's classical symbol \textit{B} will be $\mathrm{Pr}\left[\left.b\right|{\rho }_A\right]\mathrm{=}Tr\left(M^{\mathrm{\dagger }}_bM_b{\rho }_A\right)$. As had been shown by Holevo [\cref{Holevo73}], the bound for the maximal classical mutual information between Alice and Bob is
\begin{equation} \label{ZEqnNum240233} 
I\left(A\mathrm{:}B\right)\mathrm{\le }\mathrm{S}\left({\rho }_A\right)\mathrm{-}\sum_i{p_i\mathrm{S}\left({\rho }_i\right)}\mathrm{\equiv }\chi ,                                                        
\end{equation} 
where $\chi $ is called the \textit{Holevo quantity, and }\eqref{ZEqnNum240233} known as the \textit{Holevo bound}. 

 In classical information theory and classical communication systems, the mutual information $I\left(A\mathrm{:}B\right)$ is bounded only by the classical entropy of $H\left(A\right)$, hence $I\left(A\mathrm{:}B\right)\mathrm{\le }H\left(A\right)$. The mutual information $I\left(A\mathrm{:}B\right)$ is bounded by the classical entropy of $H\left(A\right)$, hence $I\left(A\mathrm{:}B\right)\mathrm{\le }H\left(A\right)$. On the other hand, for mixed states and pure non-orthogonal states the Holevo quantity $\chi $ can be greater than the mutual information $I\left(A\mathrm{:}B\right)$, however, it is still bounded by $H\left(A\right)$, which is the bound for the pure orthogonal states   
\begin{equation} \label{2.78)} 
I\left(A\mathrm{:}B\right)\mathrm{\le }\chi \mathrm{\le }H\left(A\right).                                                
\end{equation} 
The \textit{Holevo bound} highlights the important fact that one qubit can contain at most one classical bit i.e., cbit of information.

\subsubsection{Quantum Conditional Entropy}

 While the classical conditional entropy function is always takes a non negative value, the \textit{quantum conditional entropy can be negative}. The quantum conditional entropy between quantum systems \textit{A} and \textit{B} is given by  
\begin{equation} \label{2.79)} 
\mathrm{S}\left(\left.A\right|B\right)\mathrm{=S}\left({\rho }_{AB}\right)\mathrm{-}\mathrm{S}\left({\rho }_B\right).                                              
\end{equation} 
If we have two uncorrelated subsystems ${\rho }_A$ and ${\rho }_B$, then the information of the quantum system ${\rho }_A$ does not contain any information about ${\rho }_B$, or reversely, thus 
\begin{equation} \label{2.80)} 
\mathrm{S}\left({\rho }_{AB}\right)\mathrm{=}\mathrm{S}\left({\rho }_A\right)\mathrm{+S}\left({\rho }_B\right),                                        
\end{equation} 
hence we get $\mathrm{S}\left(\left.A\right|B\right)\mathrm{=}\mathrm{S}\left({\rho }_A\right)$, and similarly $\mathrm{S}\left(\left.B\right|A\right)\mathrm{=S}\left({\rho }_B\right)$. The negative property of conditional entropy $\mathrm{S}\left(\left.A\right|B\right)$ can be demonstrated with an \textit{entangled} state, since in this case, the joint quantum entropy of the joint state less than the sum of the von Neumann entropies of its individual components. For a pure entangled state, $\mathrm{S}\left({\rho }_{AB}\right)\mathrm{=0}$, while $\mathrm{S}\left({\rho }_A\right)\mathrm{=S}\left({\rho }_B\right)\mathrm{=1}$ since the two qubits are in \textit{maximally mixed} $\frac{\mathrm{1}}{\mathrm{2}}I$ state, which is classically totally unimaginable. Thus, in this case 
\begin{equation} \label{2.81)} 
\mathrm{S}\left(\left.A\right|B\right)\mathrm{=-S}\left({\rho }_B\right)\mathrm{\le }\mathrm{0},                                                
\end{equation} 
and 
$\mathrm{S}\left(\left.B\right|A\right)\mathrm{=-S}\left({\rho }_A\right)\mathrm{\le }\mathrm{0}$ and $\mathrm{S}\left({\rho }_A\right)\mathrm{=S}\left({\rho }_B\right)$.                                           \label{2.82)}

\subsubsection{Quantum Mutual Information}
The classical mutual information $I\left(\mathrm{\cdot }\right)$\textit{ }measures the information correlation between random variables \textit{A} and \textit{B. }In analogue to classical information theory, $I\left(A\mathrm{:}B\right)$ can be described by the quantum entropies of individual states and the von Neumann entropy of the joint state as follows: 
\begin{equation} \label{2.83)} 
I\left(A\mathrm{:}B\right)\mathrm{=S}\left({\rho }_A\right)\mathrm{+S}\left({\rho }_B\right)\mathrm{-}\mathrm{S}\left({\rho }_{AB}\right)\mathrm{\ge }\mathrm{0},                                             
\end{equation} 
i.e., the quantum mutual information is always a non negative function. However, there is a distinction between classical and quantum systems, since the quantum mutual information can take its value above the maximum of the classical mutual information. This statement can be confirmed, if we take into account that for an pure entangled quantum system, the quantum mutual information is 
\begin{equation} \label{2.84)} 
I\left(A\mathrm{:}B\right)\mathrm{=S}\left({\rho }_A\right)\mathrm{+S}\left({\rho }_B\right)\mathrm{-}\mathrm{S}\left({\rho }_{AB}\right)\mathrm{=1+1-0=2},                                
\end{equation} 
and we can rewrite this equation as 
\begin{equation} \label{2.85)} 
I\left(A\mathrm{:}B\right)\mathrm{=2S}\left({\rho }_A\right)\mathrm{=2S}\left({\rho }_B\right).                                              
\end{equation} 
For some pure joint system ${\rho }_{AB}$, the equation \eqref{2.85)} can be satisfied such that $\mathrm{S}\left({\rho }_A\right)\mathrm{=S}\left({\rho }_B\right)$ and $\mathrm{S}\left({\rho }_{AB}\right)\mathrm{=}0$.

If we use entangled states, the quantum mutual information could be 2, while the quantum conditional entropies could be 2. In classical information theory, negative entropies can be obtained only in the case of mutual information of three or more systems. An important property of maximized quantum mutual information: \textit{it is always additive for a quantum channel}.

The character of classical information and quantum information is significantly different. There are many phenomena in quantum systems which cannot be described classically, such as entanglement, which makes it possible to store quantum information in the correlation of quantum states. Similarly, a quantum channel can be used with pure orthogonal states to realize classical information transmission, or it can be used to transmit non-orthogonal states or even quantum entanglement. Information transmission also can be approached using the question, whether the input consists of unentangled or entangled quantum states. This leads us to say that for quantum channels many new capacity definitions exist in comparison to a classical communication channel. In possession of the general communication model and the quantities which are able to represent information content of quantum states we can begin to investigate the possibilities and limitations of information transmission through quantum channels [\cref{Lloyd04}].

\subsubsection{Quantum Relative Entropy}

 The \textit{quantum relative entropy} measures the informational distance between quantum states, and introduces a deeper characterization of the quantum states than the von Neumann entropy. Similarly to the classical relative entropy, this quantity measures the distinguishability of the quantum states, in practice it can be realized by POVM measurements. The relative entropy classically is a measure that quantifies how close a probability distribution \textit{p} is to a model or candidate probability distribution \textit{q}. For probability distributions \textit{p} and \textit{q,} the classical relative entropy is given by
\begin{equation} \label{2.88)} 
D\left(\left.p\right\|q\right)\mathrm{=}\sum_i{p_i\mathrm{log}}\left(\frac{p_i}{q_i}\right),                                                                          
\end{equation} 
while the quantum relative entropy between quantum states $\rho $ and $\sigma $ is
\begin{equation} \label{2.89)} 
\begin{split}
D\left(\left.\rho \right\|\sigma \right)&\mathrm{=}Tr\left(\rho \mathrm{log}\left(\rho \right)\right)\mathrm{-}Tr\left(\rho \mathrm{log}\left(\sigma \right)\right)\\&\mathrm{=}Tr\left[\rho \left(\mathrm{log}\left(\rho \right)\mathrm{-}\mathrm{log}\left(\sigma \right)\right)\right]. 
\end{split}
\end{equation} 
In the definition above, the term $Tr\left(\rho \mathrm{log}\left(\sigma \right)\right)$ is finite only if $\rho \mathrm{log}\left(\sigma \right)\mathrm{\ge }\mathrm{0}$ for all diagonal matrix elements. If this condition is not satisfied, then $D\left(\left.\rho \right\|\sigma \right)$ could be infinite, since the trace of the second term could go to infinity. 

The \textit{quantum informational distance} (i.e., quantum relative entropy) has some distance-like properties (for example, the quantum relative entropy function between a maximally mixed state and an arbitrary quantum state is symmetric, hence in this case it is not just a pseudo distance), however it is \textit{not commutative}, thus $D\left(\rho \left\|\sigma \right.\right)\mathrm{\ne }D\left(\sigma \left\|\rho \right.\right),$ and $D\left(\rho \left\|\sigma \right.\right)\mathrm{\ge }\mathrm{0}$ iff $\rho \mathrm{\ne }\sigma ,$ and $D\left(\rho \left\|\sigma \right.\right)\mathrm{=0}$ iff $\rho \mathrm{=}\sigma .$ Note, if $\sigma $ has zero eigenvalues, $D\left(\rho \left\|\sigma \right.\right)$ may diverge, otherwise it is a finite and continuous function. Furthermore, the quantum relative entropy function has another interesting property, since if we have two density matrices $\rho $ and $\sigma $, then the following property holds for the traces used in the expression of $D\left(\rho \left\|\sigma \right.\right)$
\begin{equation} \label{2.90)} 
Tr\left(\rho \mathrm{log}\left(\rho \right)\right)\mathrm{\ge }Tr\left(\rho \mathrm{log}\left(\sigma \right)\right).                                                                  
\end{equation} 
The symmetric Kullback-Leibler distance is widely used in classical systems, for example in computer vision and sound processing. Quantum relative entropy reduces to the classical Kullback-Leibler relative entropy for simultaneously diagonalizable matrices.

We note, the quantum mutual information can be defined by quantum relative entropy $D\left(\left.\mathrm{\cdot }\right\|\mathrm{\cdot }\right)$. This quantity can be regarded as the informational distance between the tensor product of the individual subsystems ${\rho }_A\mathrm{\otimes }{\rho }_B$, and the joint state ${\rho }_{AB}$ as follows:
\begin{equation} \label{2.91)} 
I\left(A\mathrm{:}B\right)\mathrm{=}D\left(\left.{\rho }_{AB}\right\|{\rho }_A\mathrm{\otimes }{\rho }_B\right)\mathrm{=S}\left({\rho }_A\right)\mathrm{+S}\left({\rho }_B\right)\mathrm{-}\mathrm{S}\left({\rho }_{AB}\right).                                 
\end{equation}

\subsubsection{Quantum R\'{e}nyi-Entropy}

 As we have seen, the quantum informational entropy can be defined by the $\mathrm{S}\left(\rho \right)$ von Neumann entropy function. On the other hand, another entropy function can also be defined in the quantum domain, it is called the R\'{e}nyi-entropy and denoted by $\mathrm{R}\left(\rho \right)$. This function has relevance mainly in the description of quantum entanglement. The R\'{e}nyi-entropy function is defined as follows
\begin{equation} \label{2.92)} 
\mathrm{R}\left(\rho \right)\mathrm{=}\frac{\mathrm{1}}{\mathrm{1-}r}Tr\mathrm{(}{\rho }^r\mathrm{)},                                                   
\end{equation} 
where $r\mathrm{\ge }\mathrm{0}$, while $\mathrm{R}\left(\rho \right)$ is equal to the von Neumann entropy function $\mathrm{S}\left(\rho \right)$ if
\begin{equation} \label{2.93)} 
\mathop{\mathrm{lim}}_{r\mathrm{\to }\mathrm{1}}\mathrm{R}\left(\rho \right)\mathrm{=S}\left(\rho \right).                                              
\end{equation} 
If parameter \textit{r} converges to infinity, then we have
\begin{equation} \label{2.94)} 
\mathop{\mathrm{lim}}_{r\mathrm{\to }\mathrm{\infty }}\mathrm{R}\left(\rho \right)\mathrm{=-log}\left(\left\|\rho \right\|\right).                                                         
\end{equation} 
On the other hand if $r\mathrm{=0}$ then $\mathrm{R}\left(\rho \right)$ can be expressed from the rank of the density matrix
\begin{equation} \label{2.95)} 
\mathrm{R}\left(\rho \right)\mathrm{=log}\left(rank\left(\rho \right)\right).                                                                 
\end{equation}

\subsection{Related Work}

 The field of quantum information processing is a rapidly growing field of science, however there are still many challenging questions and problems. These most important results will be discussed in further sections, but these questions cannot be exposited without a knowledge of the fundamental results of quantum information theory.

\subsubsection{Early Years of quantum information theory}

 quantum information theory extends the possibilities of classical information theory, however for some questions, it gives extremely different answers. The advanced communications and quantum networking technologies offered by quantum information processing will revolutionize traditional communication and networking methods. Classical information theory--- was founded by Claude Shannon in 1948 [\cref{Hanzo12}], [\cref{Shannon48}]. In Shannon's paper the mathematical framework of communication was invented, and the main definitions and theorems of classical information theory were laid down. On the other hand, classical information theory is just one part of quantum information theory. The other, missing part is the Quantum Theory, which was completely finalized in 1926.

 The results of quantum information theory are mainly based on the results of von Neumann, who constructed the mathematical background of quantum mechanics [\cref{Neumann96}]. An interesting---and less well known---historical fact is that quantum entropy was discovered by Neumann before the classical information theoretic concept of entropy. Quantum entropy was discovered in the 1930s, based on the older idea of entropy in classical Statistical Mechanics, while the classical information theoretic concept was discovered by Shannon only later, in 1948. It is an interesting note, since the reader might have thought that quantum entropy is an extension of the classical one, however it is not true. Classical entropy, in the context of Information Theory, is a special case of von Neumann's quantum entropy. Moreover, the name of Shannon's formula was proposed by von Neumann. Further details about the history of Quantum Theory, and the main results of physicists from the first half of the twentieth century------such as Planck, Einstein, Schr\"{o}dinger, Heisenberg, or Dirac------can be found in the works of Misner et al. [\cref{Misner09}], McEvoy [\cref{McEvoy04}], Sakurai [\cref{Sakurai94}], Griffiths [\cref{Griffiths95}] or Bohm [\cref{Bohm89}].

`\textit{Is quantum mechanics useful}'--- asked by Landauer in 1995 [\cref{Landauer95}]. Well, having the results of this paper in our hands, we can give an affirmative answer: \textit{definitely yes}. An interesting work about the importance of quantum mechanical processes was published by Dowling [\cref{Dowling03}]. Some fundamental results from the very early days of Quantum Mechanics can be found in [\cref{Broglie1924}], [\cref{Dirac82}], [\cref{Einstein1905}], [\cref{Einstein1935}], [\cref{Gerlach1922}], [\cref{Heisenberg1925}], [\cref{Planck1901}], [\cref{Schrodinger1926}], [\cref{Schrodinger1935}], [\cref{Thomson1901}]. About the early days of Information Theory see the work of Pierce [\cref{Pierce73}]. A good introduction to Information Theory can be found in the work of Yeung [\cref{Yeung02}]. More information about the connection of Information Theory and statistical mechanics can be found in work of Aspect from 1981 [\cref{Aspect81}], in the book of Jaynes [\cref{Jaynes03}] or Petz [\cref{Petz08}]. The elements of classical information theory and its mathematical background were summarized in a very good book by Cover [\cref{Cover91}]. On matrix analysis a great work was published by Horn and Johnson [\cref{Horn86}].

A very good introduction to quantum information theory was published by Bennett and Shor [\cref{Bennett98}]. The idea that the results of quantum information theory can be used to solve computational problems was first claimed by Deutsch in 1985 [\cref{Deutsch85}]. 

Later in the 90s, the answers to the most important questions of quantum information theory were answered, and the main elements and the fundamentals of this field were discovered. Details about the simulation of quantum systems and the possibility of encoding quantum information in physical particles can be found in Feynman's work from 1982 [\cref{Feynman82}]. Further information on quantum simulators and continuous-time automata can be found in the work of Vollbrecht and Cirac [\cref{Vollbrecht08}].

\subsubsection{Quantum Coding and Quantum Compression}

 The next milestone in quantum information theory is Schumacher's work from 1995 [\cref{Schumacher95a}] in which he introduced the term, `\textit{qubit}.' In [\cref{Schumacher95}, \cref{Schumacher95a}, \cref{Schumacher96b}, \cref{Schumacher96c}] the main theories of quantum source coding and the quantum compression were presented. The details of quantum data compression and quantum typical subspaces can be found in [\cref{Schumacher95a}]. In this paper, Schumacher extended those results which had been presented a year before, in 1994 by Schumacher and Jozsa on a new proof of quantum noiseless coding, for details see [\cref{Schumacher94}]. Schumacher in 1995 also defined the quantum coding of pure quantum states; in the same year, Lo published a paper in which he extended these result to mixed quantum states, and he also defined an encoding scheme for it [\cref{Lo95}]. Schumacher's results from 1995 on the compression of quantum information [\cref{Schumacher95a}] were the first main results on the encoding of quantum information------\textit{its importance and} \textit{significance in quantum information theory is similar to Shannon's noiseless channel coding theorem in classical information theory}. In this work, Schumacher also gives upper and lower bounds on the rate of quantum compression. We note, that the mathematical background of Schumacher proof is very similar to Shannon's proof, as the reader can check in [\cref{Schumacher95a}] and in Shannon's proof [\cref{Shannon48}].

 The method of sending classical bits via quantum bits was firstly completed by Schumacher et al. in their famous paper form 1995, see [\cref{Schumacher95}]. In the same year, an important paper on the encoding of information into physical particles was published by Schumacher [\cref{Schumacher95}, \cref{Schumacher95a}]. The fundaments of noiseless quantum coding were laid down by Schumacher, one year later, in 1996 [\cref{Schumacher96b}, \cref{Schumacher96c}]. In 1996, many important results were published by Schumacher and his colleges. These works cover the discussion of the relation of entropy exchange and coherent quantum information, which was completely unknown before 1996. The theory of processing of quantum information, the transmission of entanglement over a noisy quantum channel, the error-correction schemes with the achievable fidelity limits, or the classical information capacity of a quantum channel with the limits on the amount of accessible information in a quantum channel were all published in the same year. For further information on the fidelity limits and communication capabilities of a noisy quantum channel, see the work of Barnum et al. also from 1996 [\cref{Barnum96}]. In 1997, Schumacher and Westmoreland completed their proof on the classical capacity of a quantum channel, and they published in their famous work, for details see [\cref{Schumacher97}]. These results were extended in their works from 1998, see [\cref{Schumacher98a}-\cref{Schumacher98c}]. On the experimental side of fidelity testing see the work of Radmark et al. [\cref{Radmark09}].

About the limits for compression of quantum information carried by ensembles of mixed states, see the work of Horodecki [\cref{Horodecki98}]. An interesting paper about the quantum coding of mixed quantum states was presented by Barnum et al. [\cref{Barnum01}]. Universal quantum compression makes it possible to compress quantum information without the knowledge about the information source itself which emits the quantum states. Universal quantum information compression was also investigated by Jozsa et al. [\cref{Jozsa98}], and an extended version of Jozsa and Presnell [\cref{Jozsa03}]. Further information about the technique of universal quantum data compression can be found in the article of Bennett et al. [\cref{Bennett06}]. The similarity of the two schemes follows from the fact that in both cases we compress quantum information, however in the case of Schumacher's method we do not compress entanglement. The two compression schemes are not equal to each other, however in some cases------if running one of the two schemes fails------they can be used to correct the errors of the other, hence they can be viewed as auxiliary protocols of each other. Further information about the mathematical background of the processes applied in the compression of quantum information can be found in Elias's work [\cref{Elias72}].

 A good introduction to quantum error-correction can be found in the work of Gottesman, for details see [\cref{Gottesman04a}]. A paper about classical data compression with quantum side information was published by Devetak and Winter [\cref{Devetak03a}]. We note that there is a connection between the compression of quantum information and the concentration of entanglement, however the working method of Schumacher's encoding and the process of entanglement concentrating are completely different. Benjamin Schumacher and Richard Jozsa published a very important paper in 1994 [\cref{Schumacher94}]. Here, the authors were the first to give an explicit proof of the quantum noiseless coding theorem, which was a milestone in the history of quantum computation. Further information on Schumacher's noiseless quantum channel coding can be found in [\cref{Schumacher94}].

The basic coding theorems of quantum information theory were summarized by Winter in 1999 [\cref{Winter99a}]. In this work, he also analyzed the possibilities of compressing quantum information. A random coding based proof for the quantum coding theorem was shown by Klesse in 2008 [\cref{Klesse08}]. A very interesting article was presented by Horodecki in 1998 [\cref{Horodecki98}], about the limits for the compression of quantum information into mixed states. On the properties of indeterminate-length quantum coding see the work of Schumacher and Westmoreland [\cref{Schumacher01a}].

The quantum version of the well-known Huffman coding can be found in the work of Braunstein et al. from 2000 [\cref{Braunstein2000}]. Further information about the compression of quantum information and the subspaces can be found in [\cref{Fukuda10a}], [\cref{Hayden08a}], and [\cref{Hayden08b}]. The details of quantum coding for mixed states can be found in the work of Barnum et al. [\cref{Barnum01}].

\subsubsection{Quantum Entanglement}

Entanglement is one of the most important differences between the classical and the quantum worlds. An interesting paper on communication via one- and two-particle operators on Einstein-Podolsky-Rosen states was published in 1992, by Bennett [\cref{Bennett92}]. About the history of entanglement see the paper of Einstein, Podolsky and Rosen from 1935 [\cref{Einstein1935}]. In this manuscript, we did not give a complete mathematical background of quantum entanglement---further details on this topic can be found in Nielsen's book [\cref{Nielsen2000}] or by Hayashi [\cref{Hayashi06}], or in an very good article published by the four Horodeckis in 2009 [\cref{Horodecki09}]. We have seen that entanglement concentration can be applied to generate maximally mixed entangled states. We also gave the asymptotic rate at which entanglement concentration can be made, it is called the entropy of entanglement and we expressed it in an explicit form. A very important paper on the communication cost of entanglement transformations was published by Hayden and Winter, for details see [\cref{Hayden03a}]. The method of entanglement concentration was among the first quantum protocols, for details see the work of Bennett et al. from 1996 [\cref{Bennett96b}]. The method of Bennett's was improved by Nielsen in 1999, [\cref{Nielsen99}]. A very important work on variable length universal entanglement concentration by local operations and its application to teleportation and dense coding was published by Hayashi and Matsumoto [\cref{Hayashi01}]. The entanglement cost of antisymmetric states was studied by [\cref{Matsumoto04}].

The calculation of entanglement-assisted classical capacity requires a superdense protocol-like encoding and decoding strategy,------we did not explain its working mechanism in detail, further information can be found in the work of Bennett et al. [\cref{Bennett02}]. A paper about the compression of quantum-measurement operations was published by Winter and Massar in 2001 [\cref{Winter01a}]. Later, in 2004, Winter extended these results [\cref{Winter04}]. Here we note, these results are based on the work of Ahlswede and Winter [\cref{Ahlswede02}].

The definition of a conditionally typical subspace in quantum information was given by Schumacher and Westmoreland in 1997 [\cref{Schumacher97}]. Holevo also introduced it in 1998 [\cref{Holevo98}].

We did not explain in detail entanglement concentrating [\cref{Bennett96b}], entanglement transformations [\cref{Nielsen99}], or entanglement generation, entanglement distribution and quantum broadcasting,------further information can be found in [\cref{Hayashi01}], [\cref{Hayden03a}], [\cref{Hsieh08}], [\cref{Winter01}], [\cref{Yard05a}], [\cref{Yard05b}]. About the classical communication cost of entanglement manipulation see the work of Lo and Popescu from 1999 [\cref{Lo99a}]. The fact that noncommuting mixed states cannot be broadcast was shown by Barnum et al. in 1995, see [\cref{Barnum95}].

Lo and Popescu also published a work on concentrating entanglement by local actions in 2001, for details see [\cref{Lo01}]. About the purification of noisy entanglement see the article of Bennett et al. from 1996 [\cref{Bennett96a}]. The entanglement purification protocol was a very important result, since it will have great importance in the quantum capacity of a quantum channel. (However, when the authors have developed the entanglement purification scheme, this connection was still not completely cleared.)

About the quantum networks for concentrating entanglement and the distortion-free entanglement concentration, further information can be found in the paper of Kaye and Mosca from 2001 [\cref{Kaye01}]. In 2005, Devetak and Winter have shown, that there is a connection between the entanglement distillation and the quantum coherent information, which measure has tremendous relevance in the quantum capacity of the quantum channels, for details see [\cref{Devetak05a}, \cref{Devetak05a}]. An interesting paper about distortion-free entanglement concentration was published by Kohout et al. in 2009 [\cref{Kohout09}]. The method presented in that paper gives an answer to streaming universal. We did not mentioned the inverse protocol of entanglement concentration which is called entanglement dilution, for further details see the works of Lo and Popescu from 1999 [\cref{Lo99a}] and 2001 [\cref{Lo01}], and Harrow and Lo's work from 2004 [\cref{Harrow04a}]. Harrow and Lo have also given an explicit solution of the communication cost of the problem of entanglement dilution, which was an open question until 2004. Their results are based on the previous work of Hayden and Winter from 2003, for details see [\cref{Hayden03a}]. The typical entanglement in stabilizer states was studied by Smith and Leung, see [\cref{Smith06}]. The teleportation-based realization of a two-qubit entangling gate was shown by Gao et al. [\cref{Gao10}].

\subsubsection{Quantum Channels}

About the statistical properties of the HSW theory and the general HSW capacity, a very interesting paper was published by Hayashi and Nagaoka in 2003 [\cref{Hayashi03}]. As we have seen, some results of quantum information theory are similar to the results of classical information theory, however many things have no classical analogue. As we have found in this section, the Holevo theorem gives an information-theoretic meaning to the von Neumann entropy, however it does not make it possible to use it in the case of the interpretation of von Neumann entropy of physical macrosystems. Further properties of the von Neumann entropy function was studied by Audenaert in 2007 [\cref{Audenaert07}].

The concept of quantum mutual information measures the classical information which can be transmitted through a noisy quantum channel (originally introduced by Adami and Cerf [\cref{Adami96}]) however it cannot be used to measure the maximal transmittable quantum information. The maximized quantum mutual information is always additive, however this is not true for the Holevo information. In this case, the entanglement makes non-additive the Holevo information, but it has no effect on the quantum mutual information. Further information about the mathematical background of these `strange' phenomena can be found in the work of Adami from 1996 [\cref{Adami96}] or in the book of Hayashi from 2006 [\cref{Hayashi06}]. A very good book on these topics was published by Petz in 2008 [\cref{Petz08}].

For the properties of Holevo information and on the capacity of quantum channels see the works of Holevo [\cref{Holevo73}], [\cref{Holevo98}], Schumacher and Westmoreland [\cref{Schumacher94}, \cref{Schumacher95}, \cref{Schumacher95a}, \cref{Schumacher96b}, \cref{Schumacher96c}, \cref{Schumacher97}], Horodecki [\cref{Horodecki05}], Datta [\cref{Datta04a}], Arimoto [\cref{Arimoto72}]. On the geometrical interpretation of the maps of a quantum channel see the works of Cortese [\cref{Cortese02}], Petz [\cref{Petz07}-\cref{Petz10}], [\cref{Petz96}], Hiai [\cref{Hiai91}].

On physical properties of quantum communication channels the work of Levitin [\cref{Levitin69}], on the capacities of quantum communication channels see Bennett [\cref{Bennett97}], DiVincenzo [\cref{DiVincenzo98}], Schumacher [\cref{Schumacher97}], Fuchs [\cref{Fuchs97}]. In 1997, Barnum, Smolin and Terhal also summarized the actual results on quantum channel, see [\cref{Barnum97b}].

The mathematical background of distinguishing arbitrary multipartite basis unambiguously was shown by Duan et al. [\cref{Duan07}].)

In 2010, Dupis et al. [\cref{Dupuis10}] published a paper in which they described a protocol for quantum broadcast quantum channel, then Jon Yard et al. published a paper on quantum broadcast channels [\cref{Yard06}]. Before these results, in 2007, an important practical result on broadcasting was shown by Guha et al. [\cref{Guha07a}], [\cref{Guha07b}], who demonstrated the classical capacity of practical (bosonic) quantum channels. General quantum protocols---such as super-dense coding and teleportation---are not described in this article. Further information about these basic quantum protocols can be found in the book of Hayashi from 2006 [\cref{Hayashi06}], in the book of Nielsen and Chuang [\cref{Nielsen2000}], or in the paper of Bennett and Wiesner [\cref{Bennett92}], and [\cref{Bennett92a}], (both papers from 1992),, and Bennett's paper from 1993 [\cref{Bennett93}].

A very good overview of the capacity of quantum channels was presented by Smith in 2010, see [\cref{Smith10}]. About the information tradeoff relations for finite-strength quantum measurements, see the works of [\cref{Fuchs2000}]. On the mathematical background of quantum communication see the works of [\cref{Petz96}], Ruskai et al. [\cref{Ruskai01}], and [\cref{Hayashi05}], [\cref{Vedral98}]. The generalized Pauli channels are summarized by Ohno and Petz in [\cref{Ohno09}].

The relative entropy function was introduced by Solomon Kullback and Richard Leibler in 1951 [\cref{Kullback51}]. Another interpretation of the relative entropy function was introduced by Bregman, known as the class of Bregman divergences [\cref{Bregman67}]. A very important paper about the role of relative entropy in quantum information theory was published by Schumacher and Westmoreland in 2000 [\cref{Schumacher2000}]. The quantum relative entropy function was originally introduced by Umegaki, and later modified versions have been defined by Ohya, Petz and Watanbe  [\cref{Ohya97}]. Some possible applications of quantum relative entropy in quantum information processing were introduced by Vedral [\cref{Vedral2000}].

About the negativity of quantum information see the works of Horodecki et al. [\cref{Horodecki05}], [\cref{Horodecki07}]. About the use of entanglement in quantum information theory, see the work of Li et al. from 2010, [\cref{Li10}], [\cref{Li16}]. A method for measuring two-qubit entanglement by local operations and classical communication was shown by Bai et al. in 2005 [\cref{Bai05}]. About the additivity of the capacity of quantum channels see [\cref{Fujiwara02}], [\cref{King09}] and [\cref{Shor02a}]. A very good paper on the Holevo capacity of finite dimensional quantum channels and the role of additivity problem in quantum information theory was published by Shirokov [\cref{Shirokov06}]. A great summary of classical and quantum information theory can be found in the book of Desurvire from 2009 [\cref{Desurvire09}]. The bounds for the quantity of information transmittable by a quantum communication channel was analyzed by Holevo in 1973, see [\cref{Holevo73}]. About sending classical information via noisy quantum channels, see the works of Schumacher and Jozsa [\cref{Schumacher94}], Schumacher from 1996 [\cref{Schumacher96b}, \cref{Schumacher96c}], and Schumacher and Westmoreland from 1997 [\cref{Schumacher97}] and Smith's summarize [\cref{Smith10}]. The mathematical background of classical relative entropy function can be found in the works of Kullback and Leibler [\cref{Kullback51}], [\cref{Kullback59}], and [\cref{Kullback87}]. For the details of Bregman distance see [\cref{Bregman67}] and [\cref{Banerjee05}]. Further information about the Kraft-McMillan inequality can be found in [\cref{Kraft49}], [\cref{McMillan56}] and [\cref{Cover91}]. 

For research on satellite quantum communications, see [\cref{Bacsardi05}, \cref{Bacsardi07}, \cref{Bacsardi10}, \cref{Bacsardi13}], [\cref{Galambos10}]. For research results on quantum repeaters see [\cref{Azuma09}], [\cref{Bernardes10}], [\cref{Briegel98}], [\cref{Dur99}], [\cref{Jiang08}], [\cref{Ladd06}], [\cref{Loock08}], [\cref{Munro08}-\cref{Munro10}], [\cref{Sangouard09}], [\cref{VanMeter08}, \cref{VanMeter09}, \cref{VanMeter13}, \cref{VanMeter14}], and  [\cref{Yuan08}]. For some further research topic on quantum channels see [\cref{Babar15}],  [\cref{Panagiotis14}, \cref{Panagiotis14a}], [\cref{Gyongyosi11a}-\cref{Gyongyosi11d},\cref{Gyongyosi12a}-\cref{Gyongyosi12b}, \cref{Gyongyosi13}, \cref{Gyongyosi13a}, \cref{Gyongyosi14}-\cref{Gyongyosi14c}, \cref{Gyongyosi17}], [\cref{Imre13a}].

\subsubsection{Comprehensive Surveys}

 A reader who is interested in the complete mathematical background of quantum information theory can find the details for example in Nielsen and Chuang's book [\cref{Nielsen2000}]. For a general introduction to the quantum information theory and its applications see the excellent book of Hayashi [\cref{Hayashi06}]. We also suggest the book of Imre from 2005, see [\cref{Imre05}]. A very good introduction to quantum information theory was published by Bennett and Shor, for details see [\cref{Bennett98}]. Also in 1998, Preskill summarized the actual state of quantum information theory in the form of lecture notes [\cref{Preskill98}]. Preskill also summarized the conditions of reliable quantum computers, for details see [\cref{Preskill98a}]. Also in 1998, a1998, a good work on the basics of quantum computations and the mathematical formalism was published by Vedral and Plenio [\cref{Vedral98}] and by Nielsen [\cref{Nielsen98}]. On the mathematical background of quantum information processing, see the works of Shor [\cref{Shor95}, \cref{Shor96}, \cref{Shor96b}], [\cref{Shor97}], [\cref{Shor02}], and [\cref{Shor04}]. The description of classical data compression can be found in the very good book of Cover and Thomas [\cref{Cover91}], or in the book of Berger [\cref{Berger71}]. We also suggest the work of Stinespring [\cref{Stinespring55}]. A very important result regarding the compression of classical information was published by Csisz\'{a}r and K\"{o}rner in 1978 [\cref{Csiszar78}], and later the authors published a great book about coding theorems for discrete memoryless systems [\cref{Csiszar81}]. A work on the non-additivity of Renyi entropy was published by Aubrun et al. [\cref{Aubrun10}]. On the connection of quantum entanglement and classical communication through a depolarizing channel see [\cref{Bruss2000}]. Regarding the results of quantum Shannon theory, we suggest the great textbook of Wilde [\cref{Wilde11}]. The structure of random quantum channels, eigenvalue statistics and entanglement of random subspaces are discussed in [\cref{Collins09a}], [\cref{Collins09b}].  Finally, for an interesting viewpoint on `topsy turvy world of quantum computing' see [\cref{Mullins01}].

\section{Classical Capacities of a Quantum Channel}
\label{sec3}
 Communication over quantum channels is bounded by the corresponding capacities. Now, we lay down the fundamental theoretic results on \textit{classical capacities of quantum channel}s. These results are all required to analyze the advanced and more promising properties of quantum communications. 

This section is organized as follows. In the first part, we introduce the reader to formal description of a noisy quantum channel. Then we start to discuss the classical capacity of a quantum channel. Next, we show the various encoder and decoder settings for transmission of classical information. We define the exact formula for the measure of maximal transmittable classical information. Finally, we discuss some important channel maps. 

The most relevant works are included in the Related Work subsection.

\subsection{Extended Formal Model}

 The discussed model is general enough to analyze the limitations for information transfer over quantum channels. However, later we will investigate special quantum channels which models specific physical environment. Each quantum channel can be represented as a\textit{ }CPTP\textit{ }map (\textit{Completely Positive Trace Preserving}), hence the process of information transmission through a quantum communication channel can be described as a quantum operation. 

\begin{center}
\begin{figure*}[htbp]
\begin{center}
\includegraphics[angle = 0,width=0.7\linewidth]{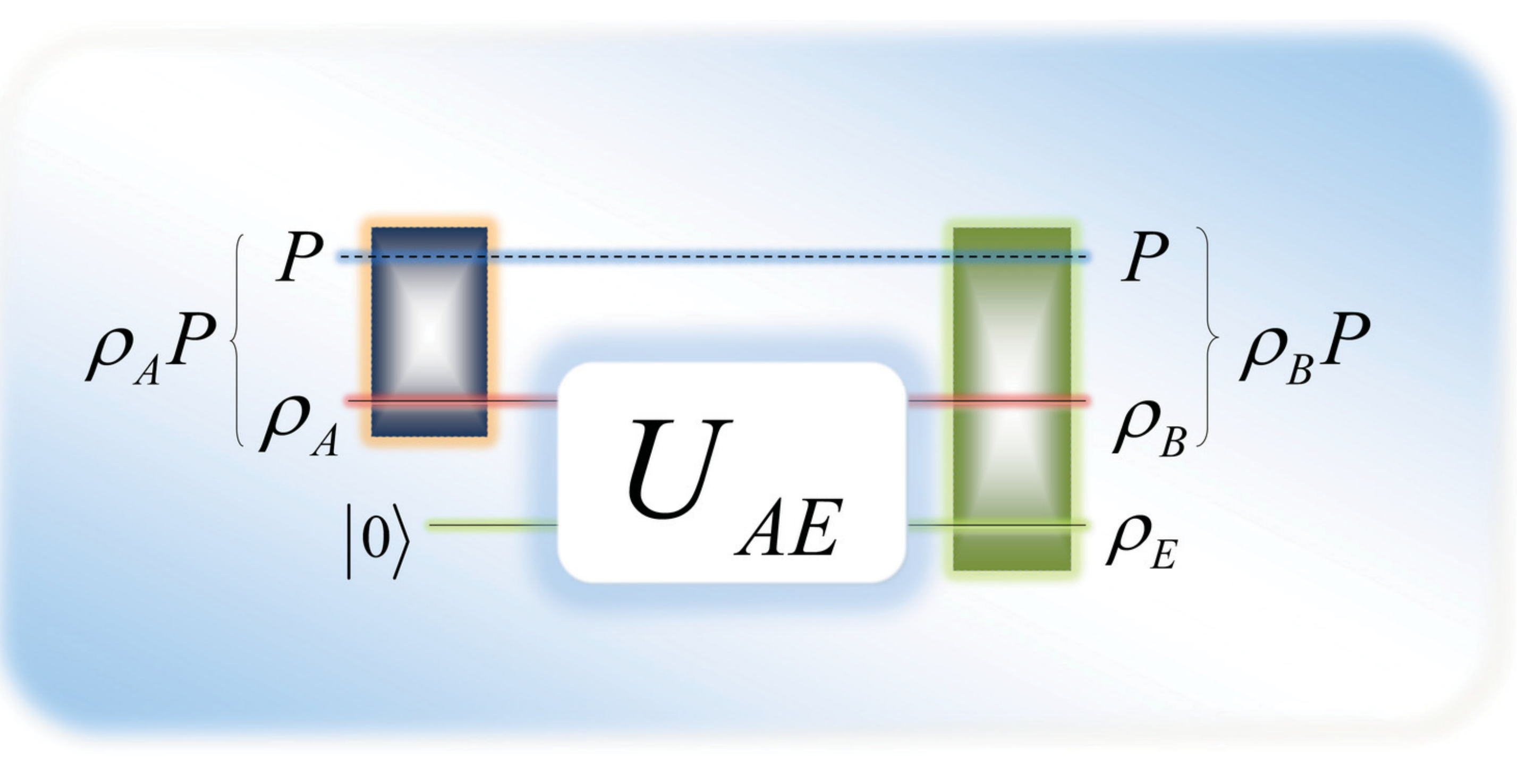}
\caption{The formal model of a noisy quantum communication channel. The output of the channel is a mixed state.} 
 \label{fig3_1}
 \end{center}
\end{figure*}
\end{center}

The general model of a quantum channel describes the transmission of an input quantum bit, and its interaction with the environment (see \fref{fig3_1}. Assuming Alice sends quantum state ${\rho }_A$ into the channel this state becomes entangled with the environment ${\rho }_E$, which is initially in a pure state $\left|\left.0\right\rangle \right.$.  For a mixed input state a so called \textit{purification} \textit{state} \textit{P} can be defined, from which the original mixed state can be restored by a partial trace (see Appendix) of the pure system ${\rho }_AP$. The unitary operation $U_{AE}$ of a quantum channel $\mathcal{N}$ entangles ${\rho }_AP$ with the environment ${\rho }_E$, and outputs Bob's mixed state as ${\rho }_B$ (and the purification state as $P$). The purification state is a reference system, it cannot be accessed, it remains the same after the transmission. 

The output of the noisy quantum channel is denoted by ${\rho }_B$, the post state of the environment by ${\rho }_E$, while the post-purification state after the output realized on the channel output is depicted by $P$.

\subsection{Capacity of Classical Channels}

 Before we start to investigate quantum channels, we survey the results of transmitting information over classical noisy channels. In order to achieve reliable (error-free) information transfer we use the so called \textit{channel codin}g which extends the payload (useful) information bits with redundancy bits so that at the receiver side Bob will be able to correct some amount of error by means of this redundancy.

The channel is given an input \textit{A, }and maps it probabilistically (it is a \textit{stochastic} mapping, not a unitary or deterministic transformation) to an output \textit{B, }and the probability of this mapping is denoted by $p\left(\left.B\right|A\right)$.

\begin{center}
\begin{figure*}[htbp]
\begin{center}
\includegraphics[angle = 0,width=0.7\linewidth]{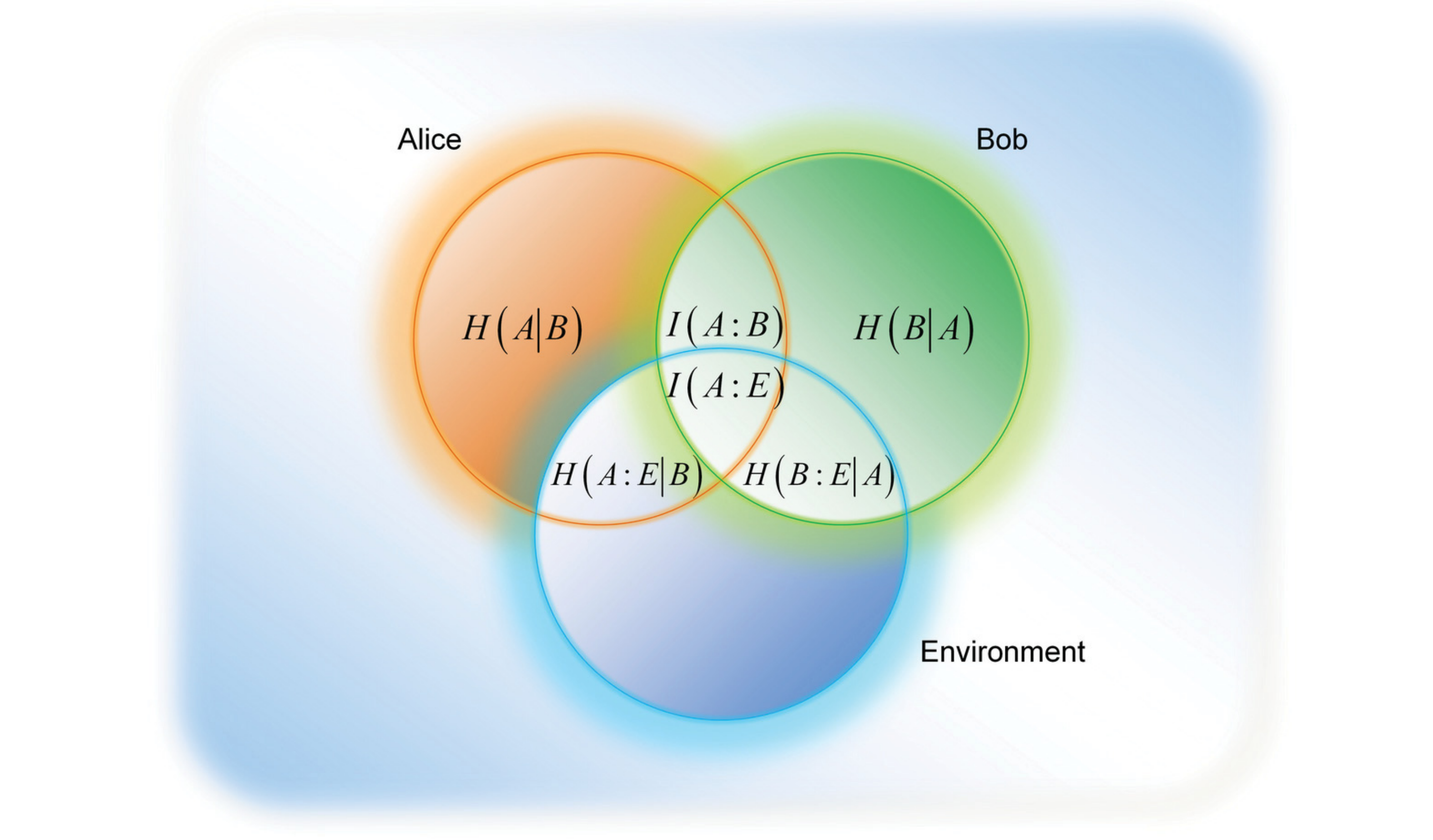}
\caption{The effects of the environment on the transmittable information and on the receiver's uncertainty.} 
\label{fig3_3}
 \end{center}
\end{figure*}
\end{center}

The \textit{channel capacity} $C\left(N\right)$ of a \textit{classical} memoryless communication channel \textit{N} gives an upper bound on the number of classical bits which can be transmitted per channel use, in reliable manner, i.e., with arbitrarily small error at the receiver.  
As it has been proven by Shannon the capacity\textit{ }$C\left(N\right)$ of a noisy classical memoryless communication channel \textit{N}, can be expressed by means of the maximum of the mutual information $I\left(A\mathrm{:}B\right)$ over all possible input distributions $p\left(x\right)$ of random variable \textit{X}
\begin{equation} \label{ZEqnNum789842} 
C\left(N\right)\mathrm{=}\mathop{\mathrm{max}}_{p\left(x\right)}I\left(A\mathrm{:}B\right).                                                            
\end{equation} 
In order to make the capacity definition more plausible let us consider \fref{fig3_3}. Here, the effect of environment \textit{E} is represented by the classical conditional entropies $H\left(\left.A\mathrm{:}E\right|B\right)\mathrm{>0}$ and $H\left(\left.B\mathrm{:}E\right|A\right)\mathrm{>0}$.                                                      

Shannon's noisy coding theorem claims that forming \textit{K} different codewords $m\mathrm{=l}\mathrm{og}K$ of length from the source bits and transmitting each of them using the channel \textit{n} times (\textit{m} to \textit{n} coding) the rate at which information can be transmitted through the channel is 
\begin{equation} \label{3.3)} 
R\mathrm{=}\frac{\mathrm{log}\left(K\right)}{n},                                                                                            
\end{equation} 
and exponentially small probability of error at this rate can be achieved only if $R\mathrm{\le }C\left(N\right)$, otherwise the probability of the successful decoding exponentially tends to zero, as the number of channel uses increases. Now, having introduced the capacity of classical channel it is important to highlight the following distinction. The asymptotic capacity of any channel describes that rate, which can be achieved if the channel can be used \textit{n} times (denoted by $N^{\mathrm{\otimes }n}$), where where $n\mathrm{\to }\mathrm{\infty }$. Without loss of generality, in case of $n\mathrm{=1}$ we speak about single-use capacity. Multiple channel uses can be implemented in consecutive or parallel ways, however from practical reasons we will prefer the latter one.

\subsection{Transmission of Classical Information over Noisy Quantum Channels}
\label{sec3c}
 As the next step during our journey towards the quantum information transfer through quantum channels (which is the most general case) we are leaving the well-known classical (macro) world and just entering into the border zone. Similar to the ancient Romans - who deployed a sophisticated wide border defense system (called \textit{the limes} which consisted of walls, towers, rivers, etc.), instead of drawing simply a red line between themselves and the barbarians -- we remain classical in terms of inputs and outputs but allow the channel operating in a quantum manner.

Quantum channels can be used in many different ways to transmit information from Alice to Bob. Alice can send classical bits to Bob, but she also has the capability of transmitting quantum bits. In the first case, we talk about the classical capacity of the quantum channel, while in the latter case, we have a different measure - the quantum capacity. The map of the channel is denoted by $\mathcal{N}$, which is trace preserving if 
\begin{equation} \label{3.4)} 
Tr\left(\mathcal{N}\left(\rho \right)\right)\mathrm{=}Tr\left(\rho \right) 
\end{equation} 
for all density matrices $\rho $, and positive if the eigenvalues of $\mathcal{N}\left(\rho \right)$ are non-negative whenever the eigenvalues of $\rho $ are non-negative. 

Compared to classical channels -- which have only one definition for capacity -- the transmittable classical information and thus the corresponding capacity definition can be different when one considers quantum channels. This fact splits the classical capacity of quantum channels into three categories, namely the (\textit{unentangled}) \textit{classical (}also known as the \textit{product-state} classical capacity, or the HSW (Holevo-Schumacher-Westmoreland) capacity\textit{) capacity }$C\left(\mathcal{N}\right)$, \textit{private classical capacity }$P\left(\mathcal{N}\right)$ and \textit{entanglement-assisted} \textit{classical} \textit{capacity }$C_E\left(\mathcal{N}\right)$.

The\textit{ }(\textit{unentangled}) \textit{classical capacity }$C\left(\mathcal{N}\right)$ is a natural extension of the capacity definition from classical channels to the quantum world. For the sake of simplicity the term \textit{classical capacity} will refer to the \textit{unentangled} version in the forthcoming pages of this paper. (The entangled version will be referred as the entanglement-assisted classical capacity. As we will see, the HSW capacity is defined for product state inputs; however it is possible to extend it for entangled input states)

The \textit{private classical capacity} $P\left(\mathcal{N}\right)$ has deep relevance in secret quantum communications and quantum cryptography. It describes the rate at which Alice is able to send classical information through the channel in secure manner. Security here means that an eavesdropper will not be able to access the encoded information without revealing her/himself.

The \textit{entanglement-assisted classical capacity} $C_E\left(\mathcal{N}\right)$ measures the classical information which can be transmitted through the channel, if Alice and Bob have already shared entanglement before the transmission. A well-known example of such protocols is `\textit{superdense coding}' [\cref{Imre05}]. Next, we discuss the above listed various classical capacities of quantum channels in detail.
 
As the first obvious generalization of classical channel capacity definition is if we maximize the quantum mutual information over all possible input ensembles 
\begin{equation} \label{3.5)} 
C\left(\mathcal{N}\right)\mathrm{=}\mathop{\mathrm{max}}_{allp_i,{\rho }_i}I\left(A\mathrm{:}B\right).                                                                            
\end{equation} 
Next, we start to discuss the classical information transmission capability of a noisy quantum channel.

\subsubsection{The Holevo-Schumacher-Westmoreland Capacity}

 The HSW (Holevo-Schumacher-Westmoreland) theorem defines the maximum of classical information which can be transmitted through a noisy quantum channel $\mathcal{N}$ if the input contains product states (i.e., entanglement is not allowed, also known as the product-state classical capacity) and the output is measured by joint measurement setting (see the \textit{second} measurement setting in subsection 3.3.2.1). In this setting, for the quantum noisy communication channel $\mathcal{N}$, the classical capacity can be expressed as follows
\begin{equation} \label{ZEqnNum611259} 
\begin{split}
C\left(\mathcal{N}\right)&\mathrm{=}\mathop{\mathrm{max}}_{allp_i,{\rho }_i}\chi \mathrm{=}\mathop{\mathrm{max}}_{allp_i,{\rho }_i}\left[\mathrm{S}\left({\sigma }_{out}\right)\mathrm{-}\sum_i{p_i\mathrm{S}\left({\sigma }_i\right)}\right] \\ 
\\&\mathrm{=}\mathop{\mathrm{max}}_{allp_i,{\rho }_i}\left[\mathrm{S}\left(\mathcal{N}\left(\sum_i{p_i{\rho }_i}\right)\right)\mathrm{-}\sum_i{p_i\mathrm{S}\left(\mathcal{N}\left({\rho }_i\right)\right)}\right] \\ 
&\mathrm{=}\chi \left(\mathcal{N}\right), 
\end{split}
\end{equation} 
where the maximum is taken over all ensembles $\left\{p_i,{\rho }_i\right\}$ of input quantum states, while for ${\sigma }_{out}$ see \eqref{ZEqnNum172273}, while $\chi \left(\mathcal{N}\right)$ is the Holevo capacity of $\mathcal{N}$. Trivially follows, that the $\chi \left(\mathcal{N}\right)$ capacity reaches its maximum for a perfect noiseless quantum channel $\mathcal{N}=I$. 

 If Alice chooses among a set of quantum codewords, then is it possible to transmit these codewords through the noisy quantum channel $\mathcal{N}$ to Bob with arbitrary small error, if
\begin{equation} \label{3.7)} 
R\mathrm{<}C\left(\mathcal{N}\right)\mathrm{=}\mathop{\mathrm{max}}_{allp_i,{\rho }_i}\left[\mathrm{S}\left(\mathcal{N}\left(\sum_i{p_i{\rho }_i}\right)\right)\mathrm{-}\sum_i{p_i\mathrm{S}\left(\mathcal{N}\left({\rho }_i\right)\right)}\right];                                                  
\end{equation} 
if Alice adjusts \textit{R} to be under $\mathop{\mathrm{max}}_{allp_i,{\rho }_i}\chi $, then she can transmit her codewords with arbitrarily small error. If Alice chooses $R\mathrm{>}C\left(\mathcal{N}\right),$then she cannot select a quantum code of arbitrary size, which was needed for her to realize an error-free communication. The HSW channel capacity guarantees an error-free quantum communication only if $R\mathrm{<}C\left(\mathcal{N}\right)\mathrm{=}\mathop{\mathrm{max}}_{allp_i,{\rho }_i}\chi $ is satisfied for her code rate \textit{R}.

\subsubsection{Various Classical Capacities of a Quantum Channel}
 The asymptotic channel capacity is the `true measure' of the various channel capacities, instead of the single-use capacity, which characterizes the capacity only in a very special case. The three classical capacities of the quantum channel of quantum channels will be discussed next. 

In the regularization step, the channel capacity is computed as a limit. In possession of this limit, we will use the following lower bounds for the single-use capacities. In Section 3.3.1 we have also seen, the \textit{Holevo-Schumacher-Westmoreland} theorem gives an explicit answer to the maximal transmittable classical information over the quantum channel. Next, we show the connection between these results. As we will see in subsection 3.3.2.1, four different measurement settings can be defined for the measurement of the \textit{classical} capacity of the quantum channel. Here we call the attention of the reader that Holevo bound \eqref{ZEqnNum240233} limits the classical information stored in a quantum bit. HSW theorem can be regarded a similar scenario but a quantum channel deployed between Alice and Bob introduces further uncertainty before extracting the classical information. Obviously if we assume an ideal channel the two scenarios become the same. 

Now, we present an example allowing the comparison of classical capacity of a simple channel model in classical and quantum context. The binary symmetric channel inverts the input cbits with probability \textit{p} and leaves it unchanged with (1-\textit{p}). The equivalent quantum bit flip channel (see \sref{sec5}) applies the Pauli \textit{X} and the identity transforms \textit{I.} 

Considering the worst case $p\mathrm{=0.5}$ all the sent information vanishes in the classical channel $C\mathrm{(}N\mathrm{)=1-}H\left(p\right)\mathrm{=0}$. However, the HSW theorem enables the optimization not only over the input probabilities but over input ensembles $\left\{p_i,{\rho }_i\right\}$. If we set ${\rho }_i$ to the eigenvectors of Pauli \textit{X} deriving them from its spectral decomposition 
\begin{equation} \label{3.8)} 
X\mathrm{=1}\left|\left.\mathrm{+}\right\rangle \right.\left\langle \left.\mathrm{+}\right|\right.\mathrm{+(-1)}\left|\left.\mathrm{-}\right\rangle \right.\left\langle \left.\mathrm{-}\right|\right.,                                   
\end{equation} 
where $\left|\left.\mathrm{\pm }\right\rangle \right.\mathrm{=}\frac{\left|\left.0\right\rangle \right.\mathrm{\pm }\left|\left.\mathrm{1}\right\rangle \right.}{\sqrt{\mathrm{2}}}$,  $C\left(\mathcal{N}\right)\mathrm{=1}$ can be achieved. This results is more than surprising, encoding into quantum states in certain cases may improve the transfer of classical information between distant points i.e., the increased degree of freedom enables reducing the uncertainty introduced by the channel.

\paragraph{Measurement Settings}

\begin{center}
\begin{figure}[htbp]
\begin{center}
\includegraphics[angle = 0,width=1\linewidth]{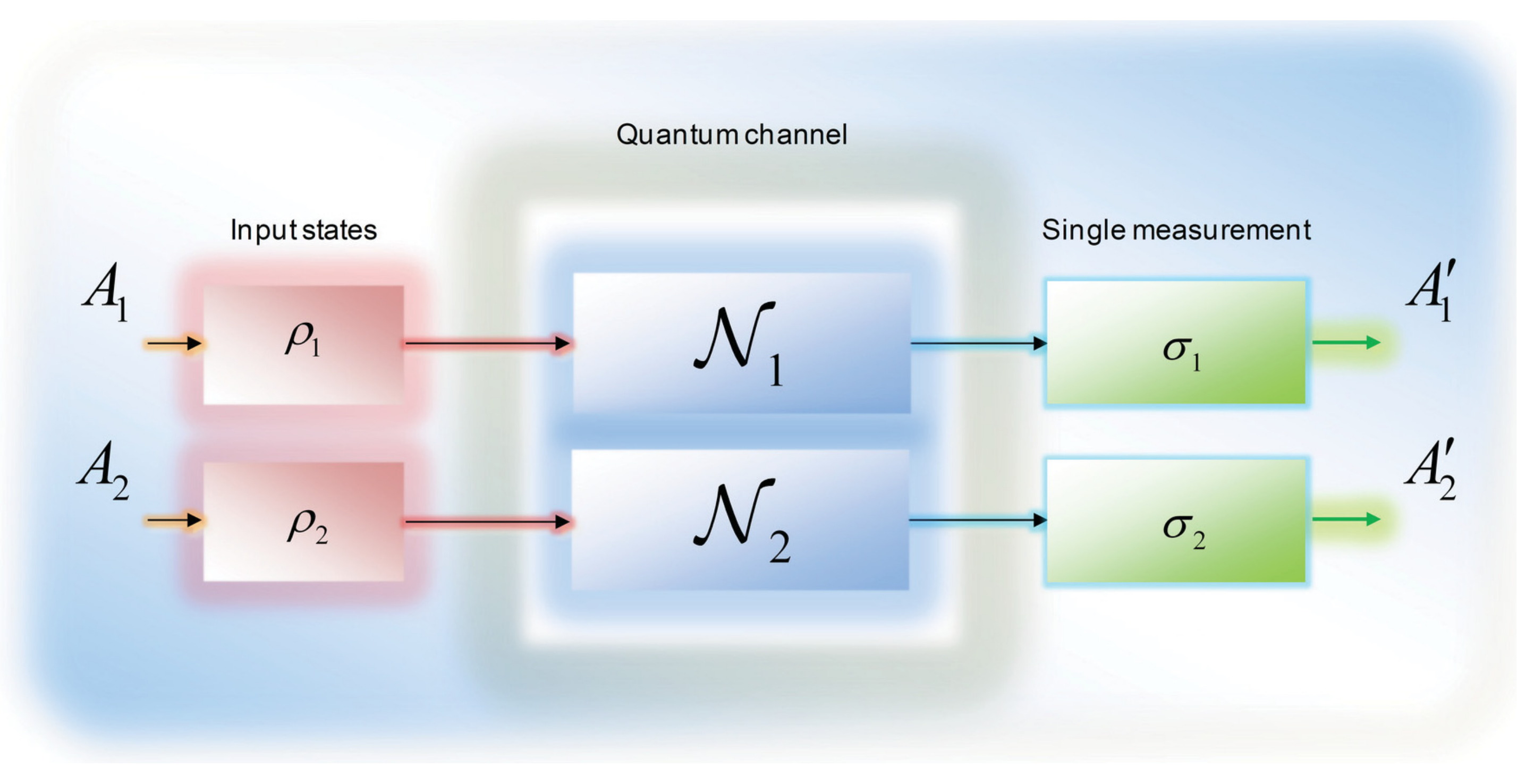}
\caption{Transmission of classical information over quantum channel with product state inputs and single measurements. Environment is not depicted.} 
 \label{fig3_5}
 \end{center}
\end{figure}
\end{center}

 Similar to classical channel encoding, the quantum states can be transmitted in codewords \textit{n} qubit of length using the quantum channel consecutively \textit{n}-times or equivalently we can send codewords over \textit{n} copies of quantum channel $\mathcal{N}$ denoted by ${\mathcal{N}}^{\mathrm{\otimes }n}$. For the sake of simplicity we use $n\mathrm{=2}$ in the figures belonging to the following explanation. In order to make the transient smoother between the single-shot and the asymptotic approaches we depicted the scenario using \textit{product input states} and \textit{single }(or independent) measurement devices at the output of the channel in \fref{fig3_5}. In that case the $C\left(\mathcal{N}\right)$ classical capacity of quantum channel $\mathcal{N}$ with input \textit{A} and output \textit{B} can be expressed by the maximization of the $I\left(A\mathrm{:}B\right)$ quantum mutual information as follows: 
\begin{equation} \label{ZEqnNum654243} 
C\left(\mathcal{N}\right)\mathrm{=}\mathop{\mathrm{max}}_{allp_i,{\rho }_i}I\left(A\mathrm{:}B\right). 
\end{equation} 

 From \eqref{ZEqnNum654243} also follows that for this setting the single-use $C^{\left(\mathrm{1}\right)}\left(N\right)$ and the asymptotic $C\left(\mathcal{N}\right)$ classical capacities are equal:
\begin{equation} \label{3.10)} 
C^{\left(\mathrm{1}\right)}\left(N\right)\mathrm{=}C\left(\mathcal{N}\right)\mathrm{=}\mathop{\mathrm{max}}_{allp_i,{\rho }_i}I\left(A\mathrm{:}B\right). 
\end{equation} 
On the other hand, if we have \textit{product state inputs} but we change the measurement setting from the single measurement setting to\textit{ joint measurement} setting, then the classical channel capacity cannot be given by \eqref{ZEqnNum654243}, hence 
\begin{equation} \label{3.11)} 
C\left(\mathcal{N}\right)\mathrm{\ne }\mathop{\mathrm{max}}_{allp_i,{\rho }_i}I\left(A\mathrm{:}B\right).                                                       
\end{equation} 
If we would like to step forward, we have to accept the fact, that the quantum mutual information cannot be used to express the asymptotic version: the\textit{ maximized }quantum mutual information is\textit{ always additive} (see \sref{sec2}) - but not the Holevo information. As follows, if we would take the regularized form of quantum mutual information to express the capacity, we will find that the asymptotic version is equal to the single-use version, since:
\begin{equation} \label{ZEqnNum465961} 
\mathop{\mathrm{lim}}_{n\mathrm{\to }\mathrm{\infty }}\frac{\mathrm{1}}{n}\mathop{\mathrm{max}}_{allp_i,{\rho }_i}I\left(A\mathrm{:}B\right)\mathrm{=}\mathop{\mathrm{max}}_{allp_i,{\rho }_i}I\left(A\mathrm{:}B\right).                                                          
\end{equation} 
From \eqref{ZEqnNum465961} follows, that if we have \textit{product inputs }and\textit{ joint measurement} at the outputs, we cannot use the $\mathop{\mathrm{max}}_{allp_i,{\rho }_i}I\left(A\mathrm{:}B\right)$ maximized quantum mutual information function to express $C\left(\mathcal{N}\right)$. If we would like to compute the classical capacity $C\left(\mathcal{N}\right)$ for that case, we have to leave the quantum mutual information function, and instead of it we have to use the maximized Holevo information $\mathop{\mathrm{max}}_{allp_i,{\rho }_i}\chi $.

 This new $C\left(\mathcal{N}\right)$ capacity (according to the \textit{Holevo-Schumacher-Westmoreland} theorem) can be expressed by the Holevo capacity $\chi \left(\mathcal{N}\right)$, which will be equal to the maximization of Holevo information of channel $\mathcal{N}$: 
\begin{equation} \label{3.13)} 
C\left(\mathcal{N}\right)\mathrm{=}\chi \left(\mathcal{N}\right)\mathrm{=}\mathop{\mathrm{max}}_{allp_i,{\rho }_i}\chi .                                                                                      
\end{equation} 
The Holevo capacity and the asymptotic channel capacity will be equal in this case.  

\begin{center}
\begin{figure}[htbp]
\vspace{-0.6cm}
\begin{center}
\includegraphics[angle = 0,width=1\linewidth]{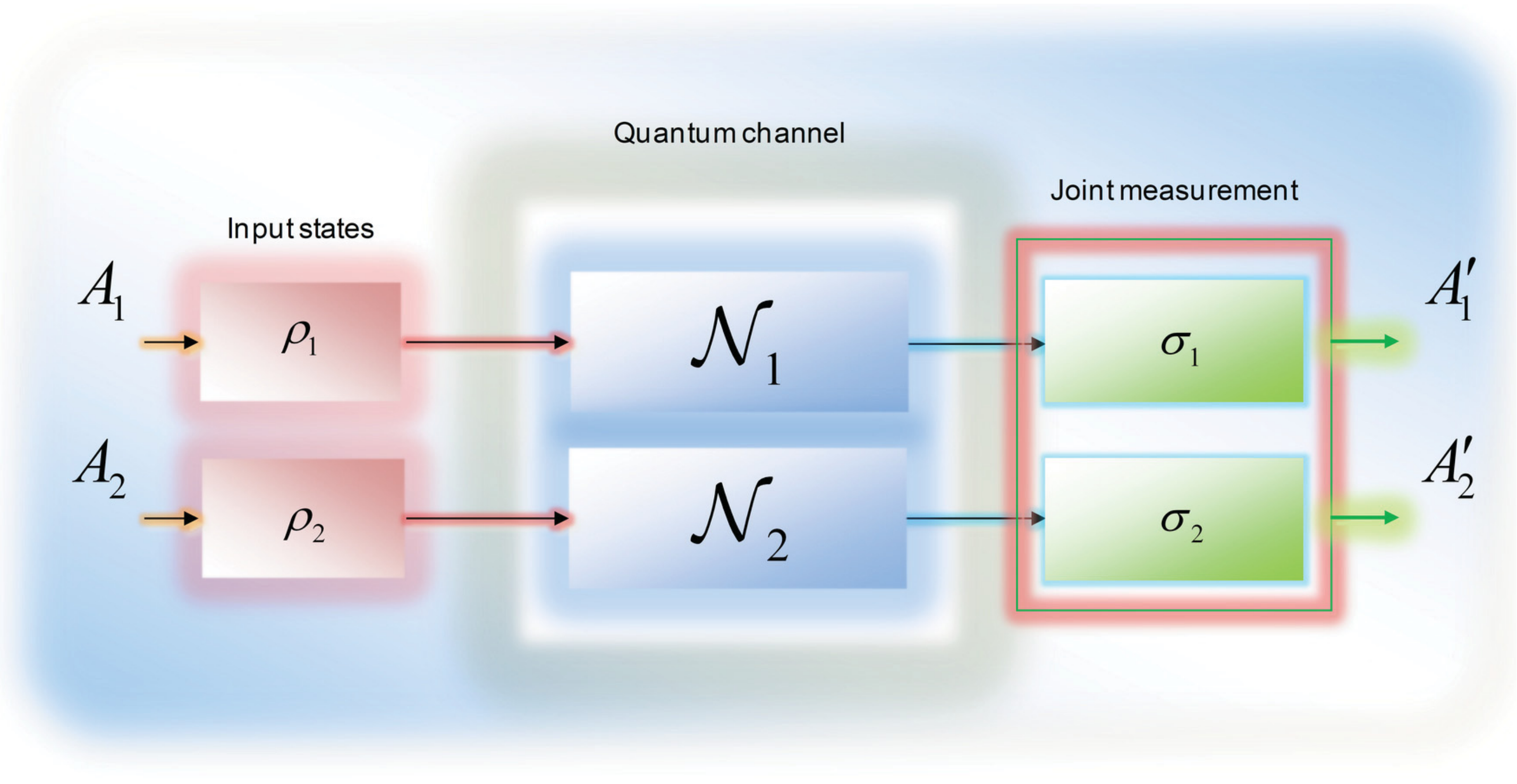}
\caption{Transmission of classical information over quantum channel with product state inputs and joint measurements. Environment is not depicted.} 
 \label{fig3_6}
 \end{center}
\end{figure}
\end{center}

The HSW theorem gives an explicit answer for the classical capacity of the \textit{product state input} with \textit{joint measurement} setting, and expresses $C\left(\mathcal{N}\right)$ as follows: 
\begin{equation} \label{ZEqnNum905584} 
\begin{split}
C\left(\mathcal{N}\right)&\mathrm{=}\chi \left(\mathcal{N}\right)\\&\mathrm{=}\mathop{\mathrm{max}}_{allp_i,{\rho }_i}\left[\mathrm{S}\left(\mathcal{N}\left(\sum_i{p_i{\rho }_i}\right)\right)\mathrm{-}\sum_i{p_i\mathrm{S}\left(\mathcal{N}\left({\rho }_i\right)\right)}\right].                                              
\end{split}
\end{equation} 
The relation discussed above holds for the restricted channel setting illustrated in \fref{fig3_6}, where the input consists of product states, and the output is measured by a joint measurement setting. 

However, if \textit{entangled inputs} are allowed with the \textit{joint measurement setting} - then this equality does not hold anymore. As a conclusion, the relation between the maximized Holevo information $\chi \left(\mathcal{N}\right)$ of the channel of the channel and the asymptotic classical channel capacity $C\left(\mathcal{N}\right)$: 
\begin{equation} \label{3.15)} 
\chi \left(\mathcal{N}\right)\mathrm{\le }C\left(\mathcal{N}\right).                                                                       
\end{equation} 
This means that we have to redefine the asymptotic formula of $C\left(\mathcal{N}\right)$ for entangled inputs and joint measurement setting, to measure the maximum transmittable classical information through a quantum channel.  

 In the 1990s, it was conjectured that the formula of \eqref{ZEqnNum905584} can be applied to describe the channel capacity for entangled inputs with the \textit{single measurement} setting; however it was an open question for a long time. Single measurement \textit{destroys} the possible benefits arising from the entangled inputs, and joint measurement is required to achieve the benefits of entangled inputs [\cref{King2000}]. 

\begin{center}
\begin{figure}[htbp]
\begin{center}
\includegraphics[angle = 0,width=1\linewidth]{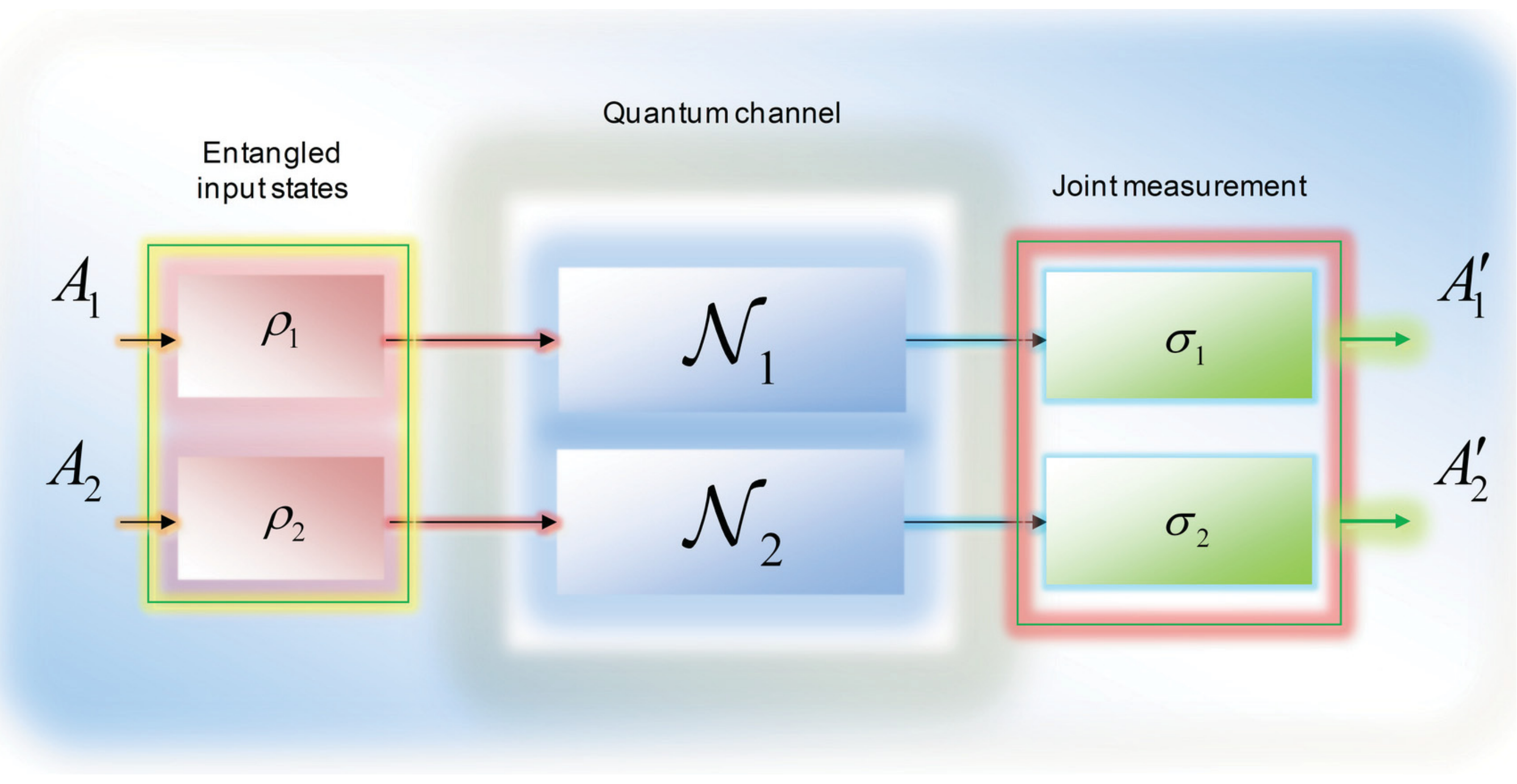}
\caption{Transmission of classical information over quantum channel with entangled inputs ${\mathrm{\Psi }}_{\mathrm{12}}$ and joint measurements. Environment is not depicted.} 
 \label{fig3_7}
 \end{center}
\end{figure}
\end{center}

 In 2009 Hastings have used \textit{entangled input} \textit{states} and showed that the entangled inputs (with the \textit{joint measurement}) can increase the amount of classical information which can be transmitted over a noisy quantum channel. In this case, $C\left(\mathcal{N}\right)\mathrm{\ne }\chi \left(\mathcal{N}\right)$ and the $C\left(\mathcal{N}\right)$ can be expressed with the help of Holevo capacity as follows, using the asymptotic formula of $\chi \left(\mathcal{N}\right)$: 
\begin{equation} \label{ZEqnNum956947} 
C\left(\mathcal{N}\right)\mathrm{=}\mathop{\mathrm{lim}}_{n\mathrm{\to }\mathrm{\infty }}\frac{\mathrm{1}}{n}\chi \left({\mathcal{N}}^{\mathrm{\otimes }n}\right).                                                                                    
\end{equation} 
The channel construction for this relation is illustrated in \fref{fig3_7}. The entangled input is formally denoted by ${\mathrm{\Psi }}_{\mathrm{12}}$.

 We also show the channel construction of the fourth possible construction to measure the classical capacity of a quantum channel. In this case, we have entangled input states, however we use a single measurement setting instead of a joint measurement setting. 

\begin{center}
\begin{figure}[htbp]
\vspace{-0.5cm}
\begin{center}
\includegraphics[angle = 0,width=1\linewidth]{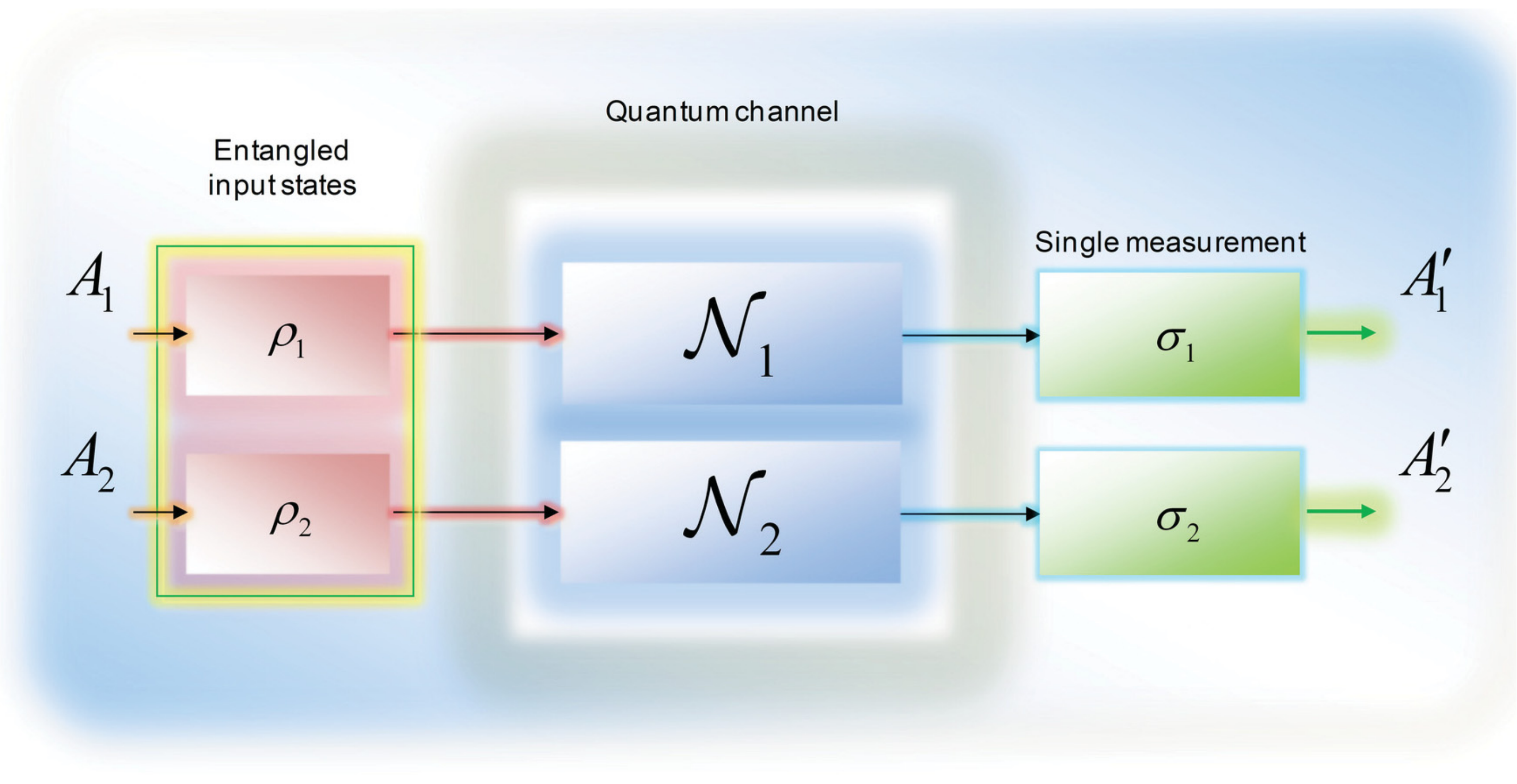}
\caption{Transmission of classical information over quantum channel with entangled inputs and single measurements. Environment is not depicted.} 
 \label{fig3_8}
 \end{center}
\end{figure}
\end{center}

To our knowledge, currently there is no quantum channel model where the channel capacity can be increased with this setting, since in this case the benefits of entanglement vanish because of the joint measurement setting has been changed into the single measurement setting. We illustrated this setting in \fref{fig3_8}.

We have seen in \eqref{ZEqnNum905584}, that if we have \textit{product input states} and we change from a single to a \textit{joint measurement} setting, then the classical capacity of $\mathcal{N}$ cannot be expressed by the maximized quantum mutual information function, because it is always additive, hence
\begin{equation} \label{3.17)} 
C\left(\mathcal{N}\right)\mathrm{\ne }\mathop{\mathrm{lim}}_{n\mathrm{\to }\mathrm{\infty }}\frac{\mathrm{1}}{n}\mathop{\mathrm{max}}_{allp_i,{\rho }_i}I\left(A\mathrm{:}B\right). 
\end{equation} 
If we allow \textit{entangled input states} and\textit{ joint measurement} (see \eqref{ZEqnNum956947}), then we have to use the $C\left(\mathcal{N}\right)$ asymptotic formula of the previously derived Holevo capacity, $\chi \left(\mathcal{N}\right)$ which yields
\begin{equation} \label{3.18)} 
C\left(\mathcal{N}\right)\mathrm{=}\mathop{\mathrm{lim}}_{n\mathrm{\to }\mathrm{\infty }}\frac{\mathrm{1}}{n}\chi \left({\mathcal{N}}^{\mathrm{\otimes }n}\right)\mathrm{\ne }\chi \left(\mathcal{N}\right).                                                                     
\end{equation} 

\subsubsection{Brief Summary}
The Holevo quantity measures the classical information, which remains in the encoded quantum states after they have transmitted through a noisy quantum channel. During the transmission, some information passes to the environment from the quantum state, which results in the increased entropy of the sent quantum state. The HSW theorem states very similar to Holevo's previous result. As in the case of the Holevo quantity, the HSW capacity measures the classical capacity of a noisy quantum channel - however, as we will see in \sref{sec4}, the Holevo quantity also can be used to express the quantum capacity of the quantum channel, which is a not trivial fact. The HSW capacity maximizes the Holevo quantity over a set of possible input states, and expresses the classical information, which can be sent through \textit{reliably} in the form of \textit{product input states} over the noisy quantum channel, hence HSW capacity is also known as \textit{product state} \textit{channel capacity}. In this case, the input states are not entangled; hence there is no entanglement between the multiple uses of the quantum channel. As we have seen in this section, if the input of the channel consists of product states and we use \textit{single measurement} setting, then the classical capacity can be expressed as the maximized of the quantum mutual information. On the other hand, if the single measurement has been changed to \textit{joint measurement}, this statement is not true anymore; - this capacity will be equal to HSW capacity, see \eqref{ZEqnNum905584}. Moreover, if we step forward, and we allow \textit{entanglement} among the input states, then we cannot use anymore the HSW capacity, which was defined in \eqref{ZEqnNum611259}. In this case we have to take its asymptotic formula, which was shown in \eqref{ZEqnNum956947}. 

Next we discuss the private classical capacity of quantum channels.

\subsection{The Private Classical Capacity}
\label{sec3d}
The private classical capacity $P\left(\mathcal{N}\right)$ of a quantum channel $\mathcal{N}$ describes the maximum rate at which the channel is able to send \textit{classical information} through the channel reliably and \textit{privately }(i.e., without any information leaked about the original message to an eavesdropper). Privately here means that an eavesdropper will not be able to access the encoded information without revealing her/himself i.e., the private classical capacity describes the maximal secure information that can be obtained by Bob on an eavesdropped quantum communication channel.

\begin{center}
\begin{figure}[htbp]
\vspace{-0.6cm}
\begin{center}
\includegraphics[angle = 0,width=\linewidth]{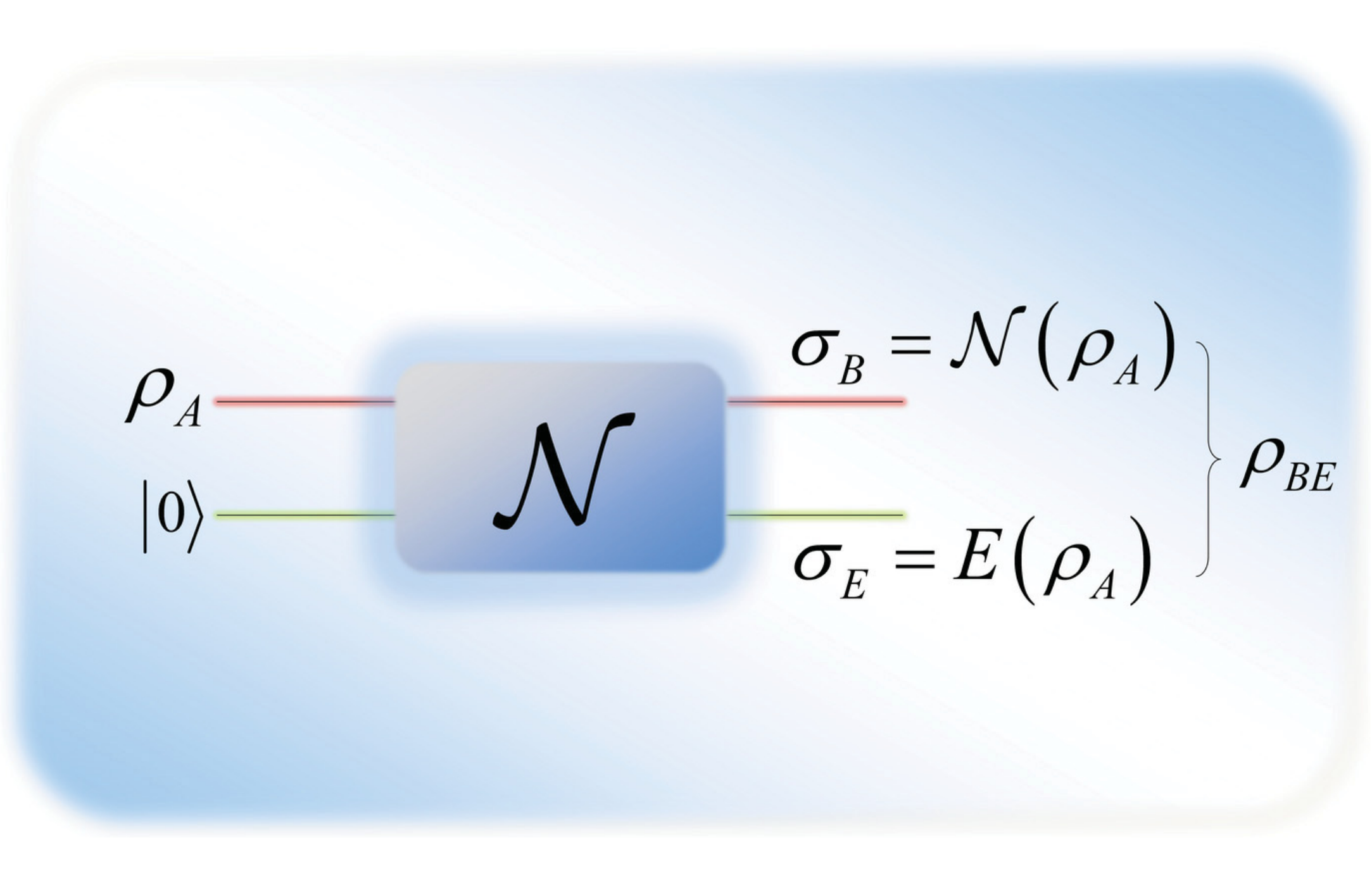}
\caption{The model of private classical communication of a quantum channel.} 
\label{fig3_11}
 \end{center}
\end{figure}
\end{center}

The generalized model of the private communication over quantum channels is illustrated in \fref{fig3_11}. The first output of the channel is denoted by ${\sigma }_B\mathrm{=}\mathcal{N}\left({\rho }_A\right)$, the second `receiver' is the eavesdropper \textit{E}, with state ${\sigma }_E$. The single-use private classical capacity from these quantities can be expressed as the maximum of the difference between two mutual information quantities. The eavesdropper, Eve, attacks the quantum channel, and she steals $I\left(A\mathrm{:}E\right)$ from the information $I\left(A\mathrm{:}B\right)$ sent by Alice to Bob, therefore the \textit{single-use} private classical capacity (or\textit{ private information}) of $\mathcal{N}$ can be determined asl
\begin{equation}  
P^{\left(\mathrm{1}\right)}\left(\mathcal{N}\right)\mathrm{=}\mathop{\mathrm{max}}_{allp_i,{\rho }_i}\left(I\left(A\mathrm{:}B\right)\mathrm{-}I\left(A\mathrm{:}E\right)\right).                                                         
\end{equation} 
while the \textit{asymptotic} private classical capacity is 
\begin{equation} \label{ZEqnNum430626} 
\begin{split}
P\left( \mathcal{N} \right)&=\underset{n\to \infty }{\mathop{\text{lim }}}\,\frac{1}{n}{{P}^{\left( 1 \right)}}\left( {{\mathcal{N}}^{\otimes n}} \right)\\&=\underset{n\to \infty }{\mathop{\text{lim }}}\,\frac{1}{n}\underset{all\text{ }{{p}_{i}},{{\rho }_{i}}}{\mathop{\text{max }}}\,\left( I\left( A:B \right)-I\left( A:E \right) \right).                              
\end{split}
\end{equation} 
The private classical capacity can be expressed as the difference of two quantum mutual information functions, see \eqref{ZEqnNum430626}. Here, we give an equivalent definition for private classical capacity $P\left(\mathcal{N}\right)$ and show, that it also can be rewritten using the Holevo quantity $\mathcal{X}$, as follows: 
\begin{equation} \label{ZEqnNum390392} 
P\left(\mathcal{N}\right)\mathrm{=}\mathop{\mathrm{lim}}_{n\mathrm{\to }\mathrm{\infty }}\frac{\mathrm{1}}{n}\mathop{\mathrm{max}}_{allp_i,{\rho }_i}\left({\mathcal{X}}_{AB}\mathrm{-}{\mathcal{X}}_{AE}\right),                                                           
\end{equation} 
where 
\begin{equation} \label{3.22)} 
{\mathcal{X}}_{AB}\mathrm{=S}\left({\mathcal{N}}_{AB}\left({\rho }_{AB}\right)\right)\mathrm{-}\sum_i{p_i\mathrm{S}\left({\mathcal{N}}_{AB}\left({\rho }_i\right)\right)} 
\end{equation} 
and 
\begin{equation} \label{3.23)} 
{\mathcal{X}}_{AE}\mathrm{=S}\left({\mathcal{N}}_{AE}\left({\rho }_{AE}\right)\right)\mathrm{-}\sum_i{p_i\mathrm{S}\left({\mathcal{N}}_{AE}\left({\rho }_i\right)\right)} 
\end{equation} 
measure the Holevo quantities between Alice and Bob, and Alice and the eavesdropper Eve, respectively, while ${\rho }_{AB}\mathrm{=}\sum_i{p_i{\rho }_i}$ and ${\rho }_{AE}\mathrm{=}\sum_i{p_i{\rho }_i}$. An important corollary from \eqref{ZEqnNum430626}, while the quantum mutual information itself is additive (see the properties of quantum mutual information function in \sref{sec2}), the difference of two quantum mutual information functions is not (i.e., we need the asymptotic version to compute the `true' private classical capacity of a quantum channel.)

\subsection{The Entanglement-assisted Classical Capacity}
\label{sec3e}
\begin{center}
\begin{figure*}[htbp]
\begin{center}
\includegraphics[angle = 0,width=0.6\linewidth]{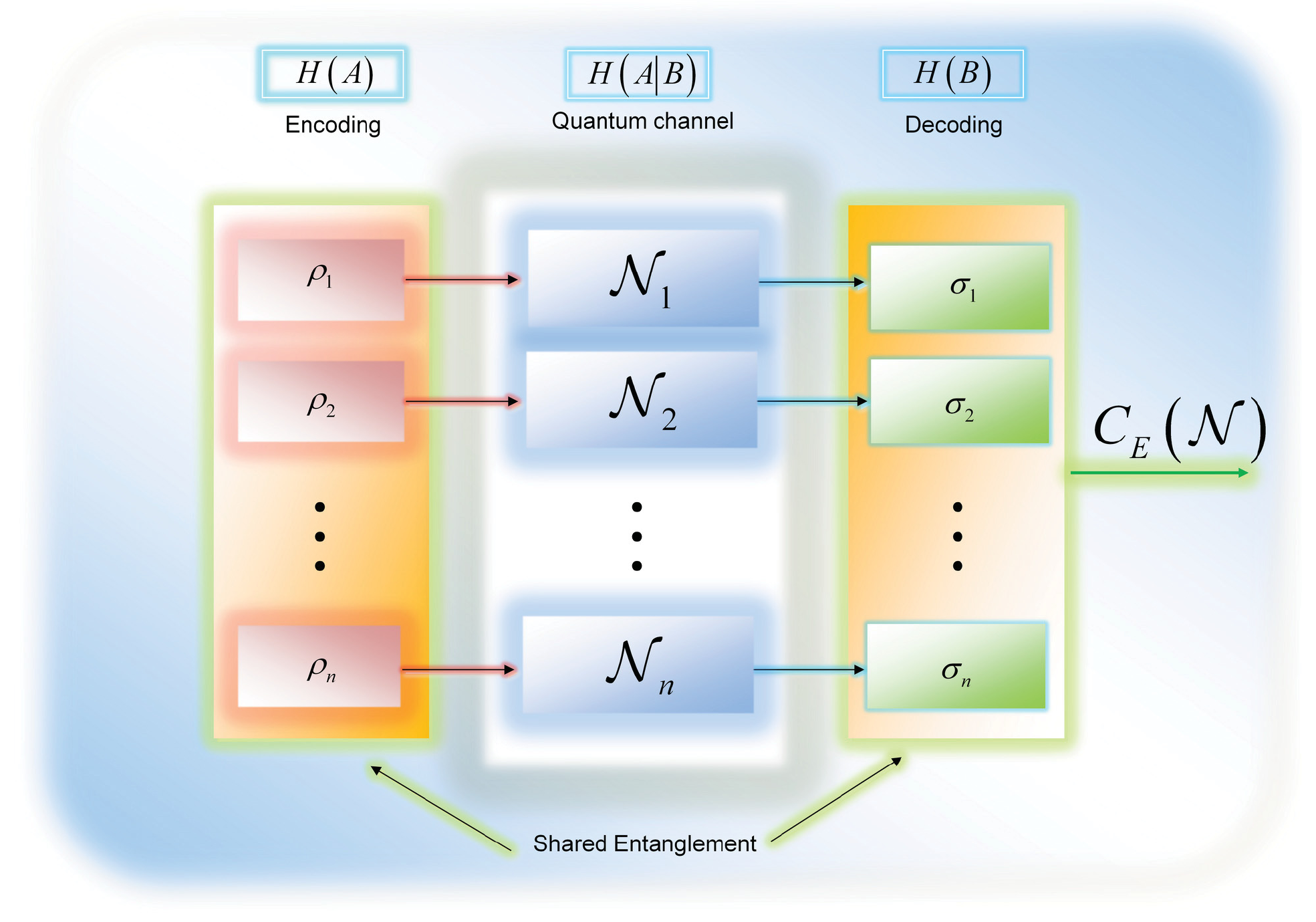}
\caption{The entanglement-assisted capacity of a quantum channel. This capacity measures the maximum of transmittable classical information through a quantum channel, if shared a priori entanglement between the parties is allowed.} 
\label{fig3_12}
 \end{center}
\end{figure*}
\end{center}

The last capacity regarding classical communication over quantum channels is called \textit{entanglement-assisted classical capacity} $C_E\left(\mathcal{N}\right)$, which measures the classical information which can be transmitted through the channel, if Alice and Bob have shared entanglement before the transmission i.e., entanglement is applied not between the input states like in case of the HSW (i.e., the product-state capacity) theorem. This capacity measures classical information, and it can be expressed with the help of the \textit{quantum mutual information function} (see \sref{sec2}) as
\begin{equation} \label{ZEqnNum643113} 
C_E\left(\mathcal{N}\right)\mathrm{=}\mathop{\mathrm{max}}_{allp_i,{\rho }_i}I\left(A\mathrm{:}B\right).                                                                        
\end{equation} 
The main difference between the classical capacity $C\left(\mathcal{N}\right)$ and the entanglement-assisted classical capacity $C_E\left(\mathcal{N}\right)$, is that in the latter case the maximum of the transmittable classical information is equal to the maximized quantum mutual information, - hence the entanglement-assisted classical capacity $C_E\left(\mathcal{N}\right)$ can be derived from the \textit{single-use} version $C^{\left(\mathrm{1}\right)}_E\left(\mathcal{N}\right)$. From \eqref{ZEqnNum643113} the reader can conclude, there is no need for the asymptotic version to express the entanglement-assisted classical capacity, i.e.:
\begin{equation} \label{3.25)} 
C_E\left(\mathcal{N}\right)\mathrm{=}C^{\left(\mathrm{1}\right)}_E\left(\mathcal{N}\right)\mathrm{=}\mathop{\mathrm{max}}_{allp_i,{\rho }_i}I\left(A\mathrm{:}B\right).                                         
\end{equation} 
It also can be concluded, that shared entanglement does not change the additivity of maximized quantum mutual information - or with other words, it remains true if the parties use shared entanglement for the transmission of classical information over $\mathcal{N}$. In \fref{fig3_12} we illustrate the general model of entanglement-assisted classical capacity $C_E\left(\mathcal{N}\right)$. 

 We note an important property of shared entanglement: while it does not provide any benefits in the improving of the classical capacity of the quantum channel, (see \eqref{ZEqnNum643113}), it can be used to increase the single-use classical capacity. It was shown, that with the help of shared entanglement the transmission of a single quantum bit can be realized with higher success probability, - this strategy is known as the CHSH (\textit{Clauser-Horne-Shimony-Holt}) game, for details see [\cref{Imre05}].

\paragraph{Brief Summary of Classical Capacities}

 Here, we give a brief summarization on the classical capacities. For the \textit{asymptotic} capacity of a quantum channel, we have
\begin{equation} \label{3.26)} 
C\left(\mathcal{N}\right)\mathrm{\ge }\chi \left(\mathcal{N}\right).                                             
\end{equation} 
According to the results of Holevo-Schumacher-Westmoreland, the asymptotic classical capacity is not equal to the single-use classical capacity. The \textit{asymptotic} formula of the classical capacity $C\left(\mathcal{N}\right)$ can be expressed by the help of the Holevo capacity $\chi \left(\mathcal{N}\right)$ as
\begin{equation} \label{3.27)} 
C\left(\mathcal{N}\right)\mathrm{=}\mathop{\mathrm{lim}}_{n\mathrm{\to }\mathrm{\infty }}\frac{\mathrm{1}}{n}\chi \left({\mathcal{N}}^{\mathrm{\otimes }n}\right).                                                      
\end{equation} 
The difference between the single-use formula and the asymptotic formula holds for the private capacity $P\left(\mathcal{N}\right)$. Unlike these capacities, in the case of entanglement-assisted classical capacity $C_E\left(\mathcal{N}\right)$, we will find something else in the expression. In this case, we have 
\begin{equation} \label{3.28)} 
C_E\left(\mathcal{N}\right)\mathrm{=}C^{\left(\mathrm{1}\right)}_E\left(\mathcal{N}\right)\mathrm{=}\mathop{\mathrm{max}}_{allp_i,{\rho }_i}I\left(A\mathrm{:}B\right),                                                     
\end{equation} 
and so we can conclude, \textit{there is no regularization}. Since there is no regularization needed, it also means that the entanglement-assisted classical capacity $C_E\left(\mathcal{N}\right)$ will always be additive. This makes it easier to compute the entanglement-assisted capacity than the other formulas, in which regularization is needed.

Originally, it was conjectured that in the general case, the Holevo information $\chi $ is additive too, for the same channels. Later, a counterexample was found by Hastings. As has been shown, in this case the additivity of the Holevo information fails. 

Similarly, for the $P\left(\mathcal{N}\right)$ private classical capacity, - which also measures classical information we have  
\begin{equation} \label{3.29)} 
P\left(\mathcal{N}\right)\mathrm{\ge }\mathop{\mathrm{max}}_{allp_i,{\rho }_i}\left(I\left(A\mathrm{:}B\right)\mathrm{-}I\left(A\mathrm{:}E\right)\right),                                                        
\end{equation} 
and finally, for the classical capacity $C\left(\mathcal{N}\right)$ of $\mathcal{N}$
\begin{equation} \label{3.30)} 
\mathop{\mathrm{max}}_{allp_i,{\rho }_i}I\left(A\mathrm{:}B\right)\mathrm{\le }C\left(\mathcal{N}\right)\mathrm{\le }\mathop{\mathrm{lim}}_{n\mathrm{\to }\mathrm{\infty }}\frac{\mathrm{1}}{n}\chi \left({\mathcal{N}}^{\mathrm{\otimes }n}\right).                                                
\end{equation} 
As can be seen, in case of the classical and private classical capacities the regularization is needed, since the asymptotic and the single-use formulas are not equal. 
\subsection{The Classical Zero-Error Capacity}
\label{sec3f}
Shannon's results on capacity [\cref{Shannon48}] guarantees transmission rate only in average when using multiple times of the channel. The zero-error capacity of the quantum channel describes the amount of (classical or quantum) information which can be transmitted \textit{perfectly} (\textit{zero probability of error}) through a noisy quantum channel. The zero-error capacity of the quantum channel could have an overriding importance in future quantum communication networks. 

 The zero-error capacity stands a very strong requirement in comparison to the standard capacity where the information transmission can be realized with asymptotically small \textit{but non-vanishing} error probability, since in the case of zero-error communication the \textit{error probability of the communication has to be zero}, hence the transmission of information has to be perfect and no errors are allowed. While in the case of classical non zero-error capacity for an \textit{n}-length code the error probabilities after the decoding process are $\mathrm{Pr}\left[error\right]\mathrm{\to }\mathrm{0}$ as $n\mathrm{\to }\mathrm{\infty }$, in case of an \textit{n}-length zero-error code, $\mathrm{Pr}\left[error\right]\mathrm{=0}$. 

 In this subsection we give the exact definitions which required for the characterization of a quantum zero-error communication system. We will discuss the classical and quantum zero-error capacities and give the connection between zero-error quantum codes and the elements of graph theory.

\subsubsection{Classical Zero-Error Capacities of Quantum Channels}

\begin{center}
\begin{figure*}[htbp]
\begin{center}
\includegraphics[angle = 0,width=0.7\linewidth]{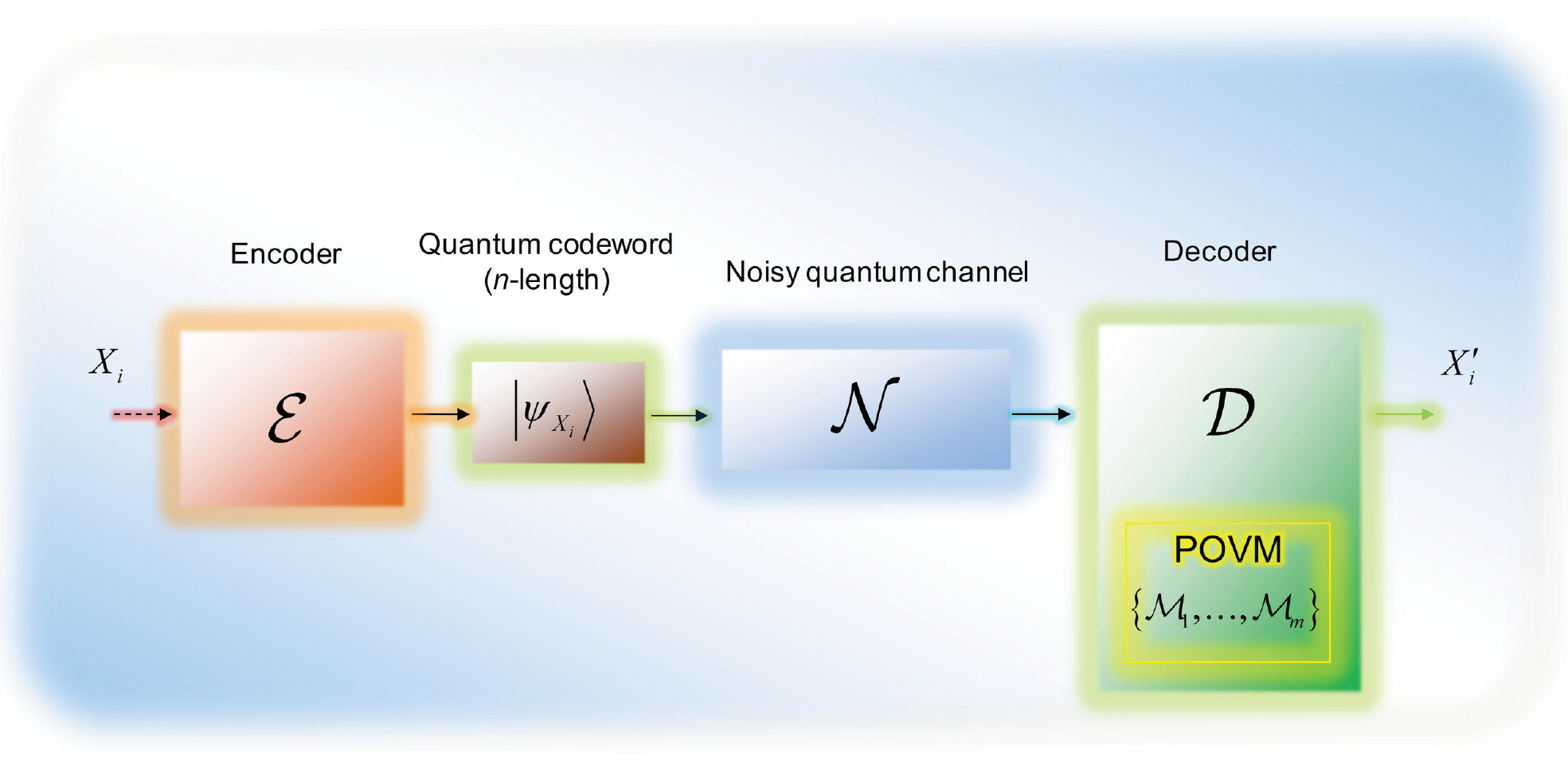}
\caption{A quantum zero-error communication system.} 
\label{fig3_13}
 \end{center}
\end{figure*}
\end{center}

 In this section we review the background of zero-error capacity $C_0\left(\mathcal{N}\right)$ of a quantum channel $\mathcal{N}$. Let us assume that Alice has information source $\left\{X_i\right\}$ encoded into quantum states $\left\{{\rho }_i\right\}$ which will be transmitted through a quantum channel $\mathcal{N}$ (see \fref{fig3_13}). The quantum states will be measured by a set of POVM operators $\mathcal{P}\mathrm{=}\left\{{\mathcal{M}}_{\mathrm{1}}\mathrm{,\dots ,}{\mathcal{M}}_k\right\}$ at the receiver (see \sref{sec2}). The classical zero-error quantum capacity $C_0\left(\mathcal{N}\right)$ for product input states can be reached if and only if the input states are\textit{ pure }states, similarly to the HSW capacity $C\left(\mathcal{N}\right)$.

 The zero-error transmission of quantum states requires perfect distinguishability. To achieve this perfect distinguishability of the zero-error quantum codewords, they have to be \textit{pairwise orthogonal}. \textit{Non-adjacent codewords can be distinguished perfectly}. Two inputs are called \textit{adjacent} if they can result in the same output. The number of possible non-adjacent codewords determines the rate of maximal transmittable classical information through $\mathcal{N}$.

In the \textit{d} dimensional Hilbert space (e.g. \textit{d}=2 for qubits) at most \textit{d} pairwise distinguishable quantum states exist, thus for a quantum system which consist of \textit{n} pieces of \textit{d} dimensional quantum states at most $d^n$ pairwise distinguishable \textit{n}-length quantum codewords are available. Obviously if two quantum codewords are not orthogonal, then they cannot be distinguished perfectly. We note, if we would like to distinguish between \textit{K} \textit{pairwise orthogonal} quantum codewords (the length of each codewords is \textit{n}) in the $d^n$ dimensional Hilbert space, then we have to define the POVM set
\begin{equation} \label{3.31)} 
\mathcal{P}\mathrm{=}\left\{{\mathcal{M}}^{\left(\mathrm{1}\right)}\mathrm{,\dots ,}{\mathcal{M}}^{\left(K\right)}\right\},                                                          
\end{equation} 
where ${\mathcal{M}}^{\left(i\right)}$ are set of \textit{d}-dimensional projectors on the individual quantum systems (e.g. qubits) which distinguish the \textit{n}-length codewords
\begin{equation} \label{3.32)} 
{\mathcal{M}}^{\left(i\right)}\mathrm{=}\left\{{\mathcal{M}}_{\mathrm{1}}\mathrm{,\dots ,}{\mathcal{M}}_m\right\} 
\end{equation} 
where $m\mathrm{=}d^n$. The probability that Bob gives measurement outcome \textit{j} from quantum state ${\rho }_i$ is
\begin{equation} \label{3.33)} 
\mathrm{Pr}\left[\left.j\right|{\rho }_i\right]\mathrm{=}Tr\left({\mathcal{M}}_j\mathcal{N}\left({\rho }_i\right)\right).                                                            
\end{equation} 
The \textit{i}-th \textit{codeword} $\left|\left.{\psi }_{X_i}\right\rangle \right.$ encodes the \textit{n}-length classical codeword $X_i\mathrm{=}\left\{x_{i\mathrm{,1}},x_{i\mathrm{,2}}\mathrm{,\dots ,}x_{i,n}\right\}$ consisting of \textit{n }product input quantum states: 
\begin{equation} \label{3.34)} 
\left|\left.{\psi }_{X_i}\right\rangle \right.\mathrm{=}\left[\left|\left.{\psi }_{i\mathrm{,1}}\right\rangle \right.\mathrm{\otimes }\left|\left.{\psi }_{i\mathrm{,2}}\right\rangle \right.\mathrm{\otimes }\left|\left.{\psi }_{i\mathrm{,3}}\right\rangle \right.\mathrm{\cdots }\mathrm{\otimes }\left|\left.{\psi }_{i,n}\right\rangle \right.\right]\mathrm{,\ }i\mathrm{=1..}K,
\end{equation} 
where ${\rho }_i\mathrm{=}\left|\left.{\psi }_{X_i}\right\rangle \right.\left\langle \left.{\psi }_{X_i}\right|\right.$.

The quantum block code consist of codewords
\begin{equation} \label{3.35)} 
 \begin{array}{l}
\left|\left.{\psi }_{X_{\mathrm{1}}}\right\rangle \right.\mathrm{=}\left[\left|\left.{\psi }_{\mathrm{1,1}}\right\rangle \right.\mathrm{\otimes }\left|\left.{\psi }_{\mathrm{1,2}}\right\rangle \right.\mathrm{\otimes }\left|\left.{\psi }_{\mathrm{1,3}}\right\rangle \right.\mathrm{\cdots }\mathrm{\otimes }\left|\left.{\psi }_{\mathrm{1,}n}\right\rangle \right.\right] \\ 
\mathrm{\vdots }\mathrm{\vdots } \\ 
\left|\left.{\psi }_{X_K}\right\rangle \right.\mathrm{=}\left[\left|\left.{\psi }_{K\mathrm{,1}}\right\rangle \right.\mathrm{\otimes }\left|\left.{\psi }_{K\mathrm{,2}}\right\rangle \right.\mathrm{\otimes }\left|\left.{\psi }_{K\mathrm{,3}}\right\rangle \right.\mathrm{\cdots }\mathrm{\otimes }\left|\left.{\psi }_{K,n}\right\rangle \right.\right], \end{array}
\end{equation} 
where \textit{K} is the number of classical (\textit{n} length) messages. 

The decoder will produce the output codeword ${{{X}'_{i}}}=\left\{ {{{{x}'_{i,1}}}},{{{{x}'_{i,2}}}},\ldots ,{{{{x}'_{i,n}}}} \right\}$ generated by the POVM measurement operators, where the POVM ${\mathcal{M}}^{\left(i\right)}$ can distinguish \textit{m} messages $\left\{ {{{{X}'_{1}}}},{{{{X}'_{2}}}},\ldots {{{{X}'_{m}}}} \right\}$  (\textit{n}-length) at the output. Bob would like to determine each message $i\mathrm{\in }\mathrm{[1,}K\mathrm{]}$ with unit probability. The zero probability of error means that for the input code $\left|\left.{\psi }_{X_i}\right\rangle \right.$ the decoder has to identify the classical output codeword ${X\mathrm{'}}_i$ with classical input codeword $X_i$ perfectly for each possible \textit{i}, otherwise the quantum channel has no zero-error capacity; that is, for the zero-error quantum communication system 
\begin{equation} \label{3.36)} 
\Pr \left[ \left. {{{{X}'_{i}}}} \right|{{X}_{i}} \right]=1.
\end{equation}

\subsubsection{Formal Definitions of Quantum Zero-Error Communication}

 In this subsection we review the most important definitions of quantum zero-error communication systems.

 The \textit{non-adjacent} elements are important for zero-error capacity, since \textit{only non-adjacent codewords can be distinguished perfectly}. Two inputs are called \textit{adjacent} if they can result in the same output, while for \textit{non-adjacent} inputs, the output of the encoder is unique. The number of possible non-adjacent codewords determines the rate of maximal transmittable classical information through quantum channels.

Formally, the \textit{non-adjacent} property of two quantum states ${\rho }_{\mathrm{1}}$ and ${\rho }_{\mathrm{2}}$ can be given as
\begin{equation} \label{ZEqnNum539102} 
Set_{\mathrm{1}}\mathrm{\cap }Set_{\mathrm{2}}\mathrm{=}\mathrm{\emptyset },                                                                          
\end{equation} 
where $Se{{t}_{i}}=\left\{ \Pr \left[ \left. {{{{X}'_{j}}}} \right|{{X}_{i}} \right]=Tr\left( {{\mathcal{M}}_{j}}\mathcal{N}\left( \left| {{\psi }_{{{X}_{i}}}} \right\rangle \langle {{\psi }_{{{X}_{i}}}}| \right) \right)>0 \right\},$ $j\in \left\{ 1,\ldots ,m \right\},i=1,2$, and $\mathcal{P}\mathrm{=}\left\{{\mathcal{M}}_{\mathrm{1}}\mathrm{,\dots ,}{\mathcal{M}}_m\right\}$  is a POVM measurement operator. In a relation of a noisy quantum channel $\mathcal{N}$, the non-adjacent property can be rephrased as follows. Two input quantum states ${\rho }_{\mathrm{1}}$ and ${\rho }_{\mathrm{2}}$ are non-adjacent with relation to $\mathcal{N}$, if $\mathcal{N}\left({\rho }_{\mathrm{1}}\right)$ and $\mathcal{N}\left({\rho }_{\mathrm{2}}\right)$ are \textit{perfectly distinguishable}. The notation ${{\rho }_{1}}\underset{\mathcal{N}}{\mathop{\bot }}\,{{\rho }_{2}}$ also can be used to denote the non-adjacent inputs of quantum channel $\mathcal{N}$.

 A quantum channel $\mathcal{N}$ has greater than zero zero-error capacity if and only if a subset of quantum states $\mathrm{\Omega }\mathrm{=}{\left\{{\rho }_i\right\}}^l_{i\mathrm{=1}}$ and POVM $\mathcal{P}\mathrm{=}\left\{{\mathcal{M}}_{\mathrm{1}}\mathrm{,\dots ,}{\mathcal{M}}_m\right\}$ exists where for at \textit{least two states} ${\rho }_{\mathrm{1}}$ and ${\rho }_{\mathrm{2}}$ from subset $\mathrm{\Omega }$, the relation \eqref{ZEqnNum539102} holds; that is, the non-adjacent property with relation to the POVM measurement is satisfied. For the quantum channel $\mathcal{N}$, the two inputs ${\rho }_{\mathrm{1}}$ and ${\rho }_{\mathrm{2}}$ are non-adjacent if and only if the quantum channel takes the input states ${\rho }_{\mathrm{1}}$ and ${\rho }_{\mathrm{2}}$  into orthogonal subspaces   
\begin{equation} \label{3.39)} 
\mathcal{N}\left({\rho }_{\mathrm{1}}\right)\mathop{\mathrm{\bot }}_{\mathcal{N}}\mathcal{N}\left({\rho }_{\mathrm{2}}\right);                                                          
\end{equation} 
that is, the quantum channel has positive classical zero-error capacity $C_0\left(\mathcal{N}\right)$ if and only if this property holds for the output of the channel for a given POVM $\mathcal{P}\mathrm{=}\left\{{\mathcal{M}}_{\mathrm{1}}\mathrm{,\dots ,}{\mathcal{M}}_m\right\}$. The previous result can be rephrased as follows. Using the trace preserving property of the quantum channe, the two quantum states ${\rho }_{\mathrm{1}}$ and ${\rho }_{\mathrm{2}}$ are non-adjacent if and only if for the channel output states $\mathcal{N}\left({\rho }_{\mathrm{1}}\right),\mathcal{N}\left({\rho }_{\mathrm{2}}\right)$, 
\begin{equation} \label{ZEqnNum384772} 
Tr\left(\mathcal{N}\left({\rho }_{\mathrm{1}}\right)\mathcal{N}\left({\rho }_{\mathrm{2}}\right)\right)\mathrm{=0},                                                                    
\end{equation} 
and if ${\rho }_{\mathrm{1}}$ and ${\rho }_{\mathrm{2}}$ are non-adjacent input states then
\begin{equation} \label{3.41)} 
Tr\left({\rho }_{\mathrm{1}}{\rho }_{\mathrm{2}}\right)\mathrm{=0}.                                                     
\end{equation} 
Let the two \textit{non-adjacent} input codewords of the $\mathcal{N}$ be denoted by $\left|\left.{\psi }_{X_{\mathrm{1}}}\right\rangle \right.$ and $\left|\left.{\psi }_{X_{\mathrm{2}}}\right\rangle \right.$. These quantum codewords encode messages $X_{\mathrm{1}}\mathrm{=}\left\{x_{\mathrm{1,1}},x_{\mathrm{1,2}}\mathrm{,\dots ,}x_{\mathrm{1,}n}\right\}$ and $X_{\mathrm{2}}\mathrm{=}\left\{x_{\mathrm{2,1}},x_{\mathrm{2,2}}\mathrm{,\dots ,}x_{\mathrm{2,}n}\right\}$. For this setting, we construct the following POVM operators for the given complete set of POVM $\mathcal{P}\mathrm{=}\left\{{\mathcal{M}}_{\mathrm{1}}\mathrm{,\dots ,}{\mathcal{M}}_m\right\}$ and the two input codewords $\left|\left.{\psi }_{X_{\mathrm{1}}}\right\rangle \right.$ and $\left|\left.{\psi }_{X_{\mathrm{2}}}\right\rangle \right.$ as follows
\begin{equation} \label{ZEqnNum745016} 
{\mathcal{M}}^{\left(\mathrm{1}\right)}\mathrm{=}\left\{{\mathcal{M}}_{\mathrm{1}}\mathrm{,\dots ,}{\mathcal{M}}_k\right\} 
\end{equation} 
and
\begin{equation} \label{ZEqnNum390187} 
{\mathcal{M}}^{\left(\mathrm{2}\right)}\mathrm{=}\left\{{\mathcal{M}}_{k\mathrm{+1}}\mathrm{,\dots ,}{\mathcal{M}}_m\right\}.                                                                      
\end{equation} 
The groups of operators, ${\mathcal{M}}^{\left(\mathrm{1}\right)}$ and ${\mathcal{M}}^{\left(\mathrm{2}\right)}$, will identify and distinguish the input codewords $\left|\left.{\psi }_{X_{\mathrm{1}}}\right\rangle \right.$ and $\left|\left.{\psi }_{X_{\mathrm{2}}}\right\rangle \right.$. Using this setting the two non-adjacent codewords $\left|\left.{\psi }_{X_{\mathrm{1}}}\right\rangle \right.$ and $\left|\left.{\psi }_{X_{\mathrm{2}}}\right\rangle \right.$ can be distinguished with probability one at the output since 
\begin{equation} \label{3.44)} 
 \begin{array}{l}
\mathrm{Pr}[{{{{X}'_{i}}}}|X_{\mathrm{1}}]\mathrm{=1,}\text{ }i\mathrm{=1,\dots ,}k, \\ 
\mathrm{Pr}[{{{{X}'_{i}}}}|X_{\mathrm{2}}]\mathrm{=1,}\text{ }i\mathrm{=}k\mathrm{+1,\dots ,}m, \end{array}
\end{equation} 
where ${{{{X}'_{i}}}}$\textit{ }is a number between $\mathrm{1}$ and \textit{m}, (according to the possible number of POVM operators) which identifies the measured unknown quantum codeword  and consequently
\begin{equation} \label{3.45)} 
 \begin{array}{l}
\mathrm{Pr}[{{{{X}'_{i}}}}|X_{\mathrm{1}}]\mathrm{=0,}\text{ }i\mathrm{=}k\mathrm{+1,\dots ,}m, \\ 
\mathrm{Pr}[{{{{X}'_{i}}}}|X_{\mathrm{2}}]\mathrm{=0,}\text{ }i\mathrm{=1,\dots ,}k. \end{array}
\end{equation} 
For input message $\left|\left.{\psi }_{X_{\mathrm{1}}}\right\rangle \right.$ and $\left|\left.{\psi }_{X_{\mathrm{2}}}\right\rangle \right.$ with the help of set ${\mathcal{M}}^{\left(\mathrm{1}\right)}$ and ${\mathcal{M}}^{\left(\mathrm{2}\right)}$ these probabilities are
\begin{equation} \label{3.46)} 
 \begin{array}{l}
\mathrm{Pr}[{X\mathrm{'_{\mathrm{1}}}}|X_{\mathrm{1}}]\mathrm{=}Tr\left({\mathcal{M}}^{\left(\mathrm{1}\right)}\mathcal{N}\left(\left|\left.{\psi }_{X_{\mathrm{1}}}\right\rangle \right.\left\langle \left.{\psi }_{X_{\mathrm{1}}}\right|\right.\right)\right)\mathrm{=1,} \\ 
\mathrm{Pr}[{X\mathrm{'_{\mathrm{2}}}}|X_{\mathrm{2}}]\mathrm{=}Tr\left({\mathcal{M}}^{\left(\mathrm{2}\right)}\mathcal{N}\left(\left|\left.{\psi }_{X_{\mathrm{2}}}\right\rangle \right.\left\langle \left.{\psi }_{X_{\mathrm{2}}}\right|\right.\right)\right)\mathrm{=1,} \end{array}
\end{equation} 
where ${\mathcal{M}}^{\left(\mathrm{1}\right)}$ and ${\mathcal{M}}^{\left(\mathrm{2}\right)}$ are orthogonal projectors, ${\mathcal{M}}^{\left(\mathrm{1}\right)}$ and ${\mathcal{M}}^{\left(\mathrm{2}\right)}$ are defined in \eqref{ZEqnNum745016} and \eqref{ZEqnNum390187}), and ${\mathcal{M}}^{\left(\mathrm{1}\right)}\mathrm{+}{\mathcal{M}}^{\left(\mathrm{2}\right)}\mathrm{+}{\mathcal{M}}^{\left(\mathrm{2+1}\right)}\mathrm{=}I$, to make it possible for the quantum channel to take the input states into orthogonal subspaces; that is, $\mathcal{N}\left(\left|\left.{\psi }_{X_{\mathrm{1}}}\right\rangle \right.\left\langle \left.{\psi }_{X_{\mathrm{1}}}\right|\right.\right)\mathrm{\bot }\mathcal{N}\left(\left|\left.{\psi }_{X_{\mathrm{2}}}\right\rangle \right.\left\langle \left.{\psi }_{X_{\mathrm{2}}}\right|\right.\right)$ has to be satisfied. The POVM measurement has to be restricted to projective measurement. As follows, the $\mathcal{P}\mathrm{=}\left\{{\mathcal{M}}^{\left(\mathrm{1}\right)},{\mathcal{M}}^{\left(\mathrm{2}\right)}\right\}$ POVM measurement can be replaced with the set of \textit{von Neumann} operators, $\mathcal{Z}\mathrm{=}\left\{{\mathcal{P}}^{\left(\mathrm{1}\right)},{\mathcal{P}}^{\left(\mathrm{2}\right)}\right\}$ , where ${\mathcal{P}}^{\left(\mathrm{1}\right)}\mathrm{+}{\mathcal{P}}^{\left(\mathrm{2}\right)}\mathrm{=}I$. This result also can be extended for arbitrarily number of operators, depending on the actual system. The non-adjacent property also can be interpreted for arbitrary length of quantum codewords. For a given quantum channel $\mathcal{N}$, the two \textit{n}-length input quantum codewords $\left|\left.{\psi }_{X_{\mathrm{1}}}\right\rangle \right.$ and $\left|\left.{\psi }_{X_{\mathrm{2}}}\right\rangle \right.$, which are tensor products of \textit{n} quantum states, then \textit{input} codewords $\left|\left.{\psi }_{X_{\mathrm{1}}}\right\rangle \right.$ and $\left|\left.{\psi }_{X_{\mathrm{2}}}\right\rangle \right.$ are non-adjacent in relation with $\mathcal{N}$ if and only if \textit{at} \textit{least one }pair of quantum states\textit{ }$\left\{\left|\left.{\psi }_{1,i}\right\rangle \right.,\left|\left.{\psi }_{2,i}\right\rangle \right.\right\}$ from the two \textit{n}-length sequences is perfectly distinguishable. Formally, at least one \textit{input} quantum state pair $\left\{\left|\left.{\psi }_{\mathrm{1,}i}\right\rangle \right.,\left|\left.{\psi }_{\mathrm{2,}i}\right\rangle \right.\right\}$ with $i$, $\mathrm{1}\mathrm{\le }i\mathrm{\le }n$, exists in $\left|\left.{\psi }_{X_{\mathrm{1}}}\right\rangle \right.$ and $\left|\left.{\psi }_{X_{\mathrm{2}}}\right\rangle \right.$, for which $\mathcal{N}\left(\left|\left.{\psi }_{\mathrm{1,}i}\right\rangle \right.\left\langle \left.{\psi }_{\mathrm{1,}i}\right|\right.\right)$ is non-adjacent to $\mathcal{N}\left(\left|\left.{\psi }_{\mathrm{2,}i}\right\rangle \right.\left\langle \left.{\psi }_{\mathrm{2,}i}\right|\right.\right)$. 
 Because we have stated that the two codewords can be distinguished at the channel output, the following relation has to be hold for their trace, according to \eqref{ZEqnNum384772}, and their non-adjacency can be verified as follows:
\begin{equation} \label{ZEqnNum472180} 
 \begin{split}
&Tr\left(\mathcal{N}\left(\left|\left.{\psi }_{X_{\mathrm{1}}}\right\rangle \right.\left\langle \left.{\psi }_{X_{\mathrm{1}}}\right|\right.\right)\mathcal{N}\left(\left|\left.{\psi }_{X_{\mathrm{2}}}\right\rangle \right.\left\langle \left.{\psi }_{X_{\mathrm{2}}}\right|\right.\right)\right) \\&
\mathrm{=}Tr\left(\left(\mathop{\mathop{\mathrm{\otimes }}_{i\mathrm{=1}}}^{n}\mathcal{N}\left(\left|\left.{\psi }_{\mathrm{1,}i}\right\rangle \right.\left\langle \left.{\psi }_{\mathrm{1,}i}\right|\right.\right)\right)\left(\mathop{\mathop{\mathrm{\otimes }}_{i\mathrm{=1}}}^{n}\mathcal{N}\left(\left|\left.{\psi }_{\mathrm{2,}i}\right\rangle \right.\left\langle \left.{\psi }_{\mathrm{2,}i}\right|\right.\right)\right)\right) \\ &
\mathrm{=}\prod^n_{i\mathrm{=1}}{Tr\left(\mathcal{N}\left(\left|\left.{\psi }_{\mathrm{1,}i}\right\rangle \right.\left\langle \left.{\psi }_{\mathrm{1,}i}\right|\right.\right)\mathcal{N}\left(\left|\left.{\psi }_{\mathrm{2,}i}\right\rangle \right.\left\langle \left.{\psi }_{\mathrm{2,}i}\right|\right.\right)\right)\mathrm{=0.}} \end{split}
\end{equation} 
As follows from \eqref{ZEqnNum472180}, a quantum channel $\mathcal{N}$ has non-zero zero-error capacity if and only if there exists at least two non-adjacent input quantum states ${\rho }_{\mathrm{1}}$ and ${\rho }_{\mathrm{2}}$. These two non-adjacent quantum states make distinguishable the two, \textit{n}-length quantum codewords at the output of quantum channel $\mathcal{N}$, and these input codewords will be called as \textit{non-adjacent quantum codewords}.
The joint measurement of the quantum states of an output codeword is \textit{necessary} and \textit{sufficient} to distinguish the input codewords with zero-error. \textit{Necessary}, because the joint measurement is required to distinguish orthogonal general (i.e., non zero-error code) tensor product states [\cref{Bennett99a}]. Sufficient, because the non-adjacent quantum states have orthogonal \textit{supports} at the output of the noisy quantum channel, i.e., $Tr\left({\rho }_i{\rho }_j\right)\mathrm{=0}$ [\cref{Medeiros05}]. (The \textit{support} of a matrix \textit{A} is the orthogonal complement of the kernel of the matrix. The \textit{kernel} of \textit{A} is the set of all vectors \textit{v}, for which $Av\mathrm{=0}$.) In the joint measurement, the $\left\{{\mathcal{M}}_i\right\},i\mathrm{=1,\dots ,}m$ projectors are $d^n\mathrm{\times }d^n$ matrices, while if we were to use a single measurement then the size of these matrices would be $d\mathrm{\times }d$.

\begin{center}
\begin{figure*}[htbp]
\begin{center}
\includegraphics[angle = 0,width=\linewidth]{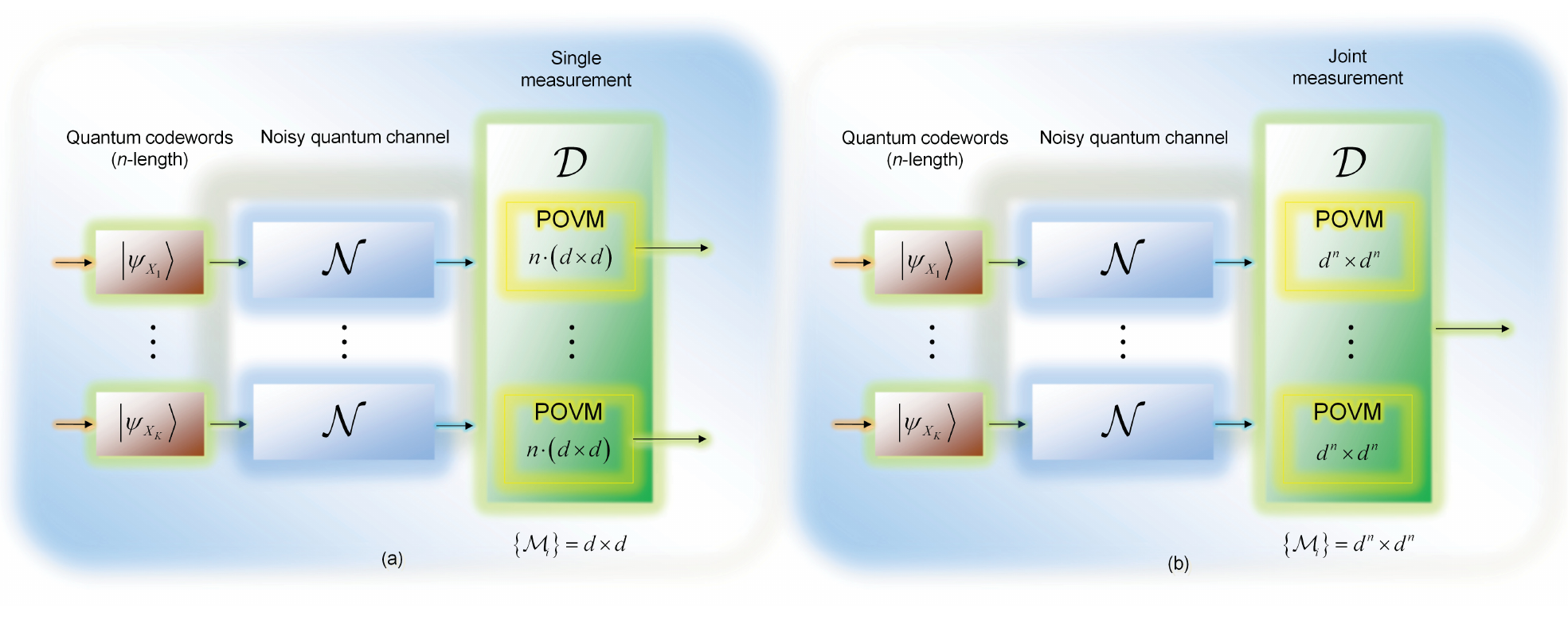}
\caption{Comparison of single (a) and joint (b) measurement settings. The joint measurement is necessary to attain the quantum zero-error communication.} 
\label{fig3_16}
 \end{center}
\end{figure*}
\end{center}

In \fref{fig3_16} we compared the difference between single and joint measurement settings for a given \textit{n}-length quantum codeword $\left|\left.{\psi }_X\right\rangle \right.\mathrm{=}\left[\left|\left.{\psi }_{\mathrm{1}}\right\rangle \right.\mathrm{\otimes }\left|\left.{\psi }_{\mathrm{2}}\right\rangle \right.\mathrm{\otimes }\left|\left.{\psi }_{\mathrm{3}}\right\rangle \right.\mathrm{\cdots }\mathrm{\otimes }\left|\left.{\psi }_n\right\rangle \right.\right]$. In the case of single measurement Bob measures each of the \textit{n} quantum states of the \textit{i}-th codeword states individually. In case of the joint measurement Bob waits until he receives the \textit{n} quantum states, then measures them together.

Next we study the achievable rates for zero error classical communication over quantum channels.

\subsubsection{Achievable Zero-Error Rates in Quantum Systems}

 Theoretically (without making any assumptions about the physical attributes of the transmission), the \textit{classical} \textit{single-use} \textit{zero-error capacity }$C^{\left(1\right)}_0\left(\mathcal{N}\right)$ of the noisy quantum channel can be expressed as 
\begin{equation} \label{3.49)} 
C^{\left(\mathrm{1}\right)}_0\left(\mathcal{N}\right)\mathrm{=log}\left(K\left(\mathcal{N}\right)\right),                                                                     
\end{equation} 
where $K\left(\mathcal{N}\right)$ is the maximum number of different messages which can be sent over the channel with a \textit{single use} of $\mathcal{N}$ (or in other words the maximum size of the set of \textit{mutually non-adjacent} inputs).

The asymptotic \textit{zero-error capacity} of the noisy quantum channel $\mathcal{N}$ can be expressed as 
\begin{equation} \label{3.50)} 
C_{0}^{{}}\left( \mathcal{N} \right)=\underset{n\to \infty }{\mathop{\text{lim }}}\,\frac{1}{n}\log \left( K\left( {{\mathcal{N}}^{\otimes n}} \right) \right),                                                                
\end{equation} 
where $K\left( {{\mathcal{N}}^{\otimes n}} \right)$ is the maximum number of \textit{n}-length classical messages that the quantum channel can transmit with zero error and ${\mathcal{N}}^{\mathrm{\otimes }n}$ denotes the \textit{n}-uses of the channel. 

 The $C_0\left(\mathcal{N}\right)$ asymptotic classical zero-error capacity of a quantum channel is \textit{upper bounded} by the HSW capacity, that is, 
\begin{equation} \label{3.51)} 
C^{\left(\mathrm{1}\right)}_0\left(\mathcal{N}\right)\mathrm{\le }C_0\left(\mathcal{N}\right)\mathrm{\le }C\left(\mathcal{N}\right).                                                                 
\end{equation} 
Next, we study the connection of zero-error quantum codes and graph theory.

\subsubsection{Connection with Graph Theory}

 The problem of finding \textit{non-adjacent} codewords for the zero-error information transmission can be rephrased in terms of graph theory. The adjacent codewords are also called \textit{confusable}, since these codewords can generate the same output with a given non-zero probability. Since we know that two input codewords $\left|\left.{\psi }_{X_{\mathrm{1}}}\right\rangle \right.$ and $\left|\left.{\psi }_{X_{\mathrm{2}}}\right\rangle \right.$ are \textit{adjacent} if there is a channel output codeword $\left|\left.{\psi }_{{X\mathrm{'}}}\right\rangle \right.$ which can be resulted by either of these two, that is $\mathrm{Pr}\left[\left.{X\mathrm{'}}\right|X_{\mathrm{1}}\right]\mathrm{>0}$ and $\mathrm{Pr}\left[\left.{X\mathrm{'}}\right|X_{\mathrm{2}}\right]\mathrm{>0}$. 

The non-adjacent property of two quantum codewords can be analyzed by the \textit{confusability graph }${\mathcal{G}}_n$, where \textit{n} denotes the \textit{length of the block code}. 

Let us take as many vertices as the number of input messages \textit{K}, and connect two vertices if these input messages are adjacent. For example, using the quantum version of the famous \textit{pentagon graph} we show how the classical zero-error capacity $C_0\left(\mathcal{N}\right)$ of the quantum channel $\mathcal{N}$ changes if we use block codes of length \textit{n}=1 and \textit{n}=2. In the pentagon graph an input codeword from the set of non-orthogonal qubits $\left\{\left|\left.0\right\rangle \right.,\left|\left.\mathrm{1}\right\rangle \right.,\left|\left.\mathrm{2}\right\rangle \right.,\left|\left.\mathrm{3}\right\rangle \right.,\left|\left.\mathrm{4}\right\rangle \right.\right\}$ is connected with two other adjacent input codewords, and the number of total codewords is 5 [\cref{Lovasz79}].

\begin{center}
\begin{figure}[htbp]
\begin{center}
\includegraphics[angle = 0,width=\linewidth]{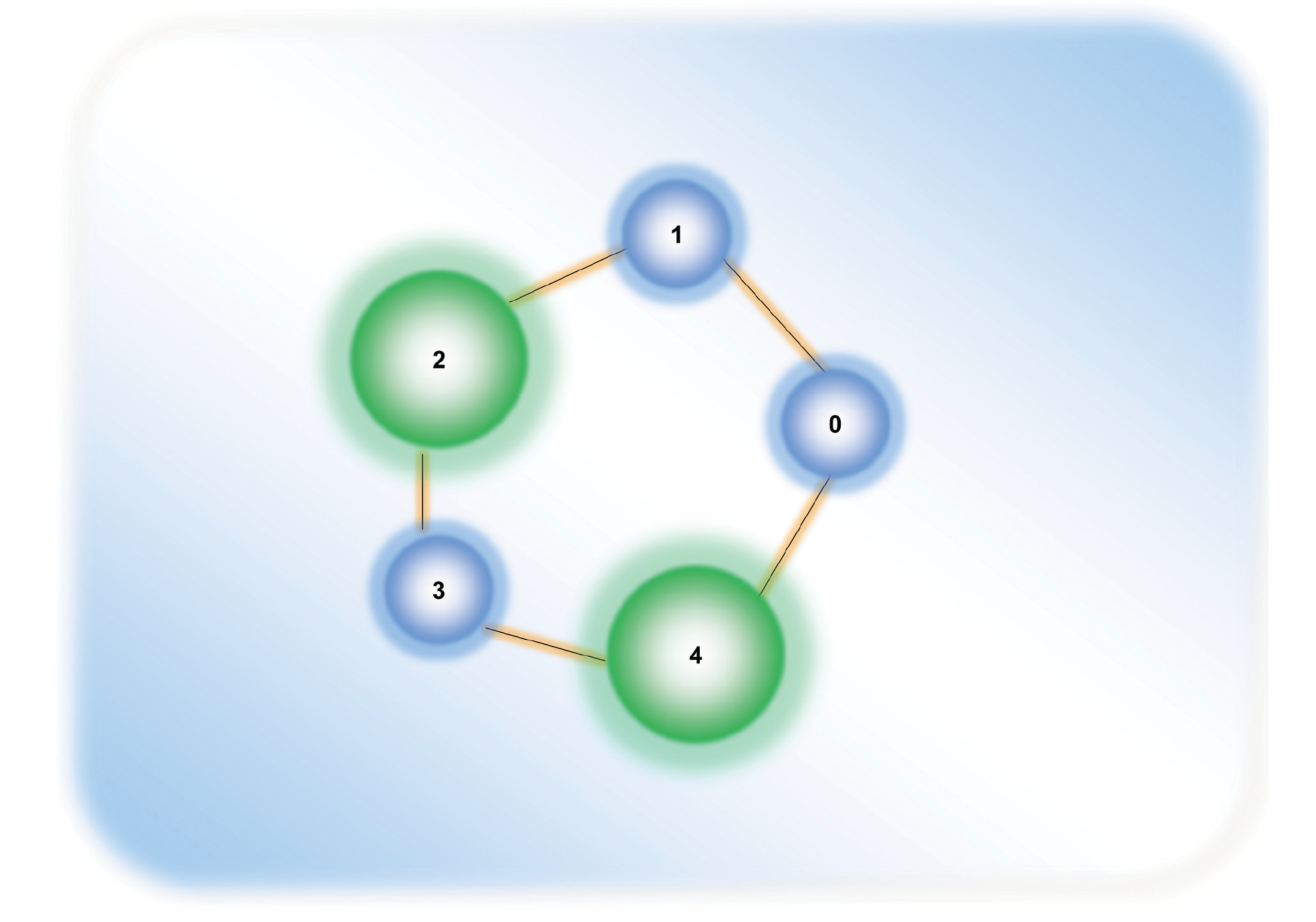}
\caption{The confusability graph of a zero-error code for one channel use. The two possible non-adjacent codewords are denoted by the large shaded circles.} 
\label{fig3_17}
 \end{center}
\end{figure}
\end{center}

 The ${\mathcal{G}}_{\mathrm{1}}$\textit{ confusability }graph of the pentagon structure for block codes of length \textit{n}=1 is shown in \fref{fig3_17}. The vertices of the graph are the possible input messages, where \textit{K = }5. The \textit{adjacent} input messages are connected by a line. The non-adjacent inputs $\left|\left.\mathrm{2}\right\rangle \right.$ and $\left|\left.\mathrm{4}\right\rangle \right.$ are denoted by gray circles, and there is no connection between these two input codewords. 

 For the block codes of length \textit{n}=1, the maximal transmittable classical information with zero error is 
\begin{equation} \label{3.52)} 
C_0\left(\mathcal{N}\right)\mathrm{=log}\left(\mathrm{2}\right)\mathrm{=1},                                                                    
\end{equation} 
since only two non-adjacent vertices can be found in the graph. We note, other possible codeword combinations also can be used to realize the zero-error transmission, in comparison with the confusability graph, for example $\left|\left.\mathrm{1}\right\rangle \right.$ and $\left|\left.\mathrm{3}\right\rangle \right.$ also non-adjacent, etc. On the other hand, the maximum number of non-adjacent vertices (two, in this case) cannot be exceeded, thus $C_0\left(\mathcal{N}\right)\mathrm{=1}$ remains in all other possible cases, too. 

\begin{center}
\begin{figure*}[htbp]
\begin{center}
\includegraphics[angle = 0,width=0.5\linewidth]{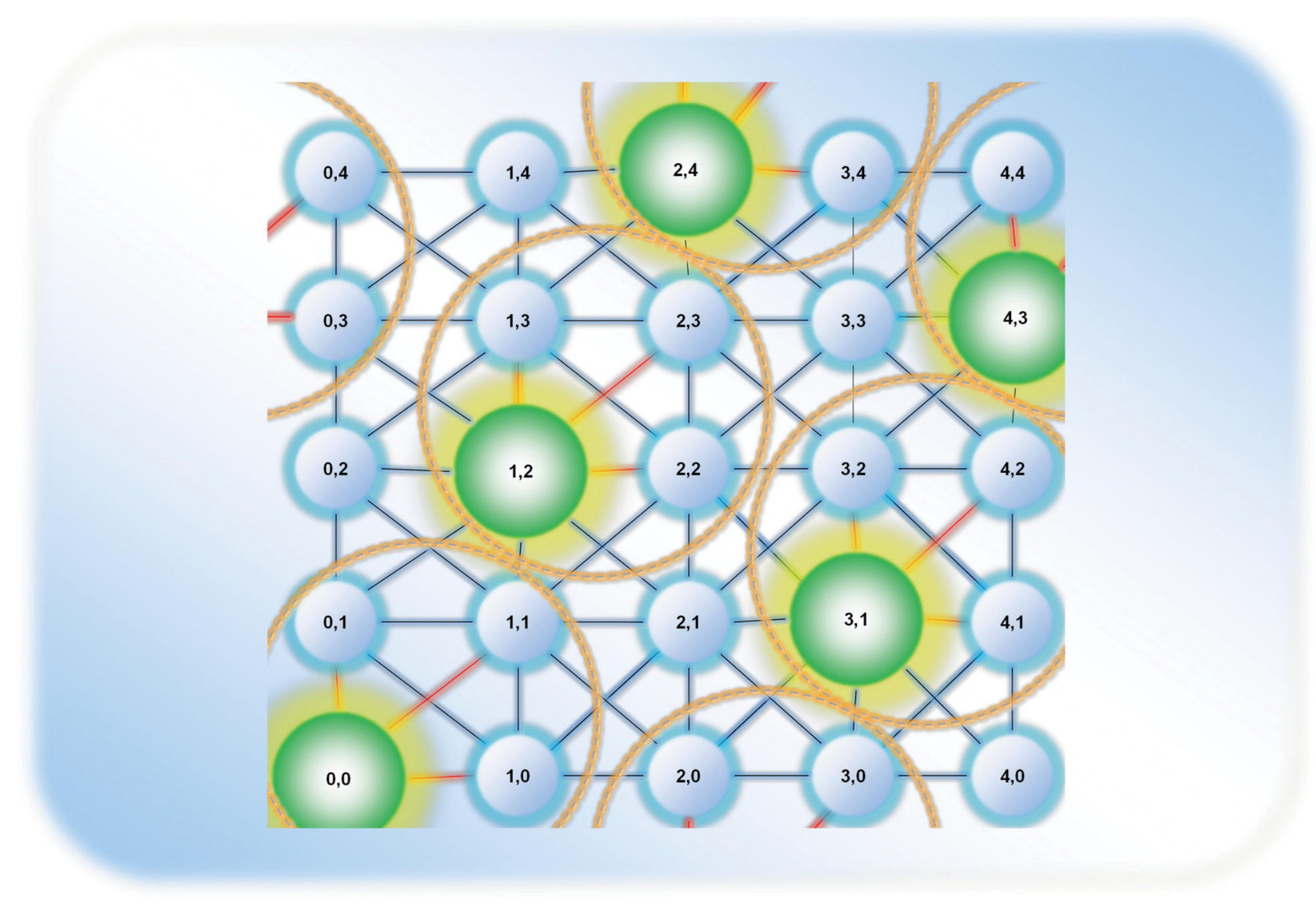}
\caption{The graph of a zero-error code for two channel uses of a quantum channel. The possible zero-error codewords are depicted by the thick lines and dashed circles.} 
\label{fig3_18}
 \end{center}
\end{figure*}
\end{center}

Let assume that we use $n\mathrm{=2}$ length of block codes. First, let us see how the graph changes. The non-adjacent inputs are denoted by the large gray shaded circles. The connections between the possible codewords (which can be used as a block code) are denoted by the thick line and the dashed circle. The confusability graph ${\mathcal{G}}_{\mathrm{2}}$\textit{ }for $n\mathrm{=2}$ length of block codes is shown in \fref{fig3_18}. The two half-circles together on the left and right sides represent one circle and the two half circles at the top and bottom of the figure also represent one circle; thus there are five dashed circles in the figure.  

It can be seen that the complexity of the structure of the graph has changed, although we have made only a small modification: we increased the lengths of the block codes from $n\mathrm{=1}$ to $n\mathrm{=2}$. The five two-length codewords and zero-error quantum block codes which can realize the zero-error transmission can be defined using the computational basis $\left\{\left|\left.0\right\rangle \right.,\left|\left.\mathrm{1}\right\rangle \right.,\left|\left.\mathrm{2}\right\rangle \right.,\left|\left.\mathrm{3}\right\rangle \right.,\left|\left.\mathrm{4}\right\rangle \right.\right\}$. 
The classical zero-error capacity which can be achieved by $n\mathrm{=2}$ length block codes is 
\begin{equation} \label{3.54)} 
C_0\left({\mathcal{N}}^{\mathrm{\otimes }\mathrm{2}}\right)\mathrm{=}\frac{\mathrm{1}}{\mathrm{2}}\mathrm{log}\left(\mathrm{5}\right)\mathrm{=1.1609}.                                                                          
\end{equation} 
From an engineering point of view this result means, that for the pentagon graph, the maximum rate at which classical information can be transmitted over a noisy quantum channel $\mathcal{N}$ with a zero error probability, can be achieved with quantum block code length of two. 

 For the classical zero-error capacities of some typical quantum channels see \sref{sec5}.

\subsection{Entanglement-assisted Classical Zero-Error Capacity}
\label{sec3g}
 In the previous subsection we discussed the main properties of zero-error capacity using product input states. Now, we add the entanglement to the picture. Here we discuss how the encoding and the decoding setting will change if we bring entanglement to the system and how it affects the classical zero-error capacity of a quantum channel. 

If entanglement allowed between the communicating parties then the single-use and asymptotic \textit{entanglement-assisted} classical zero-error capacities are defined as 
\begin{equation} \label{3.55)} 
C^{E\left(\mathrm{1}\right)}_0\left(\mathcal{N}\right)\mathrm{=log}\left(K^E\left(\mathcal{N}\right)\right) 
\end{equation} 
and
\begin{equation} \label{3.56)} 
C_{0}^{E}\left( \mathcal{N} \right)=\underset{n\to \infty }{\text{lim }}\,\frac{1}{n}\log \left( {{K}^{E}}\left( {{\mathcal{N}}^{\otimes n}} \right) \right).                                                               
\end{equation} 
where ${{K}^{E}}\left( {{\mathcal{N}}^{\otimes n}} \right)$ is the maximum number of \textit{n}-length mutually non-adjacent classical messages that the quantum channel can transmit with zero error using \textit{shared entanglement}. 

Before we start to discuss the properties of the entanglement-assisted zero-error quantum communication, we introduce a new type of graph, called the \textit{hypergraph} ${\mathcal{G}}_H$. The hypergraph is very similar to our previously shown \textit{confusability} graph ${\mathcal{G}}_n$. The hypergraph contains a set of vertices and hyperedges. The vertices represent the \textit{inputs} of the quantum channel $\mathcal{N}$, while the hyperedges contain all the channel inputs which could cause the same channel output with non-zero probability. 

We will use some new terms from graph theory in this subsection; hence we briefly summarize these definitions:
\begin{enumerate}
\item \textit{maximum independent set of }${\mathcal{G}}_n$: the maximum number of non-adjacent inputs (\textit{K}), 

\item \textit{clique of }${\mathcal{G}}_n$: ${\kappa }_i$, the set of possible inputs of a given output in a confusability graph (which inputs could result in the same output with non-zero probability),

\item \textit{complete graph}: if all the vertices are connected with one another in the graph; in this case there are no non-adjacent inputs; i.e., the channel has no zero-error capacity.
\end{enumerate}

\begin{center}
\begin{figure*}[htbp]
\begin{center}
\includegraphics[angle = 0,width=0.57\linewidth]{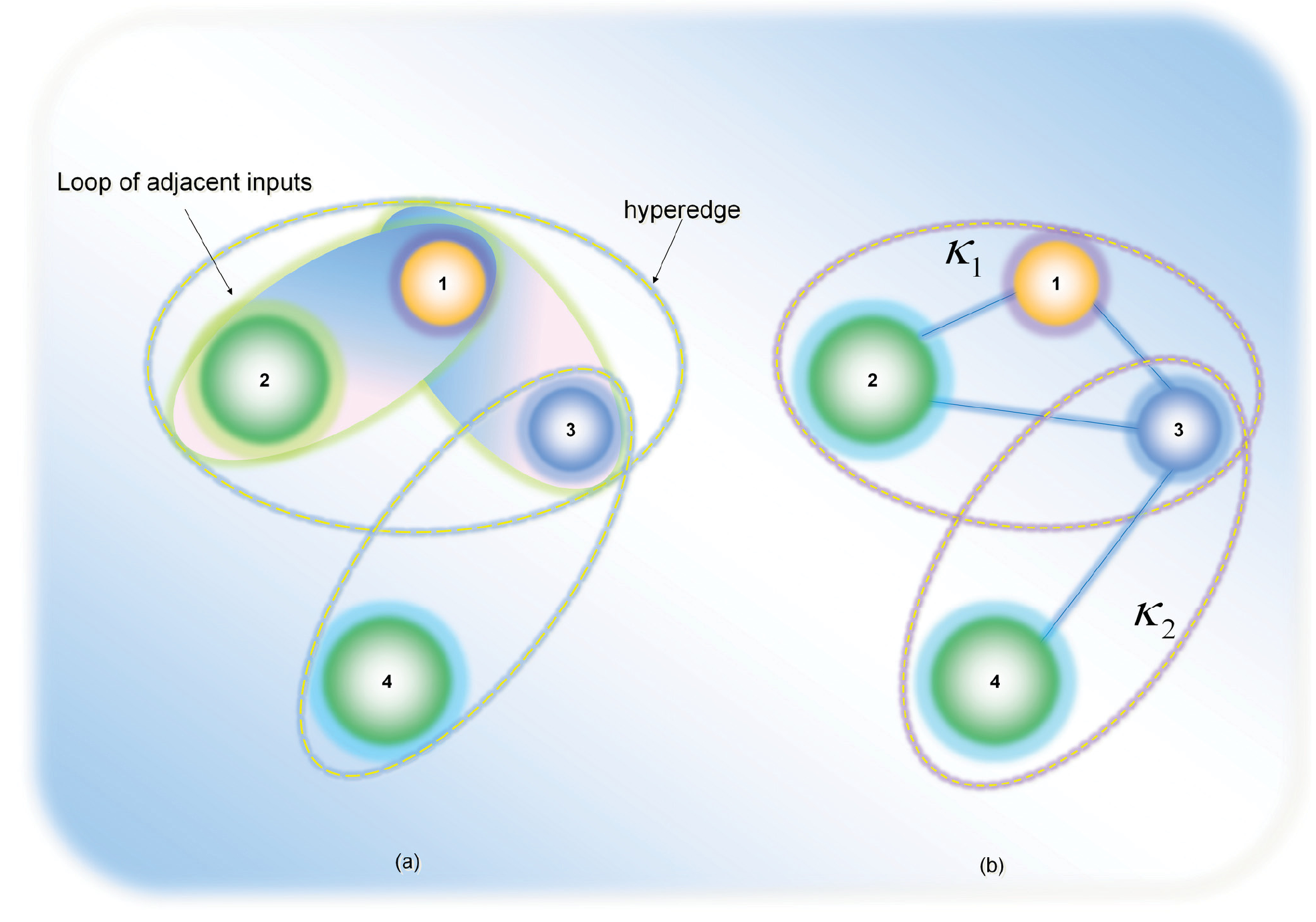}
\caption{The hypergraph and the confusability graph of a given input system with four inputs. The hyperedges of the hypergraph are labeled by the output. The number of non-adjacent inputs is two.} 
\label{fig3_19}
 \end{center}
\end{figure*}
\end{center}

 In \fref{fig3_19}(a) we show a hypergraph ${\mathcal{G}}_H$, where the inputs of the channel are the vertices and the hyperedges represent the channel outputs. Two inputs are non-adjacent if they are in a different loop. The two non-adjacent inputs are depicted by the greater grey shaded vertices. In \fref{fig3_19}(b) we give the confusability graph ${\mathcal{G}}_n$ for a single channel use ($n\mathrm{=1}$), for the same input set. The cliques in the ${\mathcal{G}}_n$ confusability graph are depicted by ${\kappa }_i$.

\begin{center}
\begin{figure*}[htbp]
\begin{center}
\includegraphics[angle = 0,width=0.6\linewidth]{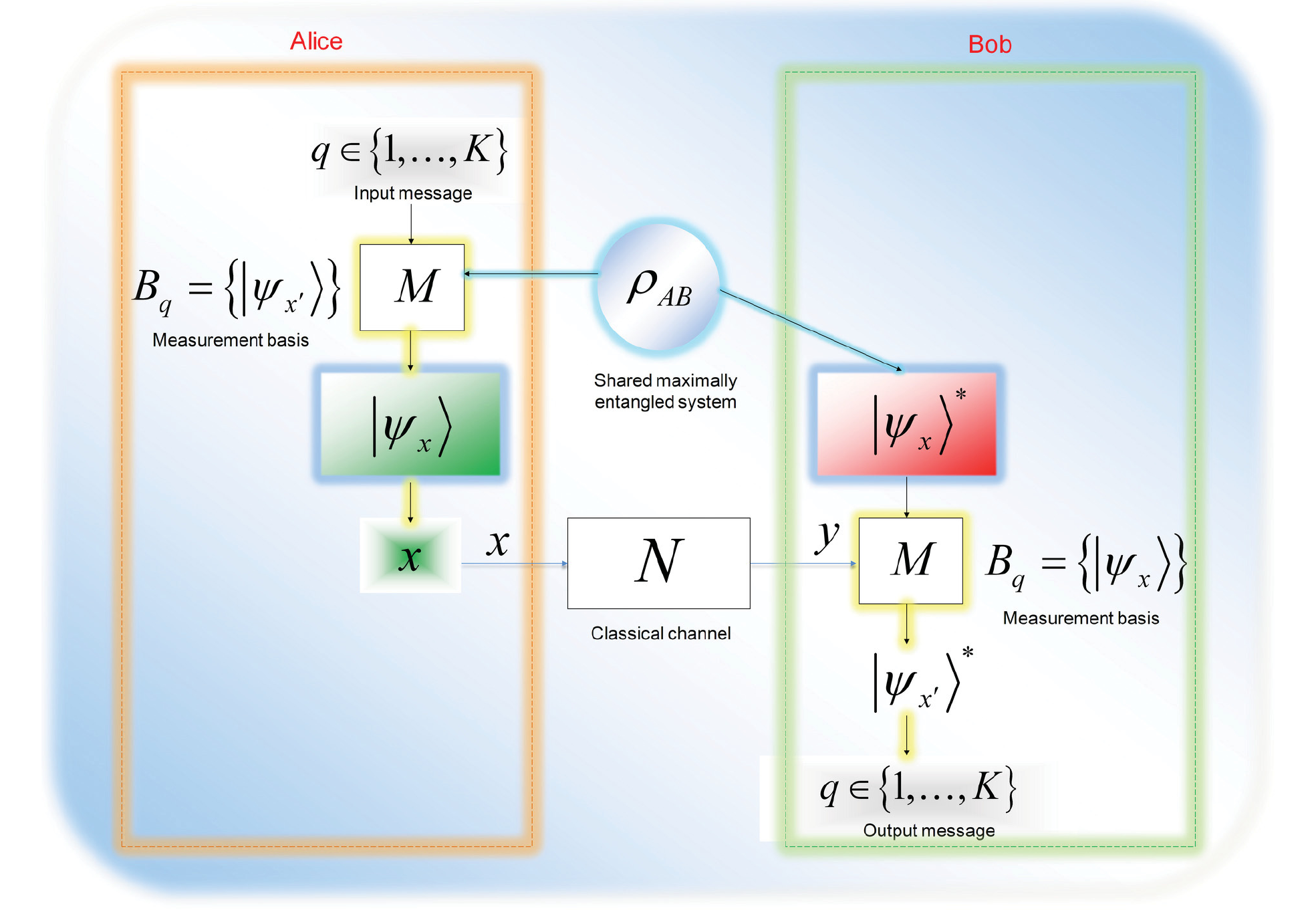}
\caption{The steps of the entanglement-assisted zero-error quantum communication protocol.} 
\label{fig3_20}
 \end{center}
\end{figure*}
\end{center}

Both the hypergraph and the confusability graph can be used to determine the non-adjacent inputs. However, if the number of inputs starts to increase, the number of hyperedges in the hypergraph will be significantly lower than the number of edges in the confusability graph of the same system. In short, the entanglement-assisted zero-error quantum communication protocol works as follows according to \fref{fig3_20} [\cref{Cubitt10}]. Before the communication, Alice and Bob share entanglement between themselves. The \textit{d}-dimensional shared system between Alice and Bob will be denoted by ${\rho }_{AB}\mathrm{=}\left|\left.{\mathrm{\Phi }}_{AB}\right\rangle \right.\left\langle \left.{\mathrm{\Phi }}_{AB}\right|\right.$, where 
\begin{equation} \label{3.57)} 
\left|\left.{\mathrm{\Phi }}_{AB}\right\rangle \right.\mathrm{=}\frac{\mathrm{1}}{\sqrt{d}}\sum^{d\mathrm{-}\mathrm{1}}_{i\mathrm{=0}}{{\left|\left.i\right\rangle \right.}_A{\left|\left.i\right\rangle \right.}_B} 
\end{equation} 
is a rank-\textit{d }maximally entangled qudit state (also called as \textit{edit}). If Alice would like to send a message $q\mathrm{\in }\left\{\mathrm{1,\dots ,}K\right\}$, where \textit{K} is the number of messages, to Bob, she has to measure her half of the entangled system using a complete orthogonal basis $B_q\mathrm{=}\left\{\left|\left.{\psi }_{x\mathrm{'}}\right\rangle \right.\right\}$, $x\mathrm{'}\mathrm{\in }{\kappa }_q$, where $x\mathrm{'}$ is a vertice in the hypergraph ${\mathcal{G}}_H$ from clique ${\kappa }_q$. The \textit{orthonormal representation of a graph is a map:} the vertice $x\mathrm{'}$ represents the unit vector $\left|\left.{\psi }_{x\mathrm{'}}\right\rangle \right.$ such that if $x$and $x\mathrm{'}$ are \textit{adjacent} then $\left\langle {\psi }_x\mathrel{\left|\vphantom{{\psi }_x {\psi }_{x\mathrm{'}}}\right.\kern-\nulldelimiterspace}{\psi }_{x\mathrm{'}}\right\rangle \mathrm{=0}$ (\textit{i.e., they are orthogonal in the orthonormal representation}) and ${\kappa }_q$ is the clique corresponding to message \textit{q} in the hypergraph ${\mathcal{G}}_H$. The hypergraph has \textit{K} cliques of size \textit{d}, $\left\{{\kappa }_{\mathrm{1}}\mathrm{,\dots ,}{\kappa }_K\right\}$ (i.e., each message $q\mathrm{\in }\left\{\mathrm{1,\dots ,}K\right\}$ is represented by a \textit{d}-size clique in the hypergraph ${\mathcal{G}}_H$.) After the measurement, Bob's state will collapse to ${{\left| {{\psi }_{x}} \right\rangle }^{*}}$. Bob will measure his state in $B_q\mathrm{=}\left\{\left|\left.{\psi }_x\right\rangle \right.\right\}$ to get the final state ${{\left| {{\psi }_{{{x}'}}} \right\rangle }^{*}}$. Bob's output is denoted by \textit{y}. Bob's possible states are determined by those vertices ${x}'$, for which $p\left( \left. y \right|{x}' \right)>0$, and these \textit{adjacent} states are \textit{mutually orthogonal}; i.e., for any two ${{{x}'_{1}}}$ and ${{{x}'_{2}}}$, $\left\langle  {{\psi }_{{{{{x}'_{1}}}}}} | {{\psi }_{{{{{x}'_{2}}}}}} \right\rangle =0$. Finally, Alice makes her measurement using ${{B}_{q}}=\left\{ \left| {{\psi }_{{{x}'}}} \right\rangle  \right\}$, then Bob measures his state ${{\left| {{\psi }_{x}} \right\rangle }^{*}}$ in ${{B}_{q}}=\left\{ \left| {{\psi }_{{{x}'}}} \right\rangle  \right\}$ to produce ${{\left| {{\psi }_{{{x}'}}} \right\rangle }^{*}}$.

In order to make the above explanations more plausible, let us provide an example. Supposed Alice's set contains $K\mathrm{=6}$ codewords and she shares a rank-four (\textit{i.e., d=}4)\textit{ }maximally entangled qudit state with Bob
\begin{equation} \label{3.58)} 
{\mathrm{\Phi }}_{AB}\mathrm{=}\frac{\mathrm{1}}{\sqrt{\mathrm{4}}}\sum^{\mathrm{3}}_{i\mathrm{=0}}{{\left|\left.i\right\rangle \right.}_A{\left|\left.i\right\rangle \right.}_B},                                                                                   
\end{equation} 
however, in the general case \textit{d} can be chosen as large as Alice and Bob would like to use. Alice measures her system from the maximally entangled state, and she chooses a basis among the \textit{K} possible states, according to which message \textit{q} she wants to send Bob. Alice's measurement outcome is depicted by \textit{x}, which is a random value. Alice sends \textit{q} and \textit{x} to the classical channel \textit{N}. In the next phase, Bob performs a projective measurement to decide which \textit{x} value was made to the classical channel by Alice. After Bob has determined it, he can answer which one of the possible \textit{K} messages had been sent by Alice with the help of the maximally entangled system. Alice makes her measurement on her side using one of the six possible bases $B_q\mathrm{=}\left\{\left|\left.{\psi }_{x\mathrm{'}}\right\rangle \right.\right\}$ on her half of the state ${{\rho }_{AB}}$. Her system collapses to $\left|\left.{\psi }_x\right\rangle \right.\mathrm{\in }B_q$, while Bob's system collapses to ${\left|\left.{\psi }_x\right\rangle \right.}^{\mathrm{*}}$, conditioned on \textit{x}. Alice makes \textit{x} to the classical channel \textit{N}; Bob will receive classical message \textit{y}. From the channel output $y\mathrm{=}N\left(x\right)$, where \textit{N} is the classical channel between Alice and Bob, Bob can determine the mutually adjacent inputs (i.e., those inputs which could produce the given output). If Bob makes a measurement in basis $B_q\mathrm{=}\left\{\left|\left.{\psi }_x\right\rangle \right.\right\}$, then he will get ${{\left| {{\psi }_{{{x}'}}} \right\rangle }^{*}}$, where these states for a given set of ${x}'$ corresponding to possible $x$ are \textit{orthogonal} \textit{states}, so he can determine \textit{x} and the original message \textit{q}. The channel output gives Bob the information that some set of mutually adjacent inputs were used on Alice's side. On his half of the entangled system, the states will be mutually orthogonal. A measurement on these mutually orthogonal states will determine Bob's state and he can tell Alice's input with certainty. 

\begin{center}
\begin{figure*}[htbp]
\begin{center}
\includegraphics[angle = 0,width=0.5\linewidth]{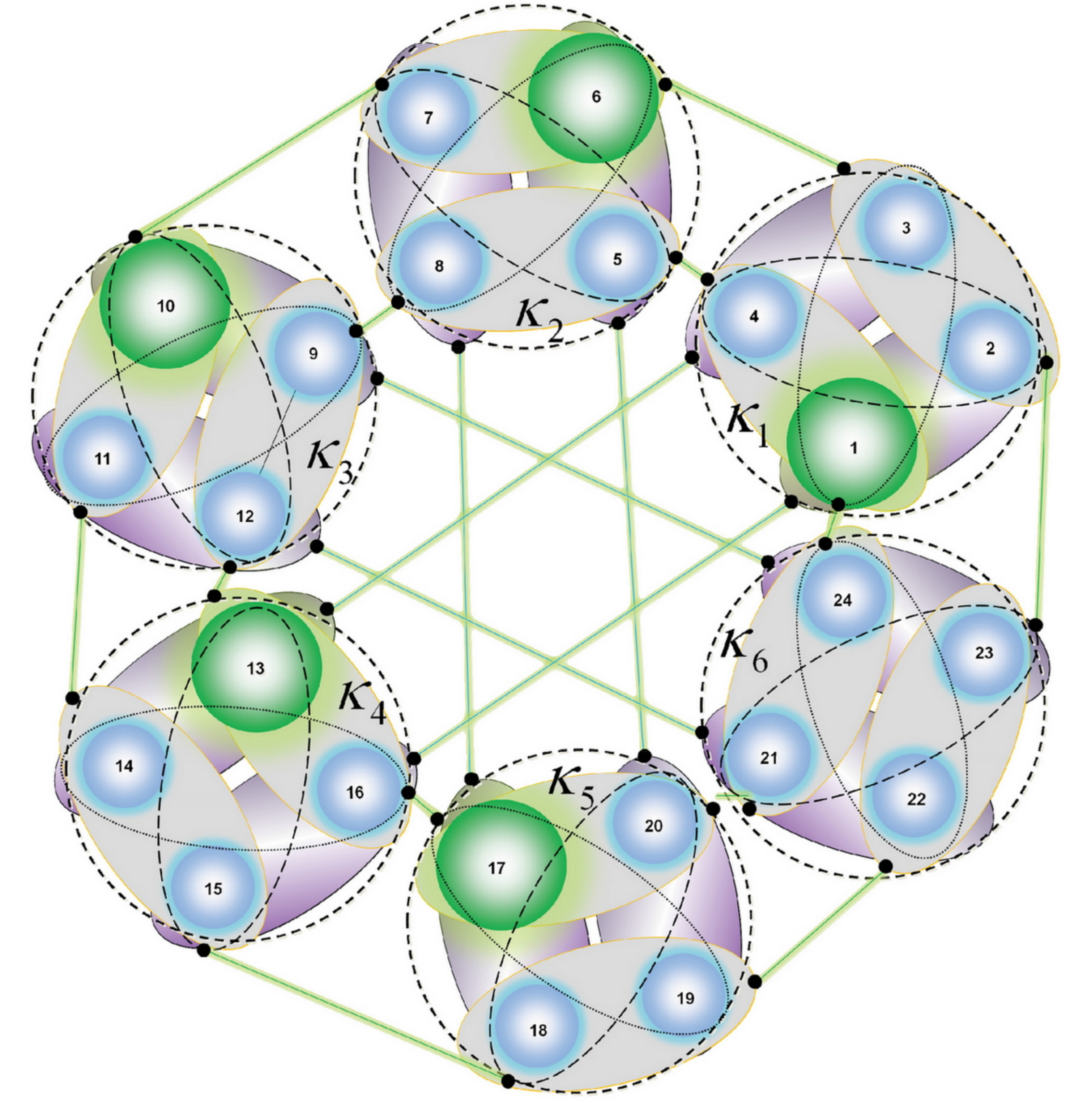}
\caption{The hypergraph of an entanglement-assisted zero-error quantum code. The non-adjacent inputs are depicted by the great shaded circles. The adjacent vertices are depicted by loops connected by the thick lines.} 
\label{fig3_21}
 \end{center}
\end{figure*}
\end{center}

Using this protocol, the number of mutually non-adjacent input messages is
\begin{equation} \label{3.59)} 
K^E\mathrm{\ge }\mathrm{6},                                                  
\end{equation} 
while if Alice and Bob would like to communicate with zero-error but without shared  entanglement, then $K\mathrm{=5}$. As follows, for the single-use classical zero-error capacities we get
\begin{equation} \label{3.60)} 
C^{\left(\mathrm{1}\right)}_0\mathrm{=log}\left(\mathrm{5}\right) 
\end{equation} 
and 
\begin{equation} \label{3.61)} 
C^{E\left(\mathrm{1}\right)}_0\mathrm{=log}\left(K^E\right)\mathrm{=log}\left(\mathrm{6}\right),                                                                     
\end{equation} 
while for the asymptotic entanglement-assisted classical zero-error capacity,
\begin{equation} \label{3.62)} 
C^E_0\mathrm{\ge }\mathrm{log}\left(K^E\right)\mathrm{=log}\left(\mathrm{6}\right).                                                                       
\end{equation} 
According to Alice's $K^E\mathrm{=6}$ messages, the hypergraph can be partitioned into six cliques of size $d\mathrm{=4}$. The adjacent vertices are denoted by a common loop. The overall system contains $\mathrm{6\times 4=24}$ basis vectors. These vectors are grouped into $K^E\mathrm{=6}$ orthogonal bases. Two input vectors are connected in the graph if they are adjacent vectors; i.e., they can produce the same output. The hypergraph ${\mathcal{G}}_H$ of this system is shown in \fref{fig3_21}. The mutually non-adjacent inputs are denoted by the great shaded circles. An important property of the entanglement-assisted classical zero-error capacity is that the number of maximally transmittable messages is not equal to the number of non-adjacent inputs. While the hypergraph has five independent vertices, the maximally transmittable messages are greater than or equal to six. The confusability graph of this system for a single use of quantum channel $\mathcal{N}$ would consist of $\mathrm{6\times 4\times 9=216}$ connections, while the hypergraph has a significantly lower number ($\mathrm{6\times 6=36}$) of hyperedges. The adjacent vertices are depicted by the loops connected by the thick lines. The six possible messages are denoted by the six, four dimensional (i.e., each contains four vertices) cliques $\left\{ {{\kappa }_{1}},\ldots ,{{\kappa }_{K}} \right\}$. The cliques (dashed circles) show the set of those input messages which could result in the same output with a given probability $p\mathrm{>0}$. 

 We note, the cliques are defined in the ${\mathcal{G}}_n$ confusability graph representation, but we also included them on the hypergraph ${\mathcal{G}}_H$. The adjacent vertices which share a loop represent mutually orthogonal input states. For these mutually orthogonal inputs the output will be the same.

 The complete theoretical background of this example, i.e., the proof of the fact, that entanglement can increase the asymptotic classical zero-error capacity $C_0\left(\mathcal{N}\right)$ of a quantum channel was described in [\cref{Cubitt10}]. 

 We have seen in this subsection that shared entanglement between Alice and Bob can help to increase the maximally transmittable classical messages using noisy quantum channels with zero error probability. According to the \textit{Cubitt-Leung-Matthews-Winter} theorem (CLMW theorem) [\cref{Cubitt10}] there exist entanglement-assisted quantum communication protocol which can send one of \textit{K} messages with \textit{zero error}; hence for the entanglement-assisted asymptotic classical zero-error capacity 
\begin{equation} \label{ZEqnNum451017} 
\begin{split}
\mathrm{log}\left(K\right)\mathrm{\le }C_0&\mathrm{=}\mathop{\mathrm{lim}}_{n\mathrm{\to }\mathrm{\infty }}\frac{\mathrm{1}}{n}\mathrm{log}\left(K\left({\mathcal{N}}^{\mathrm{\otimes }n}\right)\right)\\&\mathrm{<}C^E_0\mathrm{=}\mathop{\mathrm{lim}}_{n\mathrm{\to }\mathrm{\infty }}\frac{\mathrm{1}}{n}\mathrm{log}K^E\left({\mathcal{N}}^{\mathrm{\otimes }n}\right)\mathrm{\ge }\mathrm{log}\left(K^E\right).                  
\end{split}
\end{equation} 
Entanglement is very useful in zero-error quantum communication, since with the help of entanglement the maximum amount of perfectly transmittable information can be achieved. 

As was show by Leung et al. [\cref{Leung10}], using special input codewords (based on a special Pauli graph), entanglement can help to increase the classical zero-error capacity to the maximum achievable HSW capacity; that is, there exists a special combination for which the entanglement-assisted classical zero-error capacity $C^E_0\left(\mathcal{N}\right)$ is
\begin{equation} \label{3.64)} 
C^E_0\left(\mathcal{N}\right)\mathrm{=log}\left(\mathrm{9}\right),                                                              
\end{equation} 
while the classical zero-error capacity is
\begin{equation} \label{3.65)} 
C_0\left(\mathcal{N}\right)\mathrm{=log}\left(\mathrm{7}\right),                                                             
\end{equation} 
i.e., with the help of entanglement-assistance the number of possible input messages (\textit{K}) can be increased. 

\begin{center}
\begin{figure}[htbp]
\begin{center}
\includegraphics[angle = 0,width=\linewidth]{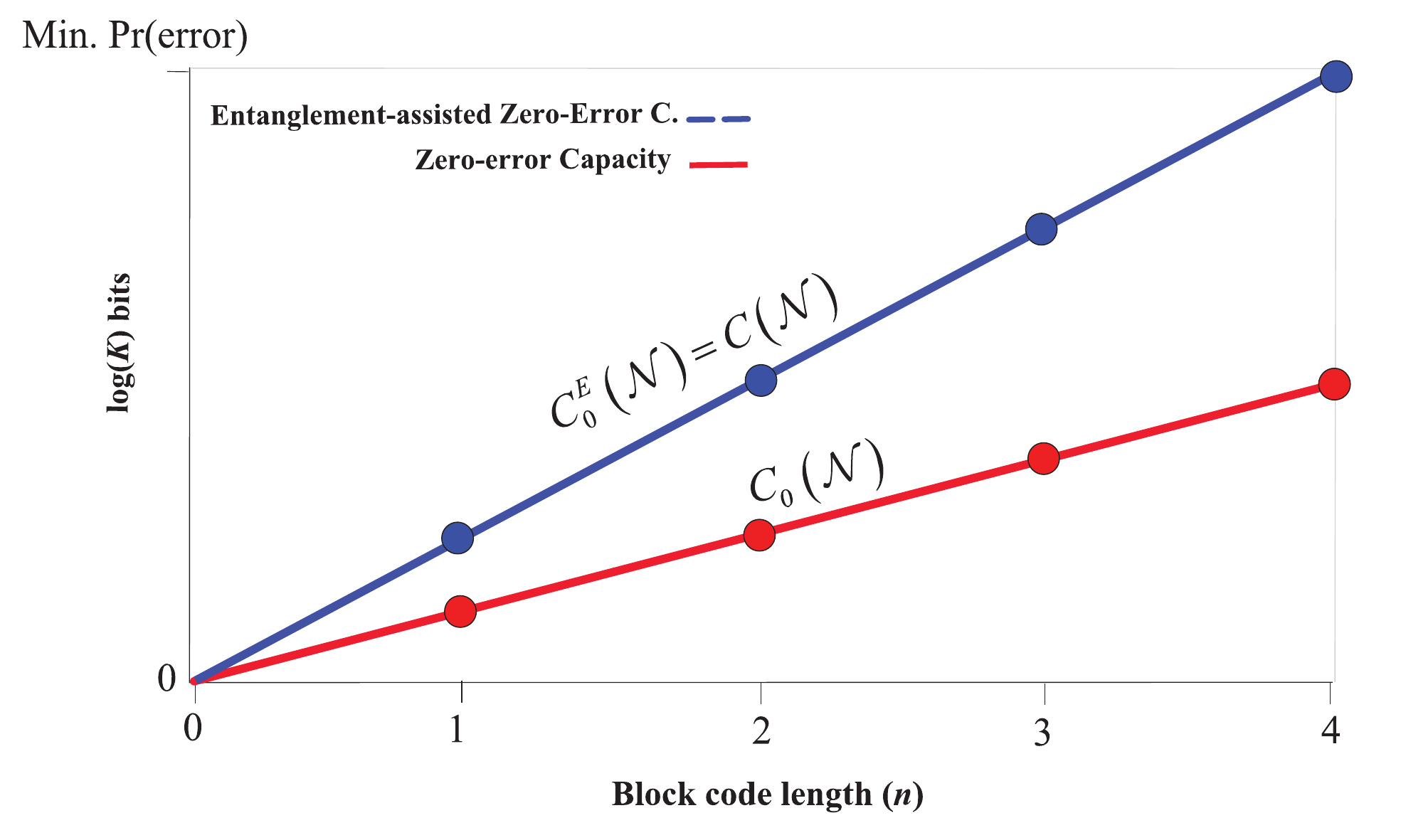}
\caption{The asymptotic classical zero-error capacities without entanglement and with entanglement assistance using a special Pauli graph.} 
\label{fig3_22}
 \end{center}
\end{figure}
\end{center}

 Another important discovery is that for this special input system the entanglement-assisted classical zero-error capacity, $C^E_0\left(\mathcal{N}\right)$, is equal to the maximal transmittable classical information over $\mathcal{N}$; that is
\begin{equation} \label{3.66)} 
C^E_0\left(\mathcal{N}\right)\mathrm{=}C\left(\mathcal{N}\right)\mathrm{=log}\left(\mathrm{9}\right).                                                                    
\end{equation} 
In the asymptotic setting the maximum achievable capacities as functions of block code length are summarized in \fref{fig3_22}.

The maximal amount of transmittable classical information which can be sent through a noisy quantum channel $\mathcal{N}$ without error increases with the length of the input block code, and with the help of EPR input states (for this special Pauli graph-based code) the classical HSW capacity can be reached, which is also the upper bound of the classical zero-error capacity. 

\subsection{Related Work}
 The classical world with the classical communication channel can be viewed as a special case of a quantum channel, since classical information can be encoded into the qubits---just as into classical bits. Classical information can also be encoded in quantum states. In this section we summarize the most important works related to the classical capacity of the quantum channels.

\paragraph{The Early Days}

 At the end of the twentieth century, the capacities of a quantum channel were still an open problem in quantum information theory. Before the several, and rather different, capacities of the quantum channel were recognized, the `academic' opinion was that quantum channels could be used only for the transmission of classical information encoded in the form of quantum states [\cref{Holevo73}], [\cref{Holevo73a}]. As has been found later, the classical capacity of the quantum channel can be measured in several different settings. It was shown that the classical capacity depends on whether the input states are entangled or not, or whether the output is measured by single or by joint measurement setting [\cref{Bennett97}], [\cref{Fuchs2000}], [\cref{King09}]. In a specified manner, the classical capacity has been defined for measuring the maximal asymptotic rate at which classical information can be transmitted through the quantum channel, with an arbitrarily high reliability [\cref{Barnum97a}], [\cref{Schumacher97}]. 

 The first proposed capacity measure was the \textit{classical capacity} of a quantum channel---denoted by $C\left(\mathcal{N}\right)$---measures the maximum transmittable classical information---in the form of product or entangled quantum states. The idea of transmitting classical information through a quantum channel was formulated in the 1970s. The Holevo bound was introduced by Holevo in 1973, however the theorem which describes the classical capacity of the quantum channel in an explicit way appeared just about\textit{ three decades later}, in the mid 1990s. 

 The maximal accessible classical information from a quantum source firstly has been characterized by Levitin [\cref{Levitin69}] and Holevo [\cref{Holevo73}], [\cref{Holevo73a}] in the early days, which were some of the first and most important results in quantum information theory regarding the classical capacity of quantum channels. More information about the connection between the Holevo bound and the accessible information (which quantifies the information of the receiver after the measurement) can be found in [\cref{Holevo73}], [\cref{Holevo73a}]. Later this result was developed and generalized by Holevo, Schumacher, and Westmoreland, and became known in quantum information theory as the \textit{HSW channel capacity} [\cref{Holevo98}], [\cref{Schumacher97}]. The HSW theorem uses the Holevo information to describe the amount of classical information which can be transmitted through a noisy quantum channel, and it makes possible to apply different measurement constructions on the sender and on the receiver's side. The proofs of the HSW theorem, such as the direct coding theorem and the converse theorem, with the complete mathematical background can be found in the work of Holevo [\cref{Holevo98}] and of Schumacher and Westmoreland [\cref{Schumacher97}]. About the efficiency problems of implementation and construction of joint POVM (Positive Operator Valued Measure) measurement setting, as it was shown in the same works of the authors. 

 One of the most important result on the mechanism of the encoding of quantum information into physical particles was discovered by Glauber in the very early years of quantum information processing [\cref{Glauber1963}] and a great summarize from more than four-decades later [\cref{Glauber05}]. Also from this era and field, important results on the encoding and decoding processes of quantum information were shown in the works of Gordon [\cref{Gordon1964}] and Helstrom [\cref{Helstrom76}]. About detection of quantum information and the process of measurement see [\cref{Fannes1973}], or the work of Helstrom from 1976 [\cref{Helstrom76}], or Herbert's work from 1982 [\cref{Herbert82}]. Before their results, Levitin published a paper about the quantum measure of the amount of information in 1969 [\cref{Levitin69}], which was a very important basis for further work.

\paragraph{Classical Capacity of a Quantum Channel}

 The amount of classical information which can be transmitted through a noisy quantum channel in a reliable form with product input states, using the quantum channel many times, was determined by the HSW theorem [\cref{Holevo98}], [\cref{Schumacher97}]. This coding theorem is an analogue to Shannon's classical channel coding theorem, however it extends its possibilities. The inventors of the HSW theorem---Holevo, Schumacher and Westmoreland---proved and concluded independently the same result. Holevo's result from 1998 can be found in [\cref{Holevo98}], Schumacher and Westmoreland's results can be found in [\cref{Schumacher97}]. They, with Hausladen et al. in 1995 [\cref{Hausladen95}], and in 1996 [\cref{Hausladen96}], have also confirmed that the maximal classical information which can be transmitted via pure quantum states is bounded by the Holevo information. 

A different approach to the proof of the HSW theorem was presented by Nielsen and Chuang in 2000 [\cref{Nielsen2000}]. An interesting connection between the mathematical background of the compressibility of quantum states and the HSW theorem was shown by Devetak in 2003 [\cref{Devetak03a}], who proved that a part of the mathematical background constructed for the compression of quantum information can be used to prove the HSW theorem. Another interesting approach for proving the HSW theorem and bounds on the error probability was presented by Hayashi and Nagaoka in 2003 [\cref{Hayashi03}]. The additivity property of qubit channels which require four inputs to achieve capacity was analyzed by Hayashi et al. in [\cref{Hayashi05}].

Very important connections regarding the transmission of classical information over noisy quantum channels was derived in the work of Schumacher and Westmoreland in 1997 [\cref{Schumacher97}], and two years later, a very important work was published on the relevance of optimal signal ensembles in the classical capacity of a noisy quantum channels [\cref{Schumacher99}]. (We also suggest their work on the characterizations of classical and quantum communication processes [\cref{Schumacher99a}].) The classical information capacity of a class of most important practical quantum channels (Gaussian quantum channels) was shown by Wolf and Eisert [\cref{Wolf05}] or the work of Lupo et al. [\cref{Lupo11}].  The generalized minimal output entropy conjecture for Gaussian channels was studied by Giovannetti et al. [\cref{Giovannetti10}].

About the role of feedback in quantum communication, we suggest the works of Bowen [\cref{Bowen04}] and 2005 [\cref{Bowen05}], the article of Bowen et al. [\cref{Bowen05a}], and the work of Harrow [\cref{Harrow04a}]. The works of Bowen provide a great introduction to the role of quantum feedback on the classical capacity of the quantum channel, it was still an open question before. As he concluded, the classical capacity of a quantum channel using quantum feedback is equal to the entanglement-assisted classical capacity, the proof was given in Bowen and Nagarajan's paper [\cref{Bowen05a}].

We have also seen that the noise of a quantum channel can be viewed as a result of the entanglement between the output and the reference system called the purification state (see purification in \eqref{ZEqnNum381325}). Some information leaks to the environment, and to the purification state, which purification state cannot be accessed. As is implicitly woven into this section, a noisy quantum channel can be viewed as a special case of an ideal quantum communication channel. The properties of the general quantum channel model and the quantum mutual information function can be found in the work of Adami and Cerf [\cref{Adami96}]. 

A great analysis of completely-positive trace preserving (CPTP) maps was published by Ruskai et al. [\cref{Ruskai01}]. Further information on the classical capacity of a quantum channel can be found in [\cref{Bennett98}], [\cref{Holevo98}], [\cref{King09}], [\cref{Nielsen2000}].

\paragraph{Entanglement-assisted Classical Capacity}

 In the early 1970s, it was also established that the classical capacity of a quantum channel can be higher with \textit{shared entanglement}---this capacity is known as the \textit{entanglement-assisted classical capacity} of a quantum channel, which was completely defined by Bennett et al. just in 1999 [\cref{Bennett99}], and is denoted by $C_E\left(\mathcal{N}\right)$. The preliminaries of the definition of this quantity were laid down by Bennett and Wiesner in 1992 [\cref{Bennett92}]. Later, in 2002 Holevo published a review paper about the entanglement-assisted classical capacity of a quantum channel [\cref{Holevo02a}]. 

Entanglement-assisted classical communication requires a super-dense protocol-like encoding and decoding strategy [\cref{Bennett02}]. About the classical capacity of a noiseless quantum channel assisted by noisy entanglement, an interesting paper was published by Horodecki et al. in 2001 [\cref{Horodecki01}]. In the same work the authors have defined the `noisy version' of the well-known superdense coding protocol, which originally was defined by Bennett in 1992 [\cref{Bennett92}] for ideal (hence noiseless) quantum channels. As can be found in the works of Bennett et al. from 1999 [\cref{Bennett99}] and from 2002 [\cref{Bennett02}], the \textit{entanglement-assisted classical capacity} opened the possibility to transmit more classical information using shared entanglement (in case of single-use capacity). As can be checked by the reader, the treatment of entanglement-assisted classical capacity is based on the working mechanism of the well-known superdense coding protocol---however, classical entanglement-assisted classical capacity used a noisy quantum channel instead of an ideal one. 

Bennett, in two papers from 1999 [\cref{Bennett99}] and 2002 [\cref{Bennett02}] showed that the \textit{quantum mutual information} function (see Adami and Cerf's work [\cref{Adami96}]) can be used to describe the classical entanglement-assisted capacity of the quantum channel i.e., the \textit{maximized quantum mutual information of a quantum channel and the entanglement-assisted classical capacity are equal.} The connection between the quantum mutual information and the entanglement-assisted capacity can be found in the works of Bennett et al. [\cref{Bennett02}] and [\cref{Bennett99}]. In the latter work, the formula of the quantum-version of the well-known classical Shannon formula was generalized for the classical capacity of the quantum channel. In these two papers the authors also proved that the entanglement-assisted classical capacity is an upper bound of the HSW channel capacity. Holevo gave an explicit upper bound on the classical information which can be transmitted through a noisy quantum channel, it is known as the Holevo-bound. The Holevo-bound states that the most classical information which can be transmitted in a qubit (i.e., two level quantum system) through a noiseless quantum channel in a reliable form, is one bit. However, as was shown later by Bennett et al. in 1999 [\cref{Bennett99}], the picture changes, if the parties use shared entanglement (known as the \textit{Bennett-Shor-Smolin-Thapliyal, or the BSST-} theorem). As follows, the BSST-theorem gives a closer approximation to the maximal transmittable classical information (i.e., to the `single-use' capacity) over quantum channels, hence it can be viewed as the \textit{true `quantum version' of the well known classical Shannon capacity formula} (since it is a maximization formula), instead of the `non entanglement-assisted' classical capacity. Moreover, the inventors of the BSST-theorem have also found a very important property of the entanglement-assisted classical capacity: \textit{its single-use version is equal to the asymptotic version}, which implies the fact that no regularization is needed. (As we have seen in this section, we are not so lucky in the case of general classical and private classical capacities. As we will show in \sref{sec4}, we are `unlucky' in the case of quantum capacity, too.) They have also found that no classical feedback channel can increase the entanglement-assisted classical capacity of a quantum channel, and this is also true for the classical (i.e., the not entanglement-assisted one) capacity of a quantum channel. These results were also confirmed by Holevo in 2002 [\cref{Holevo02a}]. It was a very important discovery in the history of the classical capacity of the quantum channel, and due to the BSST-theorem, the analogue with classical Shannon's formula \textit{has been finally} \textit{completed}. Later, it was discovered that in special cases the entanglement-assisted capacity of a quantum channel can be improved [\cref{Harrow04}], [\cref{Patron09}]. The Holevo information can be attained even with pure input states, and the concavity of the Holevo information also shown. The concavity can be used to compute the classical HSW capacity of quantum channels, since the maximum of the transmittable information can be computed by a local maximum among the input states. Moreover, as was shown by Bennett et al. in 2002, the concavity holds for the entanglement-assisted classical capacity, too [\cref{Bennett02}], [\cref{Bennett09}]--- the concavity, along with the non-necessity of any computation of an asymptotic formula, and the use of classical feedback channels to improve the capacity, \textit{makes the entanglement-assisted classical capacity the most generalized classical capacity}---and it has the same role as Shannon's formula in classical information theory [\cref{Bennett09}]. The fact that the classical feedback channel does not increase the classical capacity and the entanglement-assisted classical capacity of the quantum channel, follows from the work of Bennett et al., and the proof of the BSST-theorem [\cref{Bennett02}]. Wang and Renner's work [\cref{Wang10}] introduces the reader to the connection between the single-use classical capacity and hypothesis testing.

\paragraph{The Private Classical Capacity}
The third classical capacity of the quantum channel is the \textit{private classical capacity}, denoted by $P\left(\mathcal{N}\right)$. The concept of private classical capacity was introduced by Devetak in 2003 [\cref{Devetak03a}], and one year later by Cai et al. in 2004 [\cref{Cai04}]. Private classical capacity measures classical information, and it is always at least as large as the single-use quantum capacity (or the quantum coherent information) of any quantum channel. As shown in [\cref{Devetak05b}], for a degradable quantum channel the coherent information (see \sref{sec4}) is additive [\cref{Devetak05b}],---however for a general quantum channel these statements do not hold. The additivity of private information would also imply the fact that shared entanglement cannot help to enhance the private classical capacity for degradable quantum channels. The complete proof of the private classical capacity of the quantum channel was made by Devetak [\cref{Devetak03a}], who also cleared up the connection between private classical capacity and the quantum capacity. As was shown by Smith et al. [\cref{Smith08d}], the private classical capacity of a quantum channel is additive for degradable quantum channels, and closely related to the quantum capacity of a quantum channel (moreover, Smith has shown that the private classical capacity is equal to the quantum coherent information for degradable channels), since in both cases we have to `protect' the quantum states: in the case of private classical capacity the enemy is called Eve (the eavesdropper), while in the latter case the name of the enemy is `environment.' As was shown in [\cref{Devetak03a}], the eavesdropper in private coding acts as the environment in quantum coding of the quantum state, and vice-versa. This `gateway' or `dictionary' between the classical capacity and the quantum capacity of the quantum channel was also used by Devetak [\cref{Devetak03a}], by Devetak and Shor [\cref{Devetak05b}] and by Smith and Smolin [\cref{Smith08d}], using a different interpretation. 

About the coherent communication with continuous quantum variables over the quantum channels a work was published Wilde et al. in [\cref{Wilde07}] and [\cref{Wilde10}]. On the noisy processing of private quantum states, see the work of Renes et al. [\cref{Renes07}]. A further application of private classical information in communicating over adversarial quantum channels was shown by Leung et al. [\cref{Leung08}]. Further information about the private classical capacity can be found in [\cref{Bradler09}], [\cref{Devetak03a}], [\cref{Devetak05a}], [\cref{Li09}], [\cref{Smith08d}], [\cref{Smith09a}], [\cref{Smith09b}]. Another important work on non-additive quantum codes was shown by Smolin et al. [\cref{Smolin07}]. A great summary on the main results of Quantum Shannon Theory was published by Wilde [\cref{Wilde11}]. For further information on quantum channel capacities and advanced quantum communications see the book of Imre and Gyongyosi [\cref{Imre13}], and also [\cref{Gyongyosi13}]. We also suggest the great work of Bennett et al. on the quantum reverse Shannon theorem [\cref{Bennett09}]. A work on the connection of secure communication and Gaussian-state quantum Illumination was published by Shapiro [\cref{Shapiro09}].

\paragraph{The Zero-Error Classical Capacity}

 The properties of \textit{zero-error} communication systems are discussed in Shannon's famous paper on the zero-error capacity of a noisy channel [\cref{Shannon56}], in the work of K\"{o}rner and Orlitsky on zero-error information theory [\cref{Korner98}], and in the work of Bollob\'{a}s on modern graph theory [\cref{Bollobas98}]. We also suggest the famous proof of Lov\'{a}sz on the Shannon capacity of a graph [\cref{Lovasz79}]. The proof of the classical zero-error capacity of quantum channel can be found in Medeiros's work [\cref{Medeiros05}]. Here, he has shown, that the classical zero-error capacity of the quantum channel is also bounded above by the classical HSW capacity. The important definitions of quantum zero-error communication and the characterization of quantum states for the zero-error capacity were given by Medeiros et al., in [\cref{Medeiros06}]. On the complexity of computation of zero-error capacity of quantum channels see the work of Beigi and Shor [\cref{Beigi07}]. The fact, that the zero-error classical capacity of the quantum channel can be increased with entanglement, was shown by Cubitt et al. in 2010 [\cref{Cubitt10}]. The role of entanglement in the asymptotic rate of zero-error classical communication over quantum channels was shown by Leung et al. in 2010 [\cref{Leung10}]. For further information about the theoretical background of entanglement-assisted zero-error quantum communication see [\cref{Cubitt10}] and for the properties of entanglement, the proof of the Bell-Kochen-Specker theorem in [\cref{Bell1966}], [\cref{Kochen1967}]. 

\section{The Quantum Capacity of a Quantum Channel}
\label{sec4}
 Having discussed the general model of quantum channels and introduced various classical capacities in this section we focus on the \textit{quantum information} transfer over quantum channels. Two new quantities will be explained. By means of \textit{fidelity} \textit{F} one can describe the differences between two quantum states e.g. between the input and output states of a quantum channel. On the other hand \textit{quantum coherent information} represents the quantum information loss to the environment during quantum communication similarly as mutual information did for a classical channel \textit{N}. Exploiting this latter quantity we can define the maximal quantum information transmission rate through quantum channels -- the quantum capacity $Q\left(\mathcal{N}\right)$ analogously to Shannon's noisy channel theorem. As we have seen \sref{sec3}, the classical capacity of a quantum channel is described by the maximum of quantum mutual information and the Holevo information. The quantum capacity of the quantum channels is described by the maximum of \textit{quantum coherent information}. The concept of quantum coherent information plays a fundamental role in the computation of the \textit{LSD (Lloyd-Shor-Devetak)} channel capacity [\cref{Devetak03a}], [\cref{Lloyd97}], [\cref{Shor02}] which measures the asymptotic quantum capacity of the quantum capacity in general.  

This section is organized as follows. First, we discuss the transmission of quantum information over a nosy quantum channel. Next, we define the quantum coherent information and overview its main properties. Finally the formula for the measure of maximal transmittable quantum information over a quantum channel will be introduced. The description of the most relevant works can be found in the Related Work subsection.

\subsection{Preserving Quantum Information}

 The encoding and decoding quantum information have many similarities to the classical case, however, there exist some fundamental differences, as we will reveal in this section. In the case of quantum communication, the source is a quantum information source and the \textit{quantum information} is encoded into quantum states. When transmitting quantum information, the information is encoded into non-orthogonal superposed or entangled quantum states chosen from the ensemble $\left\{{\rho }_k\right\}$ according to a given probability $\left\{p_k\right\}$. If the states $\left\{{\rho }_k\right\}$ are pure and mutually orthogonal, we talk about classical information; that is, in this case the quantum information reduces to classical. 

\begin{center}
\begin{figure*}[htbp]
\begin{center}
\includegraphics[angle = 0,width=0.65\linewidth]{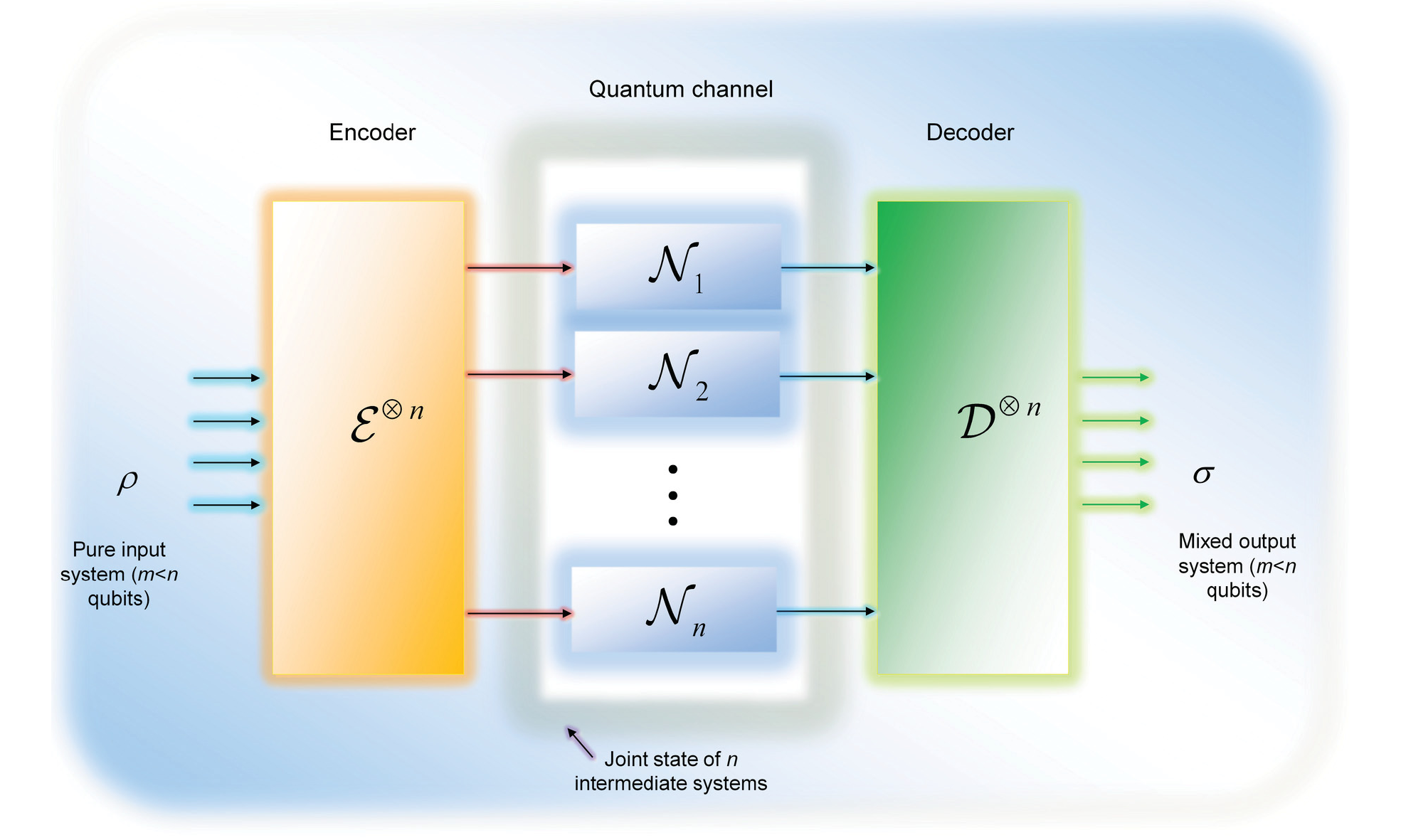}
\caption{Transmission of quantum information through the quantum channel. The encoder produces a joint state of \textit{n} intermediate systems. The encoded qubits are passed through the independent instances of the quantum channel.} 
\label{fig4_1}
 \end{center}
\end{figure*}
\end{center}

Formulating the process more precisely (see \fref{fig4_1}) the encoding and the decoding mathematically can be described by the operators $\mathcal{E}$ and $\mathcal{D}$ realized on the blocks of quantum states. The input of the encoder consists of \textit{m} pure quantum states, and the encoder maps the \textit{m} quantum states into the joint state of \textit{n} intermediate systems. Each of them is sent through an independent instance of the quantum channel $\mathcal{N}$ and decoded by the decoder $\mathcal{D}$, which results in \textit{m} quantum states again. The output of the decoder $\mathcal{D}$ is typically mixed, according to the noise of the quantum channel. The rate of the code is equal to \textit{m}/\textit{n}. 

\begin{center}
\begin{figure}[htbp]
\vspace{-0.65cm}
\begin{center}
\includegraphics[angle = 0,width=\linewidth]{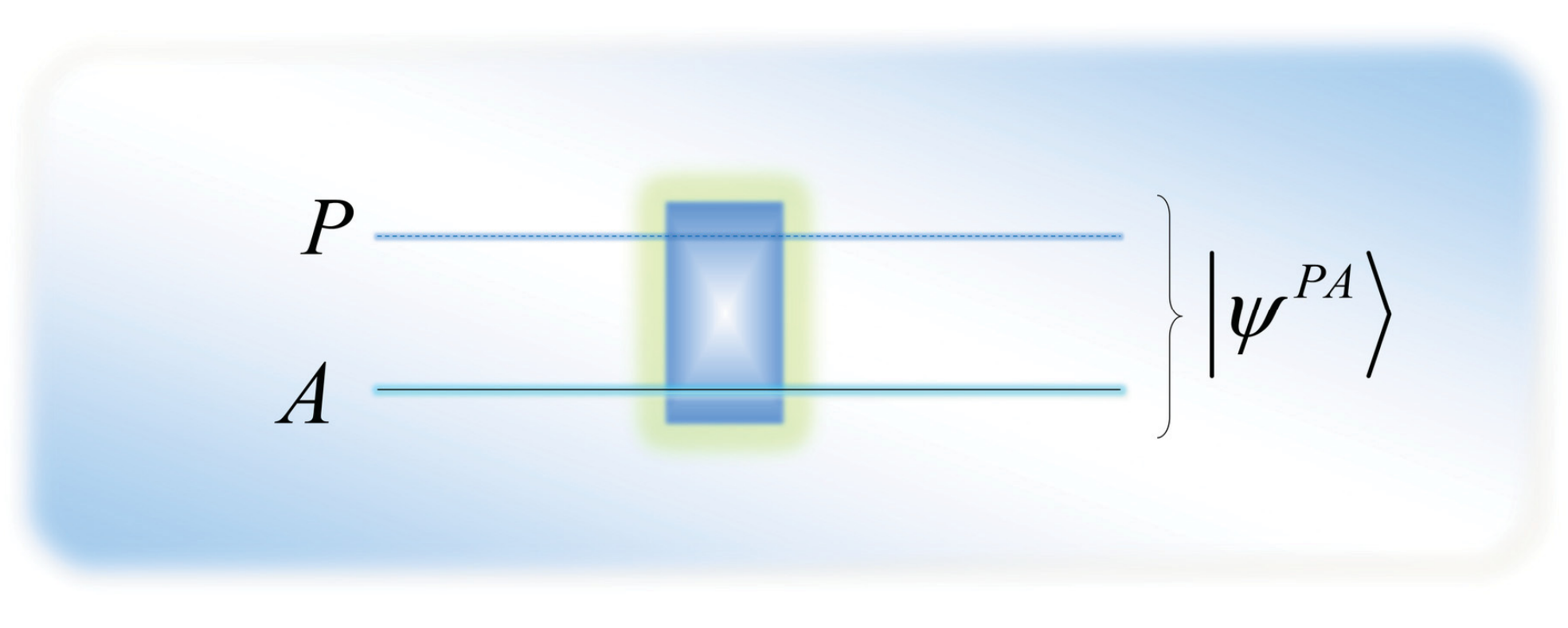}
\caption{Initially, the quantum system and the reference system are in a pure entangled state.} 
\label{fig4_2}
 \end{center}
\end{figure}
\end{center}

Theoretically quantum states have to preserve their original superposition during the whole transmission, without the disturbance of their actual properties. Practically, quantum channels are entangled with the environment which results in mixed states at the output. Mixed states are classical probability weighted sum of pure states where these probabilities appear due to the interaction with the environment (i.e., noise). Therefore, we introduce a new quantity, which is able to describe the quality of the transmission of the superposed states through the quantum channel. The fidelity (see Appendix) for two pure quantum states is defined as
\begin{equation} \label{4.1)} 
F\left({|\left.\varphi \right\rangle},\left|\left.\psi \right\rangle \right.\right)\mathrm{=}{\left|\left\langle \varphi \mathrel{\left|\vphantom{\varphi  \psi }\right.\kern-\nulldelimiterspace}\psi \right\rangle \right|}^{\mathrm{2}}.                                       
\end{equation} 
The fidelity of quantum states can describe the relation of Alice pure channel input state $\left|\left.\psi \right\rangle \right.$ and the received mixed quantum system $\sigma \mathrm{=}\sum^{n\mathrm{-}\mathrm{1}}_{i\mathrm{=0}}{p_i{\rho }_i}\mathrm{=}\sum^{n\mathrm{-}\mathrm{1}}_{i\mathrm{=0}}{p_i\left|\left.{\psi }_i\right\rangle \right.\left\langle \left.{\psi }_i\right|\right.}$ at the channel output as
\begin{equation} \label{ZEqnNum774290} 
F\left({\left| \psi  \right\rangle},\sigma  \right)=\left\langle  \psi  | \left. \sigma  \right|\psi  \right\rangle =\sum\limits_{i=0}^{n-1}{{{p}_{i}}{{\left| \left\langle  \psi  | {{\psi }_{i}} \right\rangle  \right|}^{2}}}.                                                            
\end{equation} 
Fidelity can also be defined for \textit{mixed} states $\sigma $ and $\rho $
\begin{equation} \label{4.3)} 
F\left(\rho ,\sigma \right)\mathrm{=}{\left[Tr\left(\sqrt{\sqrt{\sigma }\rho \sqrt{\sigma }}\right)\right]}^{\mathrm{2}}\mathrm{=}\sum_i{p_i}{\left[Tr\left(\sqrt{\sqrt{{\sigma }_i}{\rho }_i\sqrt{{\sigma }_i}}\right)\right]}^{\mathrm{2}}.                                                
\end{equation} 
Let us assume that we have a quantum system denoted by \textit{A} and a reference system \textit{P}. Initially, the quantum system \textit{A} and the reference system \textit{P} are in a \textit{pure} \textit{entangled} state, denoted by $\left|\left.{\psi }^{PA}\right\rangle \right.$.  The density matrix ${\rho }_A$ of system \textit{A} can be expressed by a partial trace over \textit{P}, as follows
\begin{equation} \label{4.4)} 
{\rho }_A\mathrm{=}Tr_P\left(\left|\left.{\psi }^{PA}\right\rangle \right.\left\langle \left.{\psi }^{PA}\right|\right.\right).                                                                 
\end{equation} 
The entanglement between the initial quantum system and the reference state is illustrated in \fref{fig4_2}.

\begin{center}
\begin{figure}[htbp]
\begin{center}
\includegraphics[angle = 0,width=\linewidth]{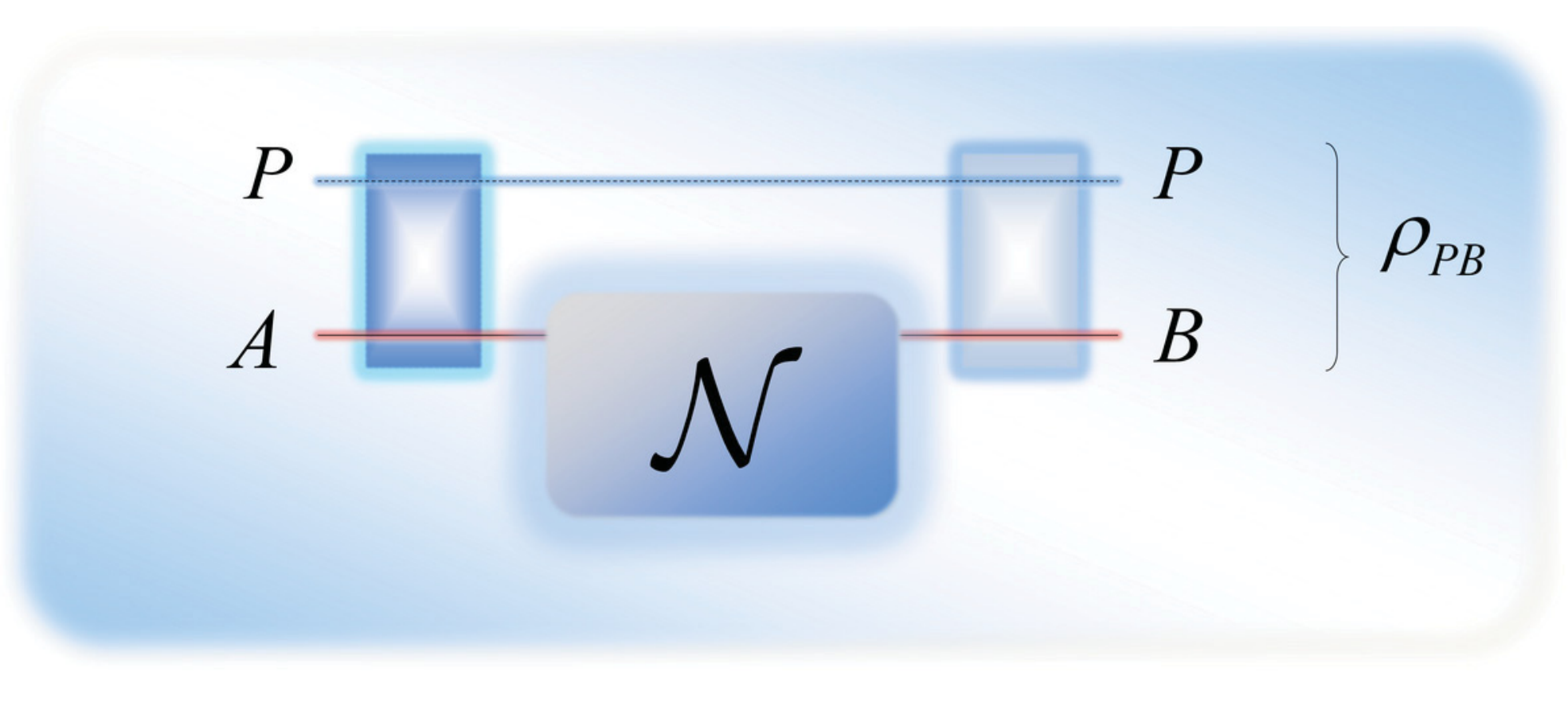}
\caption{After system \textit{A} is sent through the quantum channel $\mathcal{N}$, both the quantum system \textit{A} and the entanglement between \textit{A} and \textit{P} are affected.} 
\label{fig4_3}
 \end{center}
\end{figure}
\end{center}

In the next step, ${\rho }_A$ will be transmitted through the quantum channel $\mathcal{N}$, while the reference state \textit{P} is \textit{isolated from the environment }(see \sref{sec2}), hence it has not been not modified during the transmission. After the quantum system ${\rho }_A$ is transmitted through the quantum channel, the final state will be 
\begin{equation} \label{ZEqnNum517802} 
{\rho }^{PB}\mathrm{=}\left({\mathcal{I}}^P\mathrm{\otimes }{\mathcal{N}}^A\right)\left(\left|\left.{\psi }^{PA}\right\rangle \right.\left\langle \left.{\psi }^{PA}\right|\right.\right),                                                                      
\end{equation} 
where ${\mathcal{I}}^P$ is the identity transformation realized on the reference system \textit{P}. After the system \textit{A} is sent through the quantum channel, both the quantum system \textit{A} and the entanglement between \textit{A} and \textit{P} are affected, as we illustrated in \fref{fig4_3}. The resultant output system is denoted by $B$.

Now, we can study the preserved entanglement between the two systems \textit{A} and \textit{P}. Entanglement fidelity $F_E$ measures the fidelity between the initial pure system $\left|\left.{\psi }^{PA}\right\rangle \right.$ and the mixed output quantum system ${\rho }_{PB}$ as follows
\begin{equation} \label{4.6)} 
\begin{split}
F_E&\mathrm{=}F_E\left({\rho }_A,\mathcal{N}\right)\mathrm{=}F\left(\left|\left.{\psi }^{PA}\right\rangle \right.,{\rho }_{PB}\right)\\&\mathrm{=}\left\langle {\psi }^{PA}\mathrel{\left|\vphantom{{\psi }^{PA} \left({\mathcal{I}}^P\mathrm{\otimes }{\mathcal{N}}^A\right)\left.\left(\left|\left.{\psi }^{PA}\right\rangle \right.\left\langle \left.{\psi }^{PA}\right|\right.\right)\right|{\psi }^{PA}}\right.\kern-\nulldelimiterspace}\left({\mathcal{I}}^P\mathrm{\otimes }{\mathcal{N}}^A\right)\left.\left(\left|\left.{\psi }^{PA}\right\rangle \right.\left\langle \left.{\psi }^{PA}\right|\right.\right)\right|{\psi }^{PA}\right\rangle .                      
\end{split}
\end{equation} 
It is important to highlight the fact that $F_E$ depends on $\left|\left.{\psi }^{PA}\right\rangle \right.$ i.e., on the reference system. The sender's goal is to transmit quantum information, i.e., to preserve entanglement between \textit{A} and the inaccessible reference system \textit{P}. Alice can apply many independent channel uses of the same noisy quantum channel $\mathcal{N}$ to transmit the quantum information. 
Similar to encoding classical information into the quantum states, the quantum messages can be transmitted over copies of a quantum channel. In this case, we have \textit{n} copies of a quantum channel $\mathcal{N}$. 

\subsection{Quantum Coherent Information}

 In case of the classical capacity $C\left(\mathcal{N}\right)$, the correlation between the input and the output is measured by the Holevo information and the quantum mutual information function. In case of the quantum capacity $Q\left(\mathcal{N}\right)$, we have a completely different correlation measure with completely different behaviors: it is called the \textit{quantum coherent information}. There is a \textit{very important distinction} between the maximized quantum mutual information and maximized quantum coherent information: \textit{the maximized quantum mutual information of a quantum channel }$\mathcal{N}$ \textit{is always additive }(see \sref{sec2})\textit{, but the quantum coherent information is not.}

 The ${\mathrm{S}}_E$ \textit{entropy exchange} between the initial system $PA$ and the output system $PB$ is defined as follows. The entropy that is acquired by $PA$ when input system \textit{A} is transmitted through the quantum channel $\mathcal{N}$ can be expressed with the help of the von Neumann entropy function as follows 
\begin{equation} \label{ZEqnNum500801} 
{\mathrm{S}}_E\mathrm{=}{\mathrm{S}}_E\left({\rho }_A\mathrm{:}\mathcal{N}\left({\rho }_A\right)\right)\mathrm{=S}\left({\rho }_{PB}\right),                                                                 
\end{equation} 
or in other words the von Neumann entropy of the output system ${\rho }_{PB}$. As can be observed, the value of entropy exchange depends on ${\rho }_A$ and $\mathcal{N}$ and is independent from the purification system \textit{P}. Now, we introduce the environment state \textit{E}, and we will describe the map of the quantum channel as a unitary transformation. The environment is initially in a pure state $\left|\left.0\right\rangle \right.$. After the unitary transformation $U_{A\mathrm{\to }BE}$ has been applied to the initial system $A\left|\left.0\right\rangle \right.$, it becomes 
\begin{equation} \label{4.8)} 
U_{A\mathrm{\to }BE}\left(A\left|\left.0\right\rangle \right.\right)\mathrm{=}BE.                                                                            
\end{equation} 

From the entropy of the \textit{final state} of the environment ${\rho }_E$, the \textit{entropy exchange} ${\mathrm{S}}_E$ can be expressed as 
\begin{equation} \label{4.9)} 
\mathrm{S}\left({\rho }_{PB}\right)\mathrm{=S}\left({\rho }_E\right)\mathrm{=}{\mathrm{S}}_E.                                                                        
\end{equation} 
${\mathrm{S}}_E$ measures the increase of entropy of the environment \textit{E}, or with other words,  the entanglement between $PA$ and \textit{E,} after the unitary transformation $U_{A\mathrm{\to }BE}$ had been applied to the system. This entropy exchange ${\mathrm{S}}_E$ is analogous to the classical conditional entropy; however in this case we talk about quantum instead of classical information. 

 Using the notations of \fref{fig4_3}, the quantum coherent information can be expressed as 
\begin{equation} \label{ZEqnNum259518} 
 \begin{split}
I_{coh}\left({\rho }_A\mathrm{:}\mathcal{N}\left({\rho }_A\right)\right)\mathrm{=S}\left(\mathcal{N}\left({\rho }_A\right)\right)\mathrm{-}{\mathrm{S}}_E\left({\rho }_A\mathrm{:}\mathcal{N}\left({\rho }_A\right)\right) \\ 
\mathrm{=S}\left({\rho }_B\right)\mathrm{-}\mathrm{S}\left({\rho }_{PB}\right) \\ 
\mathrm{=S}\left({\rho }_B\right)\mathrm{-}\mathrm{S}\left({\rho }_E\right), \end{split}
\end{equation} 
where ${\mathrm{S}}_E\left({\rho }_A\mathrm{:}\mathcal{N}\left({\rho }_A\right)\right)$ is the entropy exchange as defined in \eqref{ZEqnNum500801}.

Using the definition of quantum coherent information \eqref{ZEqnNum259518}, it can be verified that quantum coherent information takes its maximum if systems \textit{A} and \textit{P} are \textit{maximally entangled} and the quantum channel $\mathcal{N}$ is \textit{completely noiseless}. This can be presented easily
\begin{equation} \label{4.11)} 
\mathrm{S}\left({\rho }_B\right)\mathrm{=S}\left({\rho }_A\right),                                      
\end{equation} 
since the input state ${\rho }_A$ is maximally mixed, and 
\begin{equation} \label{4.12)} 
\mathrm{S}\left({\rho }_{PB}\right)\mathrm{=0},                                
\end{equation} 
because $\left|\left.{\psi }^{PA}\right\rangle \right.\left\langle \left.{\psi }^{PA}\right|\right.$ will remain pure after the state has been transmitted through the ideal quantum channel. If the input system $\left|\left.{\psi }^{PA}\right\rangle \right.\left\langle \left.{\psi }^{PA}\right|\right.$ is not a maximally entangled state, or the quantum channel $\mathcal{N}$ is not ideal, then the value of quantum coherent information will decrease. 

\begin{center}
\begin{figure*}[htbp]
\begin{center}
\includegraphics[angle = 0,width=0.59\linewidth]{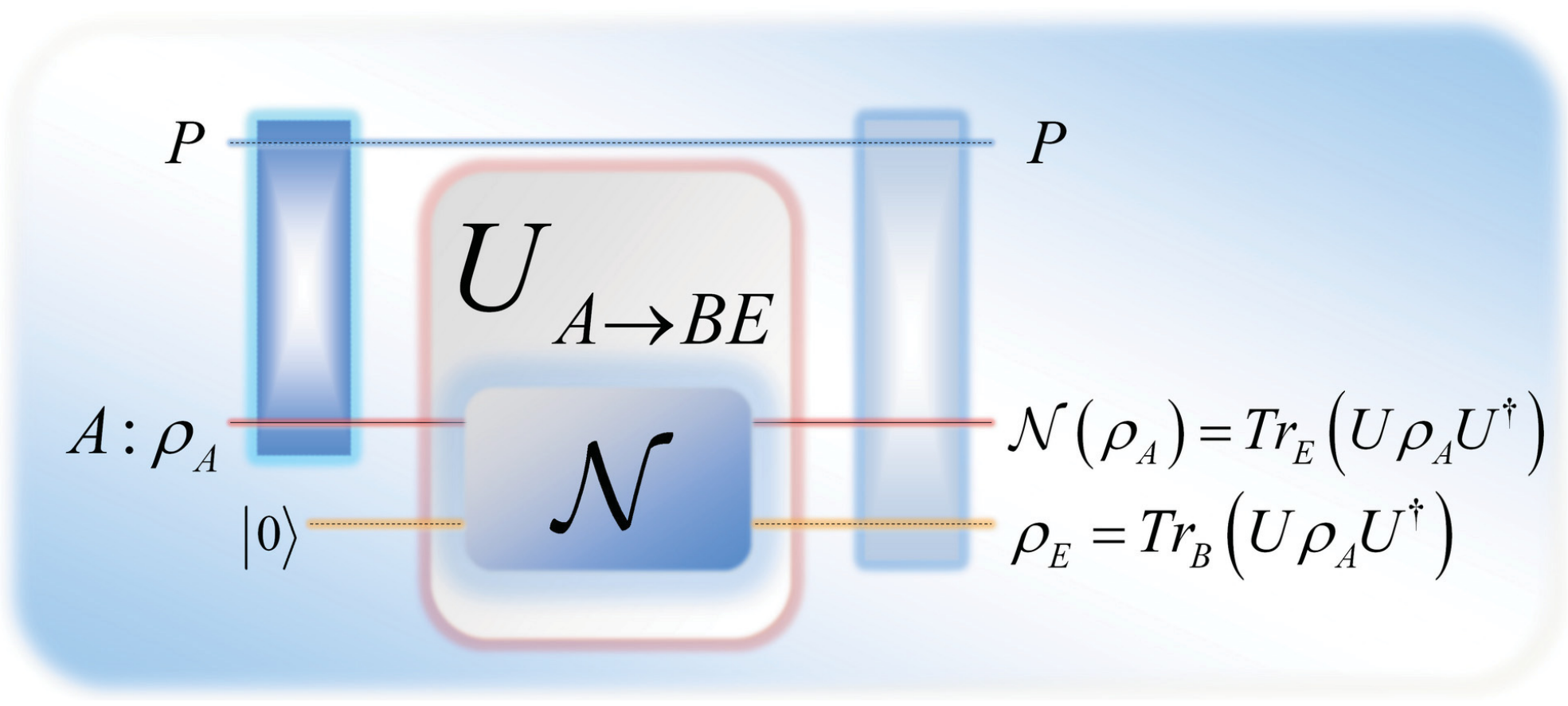}
\caption{The conceptional meaning of quantum coherent information. The unitary transformation represents the channel and the environment. The first receiver is Bob, the second is the environment. The state of the environment belonging to the unitary transformation is represented by dashed line. The outputs can be computed as the partial traces of the joint system.} 
\label{fig4_6}
 \end{center}
\end{figure*}
\end{center}

 Considering another expressive picture, quantum coherent information measures the quantum capacity as the difference between the von Neumann entropies of two channel output states. The first state is received by Bob, while the second one is received by a `second receiver' - called the environment. If we express the transformation of a quantum channel as the partial trace of the overall system, then 
\begin{equation} \label{ZEqnNum569277} 
\mathcal{N}\left({\rho }_A\right)\mathrm{=}Tr_E\left(U{\rho }_AU^{\mathrm{\dagger }}\right),                                             
\end{equation} 
and similarly, for the `effect' of the environment \textit{E}, we will get 
\begin{equation} \label{ZEqnNum431779} 
E\left({\rho }_A\right)\mathrm{=}{\rho }_E\mathrm{=}Tr_B\left(U{\rho }_AU^{\mathrm{\dagger }}\right).                                                  
\end{equation} 
The results of \eqref{ZEqnNum569277} and \eqref{ZEqnNum431779} are summarized in \fref{fig4_6}. 

It can be concluded that the quantum coherent information measures the capability of transmission of entanglement over a quantum channel. For the exact value of quantum coherent information of some important quantum channels see \sref{sec5}.

\subsection{Connection between Classical and Quantum Information}
 As it has been shown by Schumacher and Westmoreland [\cref{Schumacher2000}], the $I_{coh}$ quantum coherent information also can be expressed with the help of Holevo information, as follows 
\begin{equation} \label{ZEqnNum572683} 
I_{coh}\left({\rho }_A\mathrm{:}\mathcal{N}\left({\rho }_A\right)\right)\mathrm{=}\left({\mathcal{X}}_{AB}\mathrm{-}{\mathcal{X}}_{AE}\right),                                                          
\end{equation} 
where 
\begin{equation} \label{4.16)} 
{\mathcal{X}}_{AB}\mathrm{=S}\left({\mathcal{N}}_{AB}\left({\rho }_{AB}\right)\right)\mathrm{-}\sum_i{p_i\mathrm{S}\left({\mathcal{N}}_{AB}\left({\rho }_i\right)\right)} 
\end{equation} 
and 
\begin{equation} \label{4.17)} 
{\mathcal{X}}_{AE}\mathrm{=S}\left({\mathcal{N}}_{AE}\left({\rho }_{AE}\right)\right)\mathrm{-}\sum_i{p_i\mathrm{S}\left({\mathcal{N}}_{AE}\left({\rho }_i\right)\right)} 
\end{equation} 
measure the Holevo quantities between Alice and Bob, and between Alice and environment \textit{E}, where ${\rho }_{AB}\mathrm{=}\sum_i{p_i{\rho }_i}$ and ${\rho }_{AE}\mathrm{=}\sum_i{p_i{\rho }_i}$ are the average states. The definition of \eqref{ZEqnNum572683} also draws a very important connection: \textit{the amount of transmittable quantum information can be derived by the Holevo information}, which measures classical information. 

 As follows, the \textit{single-use} quantum capacity $Q^{\left(\mathrm{1}\right)}\left(\mathcal{N}\right)$ can be expressed as 
\begin{equation} \label{4.18)} 
 \begin{split}
Q^{\left(\mathrm{1}\right)}\left(\mathcal{N}\right)&\mathrm{=}\mathop{\mathrm{max}}_{allp_i,{\rho }_i}\left({\mathcal{X}}_{AB}\mathrm{-}{\mathcal{X}}_{AE}\right)\\ &
\mathrm{=}\mathop{\mathrm{max}}_{allp_i,{\rho }_i}\mathrm{S}\left({\mathcal{N}}_{AB}\left(\sum^n_{i\mathrm{=1}}{p_i\left({\rho }_i\right)}\right)\right)\mathrm{-}\sum^n_{i\mathrm{=1}}{p_i\mathrm{S}\left({\mathcal{N}}_{AB}\left({\rho }_i\right)\right)} \\ 
&\mathrm{-}\mathrm{S}\left({\mathcal{N}}_{AE}\left(\sum^n_{i\mathrm{=1}}{p_i\left({\rho }_i\right)}\right)\right)\mathrm{+}\sum^n_{i\mathrm{=1}}{p_i\mathrm{S}\left({\mathcal{N}}_{AE}\left({\rho }_i\right)\right)}, \end{split}
\end{equation} 
where $\mathcal{N}\left({\rho }_i\right)$ represents the \textit{i}-th output density matrix obtained from the quantum channel input density matrix ${\rho }_i$. 

 The \textit{asymptotic} quantum capacity $Q\left(\mathcal{N}\right)$ can be expressed by
\begin{equation} \label{4.19)} 
 \begin{split}
Q\left(\mathcal{N}\right)&\mathrm{=}\mathop{\mathrm{lim}}_{n\mathrm{\to }\mathrm{\infty }}\frac{\mathrm{1}}{n}Q^{\left(\mathrm{1}\right)}\left({\mathcal{N}}^{\mathrm{\otimes }n}\right) \\ &
\mathrm{=}\mathop{\mathrm{lim}}_{n\mathrm{\to }\mathrm{\infty }}\frac{\mathrm{1}}{n}\mathop{\mathrm{max}}_{allp_i,{\rho }_i}I_{coh}\left({\rho }_A\mathrm{:}{\mathcal{N}}^{\mathrm{\otimes }n}\left({\rho }_A\right)\right) \\ &
\mathrm{=}\mathop{\mathrm{lim}}_{n\mathrm{\to }\mathrm{\infty }}\frac{\mathrm{1}}{n}\mathop{\mathrm{max}}_{allp_i,{\rho }_i}\left({\mathcal{X}}_{AB}\mathrm{-}{\mathcal{X}}_{AE}\right). \end{split}
\end{equation} 
The quantum capacity $Q\left(\mathcal{N}\right)$ of a quantum channel $\mathcal{N}$ can also be expressed by ${\mathcal{X}}_{AB}$, the \textit{Holevo quantity} of Bob's output and by ${\mathcal{X}}_{AE}$,  the information leaked to the environment during the transmission. 

\subsubsection{Quantum Coherent Information and Quantum Mutual Information}
Finally let us make an interesting comparison between quantum coherent information and quantum mutual information. For classical information transmission, the \textit{quantum mutual information} can be expressed according to Section 2
\begin{equation} \label{4.20)} 
I\left(A\mathrm{:}B\right)\mathrm{=S}\left({\rho }_A\right)\mathrm{+S}\left({\rho }_B\right)\mathrm{-}\mathrm{S}\left({\rho }_{AB}\right).                                                         
\end{equation} 
However, in case of \textit{quantum coherent information }\eqref{ZEqnNum259518} the term $\mathrm{S}\left({\rho }_A\right)$ vanishes. The channel transformation $\mathcal{N}$ modifies Alice's original state ${\rho }_A$, hence Alice's original density matrix cannot be used to express $\mathrm{S}\left({\rho }_A\right)$, \textit{after Alice's qubit has been sent through} the quantum channel $\mathcal{N}$. After the channel has modified Alice's quantum state, the initially sent qubit vanishes from the system, and we will have a different density matrix, denoted by ${\rho }_B\mathrm{=}\mathcal{N}\left({\rho }_A\right)$. The coherent information can expressed as $\mathrm{S}\left({\rho }_B\right)\mathrm{-}\mathrm{S}\left({\rho }_{AB}\right)$, where ${\rho }_B$ is the transformed state of Bob, and $\mathrm{S}\left({\rho }_{AB}\right)$ is the joint von Neumann entropy. 

\begin{center}
\begin{figure*}[htbp]
\begin{center}
\includegraphics[angle = 0,width=0.62\linewidth]{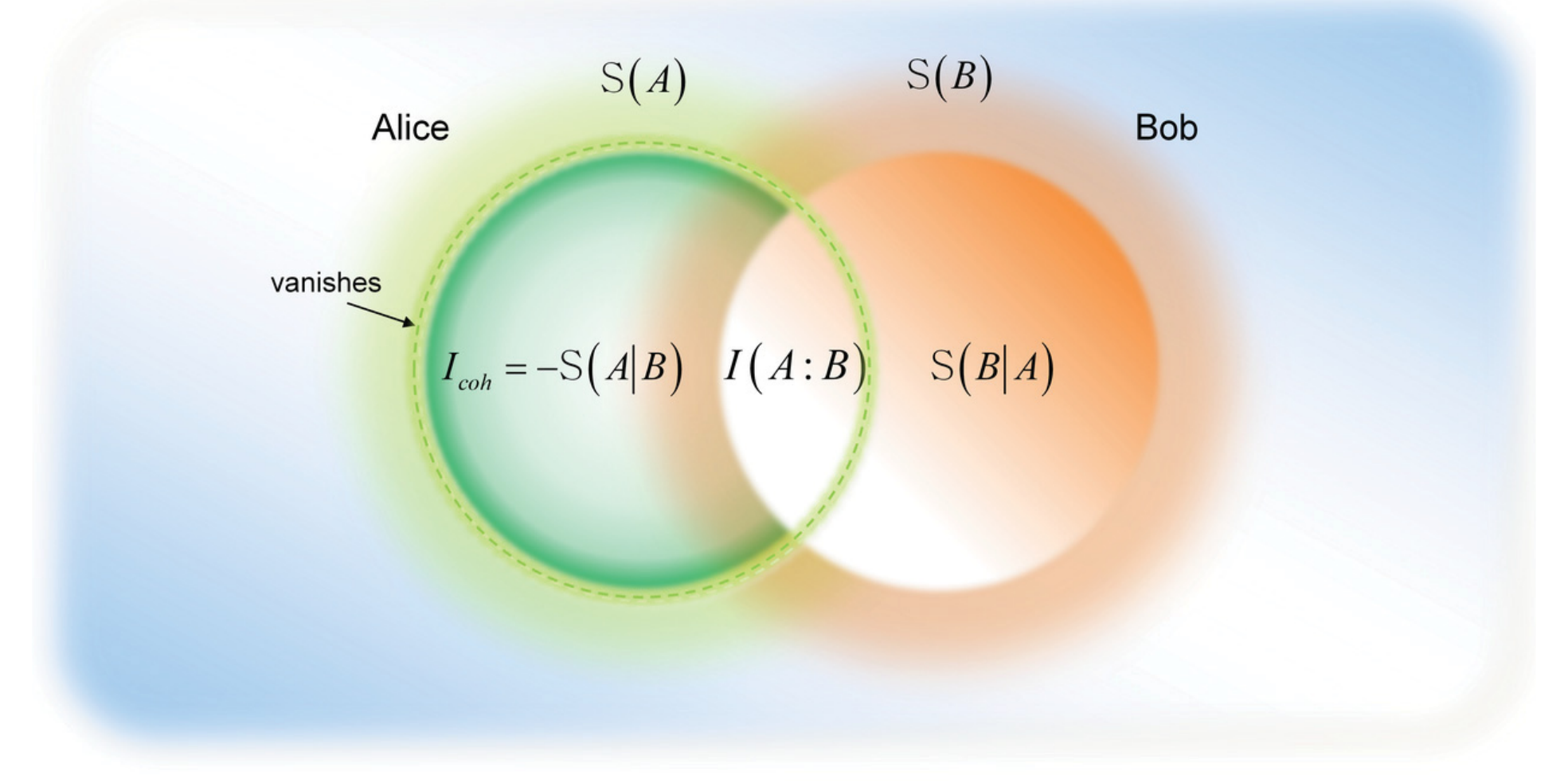}
\caption{The expression of quantum coherent information. The source entropy of Alice's state vanishes after the state is passed to Bob.} 
\label{fig4_8}
 \end{center}
\end{figure*}
\end{center}

 As follows, we will have $\mathrm{S}\left({\rho }_B\right)\mathrm{-}\mathrm{S}\left({\rho }_{AB}\right)$,  which is equal to the \textit{negative conditional entropy} $\mathrm{S}\left(\left.A\right|B\right)$, (see \sref{sec2}) thus 
\begin{equation} \label{4.21)} 
I_{coh}\left({\rho }_A\mathrm{:}\mathcal{N}\left({\rho }_A\right)\right)\mathrm{=S}\left({\rho }_B\right)\mathrm{-}\mathrm{S}\left({\rho }_{AB}\right)\mathrm{=-S}\left(\left.A\right|B\right).                                                  
\end{equation} 
This imporatnt result is summarized in \fref{fig4_8}. 

As we have seen in this section, there is a \textit{very important difference} between the maximized quantum \textit{mutual information} and the maximized \textit{quantum coherent information} of a quantum channel. While the former is always additive, it does not remain true for the latter. \textit{The quantum coherent information }is defined as follows 
\begin{equation} \label{ZEqnNum718348} 
I_{coh}\left(\mathcal{N}\right)\mathrm{=S}\left({\rho }_B\right)\mathrm{-}\mathrm{S}\left({\rho }_E\right), 
\end{equation} 
where ${\rho }_B$ refers to the output of the quantum channel $\mathcal{N}$, while ${\rho }_E$ is the state of the environment. The term $\mathrm{S}\left({\rho }_B\right)$ measures how much information Bob has, while $\mathrm{S}\left({\rho }_E\right)$ measures how much information environment has. As follows, the quantum coherent information $I_{coh}\left(\mathcal{N}\right)$ measures that `\textit{how much more information Bob has than the environment}' about the original input quantum state.

\subsubsection{Quantum Coherent Information of an Ideal Channel}
Now, we have arrived at the question of whether the $Q\left(\mathcal{N}\right)$ quantum capacity of $\mathcal{N}$, as defined previously by the $I_{coh}$ quantum coherent information, is an appropriate measure to describe the whole quantum capacity of a quantum channel. The answer is yes for an ideal channel.
If we have a completely noiseless channel, then channel ${\mathcal{N}}_{AB}\mathrm{=}I$ leads us to coherent information
\begin{equation} \label{4.23)} 
\begin{split}
Q\left(I\right)&\mathrm{=}I_{coh}\left(I\right)\\&\mathrm{=S}\left({\mathcal{N}}_{AB}\left(\rho \right)\right)\mathrm{-}\mathrm{S}\left({\mathcal{N}}_{E}\left(\left|\left.0\right\rangle \right.\left\langle \left.0\right|\right.\right)\right)\\&\mathrm{=S}\left(\rho \right).                                    
\end{split}
\end{equation} 
This equation can be used to calculate the $Q\left({\mathcal{N}}_{AB}\right)$ quantum capacity of a quantum channel (i.e., without maximization) only when we have a completely noiseless idealistic channel ${\mathcal{N}}_{AB}\mathrm{=}I$. 
It also implies the following: to achieve the maximal coherent information for an idealistic quantum channel ${\mathcal{N}}_{AB}\mathrm{=}I$, the input quantum states have to be maximally mixed states or one half of an EPR state, since in these cases, the von Neumann entropies will be maximal. 

On the other hand, if the environment of the communication system interacts with the quantum state, the quantum capacity could vanish, but not the classical capacity of the channel. In this case, the quantum channel ${\mathcal{N}}_{AB}\mathrm{=}I$ can transmit pure orthogonal states faithfully, but it cannot transmit the superposed or entangled states. Furthermore, if the interaction is more significant, it could result in an extremely noisy quantum channel for which the $C\left({\mathcal{N}}_{AB}\right)$ classical capacity of ${\mathcal{N}}_{AB}$ could also vanish.

\subsection{The Lloyd-Shor-Devetak Formula}
The concept of quantum coherent information can be used to express the \textit{asymptotic} quantum capacity $Q\left(\mathcal{N}\right)$ of quantum channel $\mathcal{N}$ called the \textit{Lloyd-Shor-Devetak (LSD) }capacity as follows 
\begin{equation} \label{ZEqnNum907582} 
 \begin{split}
Q\left(\mathcal{N}\right)&\mathrm{=}\mathop{\mathrm{lim}}_{n\mathrm{\to }\mathrm{\infty }}\frac{\mathrm{1}}{n}Q^{\left(\mathrm{1}\right)}\left({\mathcal{N}}^{\mathrm{\otimes }n}\right) \\ &
\mathrm{=}\mathop{\mathrm{lim}}_{n\mathrm{\to }\mathrm{\infty }}\frac{\mathrm{1}}{n}\mathop{\mathrm{max}}_{allp_i,{\rho }_i}I_{coh}\left({\rho }_A\mathrm{:}{\mathcal{N}}^{\mathrm{\otimes }n}\left({\rho }_A\right)\right) \\ &
\mathrm{=}\mathop{\mathrm{lim}}_{n\mathrm{\to }\mathrm{\infty }}\frac{\mathrm{1}}{n}\mathop{\mathrm{max}}_{allp_i,{\rho }_i}\left(\mathrm{S}\left({\rho }_B\right)\mathrm{-}\mathrm{S}\left({\rho }_E\right)\right), \end{split}
\end{equation} 
where $Q^{\left(\mathrm{1}\right)}\left(\mathcal{N}\right)$ represents the  \textit{single-use} quantum capacity.

 The asymptotic quantum capacity can also be expressed using the Holevo information, since as we have seen previously, the quantum coherent information can be derived from the Holevo information
\begin{equation} \label{4.25)} 
Q\left(\mathcal{N}\right)\mathrm{=}\mathop{\mathrm{lim}}_{n\mathrm{\to }\mathrm{\infty }}\frac{\mathrm{1}}{n}\mathop{\mathrm{max}}_{allp_i,{\rho }_i}\left({\mathcal{X}}_{AB}\mathrm{-}{\mathcal{X}}_{AE}\right),                                                           
\end{equation} 
where ${\mathcal{X}}_{AB}$ denotes the classical information sent from Alice to Bob, and ${\mathcal{X}}_{AE}$ describes the classical information passed from Alice to the environment during the transmission. 

 Quantum coherent information plays a fundamental role in describing the maximal amount of transmittable quantum information through a quantum channel $\mathcal{N}$, and - as the Holevo quantity has deep relevance in the classical HSW capacity of a quantum channel - the quantum coherent information will play a crucial role in the LSD capacity of $\mathcal{N}$.

\subsection{The Assisted Quantum Capacity}

 There is another important quantum capacity called \textit{assisted capacity} which measures the quantum capacity for a channel pair that contains different channel models -- and it will have relevance in the \textit{superactivation} of quantum channels [\cref{Smith08}]. If we have a quantum channel $\mathcal{N}$, then we can find a symmetric channel $\mathcal{A}$, that results in the following assisted quantum capacity
\begin{equation} \label{4.26)} 
Q_{\mathcal{A}}\left(\mathcal{N}\right)\mathrm{=}Q\left(\mathcal{N}\mathrm{\otimes }\mathcal{A}\right).                                                         
\end{equation} 
We note, that the symmetric channel has unbounded dimension in the strongest case, and this quantity cannot be evaluated in general. $Q_{\mathcal{A}}\left(\mathcal{N}\right)$ makes it possible to realize the superactivation of zero-capacity (in terms of LSD capacity) quantum channels. For example if we have a zero-capacity \textit{Horodecki channel} and a zero-capacity symmetric channel, then their combination can result in positive joint capacity [\cref{Smith08}]. 

\subsection{The Zero-Error Quantum Capacity}

 Finally, let us shortly summarize the quantum counterpart of classical zero-error capacity. In the case of quantum zero-error capacities $Q^{\left(\mathrm{1}\right)}_0\left(\mathcal{N}\right)$ and $Q_0\left(\mathcal{N}\right)$, the encoding and decoding process differs from the classical zero-error capacity: the encoding and decoding are carried out by the \textit{coherent} encoder and \textit{coherent} POVM decoder, whose special techniques make it possible to preserve the quantum information during the transmission [\cref{Harrow04}], [\cref{Hsieh08}]. 

The \textit{single-use} and \textit{asymptotic} quantum zero-error capacity is defined in a similar way
\begin{equation} \label{4.27)} 
Q^{\left(\mathrm{1}\right)}_0\left(\mathcal{N}\right)\mathrm{=log}\left(K\left(\mathcal{N}\right)\right),                                                                 
\end{equation} 
and 
\begin{equation} \label{4.28)} 
Q_{0}^{{}}\left( \mathcal{N} \right)=\underset{n\to \infty }{\mathop{\text{lim }}}\,\frac{1}{n}\log \left( K\left( {{\mathcal{N}}^{\otimes n}} \right) \right),                                                        
\end{equation} 
where $K\left( {{\mathcal{N}}^{\otimes n}} \right)$ is the maximum number of \textit{n}-length mutually non-adjacent quantum messages that the quantum channel can transmit with zero error. The quantum zero-error capacity is upper bounded by LSD channel capacity $Q\left(\mathcal{N}\right)$; that is, the following relation holds between the quantum zero-error capacities:
\begin{equation} \label{4.29)} 
Q_0\left(\mathcal{N}\right)\mathrm{\le }Q\left(\mathcal{N}\right).                                 
\end{equation} 

\subsection{Relation between Classical and Quantum Capacities of Quantum Channels}

 Before introducing some typical quantum channel maps let us summarize the main properties of various capacities in conjunction with a quantum channels. First of all, the quantum capacity of $\mathcal{N}$ cannot exceed the maximal classical capacity that can be measured with entangled inputs and joint measurement; at least, it is not possible in general. On the other hand, for some quantum channels, it is conjectured that the maximal \textit{single-use} classical capacity - hence the capacity that can be reached with \textit{product }inputs and a \textit{single} measurement setting - is lower than the \textit{quantum capacity} for the same quantum channel.

 For all quantum channels
\begin{equation} \label{4.30)} 
C\left(\mathcal{N}\right)\mathrm{\ge }Q\left(\mathcal{N}\right),                                                           
\end{equation} 
where $C\left(\mathcal{N}\right)$ is the classical capacity of the quantum channel that can be achieved with entangled input states and a joint measurement setting.

 On the other hand, it is conjectured that for some quantum channels,
\begin{equation} \label{ZEqnNum224811} 
C\left(\mathcal{N}\right)\mathrm{<}Q\left(\mathcal{N}\right) 
\end{equation} 
holds as long as the classical capacity $C\left(\mathcal{N}\right)$ of the quantum channel is measured by a classical encoder and a single measurement setting. (As we have seen in \sref{sec3}, the classical capacities of a quantum channel can be measured in different settings, and the strongest version can be achieved with the combination of entangled inputs and joint measurement decoding.) 

The fundamental differences between classical and quantum capacities are summarized in \tref{tab1}. 

\begin{table*}[htbp]
\begin{center}
\begin{tabular}{|l|l|l|l|} \hline 
\textbf{Capacity} & Type of information & Correlation measure & Capacity formula \\ \hline 
Classical & Classical information & Holevo information & HSW formula \\ \hline 
Private Classical & Private information & Private information & Li-Winter-Zou-Guo, Smith-Smolin formula \\ \hline 
Entanglement Assisted Classical & Classical information & Quantum mutual information & Bennett-Shor-Smolin-Thapliyal formula \\ \hline 
Quantum & Quantum information & Quantum Coherent Information & LSD formula \\ \hline 
\end{tabular}
\caption{The measure of classical and quantum capacities.}
\label{tab1}
\end{center}
\end{table*}

 It can be concluded from the table that in case of a quantum communication channel we have to count with so many capacities. Each of these capacities is based on different correlation measures: the \textit{classical correlation} between the input and the output is measured by the quantum mutual information and the Holevo information. The private classical capacity is measured by the private information, which is the \textit{maximization of the difference of two quantum mutual information functions. }For entanglement assisted capacity the correlation between input and output is also measured by the \textit{maximized quantum mutual information}, however in this case we do not have to compute the asymptotic version to get the true capacity. Finally, the \textit{quantum correlation }between the input and output is measured by the \textit{quantum coherent information}.

\subsection{Related Work}

 In this section we summarize the most important works regarding on the quantum capacity of the quantum channels.

 The quantum capacity is one of the most important result of quantum information theory. The classical capacity of quantum channels was discovered in early years, in the beginning of the 1970s, and the researchers from this era ---such as Holevo and Levitin---suggested that physical particles can encode only classical information [\cref{Levitin69}], [\cref{Holevo73}], [\cref{Holevo73a}]. The first step in the encoding of quantum information into a physical particle was made by Feynman, in his famous work from 1982 [\cref{Feynman82}]. However, the researchers did not see clearly and did not understand completely the importance of quantum capacity until the late 1990s. As we have shown in \sref{sec3}, a quantum channel can be used to transmit classical information and the amount of maximal transmittable information depends on the properties of the encoder and decoder setting, or whether the input quantum states are mixed or pure. Up to this point, we have mentioned just the transmission of classical information through the quantum channel---here we had broken this picture. The HSW theorem was a very useful tool to describe the amount of maximal transmittable classical information over a noisy quantum channel, however we cannot use it to describe the amount of maximal transmittable \textit{quantum information. }

\subsubsection{Quantum Coherent Information}

 The computation of quantum capacity is based on the concept of \textit{quantum coherent information}, which measures the ability of a quantum channel to preserve a quantum state. The definition of quantum coherent information (in an exact form) was originally introduced by Schumacher and Nielsen in 1996 [\cref{Schumacher96c}]. This paper is a very important milestone in the history of the quantum capacity, since here the authors were firstly shown that the concept of quantum coherent information can be used to measure the quantum information (hence not the classical information) which can be transmitted through a quantum channel. The first,---but yet not complete---definitions of the quantum capacity of the quantum channel can be found in Shor's work from 1995 [\cref{Shor95}], in which Shor has introduced a scheme for reducing decoherence in quantum computer memory, and in Schumacher's articles from one year later [\cref{Schumacher96b}, \cref{Schumacher96c}]. Shor's paper from 1995 mainly discusses the problem of implementation of quantum error correcting schemes - the main focus was not on the exact definition of quantum capacity. Later, Shor published an extended version with a completed proof in 2002 [\cref{Shor02}]. To transmit quantum information the parties have to encode and decode coherently. An interesting engineering problem is how the receiver could decode quantum states in superposition without the destruction of the original superposition [\cref{Wilde07}]. 
The quantum capacity of a quantum channel finally was formulated completely by the \textit{LSD-theorem}, named after Lloyd, Shor and Devetak [\cref{Devetak03a}], [\cref{Lloyd97}], [\cref{Shor02}], and they have shown that the rate of quantum communication can be expressed by the quantum coherent information. The LSD-channel capacity states that the asymptotic quantum capacity of the quantum channel is greater than (or equal to in some special cases) the single-use capacity; hence it is not equal to the quantum coherent information.

More information about the properties of fidelity and about the connection with other distance measures can be found in Fuch's works [\cref{Fuchs96}], [\cref{Fuchs98}]. An important article regarding the fidelity of mixed quantum states was published by Jozsa in 1994 [\cref{Jozsa94}]. Fidelity also can be measured between entangled quantum states---a method to compute the fidelity of entanglement was published by Schumacher in 1996 [\cref{Schumacher96b}]. Here, the upper bound of the quantum capacity was also mentioned, in the terms of quantum coherent information. Nielsen in 2002 [\cref{Nielsen02}] defined a connection between the different fidelity measures. 

\subsubsection{Proofs on Quantum Capacity}
The exact measure of quantum capacity was an open question for a long time. The fact that the quantum capacity cannot be increased by classical communication was formally proven by Bennett et al. [\cref{Bennett96a}], who discussed the mixed state entanglement and quantum error correction. Barnum, in 2000 [\cref{Barnum2000}], defined the connection between the fidelity and the capacity of a quantum channel, and here he also showed the same result as Bennett et al. did in 1996, namely that the quantum capacity cannot increased by classical communication [\cref{Bennett96a}]. The works of Barnum et al. [\cref{Barnum2000}] and Schumacher et al. [\cref{Schumacher98a}] from the late 1990s gave very important results to the field of quantum information theory, since these works helped to clarify exactly the maximum amount of transmittable quantum information over very noisy quantum channels [\cref{Wilde11}].

Seth Lloyd gave the first proof in 1997 on the quantum capacity of a noisy quantum channel. The details of Lloyd's proof can be found in [\cref{Lloyd97}], while Shor's results in detail can be found in [\cref{Shor02}]. On the basis of Shor's results, a proof on the quantum capacity was given by Hayden et al. in 2008 [\cref{Hayden08b}].
The next step in the history of the quantum capacity of the quantum channel was made by Devetak [\cref{Devetak03a}]. Devetak also gave a proof for the quantum capacity using the private classical capacity of the quantum channel, and he gave a clear connection between the quantum capacity and the private classical capacity of the quantum channel.
As in the case of the discoverers of the HSW-theorem, the discoverers gave different proofs. The quantum capacity of a quantum channel is generally lower than the classical one, since in this case the quantum states encode quantum information. The quantum capacity requires the transmission of arbitrary quantum states, hence not just `special' orthogonal states---which is just a subset of a more generalized case, in which the states can be arbitrary quantum states. 
On the several different encoder, decoder and measurement settings for quantum capacity see the work of Devetak and Winter [\cref{Devetak05a}], Devetak and Shor's work [\cref{Devetak05b}], and the paper of Hsieh et al. [\cref{Hsieh08}]. In this paper we have not mentioned the definition of unit resource capacity region and private unit resource capacity region, which can be found in detail in the works of Hsieh and Wilde [\cref{Hsieh10}], and Wilde and Hsieh [\cref{Wilde10}]. In 2005, Devetak and Shor published a work which analyzes the simultaneous transmission of classical and quantum information [\cref{Devetak05b}]. 

On the quantum capacities of bosonic channels a work was published by Wolf, Garcia and Giedke, see [\cref{Wolf06}]. In 2007, Wolf and P\'{e}rez-Garc\'{i}a published a paper on the quantum capacities of channels with small environment, the details can be found in [\cref{Wolf07}]. They have also determined the quantum capacity of an amplitude damping quantum channel (for the description of amplitude damping channel, see \sref{sec5}), for details see the same paper from 2007 [\cref{Wolf07}]. 
The properties of quantum coherent information and reverse coherent information were studied by Patr\'{o}n in 2009 [\cref{Patron09}]. 

The proofs of the LSD channel capacity can be found in [\cref{Devetak03a}], [\cref{Lloyd97}], [\cref{Shor02}]. The quantum communication protocols based on the transmission of quantum information were intensively studied by Devetak [\cref{Devetak04a}], and the work of the same authors on the generalized framework for quantum Shannon theory, from 2008 [\cref{Devetak08}]. 

\section{Quantum Channel Maps and Capacities}
\label{sec5}
Here, we give a brief survey of some important quantum channel maps and study some capacity formulas. For the corresponding definitions related to the state-vector description we advise to the reader to [\cref{Imre13}]. 

\subsection{Channel Maps}
\subsubsection{The Pauli Channel}
\label{Pauli channel}
The Pauli channel model having an input state $\rho$ can be formulated [\cref{Sarvepalli09}] as 
\begin{eqnarray}
\label{Pauli}
\rho \rightarrow  C_{P}(\rho) = (1-p)\rho + p_xX\rho X + p_y Y\rho Y + p_xZ\rho Z,  \
\end{eqnarray}
where that $X$, $Y$ and $Z$ are single-qubit Pauli determined by
\begin{eqnarray}
\label{X_operator}
X= \left(\begin{array} {cc}  0 & 1 \\ 
                                 1 & 0 \end{array} \right), \ 
\end{eqnarray}
\begin{eqnarray}
\label{Y_operator}
X= \left(\begin{array} {cc}  0 & -i \\ 
                                 i & 0 \end{array} \right), \ 
\end{eqnarray}
\begin{eqnarray}
\label{Z_operator}
Z= \left(\begin{array} {cc}  1 & 0 \\ 
                                 0 & -1 \end{array} \right). \ 
\end{eqnarray}
Note that the depolarizing probability $p=p_x + p_y + p_z$ is the sum of $p_x, p_y$ and $p_z$ representing the depolarizing probability of Pauli $X$, $Y$ and $Z$ errors, respectively. The probabilities of the errors at time instant $t$ are dependent to relaxation time $T_1$ and dephasing time $T_2$ as
\begin{eqnarray}
\label{pxpypz}
p_x = p_y &=& \frac{1}{4}\left( 1- \mbox{e}^{-t/T_2}\right) , \nonumber \\
p_z &=& \frac{1}{4}\left( 1+  \mbox{e}^{-t/T_1} -2\mbox{e}^{-t/T_2}\right) . \
\end{eqnarray}  

\subsubsection{The Depolarizing Channel}
 The last discussed unital channel model is the \textit{depolarizing} channel which performs the following transformation
\begin{equation} \label{ZEqnNum246685} 
\mathcal{N}\left({\rho }_i\right)\mathrm{=}p\frac{I}{\mathrm{2}}\mathrm{+}\left(\mathrm{1-}p\right){\rho }_i,                                                                            
\end{equation} 
where $p$ is the \textit{depolarizing parameter} of the channel, and if Alice uses two orthogonal states ${\rho }_0$ and ${\rho }_{\mathrm{1}}$ for the encoding then the mixed input state is
\begin{equation} \label{ZEqnNum943815} 
\rho \mathrm{=}\left(\sum_i{p_i{\rho }_i}\right)\mathrm{=}p_0{\rho }_0\mathrm{+}\left(\mathrm{1-}p_0\right){\rho }_{\mathrm{1}}.                                                                
\end{equation} 
After the unital channel has realized the transformation $\mathcal{N}$ on state $\rho $, we will get the following result
\begin{equation} \label{ZEqnNum383151} 
 \begin{split}
\mathcal{N}\left(\sum_i{p_i{\rho }_i}\right)&\mathrm{=}\mathcal{N}\left(p_0{\rho }_0\mathrm{+}\left(\mathrm{1-}p_0\right){\rho }_{\mathrm{1}}\right) \\ &
\mathrm{=}p\frac{\mathrm{1}}{\mathrm{2}}I\mathrm{+}\left(\mathrm{1-}p\right)\left(p_0{\rho }_0\mathrm{+}\left(\mathrm{1-}p_0\right){\rho }_{\mathrm{1}}\right).  
 \end{split}
\end{equation} 

\subsubsection{The Damping Channel}
\label{Damping channel}
Let us consider the influences of an environment to a single qubit of a quantum system, where for example the qubit  is realised by using a two-level atom having the ground state $\ket{0}$ and the excited state $\ket{1}$.  The atom may have a spontaneous dissipation/absorption of energy to/from the environment, which makes the atom change its state from the ground state $\ket{0}$ to the excited state $\ket{1}$ or vice versa. The transition of the state is refered to as the decoherence process.  As a result, the state of the qubit when there is no interaction with the environment is as follows [\cref{Preskill15}]
\begin{eqnarray}
\label{No}
\ket{0}\ket{0}_E \rightarrow \ket{0}\ket{0}_E, \nonumber \\
\ket{0}\ket{1}_E \rightarrow \ket{0}\ket{1}_E, \nonumber \\
\ket{1}\ket{0}_E \rightarrow \ket{1}\ket{0}_E, \nonumber \\
\ket{1}\ket{1}_E \rightarrow \ket{1}\ket{1}_E, \
\end{eqnarray}
where $\ket{0}_E$ and $\ket{1}_E$ represent the low and high basis states of the environment. Accordingly, if the dissipation/absorption occurs, we have
\begin{eqnarray}
\label{Yes}
\ket{1}\ket{0}_E \rightarrow \ket{0}\ket{1}_E, \nonumber \\
\ket{0}\ket{1}_E \rightarrow \ket{1}\ket{0}_E. \
\end{eqnarray}
The transition represented by \eref{Yes} is may be formulated as: 
\begin{eqnarray}
\label{Damping_channel}
\ket{1}\ket{0}_E &\rightarrow & \sqrt{1-p_l}\ket{1}\ket{0}_E + \sqrt{p_l}\ket{0}\ket{1}_E, \nonumber \\
\ket{0}\ket{1}_E & \rightarrow & \sqrt{1-p_o}\ket{0}\ket{1}_E + \sqrt{p_o}\ket{1}\ket{0}_E,  \
\end{eqnarray}
where $p_l$ and $p_o$ is the probability of the atom losing its energy to the environment or obtaining its energy from the environment, respectively. We may generalise the channel model of \eref{Damping_channel} by alternating the basis states by the superposition states to lead to 
\begin{eqnarray}
\begin{split}
\label{Damping_channel_general}
\left(\alpha\ket{0}+\beta\ket{1}\right)\ket{0}_E &\rightarrow \\& \left(\alpha\ket{0} + \beta\sqrt{1-p_l}\ket{1}\right)\ket{0}_E + \beta\sqrt{p_l}\ket{0}\ket{1}_E , \nonumber \\
\left(\alpha\ket{0}+\beta\ket{1}\right)\ket{1}_E &\rightarrow \\& \alpha\sqrt{p_o}\ket{1}\ket{0}_E + \left(\alpha\sqrt{1-p_o}\ket{0} + \beta\ket{1}\right)\ket{1}_E . \
\end{split}
\end{eqnarray}
It should be noted that the coefficient $\alpha$ and $\beta$ may be used representing the $(N-1)$ qubit states orthogonal to the states $\ket{0}$ and $\ket{1}$ of the considered qubit. Moreover, if it can be assumed that each qubit interacts independently with
the environment, the associated decoherence process in the $N$-qubit system may be considered as temporally and spatially uncorrelated. Accordingly, the process where the qubit loses its energy can be modelled by an amplitude damping channel $C_{AD}$ having an input state $\rho$  [\cref{Ghosh12}]:
\begin{eqnarray}
\label{Amplitude_Damping}
\rho \rightarrow  C_{AD}(\rho) = E_{AD1}~\rho ~E^{\dagger}_{AD1} + E_{AD2}~\rho ~E^{\dagger}_{AD2},  \
\end{eqnarray}
where Kraus matrices $E_{AD}$ used for characterising the amplitude damping channel are as follows:
\begin{eqnarray}
\label{AD1}
E_{AD1}= \left(\begin{array} {cc}  1 & 0 \\ 
                                 0 & \sqrt{1-p_l} \end{array} \right), \
\end{eqnarray}
\begin{eqnarray}
\label{AD2}
E_{AD2}= \left(\begin{array} {cc}  0 & \sqrt{p_l} \\ 
                                 0 & 0 \end{array} \right). \ 
\end{eqnarray}

Influences from the environment may results in random phase kicks on a single qubit. In such scenario, the decoherence process reflecting phase changes of the qubit is modelled as the phase damping channel $C_{PD}(\rho)$ as
\begin{eqnarray}
\label{Phase_Damping}
\rho \rightarrow  C_{PD}(\rho) = E_{PD1}~\rho ~E^{\dagger}_{PD1} + E_{PD2}~\rho ~E^{\dagger}_{PD2},  \
\end{eqnarray}
where we have the corresponding Kraus matrices as
\begin{eqnarray}
\label{PD1}
E_{AD1}= \left(\begin{array} {cc}  1 & 0 \\ 
                                 0 & \sqrt{1-p_l} \end{array} \right), \
\end{eqnarray}
\begin{eqnarray}
\label{PD2}
E_{AD2}= \left(\begin{array} {cc}  0 & \sqrt{p_l} \\ 
                                 0 & 0 \end{array} \right). \ 
\end{eqnarray}
In order to reflect changes of the qubit in both phase and amplitude, the combination of amplitude and phase damping channel may be used. However, in general it is not affordable to classically simulate $N$-qubit combined channel, which requires to have a $2N$-dimensional Hilbert space. For the sake of facilitating efficient classical simulations, the combined amplitude and phase damping channel may be
approximated using a Pauli channel model.
\subsubsection{The Dephasing Channel Model}
The second type of decoherence map discussed is unitary and results in relative phase differences between the computational basis states: the channel map which realizes it is called the \textit{dephasing} map. In contrast to the amplitude damping map, it realizes a unitary transformation. The unitary representation of the dephasing quantum channel for a given input $\rho \mathrm{=}\sum_{i,j}{{\rho }_{ij}\left|\left.i\right\rangle \right.}\left\langle \left.j\right|\right.$ can be expressed as 
\begin{equation} \label{3.77)} 
\mathcal{N}\left(\rho \right)\mathrm{=}\sum_i{{\rho }_{ii}}\left|\left.E_i\right\rangle \right.\left\langle \left.E_i\right|\right.,                                                                         
\end{equation} 
where $\left|\left.E_i\right\rangle \right.$ are the environment states. The dephasing quantum channel acts on the density operator $\rho $ as follows
\begin{equation} \label{3.78)} 
\mathcal{N}\left({\rho }_i\right)\mathrm{=}p{\sigma }_Z\rho {\sigma }_Z\mathrm{+}\left(\mathrm{1-}p\right){\rho }_i,                                                                       
\end{equation} 
where ${\sigma }_Z$ is the Pauli \textit{Z}-operator. The image of the dephasing channel map is similar to that of the phase flip channel map, however, the shrinkage of the original Bloch sphere is greater. The dephasing channel transforms an arbitrary superposed pure quantum state $\alpha \left|\left.0\right\rangle \right.\mathrm{+}\beta \left|\left.\mathrm{1}\right\rangle \right.$ into a mixture
\begin{equation} \label{3.79)} 
\mathcal{N}\left(\rho \right)\mathrm{\to }\rho \mathrm{'}\mathrm{=}\left[ \begin{array}{ll}
{\left|\alpha \right|}^{\mathrm{2}} & \alpha {\beta }^{\mathrm{*}}e^{\mathrm{-}\gamma \left(t\right)} \\ 
{\alpha }^{\mathrm{*}}\beta e^{\mathrm{-}\gamma \left(t\right)} & {\left|\beta \right|}^{\mathrm{2}} \end{array}
\right],                                                                   
\end{equation} 
where $\gamma \left(t\right)$ is a positive real parameter, which characterizes the coupling to the environment, using the time parameter\textit{ t}.

\subsubsection{The Pancake Map}
To give an example for physically not allowed (nonphysical, non-CP) transformations, we discuss the \textit{pancake map}. The non-CP property means, that there exists no Completely Positive Trace Preserving map, which preserves some information along the equatorial spanned by the \textit{x} and \textit{y} axes of the Bloch sphere, while it completely demolishes any information along the \textit{z} axis. This map is called the pancake map, and it realizes a physically not allowed (non-CP) transformation. The effect of the pancake map is similar to the bit-phase flip channel, however, this channel defines a non-CP transform: it `smears' the original Bloch sphere along the equatorial spanned by the \textit{x} and \textit{y} axes. On the other hand, the pancake map---besibdes the fact that is a non-physical map---can be used theoretically to transfer some information, and some information can be transmitted through these kinds of channel maps. The reason behind decoherence is\textit{ }Nature. She cannot be perfectly eliminated from quantum systems in practice. The reduction of decoherence is also a very complex task, hence it brings us on the engineering side of the problem: the quantum systems have to be designed in such a way that the unwanted interaction between the quantum states and the environment has to be minimal [\cref{Shor95}], [\cref{Shor96}]. Currently - despite the efficiency of these schemes - the most important tools to reduce decoherence are quantum error-correcting codes and decoupling methods.  

\subsection{Capacities}
Next, we study the classical and quantum capacities of the following quantum channels: 

 \begin{enumerate}
\item \textit{erasure quantum channel, }

\item \textit{phase-erasure quantum channel, }

 \item \textit{mixed erasure/phase-erasure quantum channel, }

 \item \textit{amplitude damping channel.}
\end{enumerate}

First we derive the classical capacities of these channels in closed forms. Then we give the quantum capacities and compare them. 

\subsubsection{Erasure Quantum Channel}
The \textit{erasure} quantum channel ${\mathcal{N}}_p$ erases the input state $\rho $ with probability $p$ or transmits the state unchanged with probability $\left(\mathrm{1-}p\right)$
\begin{equation} \label{5.1)} 
{\mathcal{N}}_p\left(\rho \right)\mathrm{\to }\left(\mathrm{1-}p\right)\rho \mathrm{+}\left(p\left|\left.e\right\rangle \right.\left\langle \left.e\right|\right.\right),                                           
\end{equation} 
where $\left|\left.e\right\rangle \right.$ is the erasure state. The classical capacity of the erasure quantum channel ${\mathcal{N}}_p$ can be expressed as 
\begin{equation} \label{ZEqnNum181299} 
C\left({\mathcal{N}}_p\right)\mathrm{=}\left(\mathrm{1-}p\right)\mathrm{log}\left(d\right),                                                                             
\end{equation} 
where \textit{d} is the dimension of the input system $\rho $. As follows from \eqref{ZEqnNum181299}, the classical capacity of ${\mathcal{N}}_p$ vanishes at $p\mathrm{=1}$,  while if $\mathrm{0}\mathrm{\le }p\mathrm{<1}$ then the channel ${\mathcal{N}}_p$ can transmit some classical information. 

 The quantum capacity of the erasure quantum channel ${\mathcal{N}}_p$ is 
\begin{equation} \label{ZEqnNum458810} 
Q\left({\mathcal{N}}_p\right)\mathrm{=}\left(\mathrm{1-2}p\right)\mathrm{log}\left(d\right).                                                                          
\end{equation} 
$Q\left({\mathcal{N}}_p\right)$ vanishes at $p\mathrm{=}{\mathrm{1}}/{\mathrm{2}}$, but it can transmit some quantum information if $\mathrm{0}\mathrm{\le }p\mathrm{<}{\mathrm{1}}/{\mathrm{2}}$. 

\begin{center}
\begin{figure}[htbp]
\begin{center}
\includegraphics[angle = 0,width=0.9\linewidth]{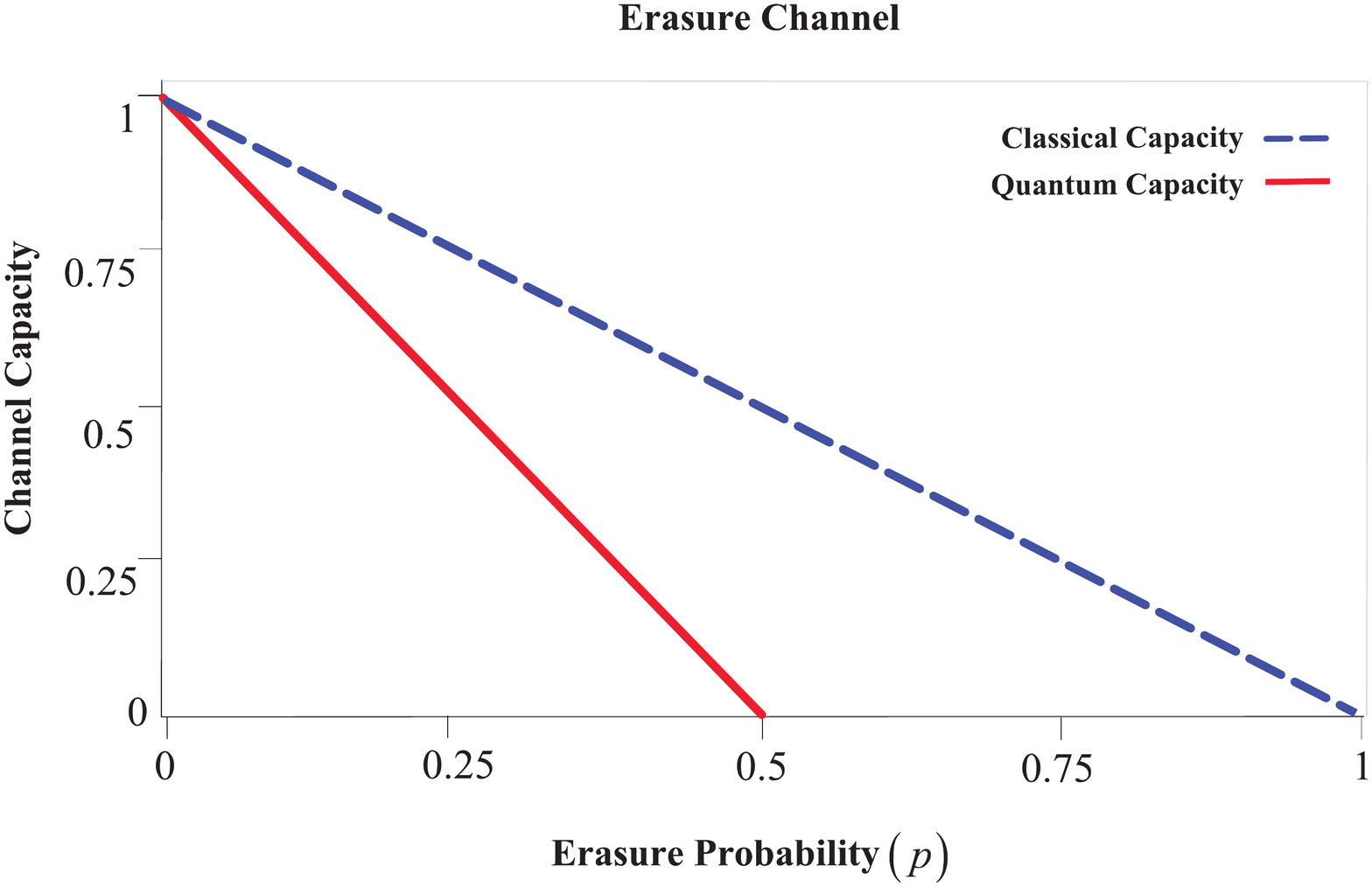}
\caption{The classical and quantum capacities of the erasure quantum channel as a function of erasure probability [Imre13].} 
 \label{fig5_1}
 \end{center}
\end{figure}
\end{center}

In \fref{fig5_1}, the classical (dashed line) and quantum capacity (solid line) of the erasure quantum channel as a function of erasure probability are shown. 

\subsubsection{Phase-Erasure Quantum Channel}
The \textit{phase-erasure} quantum channel ${\mathcal{N}}_{\delta }$ erases the phase of the input quantum state with probability \textit{p} without causing any disturbance in the amplitude. Using input density matrix $\rho $, the map of the phase-erasure quantum channel can be expressed as
\begin{equation} \label{5.4)} 
\mathcal{N}\left(\rho \right)\mathrm{\to }\left(\mathrm{1-}p\right)\rho \mathrm{\otimes }\left|\left.0\right\rangle \right.\left\langle \left.0\right|\right.\mathrm{+}p\frac{\rho \mathrm{+}Z\rho Z^{\mathrm{\dagger }}}{\mathrm{2}}\mathrm{\otimes }\left|\left.\mathrm{1}\right\rangle \right.\left\langle \left.\mathrm{1}\right|\right.,                                                  
\end{equation} 
where \textit{Z} realizes the phase transformation on the input quantum system $\rho $, while the second qubit is used as a flag qubit. 

The classical capacity of the ${\mathcal{N}}_{\delta }$ phase-erasure quantum channel using phase erasing probability \textit{q} is
\begin{equation} \label{ZEqnNum226186} 
C\left({\mathcal{N}}_{\delta }\right)\mathrm{=1},                                                               
\end{equation} 
since the phase error has no effect on the distinguishability of orthogonal input quantum states $\left|\left.0\right\rangle \right.$ and $\left|\left.\mathrm{1}\right\rangle \right.$. On the other hand, if we talk about quantum capacity $Q\left({\mathcal{N}}_{\delta }\right)$ of ${\mathcal{N}}_{\delta }$ the picture changes: 
\begin{equation} \label{ZEqnNum867646} 
Q\left({\mathcal{N}}_{\delta }\right)\mathrm{=}\left(\mathrm{1-}q\right)\mathrm{log}\left(d\right).                                                                         
\end{equation} 
\subsubsection{Mixed Erasure/Phase-Erasure Quantum Channel}
From the erasure quantum channel and the phase-erasure quantum channel a third type of quantum channel can be constructed -- the \textit{mixed erasure/phase-erasure quantum channel}. This channel erases the input quantum system with probability $p$, erases the phase with probability \textit{q}, and leaves the input unchanged with probability $\mathrm{1-}p\mathrm{-}q\mathrm{\ge }\mathrm{0}$. Using \eqref{ZEqnNum181299} and \eqref{ZEqnNum226186}, the classical capacity of the mixed erasure/phase-erasure quantum channel, ${\mathcal{N}}_{p\mathrm{+}q}$, can be expressed as 
\begin{equation} \label{5.7)} 
C\left({\mathcal{N}}_{p\mathrm{+}q}\right)\mathrm{=}\left(\mathrm{1-}p\right)\mathrm{log}\left(d\right)\mathrm{=}C\left({\mathcal{N}}_p\right).                                                               
\end{equation} 

\begin{center}
\begin{figure}[htbp]
\begin{center}
\includegraphics[angle = 0,width=0.9\linewidth]{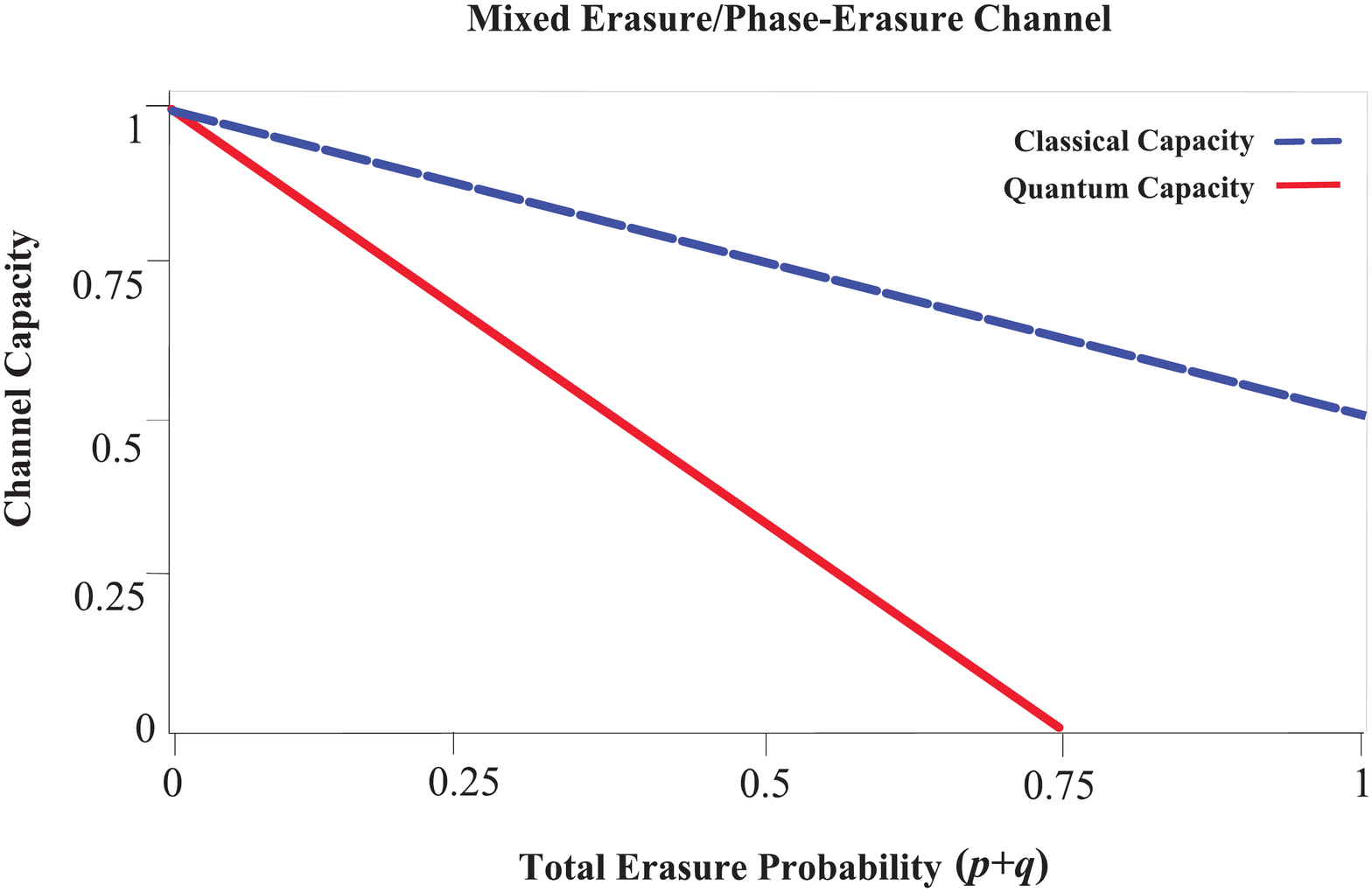}
\caption{The classical and quantum capacities of the mixed erasure/phase-erasure quantum channel as a function of total erasure probability [Imre13].} 
 \label{fig5_2}
 \end{center}
\end{figure}
\end{center}

Furthermore, combining \eqref{ZEqnNum458810} and \eqref{ZEqnNum867646}, the quantum capacity of the mixed erasure/phase-erasure quantum channel, ${\mathcal{N}}_{p\mathrm{+}q}$, we get 
\begin{equation} \label{5.8)} 
Q\left({\mathcal{N}}_{p\mathrm{+}q}\right)\mathrm{=}\left(\mathrm{1-}q\mathrm{-}\mathrm{2}p\right)\mathrm{log}\left(d\right).                                                                   
\end{equation} 
The classical (dashed line) and quantum capacities (solid line) of the mixed erasure/phase-erasure quantum channel as a function of total erasure probability $p\mathrm{+}q$ are illustrated in \fref{fig5_2}. 

\subsubsection{Amplitude damping Quantum Channel}
Finally, we give the quantum capacity of the amplitude damping channel. The classical capacity of the amplitude damping quantum channel can be expressed as 
\begin{equation} \label{ZEqnNum986768} 
C\left(A_{\gamma }\right)\mathrm{=}\mathop{\mathrm{max}}_{\tau }H\left(\tau \right)\mathrm{+}\left[\mathrm{-}H\left(\tau \left(\gamma \right)\right)\mathrm{+}H\left(\tau \left(\mathrm{1-}\gamma \right)\right)\right],                                                
\end{equation} 
where $\tau \mathrm{\in }\left[\mathrm{0,1}\right]$ is a special parameter called the \textit{population} parameter, and \textit{H} is the Shannon entropy function, and $H\left(\tau \right)\mathrm{=-}\tau \mathrm{log}\left(\tau \right)\mathrm{-}\left(\mathrm{1-}\tau \right)\mathrm{log}\left(\mathrm{1-}\tau \right)$. As follows from \eqref{ZEqnNum986768} the classical capacity $C\left(A_{\gamma }\right)$ of the amplitude damping channel completely vanishes if $\gamma \mathrm{=1}$, otherwise (if $\mathrm{0}\mathrm{\le }\gamma \mathrm{<1}$) the channel can transmit classical information. On the other hand for the quantum capacity $Q\left(A_{\gamma }\right)$ the capacity behaves differ. 

\begin{center}
\begin{figure}[htbp]
\vspace{-0.6cm}
\begin{center}
\includegraphics[angle = 0,width=0.9\linewidth]{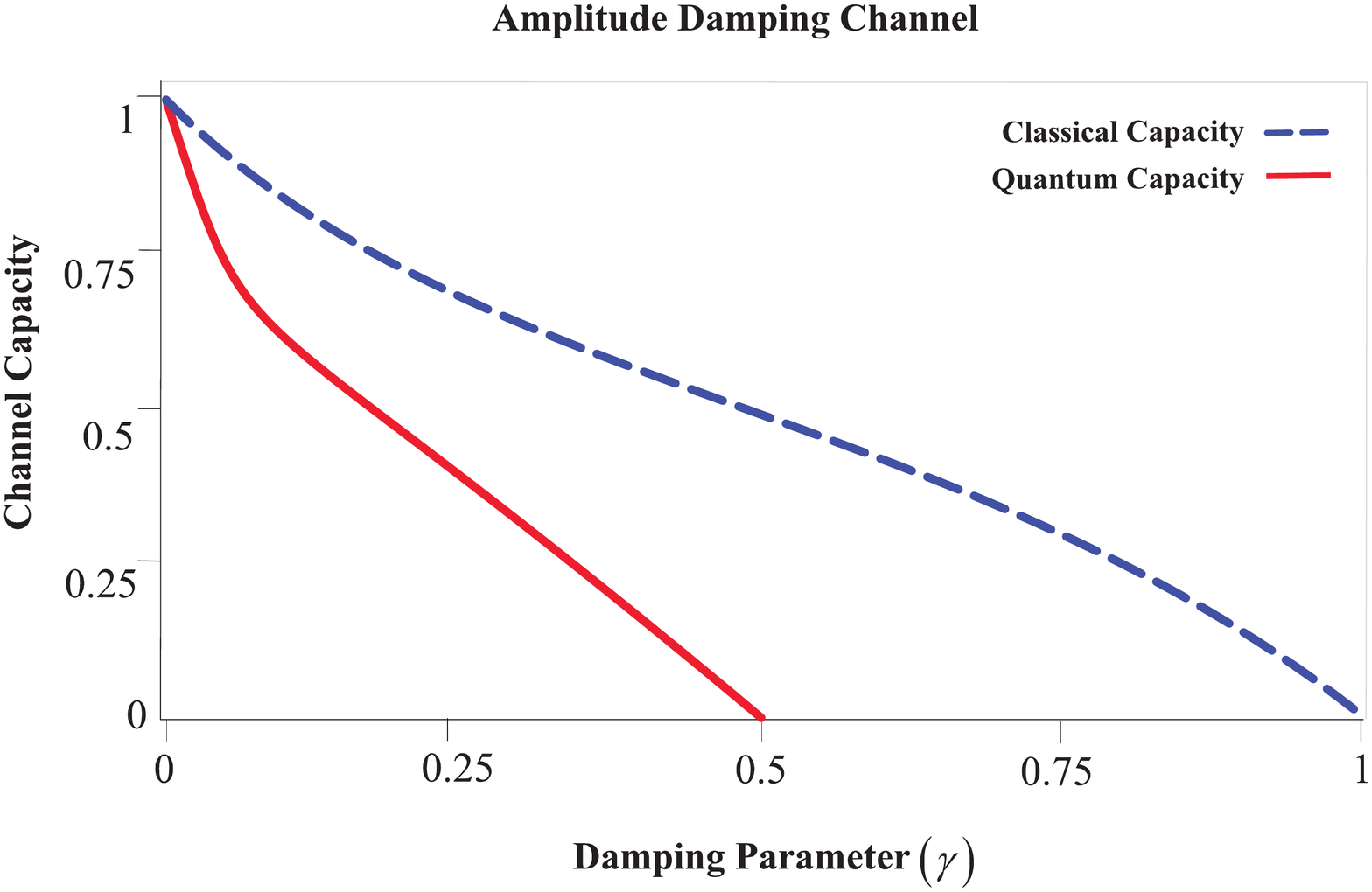}
\caption{The classical and quantum capacities of the amplitude damping quantum channel as a function of the damping parameter [Imre13].} 
 \label{fig5_3}
 \end{center}
\end{figure}
\end{center}

 The quantum capacity of this channel can be expressed as a maximization: 
\begin{equation} \label{5.10)} 
Q\left(A_{\gamma }\right)\mathrm{=}\mathop{\mathrm{max}}_{\tau }\left[H\left(\tau \left(\gamma \right)\right)\mathrm{-}H\left(\tau \left(\mathrm{1-}\gamma \right)\right)\right].                                                          
\end{equation} 
The classical (dashed line) and the quantum capacity (solid line) of the amplitude damping quantum channel as a function of the damping parameter $\gamma $ are shown in \fref{fig5_3}.

It can be concluded that the working mechanism of the amplitude damping channel is similar to the erasure channel (see \eqref{ZEqnNum181299} and \eqref{ZEqnNum458810}), since if the damping parameter value is equal to or greater than $\mathrm{0.5}$, the quantum capacity of the channel completely vanishes. We obtained the same result for the erasure channel; however in that case the erasure probability \textit{p} was the channel parameter.

\section{Practical Implementations of Quantum Channels}
\label{sec6}
In this section, we focus on the physical and experimental implications of quantum channels in different scenarios.
\subsection{Realistic Material: Asymmetric Depolarising Channel}
\label{Asymmetric deoplarising channel}
A quantum depolarizing channel characterised by the probability $p_x, p_y$ and $p_z$ can be directly used for modelling quantum systems employing diverse materials. In other words, the quantum depolarizing channel can be used for modelling the imperfections in quantum hardware, namely, qubit flips resulting from quantum decoherence and quantum gates. Furthermore, a quantum depolarizing channel can also be invoked for modelling quantum-state flips imposed by the real transmission medium, including free-space wireless channels and optical fiber links, when qubits are transmitted across these media.  For the sake of simpliciy, most recent studies of the quantum channel capacity [\cref{Bennett02}], [\cref{Bennett99}], [\cref{Bradler10}], [\cref{Wilde12}] as well as of quantum error correction (QEC) schemes considered the symmetric polarizing channel [\cref{Babar15}], [\cref{Poulin09}], [\cref{Wilde14}], where  the constituent flip probabilities obey $p_x=p_y=p_z=p/3$. By contrast, popular materials invoked for producing quantum devices often exhibit asymmetric behaviour, where a phase flip is orders of magnitude more likely than a bit flip [\cref{Loffe07}], which can be modelled by an asymmetric quantum depolarizing channel [\cref{Bertet05}], [\cref{Kaler03}], [\cref{Petta05}], [\cref{Tyryshkin06}], [\cref{Vandersypen01}]. In such asymmetric depolarizing channels, an extra parameter $\alpha$ termed as the channel's ratio of asymmetry is introduced for reflecting the ratio of the phase flip probability $p_z$ and the bit flip probability $p_x$ as [\cref{Evans07}], [\cref{Sarvepalli09}] 
\begin{eqnarray}
\label{Channel_model_1}
\alpha  &=& \frac{p_z}{p_x} = 1+ 2\frac{e^{\frac{-t}{T_1}}-e^{\left( \frac{-t}{2T_1}-\frac{2t}{T_2}\right) }}{1- e^{\frac{-t}{T_1}}} . \
\end{eqnarray}

Note that the bit flip probability $p_x$ as well as the simultaneous bit-and-phase flip probability $p_y$ may be considered to be equal [\cref{Evans07}], [\cref{Sarvepalli09}] while time instant $t$ may be interpreted as the coherent operation duration of a physical quantum gate [\cref{Williams2000}]. If the coherent operation duration $t$ is relatively short, formulated as $t<<T_1$, we can invoke the approximation of $\alpha \approx 2T_2/T_1 -1 $ [\cref{Petta05}]. As a result, the phase flip probability $p_z$ can be directly determined from the values of $\alpha$ and $p_x$. Note that in the case of having $\alpha =1$, the depolarising channel is the symmetric depolarizing channel, where the condition of having $p_x=p_y=p_z=p/3$ is satisfied. In practice the channel's ratio of asymmetry has popular values of $\alpha=10^2, 10^4, 10^6$  [\cref{Babar15}], [\cref{Poulin09}], [\cref{Wilde14}], which correspond to the typical materials of \tref{Alpha values}, which are used for producing quantum devices. 
\begin{table}[htbp]
\begin{center}
\begin{tabular}{|l|r|r|r|} \hline 
\multicolumn{1}{|c}{\bf System (Material)}{} & 
\multicolumn{1}{|c}{$T_1$}&\multicolumn{1}{|c|}{$T_2$}&\multicolumn{1}{|c|}{\textbf{$\alpha$}}\\ \hline 
P:Si [\cref{Tyryshkin06}] & 1 hour & 1ms & $10^6$ \\ \hline 
GaAs Quantum Dots [\cref{Petta05}]  & 10ms & $>1\mu$s & $10^4$\\ \hline 
Super conducting (flux qubits) [\cref{Bertet05}] & 1$\mu$s & 100 ns& $10^2$\\ \hline 
Trapped ions [\cref{Kaler03}] & 100 ms & 1 ms & $10^2$ \\ \hline 
Solid State NMR [\cref{Vandersypen01}] & $>1$ min & $>1$ s & $10^2$ \\ \hline 

\end{tabular}
\caption{Estimated asymmetric ratio $\alpha$ representing various quantum depolarizing channels associated with various quantum devices.}
\label{Alpha values}
\end{center}
\end{table}
\subsection{Acting Time in Asymmetric Channels}
\label{Acting time and asymmetric channels}
In the asymmtric depolarizing chanel, when the acting time \footnote{$t$ is the evolution time of the quantum system with the presence of decoherence, which can be considered to be equal to the duration of a physical quantum gate.} $t$ of the channels under investigation is small, the value of $\alpha$ in \eref{Channel_model_1} may be calculated by
\begin{eqnarray}
\label{Alpha_formula}
\alpha = 1+ 2\dfrac{1-e^{t/T_1(1-T_1/T_2)}}{e^{t/(T_1-1)}} \ ,
\end{eqnarray}
Then, the bit flip probability $p_x$ is calculated upon the asymmetric level $\alpha$ and the depolarizing probability of $p$ as:
\begin{eqnarray}
\label{Channel_model_2}
p_x &=& \frac{p}{\alpha + 2} \ .
\end{eqnarray}
As a result, the phase flip probability $p_z$ can be determined from the values of $\alpha$ and $p_x$. Since, the phase flip probability dominates over the bit flip one, the bit flip probability $p_x$ and the bit-and-phase flip probability $p_y$ may be considered to be equal.

We may use the precalculated $\alpha$ values in \tref{Alpha values} for characterising the quantum channel. Since this way does not take in consideration the absolute values of $t,T_1,T_2$, it may not closely characterise different systems manufactured by different materials in \tref{Alpha values} that are associated with the same value of $\alpha$. The absolute values of $t,T_1,T_2$ may be used for calculating the depolarizing probabilities of $p_x,p_z$ and $ p_y$ as follows [\cref{Evans07}]:
\begin{eqnarray}
\label{Z_actual}
p_z(t) &=& \frac{1}{4\left[ 1+e^{\frac{-t}{T_1}}-2e^{\left( \frac{-t}{2T_1}-\frac{-2t}{T_2}\right) }\right] } , \ 
\end{eqnarray}
\begin{eqnarray}
\label{X_actual}
p_x(t) &=& p_y(t) ,\\ \nonumber
       &=& \frac{1}{4(1-e^{\frac{-t}{T_1}})} .\
\end{eqnarray}

Accordingly, the encoding and decoding gate operation times pertaining to different materials are listed in \tref{Operating time}.
\begin{table*}[htbp]
\begin{center}
\begin{tabular}{|c|c|c|c|} \hline 
\multicolumn{1}{|c}{\bf Quantum Systems}{} & \multicolumn{1}{|c}{Time per gate Operation (sec)}& \multicolumn{1}{|c|}{Coherence time}& \multicolumn{1}{|c|}{Maximal number of coherence steps}\\ \hline 
Electrons from a gold atom & $10^{-14}$ & $10^{-8}$ & $10^6$ \\ \hline 
Trapped indium atoms & $10^{-14}$ & $10^{-1}$ & $10^{13}$ \\ \hline 
Optical micro cavity& $10^{-14}$ & $10^{-5}$ & $10^9$ \\ \hline 
Electron spin & $10^{-7}$ & $10^{-3}$ & $10^4$ \\ \hline 
Electron quantum dot & $10^{-6}$ & $10^{-3}$ & $10^3$ \\ \hline 

Nuclear spin & $10^{-3}$ & $10^{4}$ & $10^7$ \\ \hline 
\end{tabular}
\caption{Maximal number of computational steps that can be performed without losing coherence}
\label{Operating time}
\end{center}
\end{table*}

\subsection{Implementation of Quantum Channel in FSO-based Quantum Key Distribution}
\label{FSO_channel}
Depending on the specific form of the electromagnetic plane wave pertaining to the monochromatic laser signal generating photons, photons may be linearly polarized \bglo{LP} or elliptically polarized \bglo{EP} [\cref{Pade14}].  In the context of considering Quantum Key Distribution (QKD) systems, we only consider LP photons having polarizations of say $0^0, 90^0, -45^0, 45^0$ [\cref{Zeilinger99}]. Accordingly, the basis associated with the polarization of $0^0, 90^0$ can be characterised by:
\begin{eqnarray}
\label{photon_state1}
\ket{0^0} & = & 1\ket{0^0}+0i\ket{90^0} , \
\end{eqnarray}
\begin{eqnarray}
\label{photon_state2}
\ket{90^0} & = &  0\ket{0^0}+i\ket{90^0} . \
\end{eqnarray}

The relationship between the two bases can also be expressed by:
\begin{eqnarray}
\label{photon_state3}
\ket{0^0} & = & \frac{1}{\sqrt{2}}\ket{45^0} + \frac{i}{\sqrt{2}}\ket{-45^0} , \
\end{eqnarray}
\begin{eqnarray}
\label{photon_state4}
\ket{90^0} & = & \frac{1}{\sqrt{2}}\ket{45^0} - \frac{i}{\sqrt{2}}\ket{-45^0} . \
\end{eqnarray}

An FSO quantum transmission channel is used for carrying the photon stream to from the source (S) to the destination (D). Since the FSO channel imposes deleterious effects, such as  diffraction, atmospheric turbulence and extinction [\cref{Trinh15}], only a certain fraction $\gamma$ of the photon stream transmitted by S arrives at D. In other words, the term $\gamma$ invoked for characterising the power transfer properties of the FSO channel over a distance $L$ imposed on the QKD system's performance is approximated by [\cref{Gabay06}], [\cref{Safari09}], [\cref{Shapiro03}]
\begin{eqnarray}
\label{gamma}
\gamma &=& \mu\mbox{e}^{-\alpha L},\ 
\end{eqnarray} 
where $\mu$ represents the diffraction losses or the normalised version of the fraction $\gamma$, while $\alpha$ is the extinction coefficient.  

The value of $\mu$ depends on the Fresnel number of 
\begin{eqnarray}
\label{Dof}
D^0_f &=& \left(\frac{\pi d_1 d_2}{4 \lambda L} \right)^2,\ 
\end{eqnarray}
where $d_1$ is the transmit aperture diameter and $d_2$ is the receiver's aperture diameter, while $\lambda$ is the wavelength of the optical signal.

In the near-field region having $D^0_f>>1$, the parameter $\mu$ is bounded by [\cref{Shapiro74}], [\cref{Shapiro03}]
\begin{eqnarray}
\label{bound_near}
\mu_{NF,LB} &\leq & \mu \leq \mu_{NF,UB},\ 
\end{eqnarray} 
where the upper bound $\mu_{NF,UB}$ can be calculated by [\cref{Shapiro03}] 
\begin{eqnarray}
\label{bound_near_UB}
\mu_{NF,UB} & = & \mbox{min}(D^0_f,1),\ 
\end{eqnarray} 
while the lower bound $\mu_{NF,LB}$ is given by  [\cref{Shapiro03}]
\begin{eqnarray}
\label{bound_near_LB}
\mu_{NF,LB}&=& \frac{8\sqrt{D^0_f}}{\pi}\int^{1}_{0} \mbox{exp}\left(\frac{-D(d_2x)}{2}\right) \\ \nonumber 
       & & \times\left( \mbox{arccos}(x)-x\sqrt{1-x^2} \right)J_1\left( 4x\sqrt{D^0_f}\right) dx ,\ 
\end{eqnarray} 
where $J_1(.)$ is the first-order Bessel function. The spherical-wave structure function $D(\rho)$ of \eref{bound_near_LB} is calculated for the worse-case scenario of having $d_1=d_2$ as  [\cref{Shapiro03}]:
\begin{eqnarray}
\label{Function_D}
D(\rho) &=& 51 \sigma^2_R\left(D^0_f\right)^{5/12}\rho^{5/3},\ 
\end{eqnarray}  
where $\sigma_R^2$ is the Rytov variance [\cref{Karp88}] of 
\begin{eqnarray}
\label{Rytov_variance}
\sigma^2_R &=& 1.24\left(\frac{2\pi}{\lambda}\right)^{7/6} C^2_n L^{11/6}, \   
\end{eqnarray}
with $C^2_n$ ranging from $10^{-13}$ to $10^{-17}$ representing the altitude-dependent index of the refractive structure parameter [\cref{Zhang09}].

By contrast, in the far-field region having $D^0_f<<1$, the value of $\mu$ can be calculated by [\cref{Shapiro74}]
\begin{eqnarray}
\label{normalised_far}
\mu_{FF}&=& \frac{8\sqrt{D^0_f}}{\pi}\int^{1}_{0} \mbox{exp}\left(\frac{-D(d_2x)}{2}\right) \\ \nonumber 
       & & \times\left( \mbox{arcos}^{-1}(x)-x\sqrt{1-x^2} \right)J_1\left( 4x\sqrt{D^0_f}\right) dx ,\ 
\end{eqnarray} 
where the spherical-wave structure function $D(\rho)$ of \eref{normalised_far} can be calculated by
\begin{eqnarray}
\label{Function_D_far}
D(\rho) &=& 1.09 \left(\frac{2\pi}{\lambda}\right)^{2} C^2_nL\rho^{5/3}.\ 
\end{eqnarray} 

\begin{figure}[h!]
\begin{center}
\includegraphics[angle = 0,width=\linewidth]{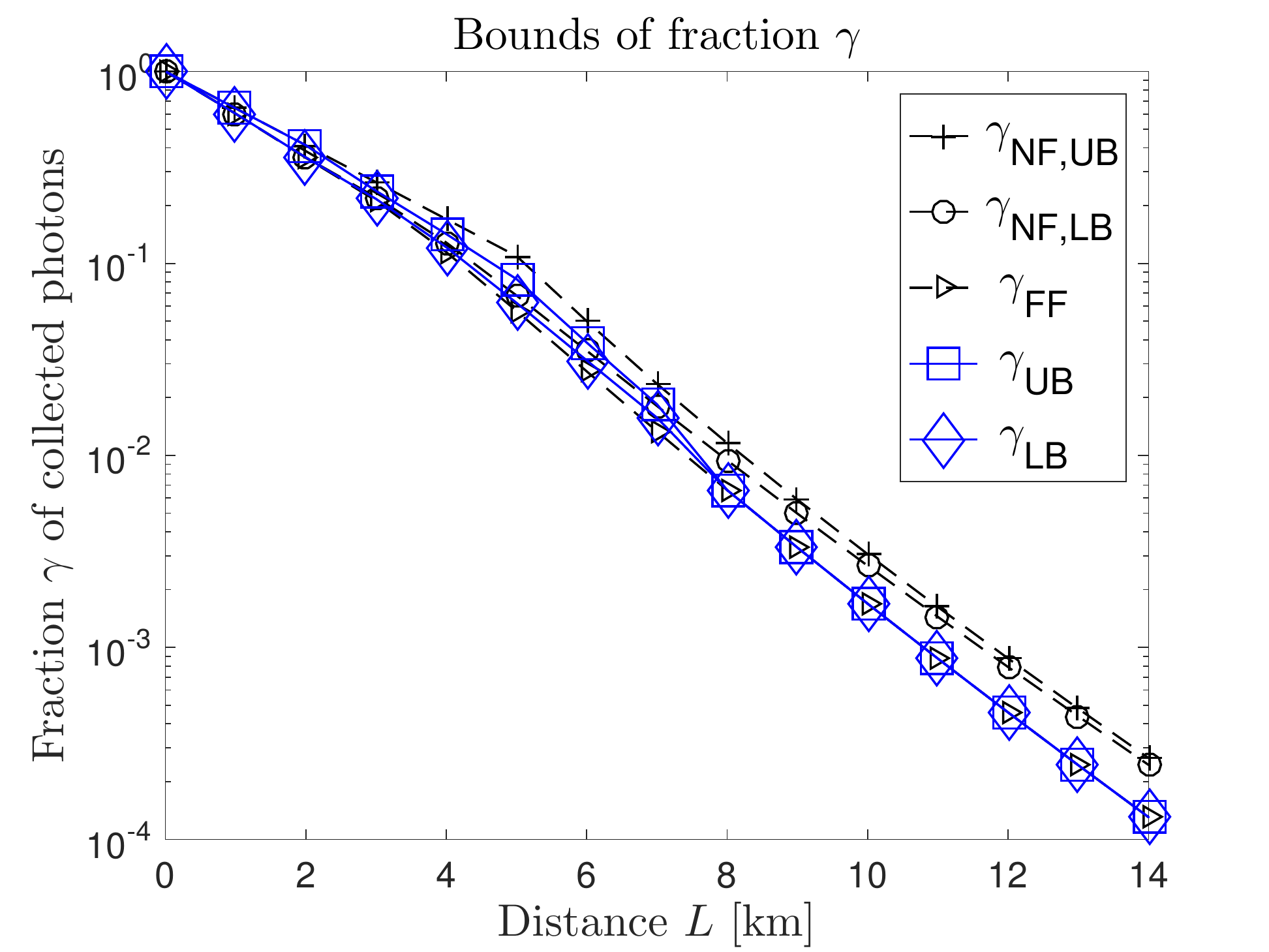}
\caption{Bounds of $\gamma$ characterised by \eref{gamma_bounds_close} for a transition region of ($T_{far}=0.5\leq D^0_f \leq T_{near}=5$).} 
 \label{figfso}
 \end{center}
\end{figure}

As a result, when a more accurate value range of $\gamma$ is sought, the following bounds should be used (see \fref{figfso})
\begin{eqnarray}
\label{gamma_bounds_close}
\gamma_{LB} &\leq & \gamma \leq \gamma_{UB},\
\end{eqnarray} 
where the upper bound $\gamma_{UB}$ is determined by:
\begin{equation}
\label{gamma_upper_close}
\gamma_{UB}  =  \left\{ \begin{array}{r@{\quad:\quad}l}
                              \gamma_{NF,UB}  & \mbox{if } D^0_f> T_{near}\\
                              (\gamma_{NF,UB}+\gamma_{FF})/2  & \mbox{if } T_{far}\leq D^0_f \leq T_{near}\\
                              \gamma_{FF} & \mbox{if }  D^0_f < T_{far}\\
\end{array}\right.\ ,
\end{equation}
while the lower bound $\gamma_{LB}$ is calculated by:
\begin{equation}
\label{gamma_lower_close}
\gamma_{LB}  =  \left\{ \begin{array}{r@{\quad:\quad}l}
                              \gamma_{NF,LB}  & \mbox{if } D^0_f> T_{near}\\
                              (\gamma_{NF,LB}+\gamma_{FF})/2  & \mbox{if } T_{far}\leq D^0_f \leq T_{near}\\
                              \gamma_{FF} & \mbox{if }  D^0_f < T_{far}\\
\end{array}\right.\ ,
\end{equation}
where the region having $T_{far}\leq D^0_f \leq T_{near}$ is the transition region between the near-field and far-field regimes.

\subsection{Quantum Channel Codes for Approaching Quantum Channel Capacity}
\label{Quantum channel codes for approaching quantum channel capacity}
The appealing parallelism of quantum computing relying on quantum bits has inspired researchers to consider various quantum-related applications in the area of quantum communications [\cref{Djordjevic08}], [\cref{Kelly15}], [\cref{Lupo11}], [\cref{Ma07}], [\cref{Paparo12}], [\cref{Schindler13}]. However, a crucial obstacle to the practical realisation of quantum communications systems is the presence of quantum perturbations. Their deleterious effects can be mitigated by Quantum Error Correction Codes [\cref{Bennett02}]. It was suggested that the employment of entanglement assistance is capable of further improving the performance of QECCs [\cref{Poulin09}], [\cref{Wilde14}] in the context of the so-called symmetric depolarizing channel, which has been routinely used in theoretical studies. In the symmetric depolarizing channel characterised by the gross depolarizing probability $p$, each transmitted qubit may independently experience either a bit flip ($X$), a phase flip ($Z$), or both ($Y$) at a probability of $p_x=p_y=p_z=p/3$. By contrast, the materials considered at the time of writing for building quantum devices, including trapped ions [\cref{Schmidt-Kaler03}] and solid state Nuclear Magnetic Resonance [\cref{Vandersypen01}], exhibit asymmetric depolarization property defined as the ratio of the phase flip probability over the bit flip probability, where the grade of asymmetry is in the range spanning from $\alpha =10^2$ to $\alpha =10^6$ [\cref{NC1}-\cref{NC5}]. QECCs designed for the asymmetric depolarizing channel were termed as asymmetric QECCs in [\cref{NC6}-\cref{NC11}], where a limited range of $\alpha$ values was assumed and no entanglement assistance was addressed. In [\cref{NC12}], a more general framework covering both symmetric and asymmetric depolarizing channels was proposed for Entanglement Assisted QECCs \bglo{EAQECCs}.

To benchmark the design of the EAQECCs, the Entanglement Assisted Quantum Channel's \bglo{EAQC} capacity was investigated in [\cref{Bradler10}], [\cref{Wilde12}]. Accordingly, the so-called Hashing bound is advocated for setting a lower limit on the achievable quantum depolarizing channel capacity, which has been used for benchmarking the performance of various QECC schemes in [\cref{Babar15}], [\cref{NC13}], [\cref{Wilde14}]. Furthermore, the powerful Extrinsic Information Transfer \bglo{EXIT} chart technique [\cref{NC14}-\cref{NC18}] that was originally introduced for analysing the convergence behaviour of iterative decoding and detection in conventional communication systems was recently further developed for analysing the iterative decoding convergence of QECCs [\cref{NC13}]. In [\cref{NC12}], entanglement assisted quantum coding schemes and the associated quantum depolarizing channel capacity were considered for both asymmetric and symmetric quantum depolarizing channels. 

\subsection{Quantum Network Coding for Entanglement Distribution}
\label{Quantum Network Coding for distribution of the Quantum Entanglement over Networks}
In the classical domain, network coding [\cref{NC19}, \cref{NC20}] is capable of increasing the throughput, while minimising the amount of energy required per packet as well as the delay of packets travelling through the network [\cref{NC21}, \cref{NC22}]. This is achieved by allowing the intermediate nodes of the network to combine multiple data packets received via the incoming links before transmission to the destination [\cref{NC23}]. Due to its merits, the concept of the network coding has been applied in diverse disciplines [\cref{NC24}]. 

Inspired by its classical counterpart [\cref{NC20}, \cref{NC25}, \cref{NC26}], the question arises if the quantum version of network coding exists. Due to the inherent nature of quantum communications, namely that cloning is impossible, negative answers to this cardinal question were suggested in  [\cref{NC27}, \cref{NC28}]. However, further studies of Quantum Network Coding \bglo{QNC} confirm that given the availability of extra resources, such as preshared entanglement  [\cref{NC29}-\cref{NC36}] or the abundance of low-cost classical communications  [\cref{NC28}, \cref{NC37}-\cref{NC39}], QNC can indeed be made feasible. 

Entanglement consitutes a valuable enabler of various quantum protocols that are essential for various applications of quantum communications, such as quantum teleportation [\cref{NC40}], remote state preparation [\cref{NC41}], quantum remote measuring [\cref{NC42}] and secret sharing [\cref{NC43}]. Entanglement refers to the fact that two or more photons have a very special connection, whereby changing for example the spin of a photon will instantaneously change that of its entangled couterpart. Anecdotally, this phenomenon is referred to as  a "spooky action at a distance" by Einstein [\cref{Einstein1935}] due to the fact that unlike in electromagnetism, interactions between entangled photons occur instantaneously, regardless of how far apart the photons are. By contrast, electromagnetic interactions are bounded by the speed of light [\cref{Imre13a}]. 

In such quantum protocols, the entangled qubits have to be distributed to distant nodes. A particularly popular application of the entanglement distribution is QKD [\cref{NC44}], which has been gradually finding its way into different practical scenarios, such as satellite communications [\cref{NC45}, \cref{NC46}], terrestrial communications [\cref{NC47}, \cref{NC48}] and over handheld communication [\cref{NC49}, \cref{NC50}]. These advances lay the foundations of the quantum Internet [\cref{NC51}-\cref{NC53}]. Entanglement distribution over a large-scale network consisting of multiple-hops and multiple-nodes can be realised by Entanglement Swapping \bglo{ES} [\cref{NC54}-\cref{NC56}] or by QNC [\cref{NC31}, \cref{NC33}, \cref{NC57}]. ES may be deemed to be similar to the classic Decode-and-Forward \bglo{DF} techniques, which is outperformed by the classical Network Coding \bglo{NC} in a number of practical scenarios [\cref{NC58}-\cref{NC60}]. This leads to another intriguing and crucial question, namely whether the QNC is similarly capable of providing a better performance than ES.
In order to answer the second question, Satoh \textit{et al.} [\cref{NC31}] provided quantitative comparisons between the QNC and the ES. Explicitly, it was shown that the fidelity-performance of the ES-based system is superior to that of the QNC-based system in a quantum communication network having $M=2$ pairs of source-to-target users that are connected via a backbone link having $N=1$ hop. However, Nguyen \textit{et al.} [\cref{NC61}] generalised the QNC of [\cref{NC31}, \cref{NC33}] to large-scale quantum communication networks, in order to demonstrate the benefits of large-scale QNC over ES.

\section{Conclusions}
\label{sec7}
Quantum channels extend the possibilities of classical communication channels allowing us to transmit classical information, entanglement assisted classical information, private classical information and quantum information. Contrary to classical channels, quantum channels can be used to construct more advanced communication primitives. Quantum entanglement or the superposed states carry quantum information, which cannot be described classically. Quantum channels can be implemented in practice easily e.g. via optical fiber networks or by wireless optical channels, and make it possible to send various types of information. The errors are a natural interference from the noisy environment, and the can be much diverse due to the extended set of quantum channel models. In the near future, advanced quantum communication and networking technologies driven by quantum information processing will revolutionize the traditional methods. Quantum information will help to resolve still open scientific and technical problems, as well as expand the boundaries of classical computation and communication systems. 

\section*{References}
\begin{enumerate}[ {[}1{]} ]
\footnotesize
\item \label{Abeyesinghe06} A. Abeyesinghe, P. Hayden, G. Smith, and A. J.Winter, ``Optimal superdense coding of entangled states,'' \textit{IEEE Transactions on Information Theory}, vol. 52, pp. 3635--3641, (2006).

 \item \label{Abramson70} Abramson, N.: The Aloha system-another alternative for computer communications. \textit{AFIPS Conf. Proc}. 36, pp. 295-298, (1970).

 \item \label{Ackermann08} M. Ackermann, J. Bl\"{o}mer, and C. Sohler. Clustering for metric and non-metric distance measures. In \textit{Proceedings of the 19th Annual ACM-SIAM Symposium on Discrete Algorithms (SODA '08),} pages 799--808. Society for Industrial and Applied Mathematics, (2008).

 \item \label{Adami96} C. Adami and N. J. Cerf, On the von Neumann capacity of noisy quantum channels, \textit{arXiv:quant-ph/9609024v3} (1996).

 \item \label{Agarwal04} P. Agarwal. Range searching. In J. E. Goodman and J. O'Rourke, editors, \textit{Handbook of Discrete and Computational Geometry}, 2nd edition., chapter 36. Chapman \& Hall/CRC, (2004).

 \item \label{Agarwal07} P. Agarwal, J. Pach, and M. Sharir. State of the union (of geometric objects): A review. In J. Goodman, J. Pach, and R. Pollack, editors, \textit{Computational Geometry}: Twenty Years Later. American Mathematical Society, (2007).

 \item \label{Aharonov97} D. Aharonov and M. Ben-Or, Fault-tolerant quantum computation with constant error. In \textit{STOC '97: Proceedings of the Twenty-Ninth Annual ACM Symposium on Theory of Computing}, pp 176-188, New York, NY, USA, ACM, (1997).

 \item \label{Ahlswede02} R. Ahlswede and A. J. Winter. Strong converse for identication via quantum channels. \textit{IEEE Transactions in Information Theory}, 48(3):569-579, arXiv:quant-ph/0012127, (2002).

 \item \label{Alicki04} R. Alicki and M. Fannes, "Note on multiple additivity of minimal entropy output of extreme SU(d)-covariant channels", \textit{Open Systems and Information Dynamics }11, 339 - 342 (2004).

 \item \label{Aluru05} S. Aluru. Quadtrees and octrees. In D. Metha and S. Sahni, editors, \textit{Handbook of Data Structures and Applications}, chapter 19. Chapman \& Hall/CRC, (2005).

 \item \label{Amari2000} S. Amari and H. Nagaoka: Methods of Information Geometry. \textit{Translations of Mathematical Monographs}, 191. AMS, Oxford University Press, Oxford, (2000).

 \item \label{Amari93} S. Amari and H. Nagaoka: Methods of Information Geometry. Translated from the 1993 Japanese original by Daishi Harada. \textit{Translations of Mathematical Monographs}, 191. AMS, Oxford University Press, Oxford, (2000).

 \item \label{Amato01} N. Amato, M. T. Goodrich, and E. A. Ramos. A randomized algorithm for triangulating a simple polygon in linear time. \textit{Discrete Comput. Geom}., 26:245-- 265, 2001.

 \item \label{Amosov04} G. Amosov. Remark on the additivity conjecture for the quantum depolarizing channel, http://arxiv.org/abs/quant-ph/0408004, (2004).

 \item \label{Amosov07} G. Amosov, "The strong superadditivity conjecture holds for the quantum depolarizing channel in any dimension", \textit{Phys. Rev. A} 75, 2104 - 2117 (2007).

 \item \label{Amosov2000} G. Amosov, A. S. Holevo and R. F. Werner, "On some additivity problems in quantum information theory", \textit{Problems in Information Transmission} 36, 305 - 313 (2000).

 \item \label{Arge06} L. Arge, G. Brodal, and L. Georgiadis. Improved dynamic planar point location. In Proc. 47th Annu. \textit{IEEE Sympos. Found. Comput. Sci}., pages 305--314, (2006).

 \item \label{Arimoto72} S. Arimoto. An algorithm for calculating the capacity of an arbitrary discrete memoryless channel. \textit{IEEE Trans. Inf. Theory}, 18:14--20, (1972).

 \item \label{Arrighi07} P. Arrighi, V. Nesme, \& R. Werner, Unitarity plus causality implies localizability, \textit{Proceedings of the Quantum Information Processing} 2010 conference (QIP2010). (2010).

 \item \label{Asano06} T. Asano, N. Katoh, N. Tamaki, and T. Tokuyama. Angular Voronoi diagram with applications. In \textit{Proceedings of 3rd International Symposium on Voronoi Diagram in Science and Engineering}, pages 32--39, Banff, Canada, (2006).

 \item \label{Asano07} T. Asano, N. Katoh, N. Tamaki, and T. Tokuyama. Voronoi diagram with respect to criteria on vision information. In \textit{Proceedings of 4th International Symposium on Voronoi Diagram in Science and Engineering}, pages 25--32, Wales, UK, (2007).

 \item \label{Asano07a} T. Asano. Aspect-ratio voronoi diagram and its complexity bounds. \textit{Information Processing Letters}, 105(1. 31):26--31, (2007).

 \item \label{Aspect81} A. Aspect, P. Grangier, and G. Roger. Experimental tests of realistic local theories via Bell's theorem, \textit{Physical Review Letters}, 47(7):460-463, (1981).

 \item \label{Aubrun10} G. Aubrun, S. Szarek, and E. Werner, ``Hastings' additivity counterexample via Dvoretzky's theorem,'' ArXiv e-prints\textit{, arXiv:1003.4925}. (2010).

 \item \label{Audenaert07} K. M. R. Audenaert. A sharp continuity estimate for the von Neumann entropy. \textit{Journal of Physics A: Mathematical and Theoretical}, 40(28):8127, (2007).

 \item \label{Aurenhammer00} F. Aurenhammer and R. Klein. Voronoi Diagrams. In J. Sack and G. Urrutia (Eds), \textit{Handbook of Computational Geometry,} Chapter V, pp. 201--290. Elsevier \textit{Science} Publishing, (2000).

 \item \label{Aurenhammer84} F. Aurenhammer and H. Edelsbrunner. An optimal algorithm for constructing the weighted Voronoi diagram in the plane. \textit{Pattern Recogn}., 17:251--257, (1984).

 \item \label{Aurenhammer87} F. Aurenhammer, ``Power diagrams: properties, algorithms and applications,'' \textit{SIAM J. Comput}., vol. 16, no. 1, pp. 78--96, (1987).

 \item \label{Aurenhammer91} F. Aurenhammer. Voronoi diagrams: A survey of a fundamental geometric data structure\textit{. ACM Comput. Surv}., 23:345--405, (1991).

 \item \label{Aurenhammer92} F. Aurenhammer and O. Schwarzkopf. A simple on-line randomized incremental algorithm for computing higher order Voronoi diagrams. \textit{Internat. J. Comput. Geom. Appl.}, 2:363--381, (1992).

 \item \label{Aurenhammer98} F. Aurenhammer, F. Hoffmann, and B. Aronov. Minkowski-type theorems and least-squares clustering. \textit{Algorithmica}, 20:61--76, (1998).

 \item \label{Azuma09} K. Azuma, N. Sota, R. Namiki, S. Kaya Ozdemir, T. Yamamoto, M. Koashi, Optimal entanglement generation for efficient hybrid quantum repeaters, \textit{Physical Review A} 80, 060303(R) (2009).

 \item \label{Babar13} Z. Babar, S. X. Ng, and L. Hanzo, ``Near-capacity code design for entanglement-assisted classical communication over quantum depolarizing channels,'' \textit{IEEE Transactions on Communications}, vol. 61, pp. 4801--4807, (2013).

 \item \label{Babar15} Z. Babar, S. X. Ng, and L. Hanzo, ``Exit-chart-aided near-capacity quantum turbo code design,'' \textit{IEEE Transactions on Vehicular Technology}, vol. 64, pp. 866--875 (2015).

 \item \label{Bacsardi05} L. Bacsardi, Using Quantum Computing Algorithms in Future Satellite Communication, \textit{ACTA ASTRONAUTICA} 57: pp. 224-229. (2005)

 \item \label{Bacsardi07} L. Bacsardi, Satellite Communication Over Quantum Channel, \textit{ACTA ASTRONAUTICA} 61:(1-6) pp. 151-159. (2007)

 \item \label{Bacsardi10} L. Bacsardi, S. Imre, Quantum Based Information Transfer in Satellite Communication, \textit{Satellite Communications}, SCIYO, 2010. pp. 421-436.

 \item \label{Bacsardi13} L. Bacsardi, On the Way to Quantum-Based Satellite Communication, \textit{IEEE COMMUNICATIONS MAGAZINE} 51:(08) pp. 50-55. (2013)

 \item \label{Badoiu02} M. Badoiu, S. Har-Peled, and P. Indyk. Approximate clustering via core-sets. In \textit{Proceedings 34th ACM Symposium on Theory of Computing}, pages 250--257, (2002).

 \item \label{Bai05} Y-K. Bai, S. Li and H. Zheng, ``Method for measuring two-qubit entanglement of formation by local operations and classical communication'', \textit{J. Phys. A}: Math. Gen. 38 8633, doi: 10.1088/0305-4470/38/40/010 (2005).

 \item \label{Banerjee05} Banerjee, Arindam; Merugu, Srujana; Dhillon, Inderjit S.; Ghosh, Joydeep ``Clustering with Bregman divergences''. \textit{Journal of Machine Learning Research} 6: 1705--1749. (2005).

 \item \label{Barnum01} H. Barnum, C. Caves, C. Fuchs, R. Jozsa, and B. Schumacher. On quantum coding for ensembles of mixed states. \textit{Journal of Physics A: Mathematical and General}, 34(35):6767, (2001).

 \item \label{Barnum2000} H. Barnum, Emanuel Knill, and Michael A. Nielsen. On quantum fidelities and channel capacities. \textit{IEEE Transactions on Information Theory}, 46:1317-1329, (2000).

 \item \label{Barnum95} H. Barnum, C. M. Caves, C. A. Fuchs, R. Jozsa, and B. Schumacher, ``Noncommuting mixed states cannot be broadcast,'' \textit{Phys. Rev. Lett}., vol. 76, pp. 2818--2821, eprint quant-ph/9511010, (1996).

 \item \label{Barnum96} H. Barnum, C. A. Fuchs, R. Jozsa, and B. Schumacher, ``A general fidelity limit for quantum channels,'' eprint \textit{quant-ph/9603014,} (1996).

 \item \label{Barnum97a} H. Barnum, M. A. Nielsen, and B. Schumacher, ``Information transmission through noisy quantum channels,'' \textit{eprint quant-ph/9702049}, (1997).

 \item \label{Barnum97b} H. Barnum, J. Smolin, and B. Terhal, ``Results on quantum channel capacity,'' submitted to Phys. Rev A (eprint \textit{quant-ph/9711032}). (1997).

 \item \label{Barrett10} S. D. Barrett, T. M. Stace, Fault tolerant quantum computation with very high threshold for loss errors, \textit{Physical Review Letters}, DOI:10.1103/PhysRevLett.105.200502, http://arxiv.org/abs/1005.2456. (2010).

 \item \label{Becerra14} F. E. Becerra, J. Fan, and A. Migdall, ``Photon number resolution enables quantum receiver for realistic coherent optical communications,'' \textit{Nature Photonics}, vol. 9, p. 48, (2014).

 \item \label{Beigi07} S. Beigi and P. W. Shor, On the Complexity of Computing Zero-Error and Holevo Capacity of Quantum Channels, \textit{arXiv:0709.2090} {[}quant-ph{]} (2007).

 \item \label{Bell1966} J. Bell, ``On the Problem of Hidden Variables in Quantum Mechanics'', \textit{Rev. Mod. Phys}. 38: 447-452 (1966).

 \item \label{Benioff80} P. Benioff, The computer as a physical system: A microscopic quantum mechanical hamiltonian model of computers as represented by Turing machines. \textit{Journal of Statistical Physics}, 22:563--591. (1980).

 \item \label{Bennett14} C. H. Bennett and G. Brassard, ``Quantum cryptography: Public key distribution and coin tossing,'' \textit{Theoretical Computer Science}, vol. 560, pp. 7--11, (2014).

 \item \label{Bennett02} C. Bennett, P. W. Shor, J. A. Smolin, and A. V. Thapliyal, ``Entanglement assisted capacity of a quantum channel and the reverse shannon theorem,'' \textit{IEEE Trans. Inf. Theory}, vol. 48, pp. 2637--2655, (2002).

 \item \label{Bennett05} C. H. Bennett, P. Hayden, D. W. Leung, P. W. Shor, and A. Winter, ``Remote preparation of quantum states,'' \textit{IEEE Transactions on Information Theory}, vol. 51, pp. 56--74, (2005).

 \item \label{Bennett06} C. Bennett, A. Harrow, and S. Lloyd. Universal quantum data compression via nondestructive tomography. \textit{Physical Review A}, 73(3):032336, (2006).

 \item \label{Bennett09} C. Bennett, I. Devetak, A. Harrow, P. Shor, A. Winter, Quantum Reverse Shannon Theorem, (2009), \textit{arXiv:0912.5537}

 \item \label{Bennett92} C. Bennett and S. Wiesner. Communication via one- and two-particle operators on Einstein-Podolsky-Rosen states.\textit{Phys.Rev.Lett}.,69:2881-2884, (1992).

 \item \label{Bennett92a} Charles H. Bennett, Gilles Brassard, Claude Crepeau, and Marie-Helene Skubiszewska, Practical Quantum Oblivious Transfer" in Advances in Cryptology. \textit{Lecture Notes in Computer Science, 576}., 351-366 (1992).

 \item \label{Bennett93} C. Bennett, G. Brassard, C. Cr\'{e}peau, R. Jozsa, A. Peres, and William K. Wootters. Teleporting an unknown quantum state via dual classical and Einstein-Podolsky-Rosen channels. \textit{Physical Review Letters}, 70(13):1895-1899, (1993).

 \item \label{Bennett95} C. Bennett. \textit{Quantum Information and Computation}. \textit{Physics Today}, 48(10):24-30, (1995).

 \item \label{Bennett96a} C. Bennett, D. P. DiVincenzo, J. A. Smolin, and W. K. Wootters, ``Mixed state entanglement and quantum error correction''. \textit{Phys. Rev. A}, 54(5):3824-- 3851, (1996).

 \item \label{Bennett96b} C. Bennett, H. Bernstein, S. Popescu, and B. Schumacher. Concentrating partial entanglement by local operations. \textit{Physical Review A}, 53(4):2046-2052, (1996).

 \item \label{Bennett97} C. Bennett, D. P. DiVincenzo, and J. A. Smolin, ``Capacities of quantum erasure channels,'' \textit{Phys. Rev. Lett}., vol. 78, no. 16, pp. 3217--3220, (1997).

 \item \label{Bennett98} C. Bennett and P. W. Shor, ``quantum information theory'', \textit{IEEE Trans. Info. Theory} 44, 2724 - 2742 (1998).

 \item \label{Bennett99} C. Bennett, P. Shor, J. Smolin, and A. Thapliyal, ``Entanglement-assisted classical capacity of noisy quantum channels,'' \textit{Physical Review Letters}, vol. 83, pp. 3081--3084, Oct. 11 (1999).

 \item \label{Bennett99a} C. H. Bennett, D. P. DiVincenzo, C. A. Fuchs, T. Mor, E. Rains, P. W. Shor, J. A. Smolin, and W. K. Wootters. Quantum nonlocality without entanglement. \textit{Phys.Rev. A}, 59(2):1070--1091, (1999).

 \item \label{Bennett83} C. H. Bennett and G. Brassard, ``Quantum cryptography and its application to provably secure key expansion, public-key distribution, and coin-tossing,'' 1983 \textit{IEEE International Symposium on Information Theory,} pp. 91--91 (1983).

 \item \label{Bengtsson06} I. Bengtsson, K. Zyczkowski, \textit{Entanglement \& geometry of quantum states}, Cambridge University Press, Cambridge, United Kingdom, (2006).

 \item \label{Berces15} Berces M, Imre S.:Extension and analysis of modifiedsuperdense-codinginmulti-userenvironment, \textit{IEEE 19th International Conference on Intelligent Engineering Systems }(INES2015), pp. 291-294. (2015).

 \item \label{Berger71} T. Berger. Rate distortion theory: \textit{A mathematical basis for data compression}. Prentice-Hall, Englewood Cliffs, New Jersey, USA, (1971).

 \item \label{Bern04} M. Bern. \textit{Triangulation and mesh generation}. In J. E. Goodman and J. O'Rourke, editors, \textit{Handbook of Discrete and Computational Geometry}, 2nd edn., chapter 25. Chapman \& Hall/CRC, (2004).

 \item \label{Bern99} M. Bern and P. Plasman. Mesh generation. In J.-R. Sack and J. Urrutia, editors, \textit{Handbook of Computational Geometry}, 2nd edn., chapter 6. Elsevier, (1999).

 \item \label{Bernardes10} N. K. Bernardes, L. Praxmeyer, P. van Loock, Rate analysis for a hybrid quantum repeater, arXiv:1010.0106v1, (2010).

 \item \label{Bertet05} P. Bertet, I. Chiorescu, G. Burkard, K. Semba, C. J. P. M. Harmans, D. P. DiVincenzo, and J. E. Mooij, ``Dephasing of a superconducting qubit induced by photon noise,''\textit{ Physical Review Lettes}, vol. 95 (2005).

 \item \label{Bohm89} D. Bohm. \textit{Quantum Theory}. Courier Dover Publications, (1989).

 \item \label{Bollobas98} B. Bollobas, \textit{Modern graph theory}. Springer-Verlag New York, Inc., New York, (1998).

 \item \label{Bonato10} C. Bonato, S. Bonora, A. Chiuri, P. Mataloni, G. Milani, G. Vallone, and P. Villoresi, Phase control of a path-entangled photon state by a deformable membrane mirror, \textit{JOSA B,} Vol. 27, Issue 6, pp. A175-A180, doi:10.1364/JOSAB.27.00A175, (2010).

 \item \label{Bowen04} G. Bowen. Quantum feedback channels. \textit{IEEE Transactions in Information Theory},50(10):2429-2434, \textit{arXiv:quant-ph/0209076}, (2004).

 \item \label{Bowen05} G. Bowen. Feedback in quantum communication. International Journal of Quantum Information, 3(1):123-127\textit{, arXiv:quant-ph/0410191}, (2005).

 \item \label{Bowen05a} G. Bowen and R. Nagarajan. On feedback and the classical capacity of a noisy quantum channel. \textit{IEEE Transactions in Information Theory}, 51(1):320-324, \textit{arXiv:quant-ph/0305176}, (2005).

 \item \label{Boyd04} S. Boyd and L. Vandenberghe. \textit{Convex Optimization}. Cambridge University Press, The Edinburgh Building, Cambridge, CB2 8RU, UK, (2004).

 \item \label{Bradler09} K. Bradler, P. Hayden, and P. Panangaden. Private information via the Unruh effect, \textit{Journal of High Energy Physics} 08, 074 (2009).

\item \label{Bradler09a} K. Bradler, An infinite sequence of additive channels: the classical capacity of cloning channels. \textit{arXiv:0903.1638}, (2009).

 \item \label{Bradler10} K. Bradler, P. Hayden, D. Touchette, and M. M. Wilde, Trade-off capacities of the quantum Hadamard channels, \textit{arXiv:1001.1732v2}, (2010).

 \item \label{Brandao10} F. Brandao and J. Oppenheim, ``Public Quantum Communication and Superactivation,'' \textit{arXiv:1005.1975}. (2010).

 \item \label{Brandao11} F. Brandao, J. Oppenheim and S. Strelchuk, "When does noise increase the quantum capacity?", \textit{arXiv:1107.4385v1} {[}quant-ph{]} (2011).

 \item \label{Braunstein2000} S. Braunstein, C. Fuchs, D. Gottesman, and H. Lo. A quantum analog of Huffman coding. \textit{IEEE Transactions in Information Theory}, 46(4):1644-1649, (2000).

 \item \label{Bregman67} L. Bregman, ``The relaxation method of finding the common points of convex sets and its application to the solution of problems in convex programming''. \textit{USSR Computational Mathematics and Mathematical Physics} 7: 200--217. doi:10.1016/0041 5553(67)90040-7. (1967).

 \item \label{Briegel98} H.-J. Briegel, W. D\"{u}r, J.I. Cirac, and P. Zoller, Quantum repeaters: the role of imperfect local operations in quantum communication. \textit{Physical Review Letters}, 81:5932 5935, (1998).

 \item \label{Brodal02} G. Brodal and R. Jacob. Dynamic planar convex hull. In Proc. 43rd Annu. \textit{IEEE Sympos. Found. Comput. Sci}., pages 617--626, (2002).

 \item \label{Broglie1924} L. de Broglie. Recherches sur la th\'{e}orie des quanta. PhD thesis, Paris, (1924).

 \item \label{Bruss2000} D. Bruss, L. Faoro, C. Macchiavello and M. Palma, "Quantum entanglement and classical communication through a depolarizing channel", \textit{J. Mod. Opt}. 47 325 (2000).

 \item \label{Buckley10} B. Buckley, G. Fuchs, L. Bassett, and D. Awschalom, Spin-Light Coherence for Single-Spin Measurement and Control in Diamond. \textit{Science}, Online October 14 2010 DOI: 10.1126/\textit{Science}.1196436. (2010).

 \item \label{Buckley88} C. Buckley. A divide-and-conquer algorithm for computing 4-dimensional convex hulls. In Lecture Note in Computer \textit{Science}, volume 333, pages 113--135. Berlin: Springer-Verlag, (1988).

 \item \label{Burbea82} J. Burbea and C. Rao, ``On the convexity of some divergence measures based on entropy functions,'' \textit{IEEE Transactions on Information Theory}, vol. 28, no. 3, pp. 489--495, (1982).

 \item \label{Bures69} D. Bures, ``An extension of kakutani's theorem on infinite product measures to the tensor product of semifinite w*-algebras,'' \textit{Transactions on American Mathematical Society}, vol. 135, pp. 199--212, (1969).

 \item \label{Cai04} N. Cai, A. Winter, and R. Yeung, ``Quantum privacy and quantum wiretap channels,'' \textit{Problems of Information Transmission}, vol. 40, no. 4, pp. 318--336, (2004).

 \item \label{Calderbank96} R. Calderbank and P. Shor, Good quantum error-correcting codes exist, \textit{Physical Review A}, 54:1098, (1996).

 \item \label{Calderbank97} R. Calderbank, E. Rains, P. Shor, and N. J. A. Sloane, Quantum error correction and orthogonal geometry. \textit{Physical Review Letters}, 78(3):405-408, (1997).

 \item \label{Calderbank98} A. R. Calderbank, E. Rains, P. Shor, and N. J. A. Sloane, Quantum error correction via codes over GF(4). \textit{IEEE Transactions on Information Theory}, 44:1369-1387, (1998).

 \item \label{Cerf97} N. Cerf and C. Adami. Negative entropy and information in quantum mechanics. \textit{Physical Review Letters}, 79:5194-5197, (1997).

 \item \label{Chan01} J. Chan. Dynamic planar convex hull operations in near-logarithmaic amortized time. \textit{J. ACM}, 48:1--12, (2001).

 \item \label{Chen06} K. Chen. On k-median clustering in high dimensions. In Proceedings of the \textit{17th Annual ACM-SIAM Symposium on Discrete Algorithms (SODA '06),} pages 1177--1185, (2006).

 \item \label{Chen06a} Chen, Shuai, Chen, Yu-Ao, Strassel, Thorsten, Yuan, Zhen-Sheng, Zhao, Bo, Schmiedmayer, Jorg, and Pan, Jian-Wei, Deterministic and Storable Single-Photon Source Based on a Quantum Memory, \textit{Physical Review Letters} 97, 173004 (2006).

 \item \label{Chen07} K. Chen. On k-median and k-means clustering in metric and Euclidean spaces and their applications. Manuscript, available at:http://ews.uiuc.edu/${\rm \sim }$kechen/, (2007).

 \item \label{Clarkson89} K. Clarkson and P. Shor. Applications of random sampling in computational geometry, II. \textit{Discrete Comput. Geom}., 4:387--421, (1989).

 \item \label{Clarkson89a} K. Clarkson, R. E. Tarjan, and C. J. Van Wyk. A fast Las Vegas algorithm for triangulating a simple polygon. \textit{Discrete Comput. Geom}., 4:423--432, (1989).

 \item \label{Collins09} B. Collins and I. Nechita, ,,Random quantum channels I: graphical calculus and the Bell state phenomenon\textit{", arXiv:0905.2313}, (2009).

 \item \label{Collins09a} B. Collins and I. Nechita, ,,Random quantum channels II: Entanglement of random subspaces, Renyi entropy estimates and additivity problems", arXiv:0906.1877, (2009).

 \item \label{Collins09b} B. Collins and I. Nechita, ,,Gaussianization and eigenvalue statistics for Random quantum channels (III\textit{)", arXiv:0910.1768}, (2009).

 \item \label{Cormen01} T. Cormen, C. E. Leiserson, R. L. Rivest, and C. Stein. Introduction to Algorithms, 2nd edn. MIT Press, Cambridge, MA, (2001).

 \item \label{Cornwell97} B. Cornwell, \textit{Group Theory In Physics}: An Introduction, Academic Press, (1997).

 \item \label{Cortese02} J. Cortese, "The Holevo-Schumacher-Westmoreland Channel Capacity for a Class of Qudit Unital Channels", LANL ArXiV e-print \textit{quant-ph/0211093}, (2002).

 \item \label{Cortese03} J. Cortese, ``\textit{Classical Communication over Quantum Channels}''. PhD Thesis by. John A. Cortese. California Institute of Technology (2003).

 \item \label{Cover91} T. Cover and J. A. Thomas,\textit{ Elements of Information Theory}, John Wiley and Sons, New York, (1991).

 \item \label{Csiszar78} I. Csiszar, and J. Korner, \textit{IEEE Trans. Inf. Theory} 24, 339. (1978).

 \item \label{Csiszar81} I. Csiszar and J. Korner.\textit{ Information Theory: Coding Theorems for Discrete Memoryless Systems}. Probability and mathematical statistics. Akademiai Kiado, Budapest, (1981).

 \item \label{Cubitt08b} T. Cubitt, M. Beth Ruskai and G. Smith, The Structure of Degradable Quantum Channels \textit{J. Math. Phys}. 49, 102104 (2008).

 \item \label{Cubitt08c} T. Cubitt and J. Cirac, Engineering Correlation and Entanglement Dynamics in Spin Systems \textit{Physical Review Letters} 100, 180406 (2008).

 \item \label{Cubitt09} T. Cubitt, J. X. Chen, and A. Harrow, Superactivation of the Asymptotic Zero-Error Classical Capacity of a Quantum Channel, \textit{arXiv: 0906.2547}. (2009).

 \item \label{Cubitt09a} T. Cubitt, G. Smith, Super-Duper-Activation of Quantum Zero-Error Capacities, \textit{arXiv:0912.2737v1}. (2009).

 \item \label{Cubitt10} T. Cubitt, D. Leung, W. Matthews and A. Winter, Improving Zero-Error Classical Communication with Entanglement, \textit{Phys. Rev. Lett}. 104, 230503, arXiv:0911.5300, (2010).

 \item \label{Czekaj08} L. Czekaj and P. Horodecki. Nonadditivity effects in classical capacities of quantum multiple-access channels. \textit{arXiv:0807.3977}, (2008).

 \item \label{Czekaj08} L. Czekaj and P. Horodecki. Nonadditivity effects in classical capacities of quantum multiple-access channels. \textit{arXiv:0807.3977}, (2008).

 \item \label{Datta04} N. Datta, A. S. Holevo, Yu. M. Sukhov, On a Sufficient Additivity Condition in quantum information theory, \textit{Probl. Peredachi Inf}., Volume 41. Issue 2, Pages 9--25 (2004).

 \item \label{Datta04a} N. Datta, A. S. Holevo, and Y. Suhov, A quantum channel with additive minimum output entropy, http://arxiv.org/abs/quant-ph/0403072, (2004).

 \item \label{Datta04b} N. Datta, A. S. Holevo, and Y. Suhov. Additivity for transpose depolarizing channels, http://arxiv.org/abs/quant-ph/0412034, (2004).

 \item \label{Datta05} N. Datta and M. B. Ruskai, ,,Maximal output purity and capacity for asymmetric unital qudit channels", \textit{J. Phys. A}: Math. Gen. 38, 9785 -- 9802 (2005).

 \item \label{Datta06} N. Datta, M. Fukuda and A. S. Holevo, ,,Complementarity and additivity for covariant channels", \textit{Quant. Info. Proc}. 5, 179 - 207 (2006).

 \item \label{Denchev16} V. S. Denchev, S. Boixo, S. V. Isakov, N. Ding, R. Babbush, V. Smelyanskiy, J. Martinis, and H. Neven, ``What is the computational value of finite-range tunneling?,'' \textit{Phys. Rev. X}, vol. 6, p. 031015, (2016).

 \item \label{Desurvire09} E. Desurvire, \textit{Classical and quantum information theory,} Cambridge University Press, The Edinburgh Building, Cambridge CB2 8RU, UK, (2009).

 \item \label{Deutsch85} D. Deutsch. Quantum theory, the Church-Turing principle and the universal quantum computer. \textit{Proceedings of the Royal Society of London A}, 400(1818):97-117, July (1985).

 \item \label{Devetak03a} I. Devetak and A. Winter. Classical data compression with quantum side information. \textit{Physical Review A}, 68(4):042301, (2003).

 \item \label{Devetak04a} I. Devetak, A. W. Harrow, and A. J. Winter. A family of quantum protocols. \textit{Physical Review Letters}, 93:239503, (2004).

 \item \label{Devetak05} I. Devetak, ``The private classical capacity and quantum capacity of a quantum channel,'' \textit{IEEE Trans. Inf. Theory}, vol. 51, pp. 44--55, arXiv:quant-ph/0304127, (2005).

 \item \label{Devetak05a} I. Devetak and A. Winter. Distillation of secret key and entanglement from quantum states. \textit{Proceedings of the Royal Society A}, 461:207-235, (2005).

 \item \label{Devetak05b} I. Devetak and P. Shor. The capacity of a quantum channel for simultaneous transmission of classical and quantum information. \textit{Communications in Mathematical Physics}, 256:287-303, (2005).

 \item \label{Devetak08} I. Devetak, A. W. Harrow, and A. Winter. A resource framework for quantum Shannon theory. \textit{IEEE Transactions on Information Theory}, 54(10):4587- 4618, (2008).

 \item \label{Devitt08} S. J. Devitt, W. J. Munro, K. Nemoto, High Performance Quantum Computing,  \textit{arXiv:0810.2444} (2008).

 \item \label{Dirac82} P. A. M. Dirac. \textit{The Principles of Quantum Mechanics} (International Series of Monographs on Physics). Oxford University Press, USA, (1982).

 \item \label{DiVincenzo98} D. DiVincenzo, P. Shor, and J. Smolin, ``Quantum-channel capacity of very noisy channels,'' \textit{Phys. Rev. A}., vol. 57, pp. 830--839, eprint quant-ph/9706061, (1998).

 \item \label{Djordjevic08} I. Djordjevic, ``Quantum LDPC Codes from Balanced Incomplete Block Designs,'' \textit{IEEE Communications Letters}, vol. 12, pp. 389--391 (2008).

 \item \label{Dowling03} J. P. Dowling and G. J. Milburn. Quantum technology: The second quantum revolution. \textit{Philosophical Transactions of The Royal Society of London Series A}, 361(1809):1655-1674, (2003).

 \item \label{Duan01} L. Duan, M. D. Lukin, J. I. Cirac, and P. Zoller, Long-distance quantum communication with atomic ensembles and linear optics, \textit{Nature} 414, 413 (2001).

 \item \label{Duan07} R. Duan, Y. Feng, Z. F. Ji, and M. S. Ying, ``Distinguishing Arbitrary Multipartite Basis Unambiguously Using Local Operations and Classical Communication'', \textit{Phys. Rev. Lett}. 98, 230502 (2007).

 \item \label{Duan09} R. Duan, Superactivation of zero-error capacity of noisy quantum channels.\textit{arXiv:0906.2527}, (2009).

 \item \label{Dupuis10} F. Dupuis, P. Hayden, and K. Li. A father protocol for quantum broadcast channels. \textit{IEEE Transactions on Information Theory}, 56(6):2946-2956, arXiv:quant ph/0612155. (2010).

 \item \label{Dur07} W. D\"{u}r and H.J. Briegel, Entanglement purification and quantum error correction. \textit{Rep. Prog. Phys}, 70:1381--1424, (2007).

 \item \label{Dur99} W. D\"{u}r, H.-J. Briegel, J. I. Cirac, and P. Zoller, Quantum repeaters based on entanglement purification. \textit{Physical Review A}, 59(1):169--181, (1999).

 \item \label{Eetimes10} Electronic Engineering Times Portal, MEMS shown to enable quantum computing, http://www.eetimes.com/electronics-news/4211424/MEMS-enables-quantum computing, Mark Saffman's group at University of Wisconsin-Madison, (2010). 

 \item \label{Einstein1905} A. Einstein. \"{U}ber einen die erzeugung und verwandlung des lichtes betreenden heuristischen gesichtspunkt. \textit{Annalen der Physik}, 17:132-148, (1905).

 \item \label{Einstein1935} A. Einstein, B. Podolsky, N. Rosen, ``Can quantum-mechanical description of physical reality be considered complete?'' \textit{Physical Review} 47(10). 777-780. (1935).

 \item \label{Eisert05} J. Eisert and M. M. Wolf, ``Gaussian quantum channels,'', arXiv:quant-ph/0505151. (2005).

 \item \label{Elias72} P. Elias. The efficient construction of an unbiased random sequence. \textit{Annals of Mathematical Statistics}, 43(3):865-870, (1972).

 \item \label{Evans07} Z. W. E. Evans, A. M. Stephens, J. H. Cole, L. C. L. Hollenberg, ``Error correction optimisation in the presence of X/Z asymmetry,'' \textit{ArXiv}, (2007).

 \item \label{Fannes1973} M. Fannes. A continuity property of the entropy density for spin lattices. \textit{Communications in Mathematical Physics}, 31:291, (1973).

 \item \label{Fedrizzi09} A. Fedrizzi, R. Ursin, T. Herbst, M. Nespoli, R. Prevedel, T. Scheidl, F. Tiefenbacher, T. Jennewein, and A. Zeilinger, ``High-fidelity transmission of entanglement over a high-loss free-space channel,'' \textit{Nature Physics}, vol. 5, pp. 389--392, (2009).

 \item \label{Feldman07} D. Feldman, M. Monemizadeh, and C. Sohler. A PTAS for k-means clustering based on weak coresets. In \textit{Proceedings of the 23rd ACM Symposium on Computational Geometry (SCG'07),} pages 11--18, (2007).

 \item \label{Feynman82} R. Feynman, Simulating physics with computers. \textit{International Journal of Theoretical Physics}, 21:467--488. (1982).

 \item \label{Fischer04} K. Fischer and B. Gartner. The smallest enclosing ball of balls: combinatorial structure and algorithms. \textit{International Journal of Computational Geometry \& Application}, 14(4-5):341--378, (2004).

 \item \label{Frahling05} G. Frahling and C. Sohler. Coresets in dynamic geometric data streams. In \textit{Proceedings of the 27th Annual ACM Symposium on Theory of Computing (STOC'05),} pages 209--217, New York, NY, USA, ACM, (2005).

 \item \label{Fuchs2000} C. Fuchs, ``Information Tradeoff Relations for Finite-Strength Quantum Measurements'', LANL ArXiV e-print \textit{quant-ph/0009101.} (2000).

 \item \label{Fuchs96} C. Fuchs. \textit{Distinguishability and Accessible Information in Quantum Theory}. PhD thesis, University of New Mexico, December 1996. arXiv:quant-ph/9601020, (1996).

 \item \label{Fuchs97} C. Fuchs, N. Gisin, R. B. Griffiths, C.-S. Niu, and A. Peres, \textit{Phys. Rev. A} 56, 1163. (1997).

 \item \label{Fuchs98} C. Fuchs and J. van de Graaf. Cryptographic distinguishability measures for quantum mechanical states. \textit{IEEE Transactions on Information Theory}, 45(4):1216-1227,. arXiv:quant-ph/9712042, (1998).

 \item \label{Fujiwara02} T. Fujiwara and T. Hashizume, ``Additivity of the capacity of depolarizing channels''. \textit{Phys. Lett. A}, 299:469--475, (2002).

 \item \label{Fukuda10} M. Fukuda, C. King, D. K. Moser, Comments on Hastings' Additivity Counterexamples, \textit{Communications in Mathematical Physics}, DOI 10.1007/s00220-010-0996-9, (2010).

 \item \label{Fukuda10a} M. Fukuda and C. King. Entanglement of random subspaces via the Hastings bound. \textit{Journal of Mathematical Physics}, 51(4):042201, (2010).

 \item \label{Gabay06} M. Gabay and S. Arnon, ``Quantum key distribution by a free-space MIMO system,'' \textit{Journal of Lightwave Technology}, vol. 24, pp. 3114-- 3120 (2006).

 \item \label{Gaitan08} F. Gaitan, \textit{Quantum Error Correction and Fault Tolerant Quantum Computing}, CRC Press, ISBN 9780849371998, p.312 (2008).

 \item \label{Galambos10} M. Galambos, L. Bacsardi, S. Imre, Modeling the superdense coding in satellite-satellite and ground-satellite communications, \textit{Proc of 10th Asian Conference on Quantum Information Science}, Tokyo, Jap\'{a}n, .08.27-2010.08.31.pp. 205-206 (2010).

 \item \label{Gao10} W. Gao, et al. ``Teleportation-based realization of an optical quantum two-qubit entangling gate.'' \textit{PNAS} Early Edition.DOI:10.1073/pnas.1005720107, (2010).

 \item \label{Gardner70} M. Gardner, Mathematical Games: The fantastic combinations of John Conway's new solitaire game "Life", \textit{Scientific American} 223: 120--123. (1970).

 \item \label{Gerlach1922} W. Gerlach and Otto Stern. Das magnetische moment des silberatoms. Zeitschrift f\"{u}r Physik, 9:353-355, (1922).

 \item \label{Ghilen17} A. Ghilen, M. Azizi, and R. Bouallegue, ``Enhancing the Security of IEEE 802.11i Standard by Integrating a Quantum Scheme for Authentication and Encryption Key Distribution,'' \textit{Wireless Personal Communications}, vol. 95, pp. 1655--1675, (2017).

 \item \label{Ghosh12} J. Ghosh, A. G. Fowler, and M. R. Geller, ``Surface code with decoherence: An analysis of three superconducting architectures,'' \textit{Phys. Rev. A}, vol. 86, p. 062318 (2012).

 \item \label{Giovannetti04} V. Giovannetti, S. Guha, S. Lloyd, L. Maccone, J H. Shapiro, B J. Yen, H P. Yuen, Classical capacity of free-space optical communication, \textit{Quantum Information \& Computation}, Volume 4 Issue 6, Dec Pages 489-499 (2004).

 \item \label{Giovannetti05} V. Giovannetti and R. Fazio. Information-capacity description of spin-chain correlations. \textit{Physical Review A}, 71(3):032314, (2005).

 \item \label{Giovannetti10} V. Giovannetti, A. S. Holevo, S. Lloyd, L. Maccone, Generalized minimal output entropy 
conjecture for one-mode Gaussian channels: definitions and some exact results, \textit{J. Phys. A: Math. Theor. 43 415305}, 8, 796--800, (2010).

 \item \label{Giovannetti14} V. Giovannetti, R. Garc\'{i}a-Patr\'{o}n, N. J. Cerf, A. S. Holevo, Ultimate classical communication rates of quantum optical channels, \textit{Nature Photonics}, 8, 796--800, (2014).

 \item \label{Glauber05} R. J. Glauber\textit{. One hundred years of light quanta.} In Karl Grandin, editor, Les Prix Nobel. The Nobel Prizes 2005, pages 90-91. Nobel Foundation, (2005).

 \item \label{Glauber1963} R. J. Glauber. The quantum theory of optical coherence. \textit{Physical Review}, 130(6):2529-2539, (1963).

 \item \label{Goodman04} J. Goodman and J. O'Rourke, editors. \textit{Handbook of Discrete and Computational Geometry}, 2nd edn. Chapman \& Hall/CRC, 2004.

 \item \label{Gordon1964} J. P. Gordon. \textit{Noise at optical frequencies; information theory}. In P. A. Miles, editor, Quantum Electronics and Coherent Light; Proceedings of the International School of Physics Enrico Fermi, Course XXXI, pages 156-181, Academic Press New York, (1964).

 \item \label{Gottesman96} D. Gottesman, Class of quantum error-correcting codes saturating the quantum Hamming bound. \textit{Physical Review A}, 54(3):1862-1868, (1996).

 \item \label{Gottesman97} D. Gottesman, \textit{Stabilizer Codes and Quantum Error Correction}. PhD thesis, California Institute of Technology (arXiv:quant-ph/9705052), (1997).

 \item \label{Gottesman04a} D. Gottesman\textit{, An Introduction to Quantum Error Correction, Quantum Computation: A Grand Mathematical Challenge for the Twenty-First Century and the Millennium}, ed. S. J. Lomonaco, Jr., pp. 221-235, (2002).

 \item \label{Grassl97} M. Grassl, T. Beth, and T. Pellizzari. Codes for the quantum erasure channel. \textit{Phys.Rev.A},56:33-38, quant-ph/9610042. (1997).

 \item \label{Griffiths95} D. Griffiths. Introduction to Quantum Mechanics. Prentice-Hall, Inc., (1995).

 \item \label{Grossing88} G. Gr\"{o}ssing, A. Zeilinger, Quantum cellular automata. \textit{Complex Syst}., 2:197--208. (1988).

 \item \label{Guha07a} S. Guha, J. Shapiro. Classical information capacity of the Bosonic broadcast channel. In \textit{Proceedings of the IEEE International Symposium on Information Theory}, pages 1896-1900, Nice, France, (2007).

 \item \label{Guha07b} S. Guha, J. Shapiro, B. Erkmen. Classical capacity of bosonic broadcast communication and a minimum output entropy conjecture. \textit{Physical Review A}, 76(3):032303, (2007).

 \item \label{Gyongyosi11a} L. Gyongyosi, S. Imre: Information Geometric Superactivation of Classical Zero-Error Capacity of Quantum Channels, \textit{Progress in Informatics}, Quantum Information Technology, Quantum Information Science Theory Group, National Institute of Informatics, Tokyo, Japan, Print ISSN : 1349-8614, Online ISSN : 1349-8606; (2011.)

 \item \label{Gyongyosi11b} L. Gyongyosi, S. Imre: Algorithmic Superactivation of Asymptotic Quantum Capacity of Zero-Capacity Quantum Channels, \textit{Information Sciences}, ELSEVIER, ISSN: 0020-0255. (2011).

 \item \label{Gyongyosi11c} L. Gyongyosi, S. Imre\textit{: Quantum Cryptographic Protocols and Quantum Security}, in "Cryptography: Protocols, Design and Applications", Nova Science Publishers, USA. (2011).

 \item \label{Gyongyosi11d} L. Gyongyosi, S. Imre: \textit{Quantum Cellular Automata Controlled Self-Organizing Networks}, in "Cellular Automata", Book Chapter, INTECH, New York, USA, ISBN 978-953-7619-X-X; (2011).

 \item \label{Gyongyosi12a} L. Gyongyosi, S. Imre: Quasi-Superactivation of Classical Capacity of Zero-Capacity Quantum Channels, \textit{Journal of Modern Optics}, Taylor \& Francis, 0950-0340 (Print), 1362-3044 (Online). 2012.

 \item \label{Gyongyosi12b} L. Gyongyosi, S. Imre: Superactivation of Quantum Channels is Limited by the Quantum Relative Entropy Function, \textit{Quantum Information Processing}, Springer, ISSN: 1570-0755 (print version), ISSN: 1573-1332 (electronic version). 2012.

 \item \label{Gyongyosi13} L. Gyongyosi, \textit{Information Geometric Superactivation of Asymptotic Quantum Capacity and Classical Zero-Error Capacity of Zero-Capacity Quantum Channels}, PhD Thesis, Budapest University of Technology and Economics, 2013.

 \item \label{Gyongyosi13a} L. Gyongyosi, The Correlation Conversion Property of Quantum Channels, \textit{Quantum Information Processing}, Springer, ISSN: 1570-0755 (print version), ISSN: 1573-1332 (electronic version). (2013).

 \item \label{Gyongyosi14} L. Gyongyosi, A Statistical Model of Information Evaporation of Perfectly Reflecting Black Holes\textit{, International Journal of Quantum Information} (IJQI), ISSN 0219-7499 (print), 1793-6918 (online), 2014.

 \item \label{Gyongyosi14a} L. Gyongyosi: The Private Classical Capacity of a Partially Degradable Quantum Channel, \textit{Physica Scripta - Special Issue on Quantum Information}, Institute of Physics (IOP), Online ISSN: 1402-4896 Print ISSN: 0031-8949, 2014.

 \item \label{Gyongyosi14b} L. Gyongyosi: The Structure and Quantum Capacity of a Partially Degradable Quantum Channel, \textit{IEEE Access}, ISSN: 2169-3536, 2014.

 \item \label{Gyongyosi14c} L. Gyongyosi: Quantum Information Transmission over a Partially Degradable Channel, \textit{IEEE Access}, ISSN: 2169-3536 (2014).

 \item \label{Gyongyosi17} Quantum Imaging of High-Dimensional Hilbert Spaces with Radon Transform, \textit{International Journal of Circuit Theory and Applications }(IJCTA), Special Issue on Quantum Circuits (Wiley), 2017.

 \item \label{Hagar11} A. Hagar, "Quantum Computing", The Stanford Encyclopedia of Philosophy (Spring 2011 Edition), Edward N. Zalta (ed.), http://plato.stanford.edu/archives/spr2011/entries/qt-quantcomp, (2011).

 \item \label{Hamada05} M. Hamada. Information rates achievable with algebraic codes on quantum discrete memoryless channels. \textit{IEEE Transactions in Information Theory}, 51(12):4263--4277, arXiv:quant-ph/0207113, (2005).

 \item \label{Hanzo12} L. Hanzo, H. Haas, S. Imre, D. O'Brien, M. Rupp, L. Gyongyosi. Wireless Myths, Realities, and Futures: From 3G/4G to Optical and Quantum Wireless, \textit{Proceedings of the IEEE}, Volume: 100, Issue: Special Centennial Issue, pp. 1853-1888. (2012).

 \item \label{Har-Peled05} S. Har-Peled and A. Kushal. Smaller coresets for k-median and k-means clustering. \textit{In Proceedings of the 21st Annual Symposium on Computational Geometry (SCG'05), }pages 126--134, New York, NY, USA, ACM, (2005).

 \item \label{Harrow04} A. Harrow. Coherent communication of classical messages. \textit{Physical Review Letters}, 92:097902, (2004).

 \item \label{Harrow04a} A. Harrow and H. Lo. A tight lower bound on the classical communication cost of entanglement dilution. \textit{IEEE Transactions on Information Theory},

 \item \label{Harrow04a} A. Harrow and H. Lo. A tight lower bound on the classical communication cost of entanglement dilution. \textit{IEEE Transactions on Information Theory}, 50(2):319-327, (2004).

 \item \label{Hastings09} M. Hastings, ``A Counterexample to Additivity of Minimum Output Entropy'' \textit{Nature Physics} 5, 255, arXiv:0809.3972, (2009).

 \item \label{Hausladen95} P. Hausladen, B. Schumacher, M. Westmoreland, and W. Wootters. Sending classical bits via quantum its. \textit{Annals of the New York Academy of Sciences}, 755:698-705, (1995).

 \item \label{Hausladen96} P. Hausladen, R. Jozsa, B. Schumacher, M. Westmoreland, and W. K. Wootters. Classical information capacity of a quantum channel. \textit{Physical Review A}, 54(3):1869-1876, (1996).

 \item \label{Hayashi01} M. Hayashi and K. Matsumoto. Variable length universal entanglement concentration by local operations and its application to teleportation and dense coding. \textit{arXiv:quant-ph/0109028}, (2001).

 \item \label{Hayashi03} M. Hayashi and H. Nagaoka, ``General formulas for capacity of classical-quantum channels,'' \textit{IEEE Transactions on Information Theory}, Vol.49, No.7, pp.1753-1768 (2003).

 \item \label{Hayashi05} M. Hayashi, H. Imai, K. Matsumoto, M. B. Ruskai, and T. Shimono. Qubit channels which require four inputs to achieve capacity: Implications for additivity conjectures. \textit{QUANTUM INF.COMPUT}., 5:13, http://arxiv.org/abs/quant-ph/0403176, (2005).

 \item \label{Hayashi06} M. Hayashi. \textit{Quantum Information: An Introduction}. Springer-Verlag, (2006).

 \item \label{Hayden03a} P. Hayden and A. Winter. Communication cost of entanglement transformations. \textit{Physical Review A}, 67(1):012326, (2003).

 \item \label{Hayden08} P. Hayden and A. Winter, ,,Counterexamples to the maximal p-norm multiplicativity conjecture for all p $>$ 1", \textit{Commun. Math. Phys}. 284, 263 -- 280 (2008).

 \item \label{Hayden08a} P. Hayden, M. Horodecki, J. Yard, and A. Winter, ``A decoupling approach to the quantum capacity,'' \textit{Open Syst. Inf. Dyn}., vol. 15, pp. 7--19, arXiv:quant-ph/0702005. (2008).

 \item \label{Hayden08b} P. Hayden, P. Shor, and A. Winter. Random quantum codes from Gaussian ensembles and an uncertainty relation. \textit{Open Systems \& Information Dynamics}, 15:71-89, (2008).

 \item \label{Heisenberg1925} W. Heisenberg. \"{U}ber quantentheoretische umdeutung kinematischer und mechanischer beziehungen. \textit{Zeitschrift fur Physik}, 33:879-893, (1925).

\item \label{Hellemans04} A. Hellemans, ``Teleporting what matters,'' \textit{IEEE Spectrum}, vol. 41, pp. 18, 20--, (2004).

 \item \label{Helstrom76} C. Helstrom, \textit{Quantum Detection and Estimation Theory }``Academic, New York". (1976).

 \item \label{Herbert82} N. Herbert. Flash - a superluminal communicator based upon a new kind of quantum measurement. \textit{Foundations of Physics}, 12(12):1171-1179, (1982).

 \item \label{Hiai91} F. Hiai and D. Petz. The proper formula for relative entropy and its asymptotics in quantum probability. \textit{Communications in Mathematical Physics}, 143(1):99--114, (1991).

 \item \label{Holevo02a} A. Holevo. On entanglement assisted classical capacity. \textit{Journal of Mathematical Physics}, 43(9):4326-4333, (2002).

 \item \label{Holevo73} A. Holevo. Bounds for the quantity of information transmitted by a quantum communication channel. Problemly Peredachi Informatsii, 9(3):3--11, (1973). English Translation: \textit{Probl. Inform. Transm}., 9, 177--183, (1975).

 \item \label{Holevo73a} A. Holevo, Information theoretical aspects of quantum measurements. \textit{Probl. Info. Transm}., 9(2):31--42, (1973).

 \item \label{Holevo98} A. Holevo, ``The capacity of the quantum channel with general signal states'', \textit{IEEE Trans. Info. Theory} 44, 269 - 273 (1998).

 \item \label{Horn86} R. Horn and C. Johnson, \textit{Matrix Analysis}, Cambridge University Press, (1986).

 \item \label{Horodecki01} M. Horodecki, P. Horodecki, R. Horodecki, D. Leung, and B. Terhal. Classical capacity of a noiseless quantum channel assisted by noisy entanglement. \textit{Quantum Information and Computation}, 1(3):70-78, arXiv:quant-ph/0106080, (2001).

 \item \label{Horodecki03} M. Horodecki, P. Shor, and M. Ruskai. Entanglement breaking channels. \textit{Reviews in Mathematical Physics}, 15(6):629-641, arXiv:quant-ph/0302031, (2003).

 \item \label{Horodecki05} M. Horodecki, J. Oppenheim and A. Winter, ``Partial quantum information'', \textit{Nature} 436, 673-676 {\textbar} doi:10.1038/nature03909. (2005).

 \item \label{Horodecki07} M. Horodecki, Jonathan Oppenheim, and Andreas Winter. Quantum state merging and negative information. \textit{Communications in Mathematical Physics}, 269:107-136, (2007).

 \item \label{Horodecki09} R. Horodecki, P. Horodecki, M. Horodecki, and Karol Horodecki. Quantum entanglement. \textit{Reviews of Modern Physics}, 81(2):865-942, (2009).

 \item \label{Horodecki98} M. Horodecki. Limits for compression of quantum information carried by ensembles of mixed states. \textit{Physical Review A}, 57(5):3364-3369, (1998).

 \item \label{Hsieh08} M.Hsieh, I. Devetak, and A. Winter. Entanglement-assisted capacity of quantum multiple-access channels. \textit{IEEE Transactions on Information Theory}, 54(7):3078-3090, (2008).

 \item \label{Hsieh10} M. Hsieh and M. Wilde. Trading classical communication, quantum communication, and entanglement in quantum Shannon theory. \textit{IEEE Transactions on Information Theory}, 56(9):4705-4730, (2010).

 \item \label{Idquantique} https://www.idquantique.com

 \item \label{Imre05} S. Imre, F. Bal\'{a}zs: \textit{Quantum Computing and Communications -- An Engineering Approach}, Published by John Wiley and Sons Ltd, (2005).

 \item \label{Imre13} S. Imre, L. Gyongyosi: \textit{Advanced Quantum Communications -- An Engineering Approach}, Published by Wiley-IEEE Press, (2013).

 \item \label{Imre13a} S. Imre, Quantum communications: explained for communication engineers, \textit{IEEE COMMUNICATIONS MAGAZINE} 51:(08) (2013).

 \item \label{Isenburg06} M. Isenburg, Y. Liu, J. Shewchuk, and J. Snoeyink. Streaming computation of delaunay triangulations. In SIGGRAPH '06: ACM SIGGRAPH 2006 Papers, pages 1049 1056, New York, NY, USA, ACM, (2006).

 \item \label{Isham99} C. Isham, \textit{Modern Differential Geometry for Physicists}, World Scientific, Second Edition, Page 187, (1999).

 \item \label{Ivanova17} A. E. Ivanova, S. A. Chivilikhin, and A. V. Gleim, ``Quantum random number generator based on homodyne detection,'' \textit{Nanosystems-Physics Chemistry Mathematics}, vol. 8, pp. 239--242, (2017).

 \item \label{Janardan04} R. Janardan and T. C. Woo. Design and manufacturing. In J. E. Goodman and J. O'Rourke, editors, \textit{Handbook of Discrete and Computational Geometry}, 2nd edn., chapter 55. Chapman \& Hall/CRC, (2004).

 \item \label{Jaynes03} E. T. Jaynes. \textit{Probability Theory: The Logic }of \textit{Science}. Cambridge University Press, (2003).

 \item \label{Jaynes57a} E. T. Jaynes. Information theory and statistical mechanics. \textit{Physical Review}, 106:620, (1957).

 \item \label{Jaynes57b} E. T. Jaynes. Information theory and statistical mechanics II. \textit{Physical Review}, 108:171, (1957).

 \item \label{Jiang08} L. Jiang, J. M. Taylor, K. Nemoto, W. J. Munro, R. Van Meter, M. D. Lukin, Quantum Repeater with Encoding, \textit{arXiv:0809.3629} (2008).

 \item \label{Jouguet13} Paul Jouguet, Sebastien Kunz-Jacques, Anthony Leverrier, Philippe Grangier, Eleni Diamanti, Experimental demonstration of long-distance continuous-variable quantum key distribution, \textit{Nature Photonics} 7, 378-381 (2013).

 \item \label{Jozsa03} R. Jozsa and S. Presnell. Universal quantum information compression and degrees of prior knowledge. \textit{Proceedings of the Royal Society A}: Mathematical, Physical and Engineering Sciences, 459(2040):3061-3077, (2003).

 \item \label{Jozsa94} R. Jozsa. Fidelity for mixed quantum states. Journal of Modern Optics, 41(12):2315-2323, (1994).

 \item \label{Jozsa98} R. Jozsa, M. Horodecki, P. Horodecki, and R. Horodecki. Universal quantum information compression. \textit{Physical Review Letters}, 81(8):1714-1717, (1998).

 \item \label{Kaler03} F. Schmidt-Kaler, S. Gulde, M. Riebe, T. Deuschle, A. Kreuter, G. Lancaster, C. Becher, J. Eschner, H. Hffner, and R. Blatt, ``The coherence of qubits based on single Ca+ ions,'' \textit{Journal of Physics B-Atomic Molecular and Optical Physics}, vol. 36, pp. 623--636 (2003).

 \item \label{Kato06} K. Kato, M. Oto, H. Imai, and K. Imai, ``Voronoi diagrams for pure 1-qubit quantum states, quant-ph/0604101, (2006).

 \item \label{Karp88} Karp, S., Gagliardi, R.M., Moran, S.E., Stotts, L.B. (Eds.), \textit{Optical channels: fibers, clouds, water and the atmosphere}. Plenum Press (1988).

 \item \label{Kaye01} P. Kaye and M. Mosca. Quantum networks for concentrating entanglement. \textit{Journal of Physics A: Mathematical and General}, 34(35):6939, (2001).

 \item \label{Kelly15} J. Kelly, R. Barends, A. G. Fowler, A. Megrant, E. Jeffrey, T. C. White, D. Sank, J. Y. Mutus, B. Campbell, Yu Chen, Z. Chen, B. Chiaro, A. Dunsworth, I.-C. Hoi, C. Neill, P. J. J. O'Malley, C. Quintana, P. Roushan, A. Vainsencher, J. Wenner, A. N. Cleland, John M. Martinis, ``State preservation by repetitive error detection in a superconducting quantum circuit,'' \textit{Nature}, vol. 519, pp. 66--69 (2015).

 \item \label{Kerckhoff10} J. Kerckhoff, et al., Designing Quantum Memories with Embedded Control: Photonic Circuits for Autonomous Quantum Error Correction, \textit{Physical Review Letters} 105, 040502 (2010).

 \item \label{Khanmohammadi15} A. Khanmohammadi, R. Enne, M. Hofbauer, and H. Zimmermanna, ``A monolithic silicon quantum random number generator based on measurement of photon detection time,'' \textit{IEEE Photonics Journal}, vol. 7, pp. 1--13, (2015).

 \item \label{Kimura03} G. Kimura. The Bloch vector for n-level systems. Physics Letter A, 314(339), (2003).

 \item \label{Kimble08} Kimble, H. J. The quantum internet. \textit{Nature} 453, 1023 - 1030 (2008).

 \item \label{King01a} C. King and M. B. Ruskai, ,,Minimal entropy of states emerging from noisy quantum channels", \textit{IEEE Trans. Info. Theory} 47, 192 - 209 (2001).

 \item \label{King01b} C. King, M. Nathanson, and M. B. Ruskai, ,,Qubit Channels Can Require More Than Two Inputs to Achieve Capacity", LANL ArXiV e-print quant-ph/0109079, (2001).

 \item \label{King02} C. King. Additivity for unital qubit channels. \textit{J. Math. Phys}., 43:4641--4653, (2002).

 \item \label{King03a} C. King, ,,Maximal p-norms of entanglement breaking channels", \textit{Quantum Info. and Comp}. 3, no.2, 186 - 190 (2003).

 \item \label{King03b} C. King, ,,The capacity of the quantum depolarizing channel", \textit{IEEE Trans. Info. Theory} 49, no.1, 221 - 229 (2003).

 \item \label{King07} C. King, K. Matsumoto, M. Nathanson, and M. B. Ruskai. Properties of conjugate channels with applications to additivity and multiplicativity. \textit{Markov Processes and Related Fields}, 13(2):391-423, J. T. Lewis memorial issue, (2007).

 \item \label{King09} C. King, Remarks on the Additivity Conjectures for Quantum. Channels, www.mathphys.org/AZschool/material/AZ09-king.pdf, (2009).

 \item \label{King2000} C. King and M. B. Ruskai, ,,Capacity of Quantum Channels Using Product Measurements", LANL ArXiV e-print quant-ph/0004062, (2000).

 \item \label{Kitaev97} A. Kitaev. U. Nauk., 52(53), (1997).

 \item \label{Klesse08} R. Klesse. A random coding based proof for the quantum coding theorem. \textit{Open Systems \& Information Dynamics}, 15:21-45, (2008).

 \item \label{Knill05} E. Knill, Quantum computing with realistically noisy devices, \textit{Nature}. March 3. (2005).

 \item \label{Knill98} E. H. Knill, R. Laflamme, and W. H. Zurek, Resilient quantum computation. \textit{Science}, 279:342-345, \textit{quant-ph/9610011}, (1998).

 \item \label{Kochen1967} S. Kochen and E. P. Specker, ``The problem of hidden variables in quantum mechanics", \textit{Journal of Mathematics and Mechanics }17, 59$\{ $87 (1967).

 \item \label{Kohout09} R. Kohout, Sarah Croke, and Daniel Gottesman. Streaming universal distortion-free entanglement concentration, \textit{arXiv:0910.5952}, (2009).

 \item \label{Korzh15} Boris Korzh, Charles Ci Wen Lim, Raphael Houlmann, Nicolas Gisin, Ming Jun Li, Daniel Nolan, Bruno Sanguinetti, Rob Thew and Hugo Zbinden, Provably secure and practical quantum key distribution over 307 km of optical fibre, \textit{Nature Photonics} 9, 163-168 (2015)

 \item \label{Korner98} J. K\"{o}rner and A. Orlitsky, Zero-error information theory. \textit{IEEE Trans. Info. Theory}, 44(6):2207--2229, (1998).

 \item \label{Kraft49} L. Kraft, ``\textit{A device for quantizing, grouping, and coding amplitude modulated pulses''}, Cambridge, MA: MS Thesis, Electrical Engineering Department, Massachusetts Institute of Technology. (1949).

 \item \label{Kullback51} S. Kullback, R.A. Leibler, ``On Information and Sufficiency''. \textit{Annals of Mathematical Statistics} 22 (1): 79--86. doi:10.1214/aoms/1177729694. MR39968. (1951).

 \item \label{Kullback59} S. Kullback \textit{Information theory and statistics} (John Wiley and Sons, NY). (1959).

 \item \label{Kullback87} S. Kullback, ``Letter to the Editor: The Kullback--Leibler distance''. The \textit{American Statistician} 41 (4): 340--341. JSTOR 2684769. (1987).

 \item \label{Kullback87} S. Kullback, ``Letter to the Editor: The Kullback--Leibler distance''. The \textit{American Statistician} 41 (4): 340--341. JSTOR 2684769. (1987).

 \item \label{Ladd06} T. Ladd, P. van Loock, K. Nemoto, W. J. Munro, and Y. Yamamoto, Hybrid quantum repeater based on dispersive CQED interactions between matter qubits and bright coherent light, \textit{New Journal of Physics} 8, 184 (2006).

 \item \label{Laflamme96} R.Laflamme, C. Miquel, J. Paz, and W. Zurek, Perfect quantum error correcting code. \textit{Physical Review Letters}, 77(1):198-201, (1996).

 \item \label{Landauer95} R. Landauer. Is quantum mechanics useful? \textit{Philosophical Transactions of the Royal Society: Physical and Engineering Sciences}, 353(1703):367-376, (1995).

 \item \label{Leung08} D. Leung, G. Smith. Communicating over adversarial quantum channels using quantum list codes. \textit{IEEE Trans. Info. Theory} 54, 2, 883-887 (2008). 

\item \label{Leung09} D. Leung and G. Smith. Continuity of quantum channel capacities, \textit{Communications in Mathematical Physics} 292, 201-215 (November 2009).

\item \label{Leung10} D. Leung, L. Mancinska, W. Matthews, M. Ozols, A. Roy, Entanglement can increase asymptotic rates of zero-error classical communication over classical channels, \textit{arXiv:1009.1195v2} (2010).

 \item \label{Levitin69} L. Levitin, ,,On the quantum measure of the amount of information", Proceedings of the Fourth all-union conference on Information Theory (in Russian), Tashkent, 111 - 115 (1969).

 \item \label{Li09} K. Li, A. Winter, X. Zou, and G. Guo, ``Private classical capacity of Quantum Channels is Not Additive,'' \textit{Physical Review Letters}, vol. 103, no. 12, p. 120501, \textit{arXiv:0903.4308}, (2009).

 \item \label{Li10} M. Li, S-M. Fei and X. Li-Jost, ``Quantum Entanglement: Separability, Measure, Fidelity of Teleportation, and Distillation'', \textit{Advances in Mathematical Physics}Volume 2010, Article ID 301072, 57 pagesdoi:10.1155/2010/301072. (2010).

 \item \label{Li15} J. Li, X.-B. Chen, G. Xu, Y.-X. Yang, and Z.-P. Li, ``Perfect quantum network coding independent of classical network solutions,'' \textit{IEEE Communications Letters}, vol. 19, pp. 115--118,(2015).

 \item \label{Li16} Z. Li, G. Xu, X. Chen, X. Sun, and Y.-X. Yang, "Multi-User Quantum Wireless Network Communication Based on Multi-Qubit GHZ State", \textit{IEEE Communications Letters}, vol. 20, no. 12, pp. 2470 - 2473, December 2016.

 \item \label{Li17} J. Li, J. Xiong, Q. Zhang, L. Zhong, Y. Zhou, J. Li, and X. Lu, ``A one-time pad encryption method combining full-phase image encryption and hiding,'' \textit{Journal Of Optics}, vol. 19, (2017).

 \item \label{Liao17sk} Sheng-Kai Liao et al., Satellite-to-ground quantum key distribution, \textit{Nature} 549, 43-47 (2017)

 \item \label{Liao17} Q. Liao, Y. Guo, and D. Huang, ``Cancelable remote quantum fingerprint templates protection scheme,'' \textit{Chinese Physics} \textit{B}, vol. 26, (2017).

 \item \label{Lloyd97} S. Lloyd, ``Capacity of the noisy quantum channel,'' \textit{Phys. Rev. A}, vol. 55, pp. 1613--1622, (1997).

 \item \label{Lloyd04} S. Lloyd, J.H. Shapiro, F.N.C. Wong, P. Kumar, S.M. Shahriar, and H.P. Yuen. In-frastructure for the quantum Internet, \textit{ACM SIGCOMM Computer Communication Review}, 34(5):9--20, (2004)

 \item \label{Lo01} H. Lo and S. Popescu. Concentrating entanglement by local actions: Beyond mean values. \textit{Physical Review A}, 63(2):022301, (2001).

 \item \label{Lo95} H. Lo. Quantum coding theorem for mixed states. \textit{Optics Communications}, 119(5-6):552-556, September (1995).

 \item \label{Lo99a} H. Lo and Sandu Popescu. Classical communication cost of entanglement manipulation: Is entanglement an interconvertible resource? \textit{Physical Review Letters}, 83(7):1459-1462, (1999).

 \item \label{Loffe07} L. Loffe, and M. Mezard, ``Asymmetric quantum error-correcting codes,'' \textit{Physical Review A}, vol. 75, no. 3, p. 032345, PRA. (2007).

 \item \label{Loock08} P. van Loock, N. L\"{u}tkenhaus, W. J. Munro, and K. Nemoto, Quantum repeaters using coherent-state communication, \textit{Physical Review A} 78, 062319 (2008).

 \item \label{Louis08} S. G. R. Louis, W. J.Munro, T. P. Spiller, and K. Nemoto, Loss in hybrid qubit-bus couplings and gates, \textit{Physical Review A} 78, 022326 (2008).

 \item \label{Lovasz79} L . Lov\'{a}sz. On the Shannon capacity of a graph. \textit{IEEE Trans. Info. Theory}, 25(1):1 -- 7, (1979).

 \item \label{Lukin10} M. Lukin, J. Taylor, Quantum physics: Quantum leaps in the solid state, \textit{Nature} 467, 278--279 (16 September 2010) doi:10.1038/467278a. (2010).

 \item \label{Lupo11} C. Lupo, S. Pirandola, P. Aniello, S. Mancini, On the classical capacity of quantum Gaussian channels, arXiv:1012.5965v2, (2011).

 \item \label{Ma07} L. Ma, A. Mink, H. Xu, O. Slattery, and X. Tang, ``Experimental demonstration of an active quantum key distribution network with over GBPS clock synchronization,'' \textit{IEEE Communications Letters}, vol. 11, pp. 1019--1021 (2007).

 \item \label{Ma08} L. Ma, T. Chang, A. Mink, O. Slattery, B. Hershman, and X. Tang, ``Experimental demonstration of a detection-time-bin-shift polarization encoding quantum key distribution system,'' \textit{IEEE Communications Letters}, vol. 12, pp. 459--461 (2008).

 \item \label{Ma16} X. Ma, X. Yuan, Z. Cao, B. Qi, and Z. Zhang, ``Quantum random number generation,'' \textit{Npj Quantum Information}, vol. 2, (2016).

 \item \label{Matsumoto04} K. Matsumoto and F. Yura, Entanglement cost of antisymmetric states and additivity of capacity of some quantum channels. \textit{Journal of Phys. A}, 37:L167--L171, (2004).

 \item \label{McEvoy04} J. P. McEvoy and Oscar Zarate. Introducing Quantum Theory. Totem Books, third edition, (2004).

 \item \label{McMillan56} McMillan, Brockway, ``Two inequalities implied by unique decipherability'', \textit{IEEE Trans. Information Theory} 2 (4): 115--116, doi:10.1109/TIT.1956.1056818. (1956).

 \item \label{Medeiros05} R. Medeiros and F. M. de Assis, ``Quantum zero-error capacity'', \textit{Int. J. Quant. Inf}. 3, 135 (2005);

 \item \label{Medeiros06} R. Medeiros, R. Alleaume, G. Cohen, and F. M. de Assis, ``Quantum states characterization for the zero-error capacity'', \textit{quant-ph/0611042}.

 \item \label{Meyer96} D. Meyer, From quantum cellular automata to quantum lattice gases. \textit{Journal of Statistical Physics}, 85:551--574. (1996).

 \item \label{Michalakis07} S. Michalakis, ,,Multiplicativity of the maximal output 2-norm for depolarized Werner-Holevo channels", \textit{J. Math. Phys}. 48, no. 12, 122102, (2007).

 \item \label{Miller06} D. Miller, D. Maslov, \& G. Dueck, Synthesis of quantum multiple-valued circuits, \textit{Journal of Multiple-Valued Logic and Soft Computing}, vol. 12, no. 5-6, pp. 431--450. (2006).

 \item \label{Misner09} C. Misner, K. Thorne, and W. Zurek. J. Wheeler, Relativity, and quantum information. \textit{Physics Today}, (2009).

 \item \label{Moore65}, Moore Gordon E. Cramming more components onto integrated circuits. \textit{Electronics}. (1965).

 \item \label{Morley10} G. Morley, M. Warner, A. Stoneham, P. Greenland, J. van Tol, C. Kay \& G. Aeppli, The initialization and manipulation of quantum information stored in silicon by bismuth dopants, \textit{Nature} \textit{Materials} 9, 725 (2010).

 \item \label{Mosonyi09} M. Mosonyi and N. Datta. Generalized relative entropies and the capacity of classical-quantum channels. \textit{Journal of Mathematical Physics}, 50(7):072104, (2009).

 \item \label{Mullins01} J. Mullins. The topsy turvy world of quantum computing. \textit{IEEE Spectrum}, 38(2):42-49, (2001).

 \item \label{Munro08} W. J. Munro, R. Van Meter, S. G. R. Louis, and K. Nemoto, High-Bandwidth Hybrid Quantum Repeater, \textit{Physical Review Letters} 101, 040502 (2008).

 \item \label{Munro09} W. J. Munro, K. A. Harrison, A. M. Stephens, S. J. Devitt, K. Nemoto, From quantum fusiliers to high-performance networks \textit{arXiv:0910.4038} (2009).

 \item \label{Munro10} W. J. Munro, K. A. Harrison, A. M. Stephens, S. J. Devitt, and K. Nemoto, From quantum multiplexing to high-performance quantum networking, \textit{Nature Photonics}, 10.1038/nphoton.2010.213, (2010).

 \item \label{NC1} M. Tyryshkin, J. J. L. Morton, S. C. Benjamin, A. Ardavan, G. A. D. Briggs, J. W. Ager, and S. A. Lyon, ``Coherence of spin qubits in silicon,'' \textit{Journal of Physics-Condensed Matter}, vol. 18, pp. S783--S794, May 31 2006.

 \item \label{NC2} J. R. Petta, A. C. Johnson, J. M. Taylor, E. A. Laird, A. Yacoby, M. D. Lukin, C. M. Marcus, M. P. Hanson, and A. C. Gossard, ``Coherent manipulation of coupled electron spins in semiconductor quantum dots,'' \textit{SCIENCE}, vol. 309, pp. 2180--2184, Sep. 30 2005.

 \item \label{NC3} P. Bertet, I. Chiorescu, G. Burkard, K. Semba, C. J. P. M. Harmans, D. P. DiVincenzo, and J. E. Mooij, ``Dephasing of a superconducting qubit induced by photon noise,'' \textit{Physical Review Letters}, vol. 95 2005.

 \item \label{NC4} F. Schmidt-Kaler, S. Gulde, M. Riebe, T. Deuschle, A. Kreuter, G. Lancaster, C. Becher, J. Eschner, H. Hffner, and R. Blatt, ``The coherence of qubits based on single Ca+ ions,'' \textit{Journal of Physics B-Atomic Molecular and Optical Physics}, vol. 36, pp. 623--636, Feb. 14 2003.

 \item \label{NC5} L. Vandersypen, M. Steffen, G. Breyta, C. Yannoni, M. Sherwood, and I. Chuang, ``Experimental realization of Shor's quantum factoring algorithm using nuclear magnetic resonance,'' \textit{Nature}, vol. 414, pp. 883--887, Dec. 20 2001.

 \item \label{NC6} L. Loffe, and M. Mezard, ``Asymmetric quantum error-correcting codes,'' \textit{Physical Review A}, vol. 75, no. 3, p. 032345, 2007. PRA.

 \item \label{NC7} Z. W. E. Evans, A. M. Stephens, J. H. Cole, L. C. L. Hollenberg, ``Error correction optimisation in the presence of X/Z asymmetry,'' ArXiv, 2007.

 \item \label{NC8} L. Wang, K. Feng, S. Ling, and C. Xing, ``Asymmetric quantum codes: Characterization and constructions,'' \textit{IEEE Transactions on Information Theory}, vol. 56, pp. 2938--2945, June 2010.

 \item \label{NC9} M. Ezerman, S. Ling, and P. Sole, ``Additive asymmetric quantum codes,'' \textit{IEEE Transactions on Information Theory}, vol. 57, pp. 5536--5550, Aug 2011.

 \item \label{NC10} M. Ezerman, S. Jitman, S. Ling, and D. Pasechnik, ``CSS-like constructions of asymmetric quantum codes,'' \textit{IEEE Transactions on Information Theory}, vol. 59, pp. 6732--6754, Oct 2013.

 \item \label{NC11} G. PersonNameProductIDLa GuardiaLa Guardia, ``On the construction of asymmetric quantum codes,'' \textit{International Journal of Theoretical Physics}, vol. 53, no. 7, pp. 2312--2322, 2014.

 \item \label{NC12} H. V. Nguyen, Z. Babar, D. Alanis, P. Botsinis, D. Chandra, S. X. Ng, and L. Hanzo, ``Exit-chart aided quantum code design improves the normalised throughput of realistic quantum devices,'' \textit{IEEE Access}.

 \item \label{NC13} Z. Babar, P. Botsinis, D. Alanis, S. X. Ng, and L. Hanzo, ``The road from classical to quantum codes: A hashing bound approaching design procedure,'' \textit{IEEE Access}, vol. 3, pp. 146--176, 2015.

 \item \label{NC14} S. ten Brink, ``Convergence of iterative decoding,'' \textit{Electronics Letters}, vol. 35, pp. 806--808, May 13, 1999.

 \item \label{NC15} S. ten Brink and G. Kramer, ``Design of repeat-accumulate codes for iterative detection and decoding,'' \textit{IEEE Transactions on Signal Processing}, vol. 51, pp. 2764--2772, Nov 2003.

 \item \label{NC16} S.-J. Lee, A. Singer, and N. Shanbhag, ``Linear turbo equalization analysis via BER transfer and EXIT charts,'' \textit{IEEE Transactions on Signal Processing}, vol. 53, pp. 2883--2897, Aug 2005.

 \item \label{NC17} J. Karjalainen, M. Codreanu, A. Tolli, M. Juntti, and T. Matsumoto, ``EXIT chart-based power allocation for iterative frequency domain MIMO detector,'' \textit{IEEE Transactions on Signal Processing}, vol. 59, pp. 1624--1641, April 2011.

 \item \label{NC18} F. Babich, A. Crismani, M. Driusso, and L. Hanzo, ``Design criteria and genetic algorithm aided optimization of three-stage-concatenated spacetime shift keying systems,'' \textit{Signal Processing Letters, IEEE,} vol. 19, pp. 543--546, Aug 2012.

 \item \label{NC19} R. Ahlswede, N. Cai, and R. Yeung, ``Network information flow theory,'' in \textit{IEEE International Symposium on Information Theory}, 1998, p. 186, Aug. 1998.

 \item \label{NC20} R. Ahlswede, N. Cai, S.-Y. Li, and R. Yeung, ``Network information flow,'' \textit{IEEE Transactions on Information Theory}, vol. 46, pp. 1204 --1216, July 2000.

 \item \label{NC21} P. Chou and Y. Wu, ``Network coding for the internet and wireless networks,'' \textit{IEEE Signal Processing Magazine}, vol. 24, pp. 77 --85, Sept. 2007.

 \item \label{NC22} E. Soljanin, ``Network multicast with network coding {[}lecture notes{]},'' \textit{IEEE Signal Processing Magazine}, vol. 25, pp. 109 --112, Sept. 2008.

 \item \label{NC23} Y. Chen and S. Kishore, ``On the tradeoffs of implementing randomized network coding in multicast networks,'' \textit{IEEE Transactions on Communications}, vol. 58, pp. 2107 --2115, July 2010.

 \item \label{NC24} C. Fragouli and E. Soljanin, ``Network coding fundamentals,'' \textit{Foundation and Trends in Networking}, vol. 2, no. 1, pp. 1--133, 2007.

 \item \label{NC25} R. W. Yeung and N. Cai, ``Network error correction, Part I: Basic concepts and upper bounds,'' \textit{Communications in Information and Systems}, vol. 6, no. 1, pp. 19--36, 2006.

 \item \label{NC26} R. W. Yeung and N. Cai, ``Network error correction, Part II: Lower bounds,'' \textit{Communications in Information and Systems}, vol. 6, no. 1, pp. 37--54, 2006.

 \item \label{NC27} M. Hayashi, K. Wama, H. Nishimura, R. Raymond, and S. Yamashita, Quantum network coding, vol. 4393 of \textit{Lecture Notes in Computer Science}, pp. 610--621. Berlin: Springer-Verlag Berlin, 2007.

 \item \label{NC28} D. Leung, J. Oppenheim, and A. Winter, ``Quantum network communication-the butterfly and beyond,'' \textit{IEEE Transactions on Information Theory}, vol. 56, no. 7, pp. 3478--3490, 2010.

 \item \label{NC29} M. Mahdian and R. Bayramzadeh, ``Perfect k-pair quantum network coding using superconducting qubits,'' \textit{Journal of Superconductivity and Novel Magnetism}, vol. 28, no. 2, pp. 345--348, 2015.

 \item \label{NC30} L. Jing, C. Xiu-Bo, X. Gang, Y. Yi-Xian, and L. Zong-Peng, ``Perfect quantum network coding independent of classical network solutions,'' \textit{Communications Letters, IEEE}, vol. 19, no. 2, pp. 115--118, 2015.

 \item \label{NC31} T. Satoh, K. Ishizaki, S. Nagayama, and R. Van Meter, ``Analysis of quantum network coding for realistic repeater networks,'' \textit{Phys. Rev. A}, vol. 93, p. 032302, Mar 2016.

 \item \label{NC32} T. Shang, X.-J. Zhao, and J.-W. Liu, ``Quantum network coding based on controlled teleportation,'' \textit{IEEE Communications Letters}, vol. 18, no. 5, pp. 865--868, 2014.

 \item \label{NC33} T. Satoh, F. Le Gall, and H. Imai, ``Quantum network coding for quantum repeaters,'' \textit{Physical Review A}, vol. 86, no. 3, 2012.

 \item \label{NC34} A. Jain, M. Franceschetti, and D. A. Meyer, ``On quantum network coding,'' \textit{Journal of Mathematical Physics}, vol. 52, no. 3, 2011.

 \item \label{NC35} W. J. Munro, K. A. Harrison, A. M. Stephens, S. J. Devitt, and K. Nemoto, ``From quantum multiplexing to high-performance quantum networking,'' \textit{Nature Photonics}, vol. 4, no. 11, pp. 792--796, 2010.

 \item \label{NC36} M. Hayashi, ``Prior entanglement between senders enables perfect quantum network coding with modification,'' \textit{Physical Review A}, vol. 76, no. 4, 2007.

 \item \label{NC37} H. Kobayashi, F. Le Gall, H. Nishimura, and M. Rotteler, ``Constructing quantum network coding schemes from classical nonlinear protocols,'' in \textit{Information Theory Proceedings }(ISIT), 2011 IEEE International Symposium on, pp. 109--113.

 \item \label{NC38} H. Kobayashi, F. L. Gall, H. Nishimura, and M. Rtteler, ``Perfect quantum network communication protocol based on classical network coding,'' in 2010 \textit{IEEE International Symposium on Information Theory}, pp. 2686--2690, June 2010.

 \item \label{NC39} H. Kobayashi, F. Le Gall, H. Nishimura, and M. Roetteler, General Scheme for Perfect Quantum Network Coding with Free Classical Communication, vol. 5555 of \textit{Lecture Notes in Computer Science}, pp. 622--633. 2009

 \item \label{NC40} R. Pakniat, M. K. Tavassoly, and M. H. Zandi, ``Entanglement swapping and teleportation based on cavity QED method using the nonlinear atom-field interaction: Cavities with a hybrid of coherent and number states,'' \textit{OPTICS COMMUNICATIONS}, vol. 382, pp. 381--385, JAN 1 2017.

 \item \label{NC41} C. H. Bennett, D. P. DiVincenzo, P. W. Shor, J. A. Smolin, B. M. Terhal, and W. K. Wootters, ``Remote state preparation,'' \textit{Phys. Rev. Lett}., vol. 87, p. 077902, Jul 2001.

 \item \label{NC42} Y.-S. Ra, H.-T. Lim, and Y.-H. Kim, ``Remote preparation of threephoton entangled states via single-photon measurement,'' \textit{Phys. Rev. A}, vol. 94, p. 042329, Oct 2016.

 \item \label{NC43} H. Lu, Z. Zhang, L.-K. Chen, Z.-D. Li, C. Liu, L. Li, N.-L. Liu, X. Ma, Y.-A. Chen, and J.-W. Pan, ``Secret sharing of a quantum state,'' \textit{Phys. Rev. Lett}., vol. 117, p. 030501, Jul 2016.

 \item \label{NC44} H. P. Yuen, ``Security of quantum key distribution,'' \textit{IEEE Access}, vol. 4, pp. 724--749, 2016.

 \item \label{NC45} J.-P. Bourgoin, B. L. Higgins, N. Gigov, C. Holloway, C. J. Pugh, S. Kaiser, M. Cranmer, and T. Jennewein, ``Free-space quantum key distribution to a moving receiver,'' \textit{Optics Express}, vol. 23, no. 26, pp. 33437--33447, 2015.

 \item \label{NC46} M. T. Gruneisen, M. B. Flanagan, B. A. Sickmiller, J. P. Black, K. E. Stoltenberg, and A. W. Duchane, ``Modeling daytime sky access for a satellite quantum key distribution downlink,'' \textit{Optics Express}, vol. 23, no. 18, pp. 23924--23934, 2015.

 \item \label{NC47} G. Vallone, D. G. Marangon, M. Canale, I. Savorgnan, D. Bacco, M. Barbieri, S. Calimani, C. Barbieri, N. Laurenti, and P. Villoresi, ``Adaptive real time selection for quantum key distribution in lossy and turbulent free-space channels,'' \textit{Physical Review A}, vol. 91, no. 4, 2015.

 \item \label{NC48} A. Carrasco-Casado, N. Denisenko, and V. Fernandez, ``Correction of beam wander for a free-space quantum key distribution system operating in urban environment,'' \textit{Optical Engineering}, vol. 53, no. 8, 2014.

 \item \label{NC49} G. Vest, M. Rau, L. Fuchs, G. Corrielli, H.Weier, S. Nauerth, A. Crespi, R. Osellame, and H. Weinfurter, ``Design and evaluation of a handheld quantum key distribution sender module,'' \textit{IEEE Journal of Selected Topics in Quantum Electronics}, vol. 21, pp. 131--137, May 2015.

 \item \label{NC50} M. Rau, T. Heindel, S. Unsleber, T. Braun, J. Fischer, S. Frick, S. Nauerth, C. Schneider, G. Vest, S. Reitzenstein, M. Kamp, A. Forchel, S. Hoefling, and H. Weinfurter, ``Free space quantum key distribution over 500 meters using electrically driven quantum dot single-photon sources-a proof of principle experiment,'' \textit{New Journal of Physics}, vol. 16, 2014.

 \item \label{NC51} W. J. Munro, K. Azuma, K. Tamaki, and K. Nemoto, ``Inside quantum repeaters,'' \textit{IEEE Journal of Selected Topics in Quantum Electronics}, vol. 21, pp. 78--90, May 2015.

 \item \label{NC52} C. Y. Chen, G. J. Zeng, F. j. Lin, Y. H. Chou, and H. C. Chao, ``Quantum cryptography and its applications over the internet,'' \textit{IEEE Network}, vol. 29, pp. 64--69, September 2015.

 \item \label{NC53} A. Hellemans, ``Two steps closer to a quantum internet {[}news{]},'' \textit{IEEE Spectrum}, vol. 53, pp. 11--13, January 2016.

 \item \label{NC54} N. L. Piparo and M. Razavi, ``Long-distance trust-free quantum key distribution,'' \textit{IEEE Journal of Selected Topics in Quantum Electronics}, vol. 21, pp. 123--130, May 2015.

 \item \label{NC55} A. Delteil, Z. Sun, W.-b. Gao, E. Togan, S. Faelt, and A. Imamoglu, ``Generation of heralded entanglement between distant hole spins,'' \textit{NATURE PHYSICS}, vol. 12, pp. 218+, MAR 2016.

 \item \label{NC56} B. T. Kirby, S. Santra, V. S. Malinovsky, and M. Brodsky, ``Entanglement swapping of two arbitrarily degraded entangled states,'' \textit{PHYSICAL REVIEW A}, vol. 94, JUL 20 2016.

 \item \label{NC57} T. Shang, J. Li, Z. Pei, and J.-w. Liu, ``Quantum network coding for general repeater networks,'' \textit{Quantum Information Processing}, vol. 14, no. 9, pp. 3533--3552, 2015.

 \item \label{NC58} H. V. Nguyen, S. X. Ng, and L. Hanzo, ``Irregular convolution and unity-rate coded network-coding for cooperative multi-user communications,''\textit{IEEE Transactions on Wireless Communications}, vol. 12,no. 3, pp. 1231--1243, 2013.

 \item \label{NC59} Q. You, Y. Li, and Z. Chen, ``Joint relay selection and network coding for error-prone two-way decode-and-forward relay networks,'' \textit{IEEE Transactions on Communications}, vol. 62, pp. 3420--3433, Oct 2014.

 \item \label{NC60} T. X. Vu, P. Duhamel, and M. D. Renzo, ``On the diversity of networkcoded cooperation with decode-and-forward relay selection,'' \textit{IEEE Transactions on Wireless Communications}, vol. 14, pp. 4369--4378, Aug 2015.

 \item \label{NC61} H. Nguyen, Z. Babar, D. Alanis, P. Botsinis, D. Chandra, A. Izhar, S. Ng, and L. Hanzo, ``Towards the quantum internet: Generalised quantum network coding for large-scale quantum communication networks,'' \textit{IEEE Access}, 8 2017.

 \item \label{Neumann66} J. von Neumann, \textit{Theory of Self-Reproducing Automata}. University of Illinois Press, Champaign, IL, USA. (1966).

 \item \label{Neumann96} J. von Neumann. \textit{Mathematical Foundations of Quantum Mechanics}. Princeton University Press, (1996).

 \item \label{Nielsen02} M. Nielsen. A simple formula for the average gate fidelity of a quantum dynamical operation. \textit{Physics Letters A}, 303(4):249 - 252, (2002).

 \item \label{Nielsen07} F. Nielsen, J.-D. Boissonnat, and R. Nock. On Bregman Voronoi diagrams. \textit{In Proceedings of the 18th Annual ACM-SIAM Symposium on Discrete Algorithms (SODA'07),} pages 746--755, Philadelphia, PA, USA, Society for Industrial and Applied Mathematics, (2007).

 \item \label{Nielsen07a} F. Nielsen, J. D. Boissonnat, R. Nock, Bregman Voronoi Diagrams: Prop erties, \textit{arXiv:0709.2196v1}, (2007).

 \item \label{Nielsen08} F. Nielsen, R. Nock: Bregman Sided and Symmetrized Centroids. ICPR 2008, ICPR'08, \textit{(arXiv:0711.3242}), (2008).

 \item \label{Nielsen08a} F. Nielsen, R. Nock: On the smallest enclosing information disk. \textit{Inf. Process. Lett}. IPL'08, 105(3): 93-97 (2008).

 \item \label{Nielsen08b} F. Nielsen, R. Nock: Quantum Voronoi Diagrams and Holevo Channel Capacity for 1-Qubit Quantum States, \textit{ISIT 2008}, (2008).

 \item \label{Nielsen09} F. Nielsen, R. Nock: Approximating Smallest Enclosing Balls with Application to Machine Learning, \textit{International Journal on Computational Geometry and Applications} (IJCGA'09), (2009).

 \item \label{Nielsen2000} M. Nielsen and I. L. Chuang, Quantum Computation and Quantum Information, Cambridge University Press, (2000).

 \item \label{Nielsen98} M. Nielsen. Quantum information theory. PhD thesis, University of New Mexico, quant-ph/0011036. (1998).

 \item \label{Nielsen99} M. Nielsen. Conditions for a class of entanglement transformations. \textit{Physical Review Letters}, 83(2):436-439, (1999).

 \item \label{Nock05} R. Nock, F. Nielsen: Fitting the Smallest Enclosing Bregman Ball. ECML 2005, ECML'05, pages 649-656, (2005).

 \item \label{Ogawa07} T. Ogawa and Hiroshi Nagaoka. Making good codes for classical-quantum channel coding via quantum hypothesis testing. \textit{IEEE Transactions on Information Theory}, 53(6):2261-2266, (2007).

 \item \label{Ohno09} H. Ohno and D. Petz,  Generalized Pauli channels,  \textit{Acta Math}. Hungary 124, 165-177. (2009). 

 \item \label{Ohya97} M. Ohya, D. Petz, and N. Watanabe. ``On capacities of quantum channels''. \textit{Prob. Math. Stats}., 17:170--196, (1997).

 \item \label{Oi01} D. Oi, The Geometry of Single-Qubit Maps, \textit{arXiv:quant-ph/0106035v1} (2001)

 \item \label{Onishi97} K. Onishi and H. Imai: Voronoi Diagram in Statistical Parametric Space by Kullback-Leibler Divergence. \textit{Proceedings of the 13th ACM Symposium on Computational Geometry}, pp.463--465, (1997).

 \item \label{Onishi97} K. Onishi and H. Imai: Voronoi Diagram in Statistical Parametric Space by Kullback-Leibler Divergence. \textit{Proceedings of the 13th ACM Symposium on Computational Geometry}, pp.463--465, (1997).

 \item \label{Onishi97} K. Onishi and H. Imai: Voronoi Diagram in Statistical Parametric Space by Kullback-Leibler Divergence. \textit{Proceedings of the 13th ACM Symposium on Computational Geometry}, pp.463--465, (1997).

 \item \label{Oppenheim08} J. Oppenheim, For quantum information, two wrongs can make a right, \textit{Science}, 321, 1783 (2008), arXiv:1004.0052v1 (2008).

 \item \label{Oto04} M. Oto, H. Imai and K. Imai: Computational Geometry on 1-qubit Quantum States. \textit{Proc. International Symposium on Voronoi Diagrams in Science and Engineering} (VD 2004), Tokyo, pp.145--151, (2004).

 \item \label{Ozawa84} M. Ozawa. Quantum measuring processes of continuous observables. \textit{Journal of Mathematical Physics}, 25(1):79-87, (1984).

 \item \label{Pade14} J. Pade. (auth.), \textit{Quantum Mechanics for Pedestrians 1: Fundamentals. Undergraduate Lecture Notes in Physics}, Springer International Publishing, 1 ed., (2014).

 \item \label{Panigrahy04} R. Panigrahy, ``Minimum enclosing polytope in high dimensions,'' \textit{arXiv cs.CG/0407020}, (2004).

 \item \label{Panagiotis14} Botsinis, Panagiotis, Alanis, Dimitrios, Ng, Soon Xin and Hanzo, Lajos Low-Complexity Soft-Output Quantum-Assisted Multi-User Detection for Direct-Sequence Spreading and Slow Subcarrier-Hopping Aided SDMA-OFDM Systems. \textit{IEEE Access}, PP, (99), doi:10.1109/ACCESS.2014.2322013 (2014).

 \item \label{Panagiotis14a} Botsinis, Panagiotis, Ng, Soon Xin and Hanzo, Lajos Fixed-complexity quantum-assisted multi-user detection for CDMA and SDMA. \textit{IEEE Transactions on Communications}, vol. 62, (no. 3), pp. 990-1000, doi:10.1109/TCOMM.2014.012514.130615 (2014).

 \item \label{Paparo12} G. D. Paparo and M. A. Martin-Delgado, ``Google in a quantum network,'' \textit{Scientific Reports}, vol. 2, (2012).

 \item \label{Patron09} R. Garc\'{i}a-Patr\'{o}n, S. Pirandola, S. Lloyd, and J. H. Shapiro. Reverse coherent information. \textit{Physical Review Letters}, 102(21):210501, (2009).

 \item \label{Pelletier05} B. Pelletier, ``Informative barycentres in statistics,'' \textit{Annals of the Institute of Statistical Mathematics}, vol. 57, no. 4, pp. 767--780, (2005).

 \item \label{Perez-Delgado05} C. Perez-Delgado, D. Cheung, Models of quantum cellular automata, \textit{Physical Review A} 76, 032320, (2005).

 \item \label{Perseguers10} S. Perseguers, M. Lewenstein, A. Ac\'{i}n and J.I. Cirac, Quantum random networks, \textit{Nature Physics}, Advanced Online Publication, DOI:10.1038/NPHYS1665, (2010).

 \item \label{Petta05} J. R. Petta, A. C. Johnson, J. M. Taylor, E. A. Laird, A. Yacoby, M. D. Lukin, C. M. Marcus, M. P. Hanson, and A. C. Gossard, ``Coherent manipulation of coupled electron spins in semiconductor quantum dots,'' \textit{SCIENCE}, vol. 309, pp. 2180--2184 (2005).

 \item \label{Petz07} D. Petz, Bregman divergence as relative operator entropy, \textit{Acta Math}. Hungar, 116, 127-131. (2007).

 \item \label{Petz08} D. Petz, \textit{Quantum information theory and Quantum Statistics}, Springer-Verlag, Heidelberg, Hiv: 6. (2008).

 \item \label{Petz09} D. Petz  and V.E. S Szab\'{o},  From quasi-entropy to skew information,   \textit{Int. J. Math}. 20, 1421-1430. (2009).

 \item \label{Petz09a} D. Petz, A. Sz\'{a}nt\'{o} and M. Weiner, Complementarity and the algebraic structure of 4-level quantum systems,  \textit{J. Infin. Dim. Anal}., Quantum Probability and Related Topics 12, 99-116. (2009).

 \item \label{Petz09b} D. Petz,  Complementarity and the algebraic structure of finite quantum systems,  \textit{J. of  Physics}: Conference Series 143, 012011. (2009). 

 \item \label{Petz09c} D. Petz and J. Pitrik,  Markov property of Gaussian states of  canonical commutation relation algebras,  \textit{J. Math. Phys}. 50, 113517 (2009).

 \item \label{Petz10} D. Petz, From  f-divergence to quantum quasi-entropies and their use, Entropy 12(2010), 304-325. (2010).  

 \item \label{Petz10a} D. Petz,  Algebraic complementarity in quantum theory,  \textit{J. Math. Phys}.  51, 015215 (2010).

 \item \label{Petz96} D. Petz and C. Sud\'{a}r, ``Geometries of quantum states,'' \textit{Journal of Mathematical Physics}, vol. 37, no. 6, pp. 2662--2673, (1996).

 \item \label{Physorg} Physorg Portal, Quantum memory for communication networks of the future, http://www.physorg.com/news/2010-11-quantum-memory-networks-future.html, (2010).

 \item \label{Physorg15} Physorg Portal, Single-photon source may meet the needs of quantum communication systems, http://www.physorg.com/news82281692.html (2006).

 \item \label{Physorg16} Physorg Portal, Researchers convert quantum signals to telecom wavelengths, increase memory times, http://www.physorg.com/news204728305.html (2010).

 \item \label{Physorg17} Physorg Portal, Quantum Communication in Random Networks, http://www.physorg.com/news194080900.html (2010).

 \item \label{Pierce73} J. Pierce. The early days of information theory. \textit{IEEE Transactions on Information Theory}, IT-19(1):3-8, (1973).

 \item \label{Planck1901} M. Planck. Ueber das gesetz der energieverteilung im normalspectrum. \textit{Annalen der Physik}, 4:553-563, (1901).

 \item \label{Poulin09} D. Poulin, J. Tillich, and H. Ollivier, ``Quantum serial turbo codes,'' \textit{IEEE Transactions on Information Theory}, vol. 55, pp. 2776--2798, (2009).

 \item \label{Preskill98} J. Preskill. Lecture notes on \textit{Quantum Information Processing}. http://www.theory.alteh.edu/people/preskill/ph229/\#leture. (1998).

 \item \label{Preskill98a} J. Preskill. Reliable quantum computers. \textit{Proceedings of the Royal Society A}: Mathematical, Physical and Engineering Sciences, 454(1969):385-410, (1998).

 \item \label{Preskill15} J. Preskill,\textit{ Lecture Notes for Physics 229:Quantum Information and Computation}. CreateSpace Independent Publishing Platform, 2015.

 \item \label{Radmark09} M. Radmark, M. Zukowski, and M. Bourennane, Experimental Test of Fidelity Limits in Six-Photon Interferometry and of Rotational Invariance Properties of the Photonic Six-Qubit Entanglement Singlet State, \textit{Phys. Rev. Lett}. 103, 150501, (2009).

 \item \label{Rajan94} V. Rajan. Optimality of the Delaunay triangulation in Rd. \textit{Discrete \& Computational Geometry}, 12:189--202, (1994).

 \item \label{Renes07} J. Renes, G. Smith. Noisy processing and the distillation of private quantum states. \textit{Phys. Rev. Lett}. 98, 020502 (2007).

 \item \label{Renka97} R. Renka. Algorithm 772: Stripack: Delaunay triangulation and Voronoi diagram on the surface of a sphere. \textit{ACM Transactions on Mathematical Software}, 23(3):416--434, (1997).

 \item \label{Richter96} S. Richter, R. Werner, Ergodicity of quantum cellular automata. \textit{Journal of Statistical Physics}, 82:963--998. (1996).

 \item \label{Ruskai01} M. Ruskai, S. Szarek, and E. Werner, "An Analysis of Completely-Positive Trace Preserving Maps on 2 by 2 Matrices", LANL ArXiV e-print \textit{quant-ph/0101003}, (2001).

 \item \label{Sadakane98} K. Sadakane, H. Imai, K. Onishi, M. Inaba, F. Takeuchi, and K. Imai. Voronoi diagrams by divergences with additive weights. In \textit{Symposium on Computational Geometry}, pages 403--404, (1998).

 \item \label{Safari09} M. Safari and M. Uysal, ``Relay-assisted quantum-key distribution over long atmospheric channels,'' \textit{Journal of Lightwave Technology}, vol. 27, pp. 4508--4515 (2009).

 \item \label{Sakurai94} J. Sakurai. \textit{Modern Quantum Mechanics} (2nd Edition). Addison Wesley, (1994).

 \item \label{Sangouard09} N. Sangouard, C. Simon, H. de Riedmatten, N. Gisin, Quantum repeaters based on atomic ensembles and linear optics \textit{arXiv:0906.2699} (2009).

 \item \label{Sarvepalli09} P. K. Sarvepalli, A. Klappenecker, and M. Rotteler, ``Asymmetric quantum codes: constructions, bounds and performance,'' \textit{Proceedings of the Royal Society of London A: Mathematical, Physical and Engineering Sciences}, vol. 465, no. 2105, pp. 1645--1672 (2009).

 \item \label{Schindler13} Schindler et al., ``A quantum information processor with trapped ions,'' \textit{New Journal of Physics}, vol. 15 (2013).

 \item \label{Schrodinger1926} E. Schr\"{o}dinger. Quantisierung als eigenwertproblem. \textit{Annalen der Physik}, 79:361-376, (1926).

 \item \label{Schrodinger1935} E. Schr\"{o}dinger. Discussion of probability relations between separated systems. \textit{Proceedings of the Cambridge Philosophical Society}, 31:555-563, (1935).

 \item \label{Schmidt-Kaler03} F. Schmidt-Kaler, S. Gulde, M. Riebe, T. Deuschle, A. Kreuter, G. Lancaster, C. Becher, J. Eschner, H. Hffner, and R. Blatt, ``The coherence of qubits based on single Ca+ ions,'' \textit{Journal of Physics B-Atomic Molecular and Optical Physics}, vol. 36, pp. 623--636, Feb. 14 2003.

 \item \label{Schumacher01a} B. Schumacher and M. Westmoreland, "Indeterminate-length quantum coding", \textit{Physical Review A} \textbf{64}, 2304-2316, (2001).

 \item \label{Schumacher02} B. Schumacher and M. Westmoreland. Approximate quantum error correction. \textit{Quantum Information Processing}, 1(1/2):5-12, (2002).

 \item \label{Schumacher2000} B. Schumacher and M. Westmoreland, "Relative Entropy in quantum information theory" 2000, LANL ArXiV e-print quant-ph/0004045, to appear in \textit{Quantum Computation and Quantum Information: A Millenium Volume} , S. Lomonaco, editor (American Mathematical Society Contemporary Mathematics series), (2000).

 \item \label{Schumacher94} B. Schumacher and R. Jozsa, "A new proof of the quantum noiseless coding theorem",\textit{ Journal of Modern Optics} \textbf{41}, 2343-2349 (1994).

 \item \label{Schumacher95} B. Schumacher, P. Hausladen, M. D. Westmoreland and W. K. Wootters "Sending classical bits via quantum bits," \textit{Annals of the New York Academy of Sciences} \textbf{755}, 698-705 (1995).

 \item \label{Schumacher95a} B. Schumacher , "Quantum coding", \textit{Physical Review A} \textbf{51}, 2738-2747 (1995).

 \item \label{Schumacher96b} B. Schumacher, "Sending entanglement through noisy quantum channels", \textit{Physical Review A} \textbf{54}, 2614-2628 (1996).

 \item \label{Schumacher96c} B. Schumacher and M. A. Nielsen, "Quantum data processing and error correction", \textit{Physical Review A} \textbf{54}, 2629-2629 (1996).

 \item \label{Schumacher97} B. Schumacher and M. Westmoreland, ``Sending classical information via noisy quantum channels,'' \textit{Phys. Rev. A}, vol. 56, no. 1, pp. 131--138, (1997).

 \item \label{Schumacher98a} B. Schumacher and M. Westmoreland, "Quantum privacy and quantum coherence", \textit{Physical Review Letters} \textbf{80}, 5695-5697 (1998).

 \item \label{Schumacher98b} B. Schumacher, H. Barnum and M. A. Nielsen) "Information transmission through a noisy quantum channel", \textit{Physical Review A} \textbf{57}, 4153-4175 (1998).

 \item \label{Schumacher98c} B. Schumacher, C. M. Caves, M. A. Nielsen, and H. Barnum, "Information theoretic approach to quantum error correction and reversible measurement", \textit{Proceedings of the Royal Society of London A }\textbf{454}, 277-304\textit{ }(1998).

 \item \label{Schumacher99} B. Schumacher and M. Westmoreland, "Optimal Signal Ensembles", LANL ArXiV e-print \textit{quant-ph/9912122,} (1999).

 \item \label{Schumacher99a} B. Schumacher and M. Westmoreland, "Characterizations of classical and quantum communication processes", \textit{Chaos, Solitons and Fractals} \textbf{10}, 1719-1736 (1999).

 \item \label{Seidel04} R. Seidel. Convex hull computations. In J. E. Goodman and J. O'Rourke, editors, \textit{Handbook of Discrete and Computational Geometry}, 2nd ed., chapter 22. Chapman \& Hall/CRC, (2004).

 \item \label{Shang14} T. Shang, X.-J. Zhao, and J.-W. Liu, ``Quantum network coding based on controlled teleportation,'' \textit{IEEE Communications Letters}, vol. 18, pp. 865--868 (2014).

 \item \label{Shannon48} C. Shannon, ,,A mathematical theory of communication", Bell System Tech. J. 27, 379 - 423, 623 - 656 (1948).

 \item \label{Shannon56} C. Shannon, ``The zero-error capacity of a noisy channel,'' \textit{IEEE Trans. Information Theory}, pp. 8--19, (1956).

 \item \label{Shapiro74} J. H. Shapiro, ``Normal-mode approach to wave propagation in the turbulent atmosphere,'' \textit{Appl. Opt.}, vol. 13, pp. 2614--2619 (1974).

 \item \label{Shapiro03} J. H. Shapiro, ``Near-field turbulence effects on quantum-key distribution,'' \textit{Phys. Rev. A}, vol. 67, p. 022309 (2003).

 \item \label{Shapiro09} Jeffrey H. Shapiro, Secure Communication using Gaussian-State Quantum Illumination, , \textit{arXiv:0903.3150} (2009).

 \item \label{Sharir04} M. Sharir. Algorithmic motion planning. In J. E. Goodman and J. O'Rourke, editors, \textit{Handbook of Discrete and Computational Geometry}, 2nd edn., chapter 47. Chapman \& Hall/CRC, (2004).

 \item \label{Sharir94} M. Sharir. Almost tight upper bounds for lower envelopes in higher dimensions. \textit{Discrete and Computational Geometry}, 12(1):327--346, (1994).

 \item \label{Shewchuck02} J. Shewchuck. Delaunay refinement algorithms for triangular mesh generation. \textit{Comput. Geom. Theory Appl}., 22:21--74, (2002).

 \item \label{Shirley05} P. Shirley, M. Ashikhmin, M. Gleicher, S. Marschner, E. Reinhard, K. Sung, W. Thompson, and P. Willemsen. \textit{Fundamentals of Computer Graphics}, 2nd edn. A.K. Peters, (2005).

 \item \label{Shirokov06} M. Shirokov, ,,The Holevo capacity of finite dimensional channels and the additivity problem", \textit{Commun. Math. Phys}. 262, 137 - 159 (2006).

 \item \label{Shor02} P. Shor, ``The quantum channel capacity and coherent information.'' lecture notes, MSRI Workshop on Quantum Computation, Available online at http://www.msri.org/publications/ln/msri/2002/quantumcrypto/shor/1/. (2002).

 \item \label{Shor02a} P. Shor. Additivity of the classical capacity of entanglement-breaking quantum channels. \textit{Journal of Mathematical Physics}, 43(9):4334-4340, arXiv:quant-ph/0201149, (2002).

 \item \label{Shor04} P. Shor. Quantum Information, Statistics, Probability (Dedicated to A. S. Holevo on the occasion of his 60th Birthday): The classical capacity achievable by a quantum channel assisted by limited entanglement. Rinton Press, Inc., \textit{arXiv:quant-ph/0402129}, (2004).

 \item \label{Shor04a} P. W. Shor. Equivalence of additivity questions in quantum information theory. \textit{Communications in Mathematical Physics}, 246(3):453-472, arXiv:quant-ph/0305035. (2004).

 \item \label{Shor95} P. Shor. Scheme for reducing decoherence in quantum computer memory. \textit{Physical Review A}, 52(4):R2493-R2496, (1995).

 \item \label{Shor96} P. Shor. Fault-tolerant quantum computation. Annual \textit{IEEE Symposium on Foundations of Computer Science}, page 56, (1996).

 \item \label{Shor96b} P. Shor and J. Smolin, Quantum Error-Correcting Codes Need Not Completely Reveal the Error Syndrome, \textit{arXiv:quant-ph/9604006v2}, (1996).

 \item \label{Shor97} P. Shor. Polynomial time algorithms for prime factorization and discrete logarithms on a quantum computer. \textit{SIAMJ. Comp}.,26(5):1484-1509, (1997).

 \item \label{Smith06} G. Smith, D. Leung. Typical entanglement in stabilizer states. quant-ph/0510232. \textit{Phys. Rev. A} 74, 062314 (2006). 

 \item \label{Smith07} G. Smith, J. Smolin. Degenerate quantum codes for Pauli channels. \textit{Phys. Rev. Lett}. 98, 030501 (2007). 

 \item \label{Smith08} G. Smith, J. Yard, Quantum Communication with Zero-capacity Channels. \textit{Science} 321, 1812-1815 (2008).

 \item \label{Smith08a} G. Smith, J. Smolin. Additive extensions of a quantum channel. \textit{IEEE Information Theory Workshop Proceedings} (2008).

 \item \label{Smith08b} G. Smith, J. Renes, J. Smolin. Structured codes improve the Bennett-Brassard-84 quantum key rate. \textit{Phys. Rev. Lett}. 100, 170502 (2008). 

 \item \label{Smith08c} G. Smith, J. Smolin, A. Winter. The quantum capacity with symmetric side channels. \textit{IEEE Trans. Info. Theory} 54, 9, 4208-4217 (2008).

 \item \label{Smith08d} G. Smith. The private classical capacity with a symmetric side channel and its application to quantum cryptography. \textit{Phys. Rev. A} 78, 022306 (2008).

 \item \label{Smith09a} G. Smith, John Smolin. Can non-private channels transmit quantum information? \textit{Phys. Rev. Lett}. 102, 010501 (2009).

 \item \label{Smith09b} G. Smith and J. A. Smolin, ``Extensive Nonadditivity of Privacy,'' \textit{Physical Review Letters}, vol. 103, no. 12, p. 120503, Sep. \textit{arXiv:0904.4050}. (2009).

 \item \label{Smith10} G. Smith: Quantum Channel Capacities, \textit{Information Theory Workshop} (ITW), 2010 IEEE, Aug. 30 2010-Sept. 3 2010. page(s): 1 - 5 arXiv:1007.2855, (2010).

 \item \label{Smith11} G. Smith, J. A. Smolin and J. Yard, Gaussian bosonic synergy: quantum communication via realistic channels of zero quantum capacity, \textit{arXiv:1102.4580v1}, (2011).

 \item \label{Smolin07} J. Smolin, G. Smith, S. Wehner. A simple family of nonadditive quantum codes. \textit{Phys. Rev. Lett}. 99, 130505 (2007).

 \item \label{Steane96} A. Steane, Error correcting codes in quantum theory. \textit{Physical Review Letters}, 77(5):793-797, (1996).

 \item \label{Steiner10} M. Steiner, P. Neumann, J. Beck, F. Jelezko, and J. Wrachtrup, Universal enhancement of the optical readout fidelity of single electron spins at nitrogen-vacancy centers in diamond, \textit{Phys. Rev. B} 81, 035205, (2010). 

 \item \label{Stephens08} A. M. Stephens, Z.W. Evans, S.J. Devitt, A.D. Greentree, A.G. Fowler, W.J. Munro, J.L. O'Brien, K. Nemoto and L.C.L. Hollenberg, A Deterministic optical quantum computer using photonic modules, \textit{Physical Review A}. 78, 032318, (2008). 

 \item \label{Stinespring55} W. F. Stinespring. Positive functions on C*-algebras. \textit{Proceedings of the American Mathematical Society}, 6:211-216, (1955).

 \item \label{Takesue15} Hiroki Takesue, Toshihiko Sasaki, Kiyoshi Tamaki, and Masato Koashi, Experimental quantum key distribution without monitoring signal disturbance, \textit{Nature Photonics} 9, 827-831 (2015).

 \item \label{Thomson1901} W. Thomson (1st Baron Kelvin). Nineteenth-century clouds over the dynamical theory of heat and light. \textit{The London, Edinburgh and Dublin Philosophical Magazine and Journal of Science}, 2(6):1, (1901).

 \item \label{Toffoli90} T. Toffoli, \& N. Margolus, Invertible cellular automata: A review. \textit{Physica D}: Nonlinear Phenomena, 45:229--253. (1990).

 \item \label{Toth01} G. Toth, \& C. Lent, Quantum computing with quantum-dot cellular automata. \textit{Physical Review A}, 63:052315. (2001).

 \item \label{Trinh15} P. V. Trinh, N. T. Dang, and A. T. Pham, ``All-optical relaying fso systems using edfa combined with optical hard-limiter over atmospheric turbulence channels,'' \textit{Journal of Lightwave Technology}, vol. 33, pp. 4132--4144 (2015).

 \item \label{Tsujino11} K. Tsujino, D. Fukuda, G. Fujii, S. Inoue, M. Fujiwara, M. Takeoka, and M. Sasaki, ``Quantum receiver beyond the standard quantum limit of coherent optical communication,'' \textit{Phys. Rev. Lett}., vol. 106, p. 250503 (2011).

 \item \label{Tyryshkin06} A. M. Tyryshkin, J. J. L. Morton, S. C. Benjamin, A. Ardavan, G. A. D. Briggs, J. W. Ager, and S. A. Lyon, ``Coherence of spin qubits in silicon,'' \textit{Journal of Physics-Condensed Matter}, vol. 18, pp. S783--S794, (2006).

 \item \label{Unruh95} W. G. Unruh, Maintaining coherence in quantum computers. \textit{Physical Review A}, 51(2):992-997, (1995).

 \item \label{Vandersypen01} L. Vandersypen, M. Steffen, G. Breyta, C. Yannoni, M. Sherwood, and I. Chuang, ``Experimental realization of Shor's quantum factoring algorithm using nuclear magnetic resonance,'' \textit{Nature}, vol. 414, pp. 883--887 (2001).

 \item \label{VanMeter08} R. Van Meter, T. Ladd, W.J. Munro, and K. Nemoto, Communication Links for Distributed Quantum Computation, \textit{IEEE Transactions on Computers}, 56(12), 1643-1653 ( 2008).

 \item \label{VanMeter09} R. Van Meter, T. D. Ladd, W. J. Munro, and K. Nemoto, System Design for a Long-Line Quantum Repeater, \textit{IEEE/ACM Transactions on Networking} 17, 1002 (2009).

 \item \label{VanMeter13} R. Van Meter, T. Satoh, T. D. Ladd, W. J. Munro, K. Nemoto, Path Selection for Quantum Repeater Networks, Networking \textit{Science}, Vol. 3, Issue 1-4, pp 82-95 (2013).

 \item \label{VanMeter14} R. Van Meter, \textit{Quantum Networking}, John Wiley and Sons Ltd, ISBN 1118648927, 9781118648926 (2014).

 \item \label{Vedral2000} V.Vedral, ``The Role of Relative Entropy in Quantum Information Theory''. \textit{Rev. Mod. Phys}, 10.1103/RevModPhys.74.197. (2000).

 \item \label{Vedral98} V.Vedral and M. B. Plenio, ``Basics of quantum computation''. \textit{Prog. Quant. Electron}., 22, 1--39. (1998).

 \item \label{Vollbrecht08} K. Vollbrecht, \& J. Cirac, Quantum simulators, continuous-time automata, and translationally invariant systems. \textit{Phys. Rev. Lett}., 100:010501, (2008).

 \item \label{Voronoi1907} G. Voronoi. Nouvelles applications des parametres continusa la theorie des formes quadratiques. Premier M'emoire: Sur quelques propri'et'es des formes quadratiques positives parfaites. \textit{J. Reine Angew. Math}., 133:97--178, (1907).

 \item \label{Voronoi1908} G. Voronoi. Nouvelles applications des parametres continusa la theorie des formes quadratiques. \textit{Deuxi`eme M'emoire: Recherches sur les parall'ello`edres primitifs. J. Reine Angew. Math}., 134:198--287, (1908).

 \item \label{Wang10} L. Wang and R. Renner. One-shot classical-quantum capacity and hypothesis testing. \textit{arXiv:1007.5456}, (2010).

 \item \label{Watrous06} J. Watrous, \textit{Lecture Notes in Quantum Computation}, University of Calgary, (2006).

 \item \label{Watrous95} J. Watrous, On one-dimensional quantum cellular automata. In: \textit{Proceedings of the 36th Annual Symposium on Foundations of Computer Science}, pp 528--537. (1995).

 \item \label{Wein07} R. Wein, J. van den Berg, and D. Halperin. The visibility-Voronoi complex and its applications. \textit{Comput. Geom. Theory Appl}., 36:66--87, (2007).

 \item \label{Welzl85} E. Welzl. Constructing the visibility graph for n line segments in O(n2) time. \textit{Inform. Process. Lett}., 20:167--171, (1985).

 \item \label{Welzl88} E. Welzl. Partition trees for triangle counting and other range searching problems. In Proc. 4th Annu. \textit{ACM Sympos. Comput. Geom}., pages 23--33, (1988).

 \item \label{Welzl91} E. Welzl. Smallest enclosing disks (balls and ellipsoids). In H. Maurer, editor, \textit{New Results and New Trends in Computer} \textit{Science}, number 555 in \textit{Lecture Notes in Computer Science}, pages 359--370, (1991).

 \item \label{Wilde07} M. Wilde, H. Krovi, and T. A. Brun. Coherent communication with continuous quantum variables. \textit{Physical Review A}, 75(6):060303(R), (2007).

 \item \label{Wilde10} M. Wilde and Min-Hsiu Hsieh. Public and private resource trade-offs for a quantum channel. \textit{arXiv:1005.3818}. (2010).

 \item \label{Wilde11} M. Wilde, \textit{From Classical to Quantum Shannon Theory}, arxiv.org/abs/1106.1445, (2011).

 \item \label{Wilde12} M. Wilde and M.-H. Hsieh, ``The quantum dynamic capacity formula of a quantum channel,'' \textit{Quantum Information Processing}, vol. 11, pp. 1431--1463 (2012).

 \item \label{Wilde14} M. Wilde, M.-H. Hsieh, and Z. Babar, ``Entanglement-assisted quantum turbo codes,'' \textit{IEEE Transactions on Information Theory}, vol. 60, pp. 1203--1222 (2014).

 \item \label{Williams2000} P. Williams, H. Clearwater, \textit{Ultimate Zero and One Computing at the Quantum Frontier}. New York, USA : COPERNICUS Springer-Verlag New York (2000).

 \item \label{Winter01} A. Winter. The capacity of the quantum multiple access channel. \textit{IEEE Transactions on Information Theory}, 47:3059-3065, (2001).

 \item \label{Winter01a} A. Winter and Serge Massar. Compression of quantum-measurement operations. \textit{Physical Review A}, 64(1):012311, (2001).

 \item \label{Winter04} A. J. Winter. ``Extrinsic'' and ``intrinsic'' data in quantum measurements: asymptotic convex decomposition of positive operator valued measures. \textit{Communications in Mathematical Physics}, 244(1):157-185, (2004).

 \item \label{Winter07} A. Winter, ,,The maximum output p-norm of quantum channels is not multiplicative for any p $>$ 2", \textit{ArXiv:0707.0402}, (2007).

 \item \label{Winter99} A. Winter. Coding theorem and strong converse for quantum channels. \textit{IEEE Transactions on Information Theory}, 45(7):2481-2485, (1999).

 \item \label{Winter99a} A. Winter. \textit{Coding Theorems of Quantum Information Theory}. PhD thesis, Universitat Bielefeld, arXiv:quant-ph/9907077, (1999).

 \item \label{Wolf05} M. Wolf and J. Eisert, Classical information capacity of a class of quantum channels. \textit{New Journal of Physics}, 7(93), (2005).

 \item \label{Wolf06} M. M. Wolf, D. P-Garcia, G. Giedke, Quantum Capacities of Bosonic Channels, arXiv:\textit{quant-ph/0606132v1}, (2006).

 \item \label{Wolf07} M. Wolf and David P\'{e}rez-Garc\'{i}a. Quantum capacities of channels with small environment. \textit{Physical Review A}, 75(1):012303, (2007).

 \item \label{Wootters82} W. Wootters and W. H. Zurek, A single quantum cannot be cloned. \textit{Nature}, 299:802--803, doi:10.1038/299802a0. (1982).

 \item \label{Worboys04} M. Worboys and M. Duckham. GIS, a Computing Perspective, 2nd edn. Chapman \& Hall/CRC, (2004).

 \item \label{Wu16} L. Jian-Wu, C. Zi, S. Jin-Jing, and G. Ying, ``Quantum secret sharing with quantum graph states,'' \textit{Acta Physica Sinica}, vol. 65, (2016).

 \item \label{WWS11} WorldWideScience.org, Sample records for quantum computer development from WorldWideScience.org, http://worldwidescience.org/ (2011).

 \item \label{Yard05a} J. Yard. \textit{Simultaneous classical-quantum capacities of quantum multiple access channels}. PhD thesis, Stanford University, Stanford, CA,arXiv:quant-ph/0506050.(2005).

 \item \label{Yard05b} J. Yard, I. Devetak, and P. Hayden. \textit{Capacity theorems for quantum multiple access channels}. In Proceedings of the International Symposium on Information Theory, pages 884-888, Adelaide, Australia, (2005).

 \item \label{Yard06} J. Yard, P. Hayden, and I. Devetak. Quantum broadcast channels, \textit{arXiv:quant-ph/0603098}., (2006).

 \item \label{Yard08} J. Yard, P. Hayden, and I. Devetak. Capacity theorems for quantum multiple-access channels: Classical-quantum and quantum-quantum capacity regions. \textit{IEEE Transactions on Information Theory}, 54(7):3091-3113, (2008).

 \item \label{Yen05} B. Yen and J. Shapiro. Multiple-access bosonic communications. \textit{Physical Review A}, 72(6):062312, (2005).

 \item \label{Yeung02} R. Yeung. \textit{A First Course in Information Theory. Information Technology: Transmission, Processing, and Storage}. Springer (Kluwer Academic/Plenum Publishers), New York, New York, USA, (2002).

 \item \label{Yin17} J. Yin, Y. Cao, Y.-H. Li, S.-K. Liao, L. Zhang, J.-G. Ren, W.-Q. Cai, W.-Y. Liu, B. Li, H. Dai, G.-B. Li, Q.-M. Lu, Y.-H. Gong, Y. Xu, S.-L. Li, F.-Z. Li, Y.-Y. Yin, Z.-Q. Jiang, M. Li, J.-J. Jia, G. Ren, D. He, Y.-L. Zhou, X.-X. Zhang, N. Wang, X. Chang, Z.-C. Zhu, N.-L. Liu, Y.-A. Chen, C.-Y. Lu, R. Shu, C.-Z. Peng, J.-Y. Wang, and J.-W. Pan, ``Satellite-based entanglement distribution over 1200 kilometers,'' \textit{Science}, vol. 356, no. 6343, pp. 1140--1144, (2017).

 \item \label{Yoshizawa99} S. Yoshizawa and K. Tanabe, ``Dual differential geometry associated with Kullback-Leibler information on the Gaussian distributions and its 2-parameter deformations,'' \textit{SUT Journal of Mathematics}, vol. 35, no. 1, pp. 113--137, (1999).

 \item \label{Yuan08} Z-S. Yuan, Y. Chen, B. Zhao, S. Chen, J. Schmiedmayer, J. Pan, Experimental demonstration of a BDCZ quantum repeater node, \textit{Nature} 454, 1098-1101, doi:10.1038/nature07241; (2008).

 \item \label{Zeilinger99} A. Zeilinger, ``Experiment and the foundations of quantum physics,'' \textit{Rev. Mod. Phys.}, vol. 71, pp. S288--S297 (1999).

 \item \label{Zeilinger12} Xiao-Song Ma, Thomas Herbst, Thomas Scheidl, Daqing Wang, Sebastian Kropatschek, William Naylor, Bernhard Wittmann, Alexandra Mech, Johannes Kofler, Elena Anisimova, Vadim Makarov, Thomas Jennewein, Rupert Ursin and Anton Zeilinger, Quantum teleportation over 143 kilometres using active feed-forward, \textit{Nature} 489, 269-273 (2012).

 \item \label{Zhang09} W. Zhang, S. Hranilovic, and C. Shi, ``Soft-switching hybrid fso/rf links using short-length raptor codes: design and implementation,'' \textit{IEEE Journal on Selected Areas in Communications}, vol. 27, pp. 1698--1708 (2009).

 \item \label{Zhang12} Zhang Y, Djordjevic IB, Gao X., On the quantum-channel capacity for orbital angular momentum-based free-space optical communications\textit{, Opt Lett}.\textit{ 37}(15):3267-9 (2012).

 \item \label{Zhang16} Q. Zhang, W. Saad, M. Bennis, and M. Debbah, "Quantum Game Theory for Beam Alignment in Millimeter Wave Device-to-Device Communications," in \textit{Proc. of the IEEE Global Communications Conference (GLOBECOM)}, Washington, DC, USA, December 2016.

 \item \label{Zhang17} W. Zhang, D.-S. Ding, Y.-B. Sheng, L. Zhou, B.-S. Shi, and G.-C. Guo, ``Quantum Secure Direct Communication with Quantum Memory,'' \textit{Physical Review Letters}, vol. 118, (2017).
\end{enumerate}

\normalsize
\appendix
\setcounter{table}{0}
\setcounter{figure}{0}
\setcounter{equation}{0}
\renewcommand{\thetable}{\Alph{section}.\arabic{table}}
\renewcommand{\thefigure}{\Alph{section}.\arabic{figure}}
\renewcommand{\theequation}{\Alph{section}.\arabic{equation}}

\subsection{Partial Trace}
If we have a density matrix which describes only a subset of a larger quantum space, then we talk about the reduced density matrix. The larger quantum system can be expressed as the tensor product of the reduced density matrices of the subsystems, if there is no correlation (entanglement) between the subsystems. On the other hand, if we have two subsystems with reduced density matrices ${\rho }_A$ and ${\rho }_B$, then from the overall density matrix denoted by ${\rho }_{AB}$ the subsystems can be expressed as 
${\rho }_A\mathrm{=}Tr_B\left({\rho }_{AB}\right)$ and ${\rho }_B\mathrm{=}Tr_A\left({\rho }_{AB}\right)$,                  \label{1)}
 where $Tr_B$ and $Tr_A$ refers to the partial trace operators. So, this partial trace operator can be used to generate one of the subsystems from the joint state ${\rho }_{AB}\mathrm{=}\left|\left.{\psi }_A\right\rangle \right.\left\langle \left.{\psi }_A\right|\right.\mathrm{\otimes }\left|\left.{\psi }_B\right\rangle \right.\left\langle \left.{\psi }_B\right|\right.$, then 
\begin{equation} \label{2)} 
 \begin{split}
{\rho }_A&\mathrm{=}Tr_B\left({\rho }_{AB}\right)\mathrm{=}Tr_B\left(\left|\left.{\psi }_A\right\rangle \right.\left\langle \left.{\psi }_A\right|\right.\mathrm{\otimes }\left|\left.{\psi }_B\right\rangle \right.\left\langle \left.{\psi }_B\right|\right.\right) \\ &
\mathrm{=}\left|\left.{\psi }_A\right\rangle \right.\left\langle \left.{\psi }_A\right|\right.Tr\left(\left|\left.{\psi }_B\right\rangle \right.\left\langle \left.{\psi }_B\right|\right.\right)\mathrm{=}\left|\left.{\psi }_A\right\rangle \right.\left\langle \left.{\psi }_A\right|\right.\left\langle {\psi }_B\mathrel{\left|\vphantom{{\psi }_B {\psi }_B}\right.\kern-\nulldelimiterspace}{\psi }_B\right\rangle . \end{split}
\end{equation} 
Since the inner product is trivially $\left\langle {\psi }_B\mathrel{\left|\vphantom{{\psi }_B {\psi }_B}\right.\kern-\nulldelimiterspace}{\psi }_B\right\rangle \mathrm{=1}$, therefore
\begin{equation} \label{3)} 
Tr_B\left({\rho }_{AB}\right)\mathrm{=}\left\langle {\psi }_B\mathrel{\left|\vphantom{{\psi }_B {\psi }_B}\right.\kern-\nulldelimiterspace}{\psi }_B\right\rangle \left|\left.{\psi }_A\right\rangle \right.\left\langle \left.{\psi }_A\right|\right.\mathrm{=}\left|\left.{\psi }_A\right\rangle \right.\left\langle \left.{\psi }_A\right|\right.\mathrm{=}{\rho }_A.                       
\end{equation} 
In the calculation, we used the fact that $Tr\left(\left|\left.{\psi }_{\mathrm{1}}\right\rangle \right.\left\langle \left.{\psi }_{\mathrm{2}}\right|\right.\right)\mathrm{=}\left\langle {\psi }_{\mathrm{2}}\mathrel{\left|\vphantom{{\psi }_{\mathrm{2}} {\psi }_{\mathrm{1}}}\right.\kern-\nulldelimiterspace}{\psi }_{\mathrm{1}}\right\rangle $. In general, if we have to systems $A\mathrm{=}\left|\left.i\right\rangle \right.\left\langle \left.k\right|\right.$ and $B\mathrm{=}\left|\left.j\right\rangle \right.\left\langle \left.l\right|\right.$, then the partial trace can be calculated as 
\begin{equation} \label{4)} 
Tr_B\left(A\mathrm{\otimes }B\right)\mathrm{=}ATr\left(B\right),                                                                         
\end{equation} 
since
\begin{equation} \label{5)} 
 \begin{split}
Tr_{\mathrm{2}}\left(\left|\left.i\right\rangle \right.\left\langle \left.k\right|\right.\mathrm{\otimes }\left|\left.j\right\rangle \right.\left\langle \left.l\right|\right.\right)&\mathrm{=}\left|\left.i\right\rangle \right.\left\langle \left.k\right|\right.\mathrm{\otimes }Tr\left(\left|\left.j\right\rangle \right.\left\langle \left.l\right|\right.\right) \\ &
\mathrm{=}\left|\left.i\right\rangle \right.\left\langle \left.k\right|\right.\mathrm{\otimes }\left\langle l\mathrel{\left|\vphantom{l j}\right.\kern-\nulldelimiterspace}j\right\rangle  \\ &
\mathrm{=}\left\langle l\mathrel{\left|\vphantom{l j}\right.\kern-\nulldelimiterspace}j\right\rangle \left|\left.i\right\rangle \right.\left\langle \left.k\right|\right., \end{split}
\end{equation} 
where $\left|\left.i\right\rangle \right.\left\langle \left.k\right|\right.\mathrm{\otimes }\left|\left.j\right\rangle \right.\left\langle \left.l\right|\right.\mathrm{=}\left|\left.i\right\rangle \right.\left|\left.j\right\rangle \right.{\left(\left|\left.k\right\rangle \right.\left|\left.l\right\rangle \right.\right)}^T$. 

In this expression we have used the fact that $\left(AB^T\right)\mathrm{\otimes }\left(CD^T\right)\mathrm{=}\left(A\mathrm{\otimes }C\right)\left(B^T\mathrm{\otimes }D^T\right)\mathrm{=}\left(A\mathrm{\otimes }C\right){\left(B\mathrm{\otimes }D\right)}^T$.  

\subsection{Quantum Entanglement}
A quantum system ${\rho }_{AB}$ is separable if it can be written as a tensor product of the two subsystems ${\rho }_{AB}\mathrm{=}{\rho }_A\mathrm{\otimes }{\rho }_B$.  Beside product states ${\rho }_A\mathrm{\otimes }{\rho }_B$ which represent a composite system consisting of several independent states merged by means of tensor product $\mathrm{\otimes }$ similarly to classical composite systems, quantum mechanics offers a unique new phenomenon called \textit{entanglement}. For example the so called \textit{Bell states} (or EPR states, named after Einstein, Podolsky and Rosen) are entangled ones:
\begin{equation} \label{6)} 
 \begin{array}{l}
\left|\left.{\beta }_{00}\right\rangle \right.\mathrm{=}\frac{\mathrm{1}}{\sqrt{\mathrm{2}}}\left(\left|\left.00\right\rangle \right.\mathrm{+}\left|\left.\mathrm{11}\right\rangle \right.\right), \\ 
\left|\left.{\beta }_{\mathrm{01}}\right\rangle \right.\mathrm{=}\frac{\mathrm{1}}{\sqrt{\mathrm{2}}}\left(\left|\left.\mathrm{01}\right\rangle \right.\mathrm{+}\left|\left.\mathrm{10}\right\rangle \right.\right), \\ 
\left|\left.{\beta }_{\mathrm{10}}\right\rangle \right.\mathrm{=}\frac{\mathrm{1}}{\sqrt{\mathrm{2}}}\left(\left|\left.00\right\rangle \right.\mathrm{-}\left|\left.\mathrm{11}\right\rangle \right.\right), \\ 
\left|\left.{\beta }_{\mathrm{11}}\right\rangle \right.\mathrm{=}\frac{\mathrm{1}}{\sqrt{\mathrm{2}}}\left(\left|\left.\mathrm{01}\right\rangle \right.\mathrm{-}\left|\left.\mathrm{10}\right\rangle \right.\right). \end{array}
\end{equation} 

\subsection{Fidelity}
Theoretically, quantum states have to preserve their original superposition during the whole transmission, without the disturbance of their actual properties. Practically, quantum channels are entangled with the environment which results in mixed states at the output. Mixed states are classical probability weighted sum of pure states where these probabilities appear due to the interaction with the environment (i.e., noise). Therefore, we introduce a new quantity, which is able to describe the quality of the transmission of the superposed states through the quantum channel. The quantity which measures this distance is called the \textit{fidelity.}
The fidelity for two pure quantum states is defined as
\begin{equation} \label{7)} 
F\left({\left| \varphi  \right\rangle},\left| \psi  \right\rangle  \right)={{\left| \left\langle  \varphi  | \psi  \right\rangle  \right|}^{2}}.                                          
\end{equation} 
The fidelity of quantum states can describe the relation of Alice pure channel input state $\left|\left.\psi \right\rangle \right.$ and the received mixed quantum system $\sigma \mathrm{=}\sum^{n\mathrm{-}\mathrm{1}}_{i\mathrm{=0}}{p_i{\rho }_i}\mathrm{=}\sum^{n\mathrm{-}\mathrm{1}}_{i\mathrm{=0}}{p_i\left|\left.{\psi }_i\right\rangle \right.\left\langle \left.{\psi }_i\right|\right.}$ at the channel output as
\begin{equation} \label{8)} 
F\left({\left| \psi  \right\rangle},\sigma  \right)=\left\langle  \psi  | \left. \sigma  \right|\psi  \right\rangle =\sum\limits_{i=0}^{n-1}{{{p}_{i}}{{\left| \left\langle  \psi  | {{\psi }_{i}} \right\rangle  \right|}^{2}}}.                                   
\end{equation} 
Fidelity can also be defined for \textit{mixed} states $\sigma $ and $\rho $
\begin{equation} \label{9)} 
F\left( \rho ,\sigma  \right)={{\left[ Tr\left( \sqrt{\sqrt{\sigma }\rho \sqrt{\sigma }} \right) \right]}^{2}}=\sum\limits_{i}{{{p}_{i}}}{{\left[ Tr\left( \sqrt{\sqrt{{{\sigma }_{i}}}{{\rho }_{i}}\sqrt{{{\sigma }_{i}}}} \right) \right]}^{2}}.                     
\end{equation} 
Next we list the major properties of fidelity
\begin{equation} \label{10)} 
\mathrm{0}\mathrm{\le }F\left(\sigma ,\rho \right)\mathrm{\le }\mathrm{1},                                         
\end{equation} 
\begin{equation} \label{11)} 
F\left(\sigma ,\rho \right)\mathrm{=}F\left(\rho ,\sigma \right),                               
\end{equation} 
\begin{equation} \label{12)} 
F\left({\rho }_{\mathrm{1}}\mathrm{\otimes }{\rho }_{\mathrm{2}},{\sigma }_{\mathrm{1}}\mathrm{\otimes }{\sigma }_{\mathrm{2}}\right)\mathrm{=}F\left({\rho }_{\mathrm{1}},{\sigma }_{\mathrm{1}}\right)F\left({\rho }_{\mathrm{2}},{\sigma }_{\mathrm{2}}\right),                                 
\end{equation} 
\begin{equation} \label{13)} 
F\left(U\rho U^{\mathrm{\dagger }},U\sigma U^{\mathrm{\dagger }}\right)\mathrm{=}F\left(\rho ,\sigma \right),                                   
\end{equation} 
\begin{equation} \label{14)} 
F\left(\rho ,a{\sigma }_{\mathrm{1}}\mathrm{+(1-}a\mathrm{)}{\sigma }_{\mathrm{2}}\right)\mathrm{\ge }aF\left(\rho ,{\sigma }_{\mathrm{1}}\right)\mathrm{+(1-}a\mathrm{)}F\left(\rho ,{\sigma }_{\mathrm{2}}\right)\mathrm{,\ }a\mathrm{\in }\left[\mathrm{0,1}\right].                                
\end{equation} 
\end{document}